\documentclass{amsart}
\numberwithin{equation}{section}
\usepackage[utf8]{inputenc}
\usepackage[T1]{fontenc}
\usepackage{lmodern}
\usepackage[mathscr]{euscript}
\usepackage[margin=1.3in]{geometry}
\usepackage{setspace}
\usepackage[usenames, dvipsnames]{xcolor}
\usepackage{mathtools}
\usepackage{amssymb}
\usepackage{amsthm}
\usepackage{spectralsequences}
\usepackage{xparse}
\usepackage{xspace}
\usepackage{tikz-cd}
\usepackage[all]{xy}
\usepackage[backref=page, bookmarks=false]{hyperref}
\usepackage[capitalize, noabbrev]{cleveref}

\DeclareDocumentCommand{\shortexact}{s O{} O{} mmmm}{
\IfBooleanTF{#1}{ 
\begin{tikzcd}[ampersand replacement=\&]
        {1} \& {#4} \& {#5} \& {#6} \& {1#7}
        \arrow[from=1-1, to=1-2]
        \arrow["#2", from=1-2, to=1-3]
        \arrow["#3", from=1-3, to=1-4]
        \arrow[from=1-4, to=1-5]
\end{tikzcd}
}{ 
\begin{tikzcd}[ampersand replacement=\&]
        {0} \& {#4} \& {#5} \& {#6} \& {0#7}
        \arrow[from=1-1, to=1-2]
        \arrow["#2", from=1-2, to=1-3]
        \arrow["#3", from=1-3, to=1-4]
        \arrow[from=1-4, to=1-5]
\end{tikzcd}
}}
\newcommand{\Z}{\mathbb Z}
\newcommand{\Q}{\mathbb Q}
\newcommand{\R}{\mathbb R}
\newcommand{\C}{\mathbb C}
\newcommand{\RP}{\mathbb{RP}}
\newcommand{\CP}{\mathbb{CP}}
\renewcommand{\O}{\mathrm O}
\newcommand{\SO}{\mathrm{SO}}
\newcommand{\Spin}{\mathrm{Spin}}
\newcommand{\Pin}{\mathrm{Pin}}
\newcommand{\Sp}{\mathit{\mathscr Sp}}
\newcommand{\bl}{\text{--}}
\newcommand{\cA}{\mathcal A}
\newcommand{\Spinc}{{\Spin^c}}
\newcommand{\Pinc}{{\Pin^c}}
\newcommand{\ko}{\mathit{ko}}
\newcommand{\ku}{\mathit{ku}}
\newcommand{\fC}{\mathscr C}
\DeclareMathOperator{\Ext}{Ext}
\DeclareMathOperator{\Hom}{Hom}
\newcommand{\Top}{\mathit{\mathscr Top}}
\newcommand{\op}{^{\mathrm{op}}}
\newcommand{\Sph}{\mathbb S}
\newcommand{\E}{\mathbb E}
\newcommand{\pt}{\mathrm{pt}}
\newcommand{\wH}{\widetilde H}
\newcommand{\GL}{\mathrm{GL}}
\newcommand{\PSL}{\mathrm{PSL}}
\newcommand{\F}{\mathbb F}
\newcommand{\surj}{\twoheadrightarrow}
\newcommand{\Ab}{\mathit{\mathscr Ab}}
\newcommand{\cB}{\mathcal B}
\newcommand{\KO}{\mathit{KO}}
\DeclareMathOperator{\coker}{coker}
\newcommand{\SU}{\mathrm{SU}}
\newcommand{\Ki}{\mathit{Ph}}
\newcommand{\Ph}{\mathit{Ph}}

\newcommand{\gathxy}[1]{%
\begin{gathered}
	\xymatrix{#1}
\end{gathered}
}

\everyentry={\displaystyle}
\allowdisplaybreaks

\newcommand{\Pinp}{\relax\ifmmode{\Pin^+}\else Pin\textsuperscript{$+$}\xspace\fi}
\newcommand{\pinp}{pin\textsuperscript{$+$}\xspace}
\newcommand{\Pinm}{\relax\ifmmode{\Pin^-}\else Pin\textsuperscript{$-$}\xspace\fi}
\newcommand{\pinm}{pin\textsuperscript{$-$}\xspace}
\newcommand{\spinc}{spin\textsuperscript{$c$}\xspace}
\newcommand{\pinc}{pin\textsuperscript{$c$}\xspace}

\DeclareMathOperator*{\hocolim}{hocolim}
\DeclareMathOperator*{\holim}{holim}

\newtheorem{thm}[equation]{Theorem}
\newtheorem*{thm*}{Theorem}

\newtheorem{lem}[equation]{Lemma}
\newtheorem{cor}[equation]{Corollary}
\newtheorem{prop}[equation]{Proposition}

\theoremstyle{definition}
\newtheorem{exm}[equation]{Example}
\newtheorem{defn}[equation]{Definition}

\theoremstyle{remark}
\newtheorem{rem}[equation]{Remark}

\crefname{thm}{Theorem}{Theorems}
\crefname{lem}{Lemma}{Lemmas}
\crefname{cor}{Corollary}{Corollaries}
\crefname{prop}{Proposition}{Propositions}
\crefname{ex}{Exercise}{Exercises}
\crefname{exm}{Example}{Examples}
\crefname{defn}{Definition}{Definitions}
\crefname{claim}{Claim}{Claims}
\crefname{rem}{Remark}{Remarks}
\crefname{fct}{Fact}{Facts}
\crefname{note}{Note}{Notes}

\newcommand{\term}{\emph} 

\DeclarePairedDelimiter\abs{\lvert}{\rvert}
\DeclarePairedDelimiter\set{\{}{\}}
\DeclarePairedDelimiter\paren{(}{)}
\DeclarePairedDelimiter\ang{\langle}{\rangle}

\makeatletter
	\let\oldparen\paren
	\def\paren{\@ifstar{\oldparen}{\oldparen*}}
\makeatother

\newcommand{\vp}{\varphi}

\newcommand{\NewThomSpectrum}[1]{\expandafter\newcommand\csname M#1\endcsname{\mathit{M#1}}}
\newcommand{\NewMTSpectrum}[1]{\expandafter\newcommand\csname MT#1\endcsname{\mathit{MT#1}}}
\newcommand{\BothThomSpectra}[1]{\NewThomSpectrum{#1}\NewMTSpectrum{#1}}
\BothThomSpectra{O}
\BothThomSpectra{SO}
\BothThomSpectra{Spin}
\BothThomSpectra{Pin}
\BothThomSpectra{U}

\usepackage{subcaption}
\usepackage{booktabs}
\usepackage{enumitem}

\usepackage{microtype}
\usepackage{hypcap}

\hypersetup{
	colorlinks,
	urlcolor={blue!80!black},
	linkcolor={red!50!black},
	citecolor={green!50!black},
}

\newcommand{\TP}{\mathit{TP}}
\newcommand{\MTH}{\mathit{MTH}}
\newcommand{\tE}{\widetilde E} 
\newcommand{\tH}{\widetilde H}
\newcommand{\tOmega}{\widetilde\Omega}
\newcommand{\wG}{\widetilde G}
\newcommand{\hG}{\widehat G}
\newcommand{\hH}{\widehat H}

\newcommand{\Map}{\mathrm{Map}}

\newcommand{\Det}{\mathrm{Det}}
\newcommand{\MTSpinc}{\mathit{MTSpin}^c}

\newcommand{\oU}{U}
\renewcommand{\Im}{\mathrm{Im}}
\DeclareRobustCommand*{\RaiseBoxByDepth}{%
    \raisebox{\depth}%
}
\newcommand{\uQ}{\RaiseBoxByDepth{\protect\rotatebox{180}{$Q$}}}

\newcommand{\cE}{\mathcal E}
\newcommand{\MTPinc}{\mathit{MTPin}^c}
\newcommand{\Th}{\mathrm{Th}}
\newcommand{\cL}{\mathcal L}

\newcommand{\id}{\mathrm{id}}
\newcommand{\T}{\mathbb T}
\newcommand{\inj}{\hookrightarrow}

\newcommand{\Sq}{\mathrm{Sq}}

\let\shortmapsto\mapsto
\renewcommand{\mapsto}{\mathchoice{\longmapsto}{\shortmapsto}{\shortmapsto}{\shortmapsto}}

\newcommand{\hooklongrightarrow}{\lhook\joinrel\longrightarrow}

\DeclareRobustCommand\longtwoheadrightarrow
	{\relbar\joinrel\twoheadrightarrow}
\DeclareRobustCommand\longtwoheadleftarrow
	{\twoheadleftarrow\joinrel\relbar}

\newcommand{\tblref}[2]{\texorpdfstring{{\hyperref[#1]{\textcolor{black}{#2}}}}{#2}}

\bgroup\catcode`\#=12\gdef\hash{#}\egroup
\newcommand{\FHrefllink}[2]{%
	\href{https://arxiv.org/pdf/1604.06527.pdf\hash #1}{#2}%
}

\theoremstyle{plain}
\newtheorem*{thmLSL}{\cref{glidethm}}
\newtheorem*{thmFCEP}{\cref{FCEP}}

\theoremstyle{definition}
\newtheorem{ans}[equation]{Ansatz}
\newtheorem*{eqans}{\cref{equivariant_ansatz}}
\newtheorem{notat}[equation]{Data}

\usepackage{tikz}
\usetikzlibrary{calc}

\def\sqtwoL (#1,#2){
  \draw (#1,#2) .. controls (#1-1,#2+1) .. (#1,#2+2);
}

\def\sqtwoR (#1,#2,#3){
  \draw[#3] (#1,#2) .. controls (#1+1,#2+1) .. (#1,#2+2);
}

\def \sqtwoCR (#1,#2){
   \draw (#1,#2) .. controls (#1+0.75,#2+.5) and (#1+1.125,#2+2) .. (#1+1.5,#2+2);
}

\def \sqtwoCL (#1,#2){
   \draw (#1,#2) .. controls (#1-0.75,#2+.5) and (#1-1.125,#2+2)  .. (#1-1.5,#2+2);
}

\def \sqone (#1,#2){
  \draw (#1, #2) -- (#1, #2+1);
}

\newcommand{\tikzpt}[4]{
	\fill[#4] (#1, #2) circle (3pt) node[anchor=east] {#3}
}

\newcommand{\tikzptR}[4]{
	\fill[#4] (#1, #2) circle (3pt) node[anchor=west] {#3}
}

\newcommand{\tikzptB}[4]{
	\fill[#4] (#1, #2) circle (3pt) node[anchor=north] {#3}
}

\newcommand{\Aone}[3]{
	\tikzpt{#1}{#2}{#3}{};
	\tikzpt{#1}{#2+1}{}{};
	\tikzpt{#1}{#2+2}{}{};
	\tikzpt{#1}{#2+3}{}{};
	\foreach \y in {3, ..., 6} {
		\tikzpt{#1+1.5}{#2+\y}{}{};
	}
	\sqone(#1, #2);
	\sqone(#1, #2+2);
	\sqone(#1+1.5, #2+3);
	\sqone(#1+1.5, #2+5);
	\sqtwoL(#1, #2);
	\sqtwoCR(#1, #2+1);
	\sqtwoCR(#1, #2+2);
	\sqtwoCR(#1, #2+3);
	\sqtwoR(#1+1.5, #2+4, );
}

\newcommand{\AoneTruncFour}[3]{
	\tikzpt{#1}{#2}{#3}{};
	\tikzpt{#1}{#2+1}{}{};
	\tikzpt{#1}{#2+2}{}{};
	\tikzpt{#1}{#2+3}{}{};
	\tikzpt{#1+1.5}{#2+3}{}{};
	\tikzpt{#1+1.5}{#2+4}{}{};
	\sqone(#1, #2);
	\sqone(#1, #2+2);
	\sqone(#1+1.5, #2+3);
	\sqtwoL(#1, #2);
	\sqtwoCR(#1, #2+1);
	\sqtwoCR(#1, #2+2);
}

\newcommand{\SpanishQnMark}[3]{
	\tikzpt{#1}{#2}{#3}{};
	\tikzpt{#1}{#2+2}{}{};
	\tikzpt{#1}{#2+3}{}{};
	\sqtwoL(#1, #2);
	\sqone(#1, #2+2);
}

\newcommand{\EoneQnMark}[3]{
	\tikzpt{#1}{#2}{#3}{};
	\tikzpt{#1}{#2+2}{}{};
	\tikzpt{#1}{#2+3}{}{};
	\qoneL(#1, #2);
	\sqone(#1, #2+2);
}

\newcommand{\Joker}[3]{
	\tikzpt{#1}{#2}{#3}{};
	\tikzpt{#1}{#2+1}{}{};
	\tikzpt{#1}{#2+2}{}{};
	\tikzpt{#1}{#2+3}{}{};
	\tikzpt{#1}{#2+4}{}{};
	\sqone(#1, #2);
	\sqtwoR(#1, #2, );
	\sqtwoL(#1, #2+1);
	\sqtwoR(#1, #2+2, );
	\sqone(#1, #2+3);
}

\newcommand{\Rtwo}[4]{
	\tikzpt{#1}{#2}{#3}{};
	\tikzpt{#1}{#2+1}{#4}{};
	\tikzpt{#1}{#2+2}{}{};
	\foreach \y in {2, ..., 5} {
		\tikzpt{#1+1.5}{#2+\y}{}{};
	}
	\sqone(#1, #2+1);
	\sqone(#1+1.5, #2+2);
	\sqone(#1+1.5, #2+4);
	\sqtwoCR(#1, #2);
	\sqtwoCR(#1, #2+1);
	\sqtwoCR(#1, #2+2);
	\sqtwoR(#1+1.5, #2+3, );
}

\newcommand{\AdamsTower}[1]{\DoUntilOutOfBounds{
	\class[#1](\lastx, \lasty+1)
	\structline[#1]
}}

\def \qoneL (#1, #2){
	\draw[dashed] (#1, #2) .. controls (#1-1.25, #2+1.5) .. (#1, #2+3);
}
\def \qoneR (#1, #2){
	\draw[dashed] (#1, #2) .. controls (#1+1.25, #2+1.5) .. (#1, #2+3);
}
\newcommand{\Eone}[3]{
	\tikzpt{#1}{#2}{#3}{};
	\tikzpt{#1}{#2 + 1}{}{};
	\tikzpt{#1}{#2 + 3}{}{};
	\tikzpt{#1}{#2 + 4}{}{};
	\sqone(#1, #2);
	\sqone(#1, #2 + 3);
	\qoneL(#1, #2);
	\qoneR(#1, #2 + 1);
}

\newcommand{\Nfive}[4]{
	\tikzpt{#1+1.5}{#2}{#3}{};
	\foreach \y in {2, ..., 5} {
		\tikzpt{#1+1.5}{#2+\y}{}{};
	}
	\tikzpt{#1}{#2+1}{#4}{};
	\tikzpt{#1}{#2+2}{}{};
	\tikzpt{#1}{#2+3}{}{};

	\sqtwoR(#1+1.5, #2, );
	\sqone(#1+1.5, #2+2);
	\sqtwoR(#1+1.5, #2+3, );
	\sqone(#1+1.5, #2+4);
	
	\sqone(#1, #2+1);
	\sqtwoL(#1, #2+1);
	\sqtwoCR(#1, #2+2);
	\sqtwoCR(#1, #2+3);
}

\begin{document}

\title[Invertible phases for mixed spatial symmetries]{Invertible phases for mixed spatial symmetries and the
fermionic crystalline equivalence principle}

\author{Arun Debray}
\address{Department of Mathematics, University of Kentucky,
719 Patterson Office Tower, Lexington, KY 40506-0027}
\email{\href{mailto:a.debray@uky.edu}{a.debray@uky.edu}}
\urladdr{\url{https://adebray.github.io}}

\date{\today}

\begin{abstract}
Freed-Hopkins~\cite{FH19} give a mathematical ansatz for classifying gapped invertible phases of matter with a
spatial symmetry in terms of Borel-equivariant generalized homology. We propose a slight generalization of this
ansatz to account for cases where the symmetry type mixes nontrivially with the spatial symmetry, such as
crystalline phases with spin-1/2 fermions. From this ansatz, we prove as a theorem a ``fermionic crystalline
equivalence principle,'' as predicted in the physics literature. Using this and the Adams spectral sequence, we
compute classifications of some classes of phases with a point group symmetry; in cases where these phases have
been studied by other methods, our results agree with the literature.
\end{abstract}

\maketitle

\tableofcontents

\setcounter{section}{-1}
\section{Introduction} 
The classification of topological phases of matter has been the subject of intensive research in condensed-matter
physics and nearby areas of mathematics for the last decade, but difficult problems still remain: for example,
there is not yet an accepted mathematical definition of a topological phase of matter, so researchers must study
these systems using ansatzes or heuristic definitions of phases. Restricting to invertible phases, also known as
\term{symmetry-protected topological (SPT) phases}, simplifies the classification question, but defining these
phases precisely is also still an open problem. Freed-Hopkins~\cite{FH16} make an ansatz modeling SPT phases using
reflection-positive invertible field theories (IFTs), then classify these IFTs using homotopy theory. This approach
has been successfully employed in several cases to study examples of SPTs, as in \cite{FH16, Cam17, WWW18, FHHT19,
GOPWW18, PW20, BCHM22}.

Condensed-matter physicists are also interested in invertible phases in more general settings, including invertible
phases on a particular space $Y$, as in~\cite{Ran10}, or invertible phases symmetric for a group $G$ acting on
space, such as phases on the plane which have a rotation symmetry and the examples in~\cite{SMJZ13}. These spatial
symmetries are often present in real-world examples of topological phases of matter (see~\cite{WACB16, MYLNS17} for
one example), and can be modeled by lattice Hamiltonian systems in which the symmetry group also acts on the
lattice, though again providing precise definitions is still open. In the case where $G$ is a crystallographic
group acting on $Y = \R^d$, these systems are called \term{crystalline SPT phases}. Freed-Hopkins' field-theoretic
approach does not directly generalize to this setting, but there is a general ansatz of Kitaev~\cite{Kit13, Kit15}
that groups of phases on $Y$ for a fixed symmetry type should define a generalized homology theory.
Freed-Hopkins~\cite{FH19} apply this to propose a classification of invertible phases in the presence of a
$G$-action on space using equivariant generalized homology.

Researchers interested in computing groups of crystalline SPT phases provide \term{crystalline equivalence
principles}, including the first such proposal of Thorngren-Else~\cite{TE18} and subsequent work in~\cite{JR17,
CW18, FH19, ZWYQG19, ZN21, ZYQG20, MCB23, RNQWG24, SY25, ZNQG25}. Crystalline equivalence principles are arguments that groups of
crystalline SPT phases are isomorphic to groups of ordinary SPT phases, where the symmetry type is modified. The
theory is well-understood for symmetry types such as $\O_n$ and $\SO_n$, corresponding to the physicists' notion of
bosonic SPT, but for fermionic SPTs, corresponding to symmetry types such as $\Spin_n$, $\Spin_n^c$, $\Pin_n^\pm$,
etc., the story is more complicated. Cheng-Wang~\cite{CW18}, Zhang-Wang-Yang-Qi-Gu~\cite{ZWYQG19},
Zhang-Ning~\cite{ZN21}, Zhang-Yang-Qi-Gu~\cite{ZYQG20},
Ren-Ning-Qi-Wang-Gu~\cite[\S 5]{RNQWG24}, and
Zhang-Ning-Qi-Gu~\cite[\S V]{ZNQG25}
study examples of fermionic crystalline SPTs, and show
cases of a fermionic crystalline equivalence principle. Crucially, their work implies any fermionic crystalline
equivalence principle must address fermionic phases in which the spatial symmetry mixes with fermion parity, which
goes beyond the scope of Freed-Hopkins' ansatz.

The purpose of this paper is to formulate and prove such a fermionic crystalline equivalence
principle (FCEP). To do so, we provide an ansatz expressing groups of invertible phases on a $G$-space $Y$ in which
the symmetry type can be merely locally constant over space and can mix with $G$, including as a special case
spatial symmetries mixing with fermion parity. Given data $\cL$ expressing this mixing and variance of the symmetry type,
we define \term{phase homology} groups of $Y$, denoted $\Ph_*^G(Y, \cL)$, and our ansatz predicts that the group of
such invertible phases is isomorphic to $\Ph_0^G(Y, \cL)$. Providing this ansatz is an additional goal of this
paper, and is necessary input for our FCEP: the ansatz reexpresses the FCEP as an isomorphism between certain phase
homology groups and groups of IFTs, as we state and prove in \cref{FCEP}. This is the first homotopy-theoretic
account of an FCEP, and to the best of our knowledge is the first fully general version of the FCEP in the
literature.\footnote{Since the first version of this paper appeared online, Manjunath-Calvera-Barkeshli~\cite[\S
III.A]{MCB23} have also proposed a general version of the FCEP, starting with a different ansatz.}

As a corollary of the FCEP, the computation of phase homology groups that represent groups of
point-group-equivariant fermionic phases reduces to computations of bordism groups; this paper's third goal is to
make these computations in several examples, both for the purpose of testing our ansatz by comparing it to
established predictions in physics, and for making additional predictions of groups of crystalline SPT phases in
as yet unstudied settings. For symmetry types that have been studied before by other methods, our computations
agree with the literature, bolstering our ansatz.

Now we go into a little more detail about these ansatzes and theorems. Freed-Hopkins~\cite{FH19} formulate an
ansatz for invertible phases of matter on a topological space $Y$ equipped with an action of a compact Lie group
$G$. First, specify the \term{symmetry type} of the theory as a map $\rho\colon H\to\O$, where
$\O\coloneqq\varinjlim_n\O_n$ is the infinite orthogonal group and $H$ is a topological group. From this data we
can form a Madsen-Tillmann spectrum $\MTH$, whose homotopy groups compute the bordism groups of manifolds with an
$H$-structure on the tangent bundle. Let $I_\Z$ denote the Anderson dual of the sphere spectrum and $E\coloneqq
\Map(\MTH, \Sigma^2I_\Z)$.
\begin{ans}[{Freed-Hopkins~\cite[Ansatz 3.3]{FH19}}]
\label{orig_FH_ansatz}
The abelian group of isomorphism classes of phases on $Y$ equivariant for a $G$-symmetry that does not mix with the
symmetry type $H$ is the Borel-equivariant Borel-Moore homology group $E_{0,\mathrm{BM}}^{hG}(Y)$.
\end{ans}
We will define equivariant Borel-Moore homology in the generality we need in \cref{equivariant_BM}.

When $G$ is trivial and $Y = \R^n$, the group of phases in \cref{orig_FH_ansatz} is naturally isomorphic to $[\MTH,
\Sigma^{d+2}I_\Z]$, which Freed-Hopkins~\cite{FH16} and Grady~\cite{Gra23} show is the classification of invertible field theories with
symmetry type $H$.
When $Y = \R^d$ and $G$ is a crystallographic group, this group of phases is expected to model the
classification of crystalline SPT phases with this symmetry type, and indeed, Freed-Hopkins~\cite[Example
3.5]{FH19} prove a version of the bosonic crystalline equivalence principle of Thorngren-Else~\cite{TE18} as a
consequence of their ansatz, matching physicists' predictions.

For fermionic phases, \cref{orig_FH_ansatz} is not the full answer, and providing the full answer is a major goal
of this paper. Physicists distinguish between phases with ``spinless fermions'' and ``spin-$1/2$ fermions'', asking
how the spatial symmetry group $G$ mixes with fermion parity. For example, one could consider phases on the plane
equivariant for a $C_4$ rotation symmetry, and either ask that fermions' spin is unaffected by the spatial
rotations, or that a full spatial rotation flips the spin on the fermion. This is reminiscent of the
better-understood dichotomy of fermionic phases with a time-reversal symmetry $T$: one may have $T^2 = 1$ or $T^2$
equal to the fermion parity operator. These two classes of phases are modeled with different symmetry types, and
similarly we use different data to model crystalline phases with spinless vs.\ spin-$1/2$ fermions.

To accommodate this mixing between the internal symmetry type $H$ and the spatial symmetry group $G$, we generalize
Freed-Hopkins' setup slightly using parametrized homotopy theory, considering local systems $f$ of symmetry types over
the base $Y$. These give rise to local systems of Thom spectra; if $Y$ has a $G$-action we obtain from $f$ a local
system $\cL'$ of Borel-equivariant Thom spectra, modeled as a functor from $Y$, thought of as an $\infty$-groupoid,
to the $\infty$-category $\Sp^G$ of Borel-equivariant spectra. Let $\cL\coloneqq\Map(\bl,\Sigma^2I_\Z)\circ\cL'$ as
maps $Y\to\Sp^G$, where $I_\Z$ has trivial $G$-action. We define the \term{equivariant phase homology} $\Ph_*^G(Y;
f)$ to be the equivariant Borel-Moore homology of the local system $\cL\colon Y\to\Sp^G$.
\begin{eqans}
The group of $G$-equivariant invertible phases on $Y$ for this data is isomorphic to the equivariant phase homology
group $\Ph_0^G(Y; f)$.
\end{eqans}
When $f$ is trivializable, this reduces to \cref{orig_FH_ansatz}; in general, though, it allows the symmetry type
to mix with the spatial symmetry, or to be merely locally constant on $Y$.

Now we specialize to the cases of spinless and spin-$1/2$ fermions. For spinless fermions, $G$ and $H$ do not mix,
so we use the data of a constant local system of symmetry types and recover Freed-Hopkins' original ansatz. For
spin-$1/2$ fermions, we specify data of an extension of $G$ by $H$
\begin{equation}
	\shortexact{H}{\tH}{G},
\end{equation}
together with a representation $\lambda\colon G\to\O_d$ dictating how $G$ acts on space.\footnote{We also specify
some additional data; see \cref{crystalline_data} in \S\ref{ferm_crys} for the full details.} In the cases we
consider in this paper, $H = \Spin$ or $H = \Spin^c$, and we specify $\tH$ by way of the central extension
\begin{equation}
	\shortexact{\mu_2}{\wG}{G}{}
\end{equation}
whose isomorphism class is picked out by $w_2(V_\lambda) + w_1(V_\lambda)^2\in H^2(BG;\mu_2)$, where $V_\lambda\to
BG$ is the associated vector bundle to the representation $\lambda$ and $\mu_2$ is the group of square roots of
unity. Then, $\tH\coloneqq H\times_{\mu_2} \wG$. Using this data, we build an equivariant local system $f$ of
symmetry types, obtaining a phase homology group $\Ph_0^G(\R^d, f)$ that we predict is isomorphic to the group of
invertible phases for this data.

The FCEP, previously studied in special cases by~\cite{CW18, TE18, ZWYQG19, ZN21, ZYQG20, RNQWG24, ZNQG25},
identifies groups of crystalline SPT phases with groups of fermionic SPT phases with an internal $G$-symmetry ---
but exchanging symmetry types: spinless crystalline phases correspond to spin-$1/2$ internal phases, and vice
versa. Freed-Hopkins~\cite{FH16} model groups of SPT phases with an internal $G$-symmetry using IFTs, and
following Freed-Hopkins~\cite{FH16} and the excellent overview by Beaudry-Campbell~\cite{BC18}, these groups of
TFTs can be expressed in terms of bordism groups of certain Thom spectra. Standard techniques in algebraic
topology, notably the Adams spectral sequence over $\cA(1)$, can be used to compute these bordism groups, so one
application of a general version of the FCEP is to provide access to tractable tools for computing groups of
crystalline SPT phases.

One of the major aims of this paper is to state and prove as a theorem a version of the FCEP, identifying phase
homology groups with groups of IFTs; then \cref{equivariant_ansatz} translates this into a statement about
crystalline SPTs and ordinary SPTs.  In \cref{internal_1,internal_2}, we define the symmetry types for spinless and
spin-$1/2$ fermions for a purely internal $G$-symmetry. In general these definitions are a little technical, but
when the spatial representation $\lambda$ factors through $\SO_d\subset\O_d$, the spinless internal symmetry type
is $H\times G\to\O$ and the spin-$1/2$ symmetry type is $H\times_{\mu_2}\wG\to\O$, with the maps induced by the
projection onto the first factor.
\begin{thmFCEP}[Fermionic crystalline equivalence principle]
Fixing data of $G$, $H$, $\lambda$, etc.\ as above, let $f_0, f_{1/2}$ denote the local systems of symmetry types
for the case of spinless, resp.\ spin-$1/2$ fermions. Then $\Ph_0^G(\R^d; f_0)$ is isomorphic to the group of
deformation classes of $d$-dimensional IFTs for the spin-$1/2$ internal symmetry type, and $\Ph_0^G(\R^d, f_{1/2})$
is isomorphic to the group of deformation classes of $d$-dimensional IFTs for the spinless internal symmetry type.
\end{thmFCEP}
The proof has two key steps.
\begin{enumerate}
	\item Phase homology groups are defined using equivariant parametrized homotopy theory.
	\Cref{reduction_to_Thom_spectra} reexpresses them using ordinary homotopy theory, as homotopy groups of a Thom
	spectrum built from a virtual vector bundle over $B\tH$. The proof uses the
	Ando-Blumberg-Gepner-Hopkins-Rezk~\cite{ABGHR14a, ABGHR14b} approach to Thom spectra.
	\item Then, in \cref{shear_D_thm,shear_A_thm}, we ``shear'' this Thom spectrum, writing down a map $\tH_n\to
	H_{n+d}\times G$ and showing that it induces a homotopy equivalence on Thom spectra, implying that phase
	homology groups are determined by $H$-bordism groups of a Thom spectrum over $BG$. Our proof is modeled on a
	fairly general shearing theorem in Freed-Hopkins~\cite[\FHrefllink{section.10}{\S 10}]{FH16}.
\end{enumerate}
After these two steps, the proof of \cref{FCEP} amounts to looking at the Thom spectra for the internal symmetry
types and noticing that we end up with equivalent Thom spectra over $BG$ in the cases we want to equate.

With this tool in hand, we can compute phase homology groups for point groups acting on $\R^d$, which are our model
for groups of fermionic phases equivariant for point group symmetries. We do these computations for many 2d and 3d
point groups, for both spinless and spin-$1/2$ fermions, and in Altland-Zirnbauer classes D and A (corresponding to
$H = \Spin$, resp.\ $\Spinc$). Our computations use two avatars of the Adams spectral sequence.  It is well-known
that low-dimensional spin bordism can be computed using connective $\ko$-homology and the Adams spectral sequence
over $\cA(1)$, and there is an excellent introduction to this technique by Beaudry-Campbell~\cite{BC18}, but we
also use a variant, computing \spinc bordism via $\ku$-homology and the Adams spectral sequence over $\cE(1)$,
e.g.\ in \S\ref{spinc_dihedral}. This is hardly a new idea, but there appear to be no examples of this specific
kind of computation in the literature before now. We hope that our computations serve as useful examples of how to
use this version of the Adams spectral sequence for \spinc bordism; this could be of independent interest.

For 2d point groups, these phases have been studied in the physics literature using very different methods. We
compare our results with those of other researchers in \S\ref{refl_compare}, \S\ref{inversion_compare},
\S\ref{cyclic_compare}, and \S\ref{dihedral_discrepancy}, and find agreement, providing evidence in favor of
Freed-Hopkins' ansatz and our generalization. However, there is not very much work on fermionic crystalline SPT phases
for most 3d point groups, so our computations are predictions. We do many computations and make many predictions,
and in \S\ref{interesting_to_study} we collect a few that we think are relatively interesting or accessible. For
example:
\begin{thm*}
Let $A_4$ act on $\R^3$ as the orientation-preserving symmetries of a tetrahedron. Then $\Ph_0^{A_4}(\R^3; f)$
vanishes, where $f$ is the local system of symmetry types for either spinless or spin-$1/2$ fermions in both
Altland-Zirnbauer classes D and A.
\end{thm*}
This is a combination of \cref{spin_bord_X,chi_tet_spinful,spinc_chi_tet_thm}. Therefore, assuming
\cref{equivariant_ansatz}, there are no nontrivial spinless nor spin-$1/2$ fermionic SPT phases equivariant for a
chiral tetrahedral symmetry in Altland-Zirnbauer classes D or A. Since the first version of this paper appeared
online, Zhang-Ning-Qi-Gu~\cite[\S III.B, \S IV.B]{ZNQG25} studied these phases with lattice methods, starting from
a different ansatz, but that the group of phases vanishes.

In \S\ref{glide_s}, we leave behind the FCEP and consider a different class of examples, SPTs equivariant for a
glide reflection symmetry, providing a test for Freed-Hopkins' ansatz for a crystallographic group that is not a
point group. Lu-Shi-Lu~\cite{LSL17} conjecture a general classification of these SPTs: that if $\TP_d(H)$ denotes
the group of $d$-dimensional SPT phases with symmetry type $H$, then the group of $d$-dimensional glide SPTs is
isomorphic to $\TP_{d-1}(H)\otimes\Z/2$. Xiong-Alexandradinata~\cite{XA18} derive this classification using
physics-based arguments. We use Freed-Hopkins' ansatzes~\cite{FH16, FH19} to translate Lu-Shi-Lu's conjecture into
a statement about phase homology groups and prove it.

Recall $E\coloneqq\Map(\MTH, \Sigma^2I_\Z)$ and let $\widehat{\Ph}{}_*^\Z(\R^d, \underline E)$ denote the kernel of
the forgetful map from $\Z$-equivariant phase homology to nonequivariant phase homology, where $\Z$ acts on $\R^d$
by glide translations, and $\underline E\to\R^d$ is the constant local system.  This kernel models Lu-Shi-Lu's
group of glide SPTs, as they require glide SPTs to be trivial in the absence of the glide symmetry.
\begin{thmLSL}
There is a natural isomorphism $\widehat{\Ph}{}_0^\Z(\R^d;\underline E)\cong E_{-(d-1)}\otimes\Z/2$.
\end{thmLSL}
This provides additional evidence in favor of the ansatz.

We want to mention that there are other homotopy-theoretic approaches to the study of phases of matter with a
spatial symmetry, including those of Antolín Camarena, Sheinbaum, and collaborators~\cite{AACSS16, SC20, SAC24,
SAC25}, Cornfeld-Carmeli~\cite{CC20}, and Sati-Schreiber~\cite{SS25}. These authors mostly deal with free fermion phases, which are out of scope
of this paper, though see \S\ref{free_ferm}.
\subsection{Reader's guide to the different sections}
Overview:
\begin{itemize}
	\item In \S\S\ref{gen_ans_s}--\ref{ferm_crys} we discuss general aspects of our model for phases on a $G$-space
	$Y$ and prove the FCEP. These sections involve the most homotopy theory.
	\item In \S\S\ref{s_generalities}--\ref{3d_pt} we make phase homology calculations which according to
	\cref{equivariant_ansatz} calculate groups of fermionic crystalline SPT phases for which the symmetry group is
	a point group. We collect the results of these computations in
	\cref{reflections_table,,inversions_table,,spin_table,,dihedral_table,,para_n_gonal_table,,TOI_table}, and
	summarize the methods of computation in \S\ref{s_sseq_summary}.
	\item In \S\ref{glide_s} we consider phases on $\R^d$ with a glide symmetry, and prove a theorem computing the
	corresponding phase homology classification.
\end{itemize}
Now a little more detail. In \S\ref{gen_ans_s}, we use Borel-equivariant parametrized homotopy theory to state a
mild generalization of Freed-Hopkins' ansatz on invertible phases with spatial symmetry. In \S\ref{noneq_param}, we
consider phases on a space $Y$ without a group action, using local systems of symmetry types (\cref{param_ST}). We
define phase homology and in \cref{noneq_ansatz} express the group of invertible phases for such a local system in
terms of phase homology. This is a slight generalization of~\cite[Ansatz 2.1]{FH19}. In \S\ref{eq_param}, we allow
group actions, defining equivariant local systems of symmetry types and equivariant Borel-Moore homology for a
local system for the purpose of formulating \cref{equivariant_ansatz} expressing groups of invertible phases for a
spatial symmetry in terms of equivariant phase homology. This is a minor generalization of Freed-Hopkins'
ansatz~\cite[Ansatz 3.3]{FH19} to the parametrized setting. Then, in \S\ref{s_mixing}, we specialize to the case
relevant to the FCEP, defining the local systems of symmetry types for spatial symmetries that mix with fermion
parity. We prove \cref{reduction_to_Thom_spectra} expressing the phase homology groups for this data in terms of
nonequivariant, nonparametrized homotopy theory, and do not need equivariant or parametrized homotopy theory in the
rest of the paper.

Next, \S\ref{ferm_crys}, whose goal is to state and prove the FCEP. We begin in
\cref{eq_type,,internal_1,,internal_2} by defining the spinless and spin-$1/2$ local systems of symmetry types for
both equivariant (i.e.\ $G$ acting on space) and internal ($G$ not acting on space) symmetries, and use these
definitions to state our FCEP theorem in \cref{FCEP}, identifying phase homology groups for these local systems in
terms of groups of IFTs. As mentioned, the nontrivial part of the proof runs a
shearing argument to simplify a Thom spectrum over $B\tH$ into a smash product of $\MTSpin$ and a Thom spectrum
over $BG$. In \S\ref{d_shear_s}, we prove \cref{shear_D_thm} accomplishing this in class D, for which $H = \Spin$.
Then, in \S\ref{a_shear_s}, we prove \cref{shear_A_thm}, which is the analogous theorem in class A, i.e.\ for $H =
\Spinc$, via a similar proof. Finally, in \S\ref{where_we_prove_FCEP}, we combine these arguments to prove
\cref{FCEP}.

In \S\ref{s_generalities}, we address a few generalities related to the FCEP before studying it in examples. First,
in \S\ref{interesting_to_study}, we provide a summary of some phases or phenomena newly predicted by our
computations which might be interesting to investigate further. In \S\ref{s_sseq_summary}, we introduce
and review the tools from algebraic topology we need to make these computations: the Adams and Atiyah-Hirzebruch
spectral sequences. In \S\ref{s_secretly_bosonic}, we discuss how to use the Adams filtration to detect when an
invertible TFT of $\tH$-manifolds only depends on the underlying $\SO\times G$-structure, which is believed to
correspond to detecting which fermionic phases are really bosonic phases that are fermionic in a trivial way.
Finally, in \S\ref{s_util_lem}, we state and prove several lemmas needed in the computations in the next sections.

Then, in \S\S\ref{rot_refl}--\ref{3d_pt}, we implement this in examples, computing phase homology groups of $\R^d$
equivariant for two- and three-dimensional point-group symmetries, which in \cref{equivariant_ansatz} are
interpreted as groups of point group equivariant fermionic phases on $\R^d$. In all cases we consider
Altland-Zirnbauer classes D and A (corresponding to symmetry types spin and \spinc, respectively), and consider
phases with spinless fermions and spin-$1/2$ fermions. These computations amount to computing spin and \spinc
bordism groups of Thom spectra of vector bundles over $BG$, where $G$ is the point group of interest; we use the
Adams and Atiyah-Hirzebruch spectral sequences to determine these bordism groups.

In \S\ref{rot_refl}, we consider $\Z/2$ acting by a reflection (\S\ref{reflection}) and by an inversion
(\S\ref{s_inv}), as well as $C_n$ acting by rotations (\S\ref{s_rotations}) and $D_{2n}$ acting by rotations and
reflections on $\R^2$ (\S\ref{s_dihedral}) or purely by rotations on $\R^3$ (\S\ref{para_symm}). The results of
these computations can be found in
\cref{reflections_table,,inversions_table,,spin_table,,dihedral_table,,para_n_gonal_table}. Most of these symmetry
types have been studied in the physics literature, and we compare our results with other researchers'.

In \S\ref{3d_pt}, we study many 3d point groups, including chiral tetrahedral symmetry (\S\ref{s_chiral_tet}),
pyritohedral symmetry (\S\ref{pyrit}), full tetrahedral symmetry (\S\ref{s_full_tet}), chiral octahedral symmetry
(\S\ref{s_chiral_oct}), full octahedral symmetry (\S\ref{s_full_oct}), chiral icosahedral symmetry
(\S\ref{s_chiral_ico}), and full icosahedral symmetry (\S\ref{s_full_ico}). In all cases, we study phases with
spinless and spin-$1/2$ fermions in Altland-Zirnbauer types D and A. Our predictions in this section are new as far
as we can determine. See \cref{TOI_table} for the results of the computations.

In \S\ref{glide_s}, we discuss phases equivariant for a glide reflection symmetry. Lu-Shi-Lu~\cite{LSL17}
conjecture a general classification of such phases, and we translate their conjecture into a statement on phase
homology groups using Freed-Hopkins' ansatz, then prove that statement. Finally, in \S\ref{conclusion}, we suggest some
directions for further research.
\subsection*{Acknowledgments}
	I gratefully thank my advisor, Dan Freed, for his constant help, guidance, and patience.

In addition, this paper benefited from conversations with
Vladimir Calvera,
Dexter Chua,
Markus Dierigl,
Zheng-Cheng Gu,
Cameron Krulewski,
Naren Manjunath,
Miguel Montero,
Riccardo Pedrotti,
Yang Qi,
Oscar Randal-Williams,
Daniel Sheinbaum,
Luuk Stehouwer,
Ryan Thorngren,
Juven Wang,
Qing-Rui Wang,
and
Weicheng Ye. Thank you to all.

A portion of this work was supported by the National Science Foundation under Grant No.\ 1440140 while the author
was in residence at the Mathematical Sciences Research Institute in Berkeley, California, during January--March
2020.

\section{Phases on a $G$-space: the general principle}
	\label{gen_ans_s}
We reprise the ansatz of Freed-Hopkins~\cite[Ansatzes 2.1, 3.3]{FH19} on invertible phases on a $G$-space, though
we need to generalize it: physicists often consider crystalline phases in which the symmetry acting on spacetime
mixes with the internal symmetry (e.g.\ a reflection squaring to $(-1)^F$), leading us to generalize from homology
to twisted homology.

What we do not do is define a phase of matter. Precisely defining topological phases of matter, even in the absence
of spatial symmetries, is a difficult open question. Our ansatz is a heuristic that these objects can be classified
with what we call \term{phase homology}, which we do define.
\subsection{Invertible phases on a space}
\label{noneq_param}
Let $Y$ be a locally compact topological space and $\fC$ an $\infty$-category.\footnote{There are different
definitions of $\infty$-categories; we work with \term{quasicategories} as developed by Joyal~\cite{Joy02} and
Lurie~\cite{Lur09b}, so as to follow~\cite{ABG10, ABG18}. However, this paper does not depend on
implementation-specific details. See~\cite[\S 2]{ABG18} for more information and some useful references.} Following
Ando-Blumberg-Gepner~\cite{ABG10, ABG18}, we say a \term{$\fC$-valued local system} on $Y$ is a functor $\cL\colon
\pi_{\le\infty} Y\to\fC$; here $\pi_{\le\infty}Y$ is the fundamental $\infty$-groupoid of $Y$.\footnote{This is not
the only approach to parametrized homotopy theory; see also May-Sigurdsson~\cite{MS06} and
Braunack-Mayer~\cite{BM19}.} If $\cL\colon Y\to\Sp$ is a local system of spectra, the homology of $Y$ valued
in $\cL$ is $\cL_*(Y)\coloneqq \pi_*(\hocolim\cL)$, and the cohomology of $Y$ valued in $\cL$ is
$\cL^*(Y)\coloneqq\pi_*(\holim\cL)$; this generalizes (co)homology with local coefficients.

Given a subspace $j\colon Y'\inj Y$, we also define relative homology groups: $j$ induces a map
$j_*\colon\hocolim_{Y'}\cL|_{Y'}\to\hocolim_Y\cL$, and we define $\cL(Y, Y') \coloneqq
\pi_*(\mathrm{cofib}(j_*))$. Relative cohomology is analogous.
\begin{defn}
\label{noneqBM}
Assume that the one-point compactification $\overline Y$ of $Y$ is a finite CW complex and $\cL$ extends to a local
system $\overline\cL\colon \overline Y\to\Sp$. Choose such an extension $\overline\cL$ over the basepoint $*$. The
\term{Borel-Moore homology} of $Y$ valued in $\cL$ is
\begin{equation}
	\cL_{\mathrm{BM}, *}(Y) \coloneqq \overline\cL_*(\overline Y, *).
\end{equation}
\end{defn}
\Cref{noneqBM} appears to depend on the choice of extension of $\cL$ to $\overline Y$,
but given two choices of extension, the cofibers of the induced maps $\hocolim
\overline{\cL}|_*\to\hocolim\overline{\cL}$ are equivalent, hence compute the same Borel-Moore homology groups.

When $\cL$ is constant, this recovers the usual notion of Borel-Moore (generalized) homology~\cite{BM60, Mil93}.

Recall that a \term{symmetry type} is a space $B$ with a map $f\colon B\to B\O$. This is also referred to as a
\term{tangential structure}; we use the name ``symmetry type'' to emphasize the connection with the symmetries of a
field theory in physics.
\begin{defn}
\label{param_ST}
A \term{local system of symmetry types} over the space $Y$ is a local system on $Y$ valued in the $\infty$-category
of spaces with a map to $B\O$.
\end{defn}
This is closely related to Raptis-Steimle's definition of parametrized tangential structures~\cite[\S 2]{RS17}.

Symmetry types often arise as the stabilizations in $n$ of maps $B\rho_n\colon BH_n\to B\O_n$ induced from
representations $\rho_n\colon H_n\to\O_n$; see~\cite[\FHrefllink{section.2}{\S 2}]{FH16} for a general discussion.
Likewise, the local systems of symmetry types we consider arise from $BH$-bundles over $Y$.

We repeatedly use the notion of \term{Thom spectra}; the definition given by
Freed-Hopkins~\cite[\FHrefllink{page.40}{\S 6.1.4}]{FH16} covers the cases we need.\footnote{Thom spectra have been
heavily studied in homotopy theory; key references include Thom~\cite{ThomThesis}, Atiyah~\cite{Ati61},
May-Quinn-Ray-Tornehave~\cite{MQRT77}, and Ando-Blumberg-Gepner-Hopkins-Rezk~\cite{ABGHR14a, ABGHR14b}.}
\begin{defn}
Given a representation $\rho_n\colon H_n\to\O_n$ or $\rho\colon H\to\O$, where $\O\coloneqq\varinjlim_n\O_n$, we
introduce notation for several Thom spectra. Let $V_n\to B\O_n$ and $V\to B\O$ denote the tautological vector
bundle, resp.\ the tautological rank-zero virtual vector bundle. 
\begin{enumerate}
	\item The Thom spectra $\mathit{MH}_n$, resp.\ $\mathit{MH}$, are the Thom spectra of $(B\rho_n)^*V_n\to BH_n$,
	resp.\ $(B\rho)^* V\to BH$.
	\item The \term{Madsen-Tillmann spectra}~\cite{MT01, MW07} $\MTH_n$, resp.\ $\MTH$, are the Thom spectra of
	$(B\rho_n)^*(-V)\to BH_n$, resp.\ $(B\rho)^*(-V)\to BH$.
\end{enumerate}
\end{defn}
We will use $H \in\set{\O, \SO, \Spin, \Spinc, \Pin^\pm, \Pinc}$; in all of these cases, $\rho$ is
the usual map $H\to \O$ used in, e.g.,~\cite{FH16}.
\begin{rem}
Some Thom spectra go by many names. The notation $\RP_n^\infty$ denotes $(B\O_1)^{nV_1}$, and similarly
$\CP_n^\infty\coloneqq (B\SO_2)^{nV_2}$. Thus, for example, $\Sigma^2\MTSO_2$, $\Sigma^2\MTU_1$, and
$\Sigma^2\CP_{-1}^\infty$ all refer to $(B\SO_2)^{2-V_2}$.
\end{rem}
\begin{defn}
The \term{Anderson dual of the sphere spectrum}~\cite{And69, Yos75} is a spectrum $I_\Z$ satisfying the universal
property that for any spectrum $X$, there is a natural short exact sequence
\begin{equation}
\label{IZproperty}
\shortexact{\Ext(\pi_{n-1}(X), \Z)}{[X, \Sigma^n I_\Z]}{\Hom(\pi_n(X), \Z)}.
\end{equation}
As all such spectra are equivalent, we refer to ``the'' Anderson dual of the sphere spectrum to mean any particular
choice of $I_\Z$.
\end{defn}
\eqref{IZproperty} splits, but not naturally, implying a non-natural isomorphism from $[X, \Sigma^nI_\Z]$ to the
direct sum of the torsion summand of $\pi_{n-1}(X)$ and the free summand of $\pi_n(X)$. We often use this fact
implicitly, calculating $\pi_*(X)$ but depending on the reader to rearrange it into $[X, \Sigma^* I_\Z]$. For more
on $I_\Z$ and its appearance in this context, see Freed-Hopkins~\cite[\FHrefllink{page.33}{\S 5.3},
\FHrefllink{page.35}{\S 5.4}]{FH16}.

Let $\Th\colon\Top_{/B\O}\to\Sp$ denote the Thom spectrum functor and $\mathcal I\colon\Sp\op\to\Sp$
denote the functor $\Map(\bl, \Sigma^2 I_\Z)$.
\begin{defn}
\label{noneq_kitaev_homology_defn}
Let $Y$ be a locally compact space and $f\colon Y\to\Top_{/B\O}$ be a parametrized symmetry type on $Y$. The
\term{phase homology} of this data, denoted $\Ki_*(Y; f)$, is the Borel-Moore homology
\begin{equation}
	\Ki_*(Y; f) \coloneqq (\mathcal I\circ\Th\circ f)_{\mathrm{BM}, *}(Y).
\end{equation}
\end{defn}
\begin{ans}
\label{noneq_ansatz}
With $Y$ and $f$ as in \cref{noneq_kitaev_homology_defn}, the group of invertible topological phases on $Y$ for the
local system of symmetry types $f$ is the phase homology group $\Ki_0(Y; f)$.
\end{ans}
Again, this is not a mathematical definition, but rather a heuristic.
\begin{rem}
When $f$ is constant, \cref{noneq_ansatz} is the original ansatz of Freed-Hopkins~\cite[Ansatz 2.1]{FH19}. In
that case, the ansatz builds on the idea that invertible phases on $Y$ are related to families of
reflection-positive invertible field theories on $Y$. The generalization to nonconstant $f$ allows one to
prescribe how the symmetry type of the family varies along $Y$. For example, one might want to consider families of
phases in which the monodromy around a loop in $Y$ acts by orientation reversal.
\end{rem}
\subsection{Invertible phases on a $G$-space}
\label{eq_param}
Our model for invertible crystalline phases requires considering the case where a compact Lie group $G$ acts on
$Y$. Again we closely follow Freed-Hopkins~\cite[\S 3]{FH19} but using twisted Borel-Moore homology.

Throughout this section, $G$ is a Lie group; unlike in~\cite{FH16, FH19}, we do not need $G$ to be compact. Indeed,
in the study of crystalline phases, $G$ is often an infinite discrete subgroup of $\mathrm{Isom}(\mathbb E^n)$, and
we will consider one such example in \S\ref{glide_s}. We work with the $\infty$-category $\Sp^G$ of
\term{Borel $G$-equivariant spectra}, whose objects can be modeled by data of a sequence of $G$-spaces $X_n$
together with $G$-equivariant maps $\Sigma X_n\to X_{n+1}$.\footnote{There are a few different notions of
$G$-spectra in the equivariant homotopy theory literature, and their names can be confusing. Borel $G$-equivariant
spectra can be thought of as ``spectra with a $G$-action'' or ``spectra living over $BG$,'' and are different from
\term{genuine $G$-spectra}, which have a richer structure. To a geometer, ``equivariant (generalized) cohomology''
usually means the Borel theory, but to a homotopy theorist, it means the genuine theory. See~\cite[\S 2.1]{Sul20}
for a detailed introduction into the different names and notions of $G$-spaces and $G$-spectra.} Notions of
homotopy equivalence, etc., are as in~\cite[\FHrefllink{section.6}{\S 6.1}]{FH16},
and do not require the compactness assumption on $G$ in \textit{loc.\ cit.}
\begin{defn}
Suppose $G$ admits a finite-dimensional, real orthogonal representation $\lambda\colon G\to\O_d$. The one-point
compactification of $\R^d$ with this $G$-action is a $G$-space denoted $S^\lambda$ and called a
\term{representation sphere}.
\end{defn}
The suspension functor $\Sigma^\lambda\coloneqq S^\lambda\wedge\bl$ is not invertible in $G$-spaces, but upon
stabilization is invertible in Borel $G$-spectra; we denote its inverse by $\Sigma^{-\lambda}$. Given a virtual
$G$-representation $V = \lambda - \mu$ (i.e.\ a formal difference of two finite-dimensional real orthogonal
representations), we define the Borel $G$-spectrum $\Sph^V\coloneqq\Sigma^{-\mu}\Sigma^\infty S^\lambda$. We will
let $\Sph$ denote the sphere spectrum with trivial $G$-action.

\begin{defn}
Let $Y$ be a $G$-space and $\cL\colon Y\to\Sp^G$ be a local system. The \term{(Borel-)equivariant homology} of $Y$
with respect to $\cL$ is denoted $\cL_*^G(Y)$ and defined to be
\begin{equation}
	\cL^G_*(Y)\coloneqq \pi_*(\Map_{\Sp^G}(\Sph, \hocolim\cL)^{hG}),
\end{equation}
where $(\bl)^{hG}\colon\Sp^G\to\Sp$ denotes the homotopy fixed-points functor.

If $j\colon Y'\inj Y$ is an inclusion of $G$-spaces, it induces a map
\begin{equation}
	j_*\colon \Map_{\Sp^G}(\Sph, \hocolim_{Y'}\cL|_{Y'})^{hG}\longrightarrow \Map_{\Sp^G}(\Sph,
	\hocolim_Y\cL)^{hG},
\end{equation}
and we define the \term{relative (Borel-)equivariant homology}
\begin{equation}
	\cL_*^G(Y, Y')\coloneqq\pi_*(\mathrm{cofib}(j_*))
\end{equation}
as in the nonequivariant case.
\end{defn}
\begin{defn}
\label{equivariant_BM}
Let $Y$ be a $G$-space and $\cL\colon Y\to \Sp^G$ be an $\Sp^G$-valued local system. Assume that the one-point
compactification $\overline Y$ of $Y$ is a CW complex and $\cL$ extends to a local system $\overline\cL\colon
\overline Y\to\Sp^G$. Choose such an extension $\overline\cL$. The \term{equivariant Borel-Moore homology} of $Y$ valued in $\cL$ is
\begin{equation}
	\cL^G_{\mathrm{BM},*}(Y) \coloneqq \overline\cL{}^G_*(\overline Y, *).
\end{equation}
\end{defn}
Just like \cref{noneqBM}, this does not actually depend on the choice of extension.
\begin{defn}
Let $Y$ be a $G$-space. A \term{$G$-equivariant local system of symmetry types} is a $G$-space $B$ and a
$G$-equivariant map $f\colon B\to Y\times B\O$, where $B\O$ has a trivial $G$-action.
\end{defn}
Taking the Thom spectrum of the map to $B\O$ defines a local system $\Th\circ f\colon Y\to \Sp^G$.
\begin{defn}
\label{equivariant_kitaev_homology_defn}
Let $Y$ be a $G$-space whose one-point compactification is a finite CW complex, and let $f\colon B\to Y\times B\O$
be a $G$-equivariant local system of symmetry types for $Y$. The \term{$G$-equivariant phase homology} of this
data, denoted $\Ki_*^G(Y; f)$, is the equivariant Borel-Moore homology
\begin{equation}
	\Ki_*^G(Y; f)\coloneqq (\mathcal I\circ\Th\circ f)_{\mathrm{BM},0}^G(Y).
\end{equation}
\end{defn}
\begin{ans}
\label{equivariant_ansatz}
With $Y$ and $f$ as in \cref{equivariant_kitaev_homology_defn}, the group of invertible topological phases on $Y$
for the equivariant local system of symmetry types $f$ is the $G$-equivariant phase homology group $\Ki_0^G(Y; f)$.
\end{ans}
Again, this is a heuristic and not a definition. When $G$ is a discrete subgroup of $\mathrm{Isom}(\E^n)$ (e.g.\ a
wallpaper or space group) acting on $Y = \E^n$, these phases are called \term{crystalline SPT phases} in the
physics literature.
%
%
\subsection{Mixing internal and crystalline symmetries}
	\label{s_mixing}
The fermionic crystalline equivalence principle is about invertible topological phases in which an internal
symmetry mixes with the symmetry group acting on space. In this section, we construct the equivariant local systems
of symmetry types for these phases. First, we review how mixing of symmetries is handled in the purely internal
case in \cref{internal_mixing}; then we address the case of spatial symmetries in \cref{reduction_to_Thom_spectra},
showing how to reduce the computation of the relevant equivariant phase homology groups to a nonparametrized
question. We will simplify these computations further in \S\ref{ferm_crys} when we discuss the FCEP in more detail,
then study several examples in \S\S\ref{rot_refl}--\ref{3d_pt}.
\begin{exm}[Mixing for internal symmetries]
\label{internal_mixing}
In the study of SPTs, one commonly encounters symmetry types where there are two different symmetries present, such
as time reversal and fermion parity, but they mix, meaning the group they generate is not a product of the
individual symmetry groups, but rather an extension. For example, we could ask for a generator $T$ of the group of
time-reversal symmetries to square to the fermion parity $(-1)^F$, via the extension $0\to \Z/2\to\Z/4\to\Z/2\to
0$, rather than considering phases where $T^2 = 1$, corresponding to the split extension $0\to \Z/2\to
\Z/2\times\Z/2\to\Z/2\to 0$.

Freed-Hopkins~\cite{FH16} make the ansatz that SPT phases are classified up to equivalence by their low-energy
limits, which are invertible field theories. The symmetry type is expressed as an $H_n$-structure, where $H_n$ is a
group with a map to $\O_n$; mixing manifests as an extension involving the base symmetry type (e.g.\ $\Spin_n$ for
fermionic phases) and the additional symmetry. For example, the two cases of time-reversal symmetry squaring to the
identity or to fermion parity are represented by the extensions
\begin{subequations}
\begin{gather}
	\shortexact*{\Spin_n}{\Pin_n^+}{\Z/2}{}\\
	\shortexact*{\Spin_n}{\Pin_n^-}{\Z/2},
\end{gather}
\end{subequations}
respectively, together with the standard maps $\Pin_n^\pm\to\O_n$ as defined in~\cite[Theorem 3.11]{ABS}.
\end{exm}
When one of the groups we want to mix acts on space, we can specify a mixed symmetry type by the following data:
\begin{itemize}
	\item a symmetry type $\rho_n\colon H_n\to\O_n$, called the \term{base symmetry type},
	\item the point group symmetry $\lambda\colon G\to\O_d$,
	\item an extension
	\begin{equation}
	\label{the_mixing_extension}
		\shortexact*{H_n}{\widetilde H_n}{G}{}
	\end{equation}
	specifying how they mix, and
	\item an extension $\widetilde\rho_n\colon \wH_n\to\O_n$ of $\rho_n\colon H_n\to\O_n$.
\end{itemize}
Freed-Hopkins~\cite[\FHrefllink{page.77}{\S 9.2}]{FH16} relate Altland-Zirnbauer's symmetry classes of
condensed-matter systems~\cite{Zir96, AZ97} to ten symmetry types in topology.\footnote{This ``tenfold way'' is a
relativistic version of Dyson's threefold way~\cite{Dys62}, and appears in many contexts in physics,
including~\cite{Kit09, RSFL10, FM13, WS14, FH16, KZ16, GM20, IT20}.} Using this, we call the case $H = \Spin$ the
\term{class D case} and $H = \Spinc$ the \term{class A case}.

Let $Y$ be a $G$-space. Then the map
\begin{equation}
	Y\times E\widetilde H_n/H_n\longrightarrow Y
\end{equation}
is a $G$-equivariant fiber bundle with fiber $BH_n$, and the total space maps to $B\O_n$ as specified by the
virtual vector bundle
\begin{equation}
\label{an_EPST}
	f\colon{ - (Y\times(E\widetilde H_n\times_{H_n}\R^n))}\longrightarrow Y\times E\widetilde H_n/H_n.
\end{equation}
After stabilizing (i.e.\ letting $n\to\infty$), this is an equivariant local system of symmetry types over $Y$, so
has equivariant phase homology groups $\Ph_*^G(Y; f)$. Under \cref{equivariant_ansatz}, $\Ph_0^G(Y; f)$ models
the group of invertible topological phases on $Y$ in which fermion parity mixes with the spatial symmetry as
specified by~\eqref{the_mixing_extension}.  The notion of $G$-equivariant phases for this symmetry type (without a
reference space $Y$) is taken to mean $G$-equivariant phases on $\R^d$, where $G$ acts on $\R^d$ through $\lambda$.
\begin{rem}[Change of symmetry type]
\label{change_symmetry}
We would like to be able to move information between instances of this data: for example, there should be forgetful
maps from equivariant phases on a space to nonequivariant ones, and we model them with maps between phase homology
groups for the two local systems of symmetry types.

Suppose we are given two instances of the data above. That is, we ask for a commutative diagram of Lie groups
\begin{equation}
\label{change_of_ST_diagram}
\begin{tikzcd}
	1 & {H_n} & {\wH_n} & G & 1 \\
	1 & {H_n'} & {\wH_n'} & {G'} & 1
	\arrow[from=1-1, to=1-2]
	\arrow[from=2-1, to=2-2]
	\arrow[from=1-2, to=1-3]
	\arrow[from=2-2, to=2-3]
	\arrow[from=1-3, to=1-4]
	\arrow[from=2-3, to=2-4]
	\arrow[from=1-4, to=1-5]
	\arrow[from=2-4, to=2-5]
	\arrow["{\vp_G}", from=1-4, to=2-4]
	\arrow["\widetilde\vp", from=1-3, to=2-3]
	\arrow["\vp", from=1-2, to=2-2]
\end{tikzcd}
\end{equation}
together with maps $\rho_n\colon H_n\to\O_n$ and $\rho_n'\colon H_n\to\O_n$, $\lambda\colon G\to\O_d$ and
$\lambda'\colon G'\to\O_d$, and $\widetilde\rho_n\colon \wH_n\to\O_n$ and $\widetilde\rho_n'\colon\wH_n'\to\O_n$
which commute with the vertical maps in~\eqref{change_of_ST_diagram}. Fix a $G'$-space $Y$; then through~\eqref{an_EPST}
this defines equivariant local systems of symmetry types $f$ for $G$, resp.\ $f'$ for $G'$. The maps between the
data induce a pullback or forgetful map $\vp^*\colon \Ki_*^{G'}(Y; f')\to\Ki_*^G(Y; f)$, where $G$ acts on $Y$
through $\vp_G$. Using \cref{equivariant_ansatz}, we interpret this pullback map realizing an invertible phase on
$Y$ with a $G'$-symmetry to a phase with a $G$-symmetry.

The construction of $\vp^*$ amounts to checking that diagrams you would expect to commute do in fact
commute. The data we gave induces a commutative diagram
\begin{equation}
\gathxy{
	- (Y\times (E\tH_n\times_{H_n} \R^n))\ar[r]\ar[d] & Y\times E\tH_n/H_n\ar[d]\\
	- (Y\times (E\tH_n'\times_{H_n'} \R^n))\ar[r] & Y\times E\tH_n'/H_n'.
}
\end{equation}
The rows define equivariant local systems of symmetry types; then $f$ and $f'$ are the maps to $Y\times B\O$.  Let
$\vp^\circ\colon \Sp^{G'}\to\Sp^G$ be the map in which $G$ acts on Borel $G'$-spectra through $\vp$; then, upon
applying $\mathcal I\circ\Th$, we obtain local systems $\cL$, resp.\ $\cL'$ of Borel $G$-, resp.\ $G'$-spectra.  To
define phase homology, we assumed that an extension $\overline{\cL}$ of $\cL$ to $\overline Y$ exists, so choose
such an extension; then $\overline\cL{}'\coloneqq \overline{\cL}\circ\vp^\circ$ is an extension of $\cL$. We obtain
from the inclusion $*\inj\overline Y$ a commutative diagram of spectra
\begin{equation}
\gathxy{
	\Map_{\Sp^{G'}}(\Sph, \hocolim_{*}\overline\cL{}'|_{*})^{hG'}\ar[r]\ar[d]
	& \Map_{\Sp^{G}}(\Sph, \hocolim_{*}\overline\cL|_{*})^{hG}\ar[d] \\
	\Map_{\Sp^{G'}}(\Sph, \hocolim_{\overline Y} \overline\cL{}')^{hG'}\ar[r] & \Map_{\Sp^G}(\Sph,
	\hocolim_{\overline Y}\overline\cL)^{hG}.
}
\end{equation}
Thus, we get a map between the cofibers of the vertical arrows, and $\pi_*$ of that map is the desired map on phase
homology.

For us there are two particularly important examples.
\begin{enumerate}
	\item Let $H'_n = H_n$ and $G = 1$, which forces $\widetilde\vp\colon\wH_n\to\wH_n'$ to be the inclusion
	$H_n'\to\tH_n'$. The above construction produces a map from $H$-equivariant phase homology to nonequivariant
	phase homology on $Y$, which we interpret as modeling the forgetful map from phases with a $G$-symmetry to
	phases without a $G$-symmetry.
	\item Let $G' = G$, $H_n' = \SO_n$, and $H_n$ be either $\Spin_n$ or $\Spin^c_n$, with $\vp$ the usual map. In
	this case the pullback map goes from equivariant phase homology where the base symmetry type is $\SO$ to
	equivariant phase homology where the base symmetry type is $\Spin$ or $\Spin^c$. We interpret this in physics
	as modeling the procedure that regards a bosonic phase as a fermionic phase by adding some trivial fermionic
	degrees of freedom. This is analogous to the procedure which regards an oriented TFT as a spin TFT that does
	not depend on the spin structure.
\end{enumerate}
\end{rem}
Crucially for computations, we can simplify the equivariant phase homology groups for the symmetry types
in~\eqref{an_EPST} into a description not requiring equivariant or parametrized stable homotopy theory.
\begin{prop}
\label{reduction_to_Thom_spectra}
There is an isomorphism
\begin{equation}
	\Ph_0^G(\R^d; f) \overset\cong\longrightarrow [(B\widetilde
	H)^{d-\lambda-\widetilde\rho}, \Sigma^{d+2}I_\Z]
\end{equation}
natural for changing the symmetry type in the sense of \cref{change_symmetry}.
\end{prop}
\begin{proof}
We want to compute the twisted equivariant Borel-Moore homology for this equivariant local system of symmetry
types, where $Y = \R^d$ with $G$ acting through $\lambda$. This amounts to the following: one-point compactify to a
local system over $S^\lambda$; take the colimit of the local system and call it $E$; then compute $[\Sph, E]^G$ (in
the notation of \cite{FH19}; this means $\pi_0(\Map(\Sph, E)^{hG})$). Now, the local system $(\mathcal
I\circ\Th\circ f)\colon S^\lambda\to\Sp^G$ is nonequivariantly the trivial local system with fiber $\Map(\MTH,
\Sigma^2 I_\Z)$, so $E\simeq S^\lambda\wedge \Map(\MTH, \Sigma^2 I_\Z)$; in general, $G$ can act nontrivially on
both $S^\lambda$ and $\MTH$, but always acts trivially on $\Sigma^2I_\Z$.  Therefore we may
follow~\cite[(3.6)]{FH19} and identify
\begin{equation}
\label{unnatural_G_action}
	\Map(\Sph, S^\lambda\wedge \Map(\MTH, \Sigma^2 I_\Z))\simeq \Map(\Sph^{d-\lambda}\wedge\MTH, \Sigma^{d+2}I_\Z),
\end{equation}
though the $G$-action on $\Sph^{d-\lambda}\wedge\MTH$ is not the diagonal action, but rather the induced $G$-action
on the Thom spectrum of the $G$-equivariant virtual bundle $(d-\underline\lambda- \rho)\to BH$
(see~\cite[\FHrefllink{page.44}{\S 6.2.2}]{FH16}).

Since $G$ acts trivially on $\Sigma^{d+2}I_\Z$,
\begin{equation}
	\Map(\Sph^{d-\lambda}\wedge\MTH, \Sigma^{d+2}I_\Z)^{hG} \simeq \Map((\Sph^{d-\lambda}\wedge\MTH)_{hG},
	\Sigma^{d+2}I_\Z).
\end{equation}
It now suffices to show that
\begin{equation}
\label{was_a_lemma}
	(\Sph^{d-\lambda}\wedge\MTH)_{hG}\simeq (B\widetilde H)^{-\widetilde\rho - \lambda + d}.
\end{equation}
%
Ando-Blumberg-Gepner-Hopkins-Rezk~\cite[Proposition 1.20]{ABGHR14a} show that the Thom spectrum of a virtual bundle
$V\to X$, identified with a map $V\colon X\to B\O$, is the homotopy colimit
\begin{equation}
\label{ABGHR_Thom_defn}
	X^V \simeq \hocolim\paren*[\big]{\xymatrix@1{X\ar[r]^V &B\O\ar[r]^-{BJ} & B\GL_1(\Sph)\ar[r] & \Sp}},
\end{equation}
where the notation means to interpret $X$ as, through its fundamental $\infty$-groupoid, providing a diagram in the
$\infty$-category $\Sp$ of spectra. Here $B\GL_1(\Sph)$ is the classifying space of stable spherical
fibrations~\cite{Sta63, MQRT77} and $BJ\colon B\O\to B\GL_1(\Sph)$ is a form of the $J$-homomorphism~\cite{Whi42,
MQRT77}. Heuristically,~\eqref{ABGHR_Thom_defn} says that the virtual vector bundle $V$ defines a local system of
$\wedge$-invertible spectra, with the fiber at a point $x\in X$ given by $\Sph^{V_x}$, and that the Thom spectrum
is obtained from an associated bundle construction. See~\cite{ABGHR14a, ABGHR14b} for more detail on this approach
to Thom spectra.

Homotopy quotients are also homotopy colimits, meaning
\begin{subequations}
\begin{align}
\label{double_hocolim}
	(\Sph^{d-\lambda}\wedge\MTH)_{hG} &= \hocolim_{\pt/G} \paren*[\Big]{\hocolim\paren*[\big]{
	\xymatrix@1{BH\ar[r]^-{d-\underline \lambda-\rho} &B\O\ar[r]^-{BJ} & B\GL_1(\Sph)\ar[r] & \Sp}}},
\intertext{where $G$ acts on the spectra in the diagram through its action on $\lambda$, as well as on $BH$, as
prescribed by the extension~\eqref{the_mixing_extension}. This in particular implies the double homotopy colimit
above simplifies into a single homotopy colimit over a $B\widetilde H$-shaped diagram:}
	&\simeq \hocolim\paren*[\Big]{\xymatrix@1{B\widetilde H\ar[r]^-{d-\lambda-\widetilde\rho} &B\O\ar[r]^-{BJ} &
	B\GL_1(\Sph)\ar[r] & \Sp}},
\end{align}
\end{subequations}
which by~\eqref{ABGHR_Thom_defn} is the Thom spectrum for $d-\lambda-\widetilde\rho\to B\widetilde H$,
proving~\eqref{was_a_lemma}.
\end{proof}
%
Our next step in \S\ref{ferm_crys} is to simplify $(B\tH)^{d - \lambda-\widetilde\rho}$. This allows both for a
general formulation of the fermionic crystalline equivalence principle as well as explicit calculations.

The following lemma will be helpful for simplifying Thom spectra.
\begin{thm}[Relative Thom isomorphism]
\label{relative_Thom}
Let $\rho\colon H\to\O$ be a symmetry type with the two-out-of-three property, i.e.\ an $H$-structure on any two of
$E$, $F$, or $E\oplus F$ induces one on the third. If $V,W\to X$ are virtual vector bundles such that $V$ has an
$H$-structure, then there is an equivalence
\begin{equation}
	\MTH\wedge X^W \overset\simeq\longrightarrow \MTH\wedge X^{V\oplus W}.
\end{equation}
\end{thm}
\begin{proof}
The two-out-of-three property gives $\MTH$ an $E_\infty$-ring structure, which is needed for some of the
constructions we employ from~\cite{ABGHR14a, ABGHR14b} below.

Up to equivalence, the Thom spectrum of a virtual vector bundle $E\to X$ depends only on the homotopy class of the
map $f_E\colon X\to B\O\to B\GL_1(\Sph)$, where the first map is given by $E$, and the second map is the
$J$-homomorphism, as in~\eqref{ABGHR_Thom_defn}. Smashing with $\MTH$ corresponds to composing $f_E$ with the map
$B\GL_1(\Sph)\to B\GL_1(\MTH)$ induced by the Hurewicz map $\Sph\to\MTH$~\cite[\S 1.4]{ABGHR14b}, and in
particular, up to equivalence, $\MTH\wedge X^E$ only depends on the homotopy type of the map $X\to B\GL_1(\MTH)$.

Because $\MTH$ is an $E_\infty$-ring spectrum, $B\GL_1(\MTH)$ is a grouplike $E_\infty$-space, and the composition
$\psi\colon B\O\to B\GL_1(\Sph)\to B\GL_1(\MTH)$ is a map of grouplike $E_\infty$-spaces, where $B\O$ carries the
$E_\infty$ structure coming from direct sum. This means that $[X, B\GL_1(\MTH)]$ is naturally an abelian group, and
that if we define classes in this group using virtual vector bundles $V,W\to X$ to map to $B\O$ then composing with
$\psi$, the class of $V\oplus W$ is the sum of the classes of $V$ and $W$.

An $H$-structure on $V$ trivializes the map $X\to B\O\overset\psi\to B\GL_1(\MTH)$ defined by $V$, so the class of
the map defined by $V\oplus W$ is equal to the class of the map defined by $W$ in the abelian group $[X,
B\GL_1(\MTH)]$.
\end{proof}

\section{The fermionic crystalline equivalence principle}
	\label{ferm_crys}
In this section, our goal is to state and prove the FCEP, \cref{FCEP}, identifying phase homology groups in
classes D and A with groups of deformation classes of invertible field theories. Assuming
\cref{equivariant_ansatz}, this leads to the more familiar version of the FCEP. Crystalline equivalence principles
are first introduced by Thorngren-Else~\cite{TE18}: the idea is to equate the classification of crystalline
topological phases of matter for some group $G$ acting on spacetime with a classification of a different kind of
topological phases of matter, in which $G$ is part of the internal symmetry group. Then one may use preexisting
techniques for phases without a spatial symmetry to classify phases with the specified $G$-action on space.

The best-understood crystalline equivalence principles are for bosonic SPTs, as first considered by
Thorngren-Else~\cite{TE18}. ``Bosonic'' does not have a precise mathematical translation here; these are phases for
which the symmetry type is built using $\SO$ or $\O$ rather than $\Spin$, $\Spinc$, $\Pin^\pm$, and so on. If a
group $G$ acts on space by orientation-preserving symmetries and $H$ is $\SO$ or $\O$, the classification of
crystalline SPTs in dimension $n$ with symmetry type $H$ and this $G$-action is identified with the classification
of SPTs for $H = \SO\times G$. To what extent this is an ansatz or a theorem depends on one's model
for crystalline SPTs: Freed-Hopkins~\cite[Example 3.5]{FH19} derive it as a corollary of their ansatz.\footnote{If
$G$ acts by reflections, almost as nice of a story is still true, but the internal $G$-symmetry mixes with $H$.
Thorngren-Else~\cite{TE18} and Freed-Hopkins~\cite[Example 3.5]{FH19} discuss this case too.} For other derivations
of the bosonic crystalline equivalence principle from different ansatzes, see Jiang-Ran~\cite{JR17} and
Thorngren-Else~\cite{TE18, ET19}.

The fermionic analogue of this statement is more complicated because there are more ways for $G$ to mix with the
symmetry type. Thorngren-Else~\cite[\S VII.B]{TE18}, Cheng-Wang~\cite{CW18}, Zhang-Wang-Yang-Qi-Gu~\cite{ZWYQG19},
Zhang-Ning~\cite{ZN21}, Zhang-Wang-Yang-Gu~\cite[\S V]{ZYQG20}, Cheng-Wang-Yang~\cite{CWY24},
Ren-Ning-Qi-Wang-Gu~\cite[\S V]{RNQWG24}, and
Zhang-Ning-Qi-Gu~\cite[\S V]{ZNQG25}
all study examples in which an FCEP holds, and each paper discusses that
such a principle would have to account for the different ways in which $G$ mixes with $H$: crystalline phases for
which the spatial $G$-symmetry does not mix with fermion parity correspond to phases with an internal $G$-symmetry
that does mix with fermion parity, and vice versa. Examples of this
twisted correspondence also appear in work of Freed-Hopkins~\cite[Example 3.5]{FH19},
Guo-Ohmori-Putrov-Wan-Wang~\cite{GOPWW18}, and Mao-Wang~\cite{MW20}, though until now there was no precise general
version of the FCEP.\footnote{Since this paper first appeared online, Manjunath-Calvera-Barkeshli~\cite[\S
III.A]{MCB23} proposed another general FCEP, starting from a different ansatz and using different methods.}

Our version of the FCEP applies in Altland-Zirnbauer classes A and D (i.e.\ $H = \Spin$ or $H = \Spinc$), for all
compact Lie groups $G$ acting on faithfully on space, and all ways in which $G$ may mix with fermion
parity. The slogan ``mixed crystalline goes to unmixed internal, and vice versa'' is a little hard to glean from
the result when the $G$-action includes reflections, but we obtain an equivalence from phase homology groups for
certain equivariant local systems of symmetry types, which under \cref{equivariant_ansatz} stands in for groups of
crystalline SPT phases, to groups of deformation classes of IFTs, which under Freed-Hopkins' ansatz~\cite{FH16}
model groups of phases without spatial symmetries.

To precisely state our FCEP, we must fix some data.
\begin{notat}\hfill
\label{crystalline_data}
\begin{itemize}
	\item Let $H$ denote the base symmetry type, which today is either of the infinite-dimensional topological
	groups $\Spin$ or $\Spinc$.
	\item Let $G$ be a compact Lie group, $\lambda\colon G\to\O_d$ be a faithful representation, and
	$V_\lambda\coloneqq EG\times_G \R^d\to BG$ be the associated vector bundle.
	\item Let $\xi\colon G\to\O_{d'}$ be another faithful representation and $V_\xi\to BG$ be the associated
	vector bundle. Let $1\to\mu_2\to\wG\to G\to 1$ be
	the central extension classified by $w_2(V_\lambda) + w_1(V_\lambda)^2\in H^2(BG;\mu_2)$. Here $\mu_2$ denotes
	the group of square roots of unity.
	\item Let $\tH\coloneqq H\times_{\mu_2} \wG$. Let $\rho$ be the composition $\tH\to H\to\O$ and $V\to B\tH$ be
	the associated tautological vector bundle.
\end{itemize}
\end{notat}
For us, $\xi$ and $\lambda$ are usually the same, but they differ when $G = \Z/2$ acts on $\R^d$ by inversion in
the case of spin-$1/2$ fermions: here $\xi$ is the sign representation $\sigma\colon\Z/2\to\O_1$, but $\lambda =
d\sigma$.  See~\S\ref{s_inv} for more detail.
\begin{defn}
\label{eq_type}
The \term{spin-$1/2$ equivariant local system of symmetry types} for the above data is the $G$-equivariant
parametrized symmetry type $f_{1/2}\colon B\tH\to \R^{d'}\times B\O$ which sends $x\mapsto (0, B\rho(x))$, and in
which $G$ acts on $\R^d$ through $\lambda$. The \term{spinless equivariant local system of symmetry types} $f_0$ is
defined in the same way, except using $H\times G$ instead of $\tH$.
\end{defn}
\begin{defn}
\label{internal_1}
Recall that $H$ is either $\Spin$ or $\Spinc$. Let $\dagger\in\set{-,c}$ be $-$ if $H = \Spin$ and $c$ otherwise.
The \term{spinless internal symmetry type} is the symmetry type
\begin{itemize}
	\item $(-V, d-V_\lambda)\colon BH\times BG\to B\O$, if $\lambda$ is pin\textsuperscript{$\dagger$}, or
	\item $(-V, V_\xi + \Det(V_\xi) - V_\lambda)\colon BH\times BG\to B\O$, if $\lambda$ is not
	pin\textsuperscript{$\dagger$}.
\end{itemize}
For shorthand, we denote this symmetry type $\rho(0)\colon BH\times BG\to B\O$.
\end{defn}
\begin{defn}
\label{internal_2}
The \term{spin-$1/2$ internal symmetry type} is the symmetry type
\begin{equation}
	(-V, d - V_\lambda)\colon BH\times BG\to B\O.
\end{equation}
\end{defn}
For shorthand, we denote this symmetry type $\rho(1/2)\colon BH\times BG\to B\O$.
\begin{rem}
The internal symmetry types probably look pretty arbitrary. This is because of the generality of our setup: in some
cases of interest, we can rewrite these symmetry types in ways which more closely resembles the proposals of
Thorngren-Else~\cite[\S VII.B]{TE18}, Zhang-Wang-Yang-Qi-Gu~\cite{ZWYQG19}, Zhang-Ning~\cite{ZN21}, and Cheng-Wang~\cite{CW18} for the FCEP
in specific cases.

Suppose $\lambda = \xi$ and $\Im(\lambda)\subset\SO_d$ but does not lift across $\Spin_d\surj\SO_d$. Then, the
spinless internal symmetry type simplifies to $BH\times BG\to B\O$, where the map is just projection onto the first
factor followed by the usual map $BH\to B\O$. That is, for representations with image contained in $\SO_d$, the
FCEP switches the ``unmixed'' (i.e.\ $BH\times BG$) and ``mixed'' (i.e.\ $B(H\times_{\mu_2}\wG)$) symmetry types
when passing between crystalline and internal phases. This matches predictions by Thorngren-Else~\cite{TE18} and
Cheng-Wang~\cite{CW18}.
\end{rem}
Freed-Hopkins~\cite[\FHrefllink{page.65}{Corollary 8.21}]{FH16} show that the group of deformation classes of
reflection-positive IFTs with symmetry type $\rho'\colon H'\to\O$ in (space) dimension $n$ is naturally isomorphic
to
\begin{equation}
\label{FHIFT}
        [\MTH', \Sigma^{d+2}I_\Z].
\end{equation}
Freed-Hopkins (\textit{ibid.}) conjectured, and Grady~\cite{Gra23} proved, that the full group~\eqref{FHIFT}
classifies all reflection-positive invertible field theories, topological or not.
\begin{thm}[Fermionic crystalline equivalence principle]
\label{FCEP}
There are isomorphisms
\begin{subequations}
\begin{align}
	\Ph_k^G(\R^d; f_0) &\overset\cong\longrightarrow [\mathit{MT\!\rho}(1/2), \Sigma^{d+k+2}I_\Z]\\
	\Ph_k^G(\R^d; f_{1/2}) &\overset\cong\longrightarrow [\mathit{MT\!\rho}(0), \Sigma^{d+k+2}I_\Z].
\end{align}
\end{subequations}
\end{thm}
Assuming \cref{equivariant_ansatz}, the physics implication of this theorem is that the abelian group of
crystalline SPT phases for the spinless equivariant local system of symmetry types is naturally isomorphic to the
abelian group of deformation classes of IFTs for the spin-$1/2$ internal symmetry type; and the classification of
crystalline SPT phases for the spin-$1/2$ equivariant local system of symmetry types is naturally isomorphic to the
abelian group of deformation classes of IFTs of the spinless internal symmetry type.
We break the proof of \cref{FCEP} down into a few steps. First, \cref{reduction_to_Thom_spectra} simplifies the
question into one of ordinary stable homotopy theory.\footnote{For the spinless equivariant symmetry type, this
is just~\cite[Example 3.5]{FH19}.} We obtain Thom spectra for vector bundles over $B\tH$, and to finish we must
compare these spectra to $\MTH\wedge (BG)^E$, where $E\to BG$ is some rank-zero virtual vector bundle.  This
comparison, in the form of \term{shearing arguments}, is the core of the proof: we prove \cref{shear_D_thm} ($H =
\Spin$) and \cref{shear_A_thm} ($H = \Spinc$) establishing the homotopy equivalences we need, and after that
proving \cref{FCEP} amounts to verifying that the outputs of \cref{shear_D_thm,shear_A_thm} simplifying the
crystalline symmetry types match the Thom spectra for the internal symmetry types in \cref{internal_1,internal_2}.

The proofs of \cref{shear_D_thm,shear_A_thm} resemble the proofs of the more standard equivalences
\begin{subequations}
\begin{align}
  \label{pinpsplitting}
	\MTPin^+ &\simeq \MTSpin\wedge (B\Z/2)^{1 - \sigma}\\
  \label{pinmsplitting}
	\MTPin^- &\simeq \MTSpin\wedge (B\Z/2)^{\sigma-1}\\
  \label{pincsplitting}
	\MTPinc &\simeq \MTSpinc\wedge (B\Z/2)^{\pm(1 - \sigma)}\\
	\MTSpin^c &\simeq \MTSpin\wedge (B\SO_2)^{\pm(2- V_2)},
\end{align}
where $\sigma\to B\Z/2$ and $V_2\to B\SO_2$ denote the respective tautological line bundles. These decompositions
were first proven by Kirby-Taylor~\cite[Lemma 6]{KT90Pinp} (\pinp), Peterson~\cite[\S7]{Pet68} (\pinm), and
Bahri-Gilkey~\cite{BG87a, BG87b} (\spinc and \pinc). For a unified proof of all of these equivalences, see
Freed-Hopkins~\cite[\FHrefllink{section.10}{\S 10}]{FH16}.
\end{subequations}
\subsection{Case $H = \Spin$}
\label{d_shear_s}
%
\begin{thm}[Shearing, class D]\label{shear_D_thm}
Let $V\to B\tH$ be the tautological bundle.
\begin{enumerate}
	\item\label{is_pinm} Suppose $V_\xi$ admits a \pinm structure. Then there is an equivalence
	\begin{equation}
		(B\tH)^{d - V_\lambda - V}\overset\simeq\longrightarrow \MTSpin \wedge (BG)^{d- V_\lambda}.
	\end{equation}
	\item\label{isnt_pinm} If $V_\xi$ does not admit a \pinm structure, there is an equivalence
	\begin{equation}
		(B\tH)^{d - V_\lambda - V} \overset\simeq\longrightarrow \MTSpin\wedge (BG)^{V_\xi + \Det(V_\xi) -
		V_\lambda - d'-1+d}.
	\end{equation}
\end{enumerate}
\end{thm}
We will most often consider case~\eqref{isnt_pinm} with $\lambda = \xi$, in which case we learn
$(B\tH)^{d-V_\lambda-V}\simeq \MTSpin\wedge (BG)^{\Det(V_\lambda)-1}$.
\begin{proof}
Case~\eqref{is_pinm} is by far the easier of the two: $V_\xi$ admits a \pinm structure iff $w_2(V_\xi) +
w_1(V_\xi)^2 = 0$ iff the extension $1\to\mu_2\to\wG\to G\to 1$ splits. Since $\mu_2\subset\wG$ is central, a
splitting induces isomorphisms $\wG\cong \mu_2\times G$ and $\tH_n\cong \Spin_n\times G$. Passing to classifying
spaces, this identifies $d - V_\lambda - V\colon B\tH\to B\O$ with $-V \boxplus (d-\lambda)\colon B\Spin\times
BG\to B\O$; then take Thom spectra.

On to case~\eqref{isnt_pinm}. In this case, in $H^2(B\tH; \mu_2)$, $w_2(V_\xi) + w_1(V_\xi)^2 = w_2(V)$, so the map
$V + V_\xi + \Det(V_\xi)\colon B\tH\to B\SO$ lifts across $B\Spin\to B\SO$. Choose such a lift.
\begin{prop}
\label{pi_p_prop}
The induced map
\begin{equation}
\label{pi_p_map}
	(V + V_\xi + \Det(V_\xi), \xi)\colon B\tH\longrightarrow B\Spin\times BG
\end{equation}
is a homotopy equivalence commuting with the maps to $B\SO$.
\end{prop}
The proof is due to Freed-Hopkins~\cite[\FHrefllink{section.10}{\S 10}]{FH16}.
\begin{proof}
We will show that the commutative square
\begin{subequations}
\begin{equation}
\label{first_pullback}
	\gathxy{
		B\tH\ar[r]\ar[d]_{B(\pi_1\oplus \pi_2)} & B\Spin\ar[d]\\
		B\SO\times BG\ar[r]^-{B(\id\oplus \xi)} & B\SO
	}
\end{equation}
is homotopy Cartesian. Any two homotopy pullbacks of the same diagram are weakly equivalent, with the weak
equivalence intertwining the maps to $B\SO$. Since there is also a homotopy pullback square
\begin{equation}
\label{second_pullback}
	\gathxy{
		B\Spin\times BG\ar[r]\ar[d] & B\Spin\ar[d]\\
		B\SO\times BG\ar[r]^-{B(\id\oplus \xi)} & B\SO,
	}
\end{equation}
\end{subequations}
then $B\tH\simeq B\Spin\times BG$; this equivalence is realized by~\eqref{pi_p_map} because that is the only
possibility that intertwines the maps in~\eqref{first_pullback} and~\eqref{second_pullback}.

To fulfill the promise that~\eqref{first_pullback} is a homotopy pullback square, begin with the commutative
diagram of short exact sequences
\begin{equation}
\label{before_fib}
	\gathxy{
		1\ar[r] & \mu_2\ar@{=}[d]\ar[r] & \tH_n\ar[r]^-{(\pi_1, \pi_2)}\ar[d] &
		\SO_n\times G\ar[r]\ar[d]^{\id\oplus\xi} &1\\
		1\ar[r] & \mu_2\ar[r] &  \Spin_{n+d}\ar[r] & \SO_{n+d}\ar[r] & 1.
	}
\end{equation}
This induces a map of fiber sequences
\begin{equation}
\label{fiber_seq_pullback}
	\gathxy{
		B\tH\ar[r]^-{B(\pi_1, \pi_2)}\ar[d] &B\SO\times
		BG\ar[d]^{B(\id\oplus\xi)} \ar[r]^-{w_2} & K(\mu_2, 2)\ar@{=}[d]\\
		B\Spin\ar[r] & B\SO\ar[r]^-{w_2} & K(\mu_2, 2),
	}
\end{equation}
e.g.\ $B\tH$ is the fiber of $w_2\colon B\SO\times BG\to K(\mu_2, 2)$. The left square in such a pullback is always
homotopy Cartesian, and in~\eqref{fiber_seq_pullback} the left square can be identified
with~\eqref{first_pullback}.
\end{proof}
Including the maps down to $B\SO$ produces the commutative diagram
\begin{equation}
\label{shearing_diagram}
\gathxy{
	B\tH\ar[rr]^-{(V + V_\xi + \Det(V_\xi), \xi)}_-\simeq\ar[dr]_-{-V} && B\Spin\times
	BG\ar[dl]^-{-V + V_\xi + \Det(V_\xi)}\\
	& B\SO.
}
\end{equation}
Taking Thom spectra of the vertical maps, the shearing map induces a homotopy equivalence
\begin{equation}
\label{just_tH_shear}
	(B\tH)^{-V} \overset\simeq\longrightarrow \MTSpin\wedge (BG)^{V_\xi + \Det(V_\xi) - d'-1}.
\end{equation}
To finish, we subtract $V_\lambda$ from the vertical arrows in~\eqref{shearing_diagram}, then take Thom spectra again.
\end{proof}
\subsection{Case $H = \Spinc$}
\label{a_shear_s}
Let $\tH_n\coloneqq\Spin^c_n\times_{\mu_2}\wG$, and define $\tH$ similarly. The shearing argument is scarcely
different than for \cref{shear_D_thm}, but it will be useful to rephrase $\tH_n$ using the circle group $\T$
instead of $\mu_2$.

The extension of $G$ by $\mu_2$ defines an extension of $G$ by $\T$ by pushing forward along the inclusion
$\mu_2\inj\T$:
\begin{equation}
\gathxy{
	1\ar[r] & \mu_2\ar@{^(->}[d]\ar[r] & \wG\ar[d]\ar[r] & G\ar@{=}[d]\ar[r] & 1\\
	1\ar[r] & \T\ar[r] & \hG\ar[r] & G\ar[r] &1.
}
\end{equation}
In cohomology, this construction is classified by the Bockstein map $H^2(BG;\mu_2)\to H^3(BG;\Z)$. Let
$\hH_n\coloneqq \Spin_n^c\times_\T \hG$ and $\hH\coloneqq \Spin^c\times_\T\hG$. The map $\wG\to\hG$ induces maps
$\vp_n\colon\wH_n\to\hH_n$ and $\vp\colon\wH\to\hH$; $\vp$ is the colimit of the $\vp_n$s.
\begin{lem}
The maps $\vp_n\colon\wH_n\to\hH_n$ are isomorphisms of Lie groups.
\end{lem}
\begin{proof}
Write down the commutative diagram
\begin{equation}
\gathxy{
	1\ar[r] & \mu_2\ar@{=}[d]\ar[r] & \tH_n\ar[d]^\vp\ar[r] & \SO_n\times\T\times G\ar@{=}[d]\ar[r] &1\\
	1\ar[r] & \mu_2\ar[r] & \hH_n\ar[r] & \SO_n\times\T\times G\ar[r] & 1
}
\end{equation}
and apply the five lemma.
\end{proof}
And now we shear. Recall our notation from \cref{crystalline_data}.
\begin{thm}[Shearing, class A]\hfill
\label{shear_A_thm}
\begin{enumerate}
	\item Suppose $V_\xi$ admits a \pinc structure. Then there is an equivalence
	\begin{equation}
		(B\hH)^{d - V_\lambda -V} \overset\simeq\longrightarrow \MTSpinc\wedge (BG)^{d- V_\lambda}.
	\end{equation}
	\item\label{non_pinc_split} If $V_\xi$ does not admit a \pinc structure, there is an equivalence
	\begin{equation}
		(B\hH)^{d - V_\lambda - V} \overset\simeq\longrightarrow \MTSpinc\wedge (BG)^{V_\xi + \Det(V_\xi) -
		V_\lambda -d' - 1 + d}.
	\end{equation}
\end{enumerate}
\end{thm}
Again, we most often use case~\eqref{non_pinc_split} when $\lambda = \xi$, in which case the right-hand side
simplifies to $\MTSpinc\wedge (BG)^{\Det(V_\lambda)-1}$.
\begin{proof}
The proof is barely different than that of \cref{shear_D_thm}; we indicate only the differences. In that theorem,
the engine of the proof when $V_\xi$ was not \pinm was the map~\eqref{pi_p_map} from
$B(\Spin\times_{\mu_2}\wG)\to B\Spin\times BG$. Here, $V_\xi$ is not \pinc, so $V_\xi\oplus\Det(V_\xi)$ is oriented
but not \spinc. We have that if $\beta\colon H^2(B\hH;\mu_2)\to H^3(B\hH;\Z)$ is the Bockstein, $\beta(w_2(V_\xi) +
w_1(V_\xi)^2 + w_2(V)) = 0$, so $V + V_\xi + \Det(V_\xi)$, interpreted as a map $B\hH\to B\SO$, lifts to $B\Spinc$.
Our analogue of~\eqref{pi_p_map} is
\begin{equation}
	(V + V_\xi + \Det(V_\xi), \xi)\colon B\hH \longrightarrow B\Spinc\times BG.
\end{equation}
As in \cref{pi_p_prop}, this is a homotopy equivalence commuting with the maps down to $B\SO$. The proof is almost
the same, though we replace $\Spin$ with $\Spinc$ in~\eqref{first_pullback} and~\eqref{second_pullback}, $\mu_2$
with $\T$ in~\eqref{before_fib}, and $K(\mu_2, 2)$ with $K(\Z, 3)$ in~\eqref{fiber_seq_pullback}.
\end{proof}

\subsection{Putting it together}
	\label{where_we_prove_FCEP}
The hard work of the proof is already done.
\begin{proof}[Proof of \cref{FCEP}]
By \cref{reduction_to_Thom_spectra},
\begin{equation}
	\Ph_0^G(\R^d; f_{1/2})\cong[X, \Sigma^{d+1}I_\Z],
\end{equation}
where $X\coloneqq (B\tH)^{d - V_\lambda - V}$. Then \cref{shear_D_thm} ($H = \Spin$) and
\cref{shear_D_thm} ($H = \Spinc$) split this into $\MTH\wedge (BG)^E$ for some rank-zero virtual vector bundle $E$.
For $f_0$, because $\tH \cong H\times G$, \cref{reduction_to_Thom_spectra} gets us to $\MTH\wedge (BG)^E$ without
having to shear. The only thing left to do is compare these Thom spectra to \cref{internal_1,internal_2}, and sure
enough, they match.
\end{proof}

	\section{Computations in examples: summary of results and some generalities}
	\label{s_generalities}

In the next two sections, we study the fermionic crystalline equivalence principle in many examples where the
symmetry is given by a two- or three-dimensional point group. Here, we summarize the results and some takeaways for
researchers interested in crystalline phases; for more detailed results of computations of groups of phases, see
\cref{reflections_table,,inversions_table,,spin_table,,dihedral_table,,para_n_gonal_table,,TOI_table}.

In \S\ref{interesting_to_study}, we indicate some example phases predicted by our phase homology calculations that
have not been previously studied to our knowledge, and which might have accessible or interesting lattice
realizations. We also summarize which of our calculations correspond to phases already studied in the literature.
In \S\ref{s_sseq_summary}, we briefly review the computational techniques we use to study phase homology groups,
namely the Adams and Atiyah-Hirzebruch spectral sequences. In \S\ref{s_secretly_bosonic}, we use the Adams
filtration to characterize which invertible field theories with $\tH$-structure actually only require weaker
structure, such as an $\SO\times G$-structure; this is believed to model the phenomenon in physics of phases which
appear to be fermionic, but are in fact bosonic phases that are not fermionic in an interesting way. Finally, in
\S\ref{s_util_lem}, we gather some lemmas we use repeatedly in the coming sections. The reader interested in the
computations can read~\S\ref{interesting_to_study} and~\S\ref{s_sseq_summary}, returning to the other sections
later.

\subsection{Some interesting phases to study}
\label{interesting_to_study}
In \S\S\ref{rot_refl}--\ref{3d_pt}, we compute equivariant phase homology groups for many $2$- and $3$-dimensional
point groups. Using \cref{equivariant_ansatz}, these computations yield predictions of groups of invertible
topological phases. This is a lot of data, so we take the opportunity here to highlight which of our predictions would
be interesting to study by other means, e.g.\ by arguing on the lattice.

We first study some cases already present in the literature and find agreement, including
reflections in Altland-Zirnbauer classes D and A (\S\ref{reflection}), inversions in classes D and A
(\S\ref{s_inv}), cyclic groups acting by rotations in classes D and A (\S\ref{s_rotations}), and dihedral groups
acting by rotations and reflections in class D (\S\ref{s_dihedral}). In all cases we consider both spinless and
spin-$1/2$ fermions.

In addition, we study rotations in class A and many three-dimensional point group symmetries in classes D and A:
dihedral groups acting by rotations, pyritohedral symmetry, and chiral and full tetrahedral, octahedral, and
icosahedral symmetries. We consider symmetry types with both spinless and spin-$1/2$ fermions. At the time this
paper first appeared online,
these symmetry types had not been studied in the literature; since then, Zhang-Ning-Qi-Gu~\cite{ZNQG25} studied
invertible phases equivariant for all three-dimensional point groups in classes D and A, with spinless and
spin-$1/2$ fermions. They use a different formalism to calculate these groups of invertible phases, and their
calculations agree with ours.  We indicate some of the predictions of phases, implied by the calculations in this
paper and Zhang-Ning-Qi-Gu~\cite{ZNQG25}, that might be interesting to study further.
\begin{enumerate}
	\item In \S\ref{spinc_dihedral} and \S\ref{s_spinc_mixed_dih}, we compute phase homology groups for the local
	systems of symmetry types corresponding to class A phases in which the dihedral group $D_{2n}$ acts by
	rotations and reflections.
	\begin{enumerate}
		\item In dimension $d = 2$, we predict using \cref{dih_spinc_2mod4,dih_spinc_0mod4} a phase generating a
		$\Z/2n$ for even $n$ with spinless fermions.\footnote{Since this paper first appeared online,
		Herzog-Arbeitman-Bernevig-Song~\cite[\S B.2]{HABS24} also studied $d = 2$ invertible phases with this
		symmetry type, starting from a different ansatz; their results agree with ours.}
		\item In dimension $d = 3$, we would be interested in the predicted $\Z/8\oplus\Z/2$ for $n$ odd, with
		either spin-$1/2$ or spinless fermions (based on \eqref{dih_spinc_odd}, \cref{spinc_mixed_odd_thm}), as
		well as a phase generating a $\Z/4$ for $n$ even with spin-$1/2$ fermions (based on
		\cref{bord_gps_spinc_dih_mixed_2mod4,bord_gps_spinc_dih_mixed_0mod4}).
	\end{enumerate}
	\item We predict using \S\ref{pyrit} a $\Z/2\oplus\Z/2$ of 3d class D phases with a pyritohedral symmetry and
	spinless fermions. In class A, we predict a phase generating a $\Z/4$ subgroup, again with spinless fermions.
	\item In \S\ref{s_chiral_tet}, we calculate equivariant phase homology groups on $\R^3$ for $A_4$ acting by
	tetrahedral symmetry and find that for classes A and D and the spinless and spin-$1/2$ cases, the zeroth phase
	homology groups all vanish. Under our ansatz, this predicts there are no nontrivial fermionic phases
	equivariant for a tetrahedral symmetry in these cases. In the first version of this paper, we asked whether
	this be seen using a lattice argument; the work of
	Zhang-Ning-Qi-Gu~\cite{ZNQG25} answers: yes!
\end{enumerate}
Our computations predict plenty of other phases, but many of them either have Adams filtration zero (see
\S\ref{s_secretly_bosonic}) and therefore are not predicted to be intrinsically fermionic, or have more complicated
symmetry types, such a full octahedral symmetry, that would be harder to study on the lattice. See
Zhang-Ning-Qi-Gu~\cite{ZNQG25} for more on lattice-based approaches to these symmetry groups.

\begin{rem}
In the computations we make in the next several sections, we generally report more bordism groups than we need to
determine the phase homology groups corresponding to groups of invertible phases: to compute the group of
$n$-dimensional invertible field theories with symmetry type $H\to\O$, we need the torsion subgroup of
$\pi_n(\MTH)$ and the free summand in $\pi_{n+1}(\MTH)$. Bordism has other applications in geometry and physics, so
we usually report all bordism groups $\pi_k(\MTH)$ that follow from the calculations that we need for crystalline
phases. When $k\ge n+1$, these provide information about higher-dimensional crystalline phases; for $k <
\dim(\lambda)$, though, it is not clear what a crystalline phase could mean when there are not enough space
dimensions for $G$ to act by $\lambda$, and we do not give a physical meaning to these computations.
See~\cite{GOPWW18} for some discussion when space\emph{time} is $\dim(\lambda)$-dimensional.
\end{rem}
\subsection{Methods of computation}
	\label{s_sseq_summary}
In this section, we summarize the techniques we use to make these computations, and gather a few auxiliary lemmas
we need along the way. Most of our computations can be reframed as computing certain twisted $\ko$- and
$\ku$-homology groups of finite groups in low degrees; the reader interested in learning how to perform such
computations is encouraged to refer to the monographs of Bruner-Greenlees~\cite{BG03, BG10} on connective $\ko$-
and $\ku$-theory, as well as Beaudry-Campbell's article~\cite{BC18} on using the Adams spectral sequence to compute
$\ko$-theory.
\begin{description}
	\item[Computing spin bordism]
		Let $\ko$ denote connective real $K$-theory, which is a generalized homology theory.
		Anderson-Brown-Peterson~\cite{ABP67} show that the Atiyah-Bott-Shapiro map~\cite{ABS} is $7$-connected,
		meaning that for any space or connective spectrum $X$, the induced map $\Omega_k^\Spin(X)\to \ko_k(X)$ is
		an isomorphism for $k \le 7$. We often pass between spin bordism and $\ko$-theory without comment. We
		compute the free and $2$-torsion summands of $\ko_*(X)$ using the Adams spectral sequence; see below. The
		forgetful map $\Omega_*^\Spin(\bl)\to\Omega_*^\SO(\bl)$ induces an equivalence on odd-primary torsion (in
		fact, it lifts to an equivalence $\MTSpin[1/2]\to\MTSO[1/2]$), so to compute odd-primary torsion, we
		typically compute $\Omega_*^\SO(X)$ via the Atiyah-Hirzebruch spectral sequence, which we also discuss
		below.
	\item[Computing spin\textsuperscript{\itshape c} bordism]
		Let $\ku$ denote connective complex $K$-theory. Anderson-Brown-Peterson~\cite{ABP67} also produce a
		$7$-connected map $\MTSpin^c\to\ku\vee\Sigma^4\ku$, which for homology theories gives a natural
		transformation $\Omega_k^\Spinc(\bl)\to\ku_k(\bl)\oplus\ku_{k-4}(\bl)$, which is an isomorphism for $k\le
		7$ on spaces and connective spectra. We will again use the Adams spectral sequence to determine the free
		and $2$-torsion summands of $\ku_*(X)$, as described below. The forgetful map
		$\Omega_*^\Spinc(\bl)\to\Omega_*^\SO(B\T\times\bl)$ induces an equivalence on odd-primary torsion, and in
		fact an equivalence of spectra $\MTSpin^c[1/2]\overset\simeq\to\MTSO[1/2]\wedge (B\T)_+$, so we
		compute $\Omega_*^\SO(X\times B\T)$, typically with the Atiyah-Hirzebruch spectral sequence.
\end{description}
Now we briefly introduce the Adams and Atiyah-Hirzebruch spectral sequences in the ways that we use them.
\subsubsection{The Adams spectral sequence}
The ($2$-primary) Adams spectral sequence~\cite[Theorems 2.1, 2.2]{Ada58} computes the $2$-completion of the
homotopy groups of a pointed space or spectrum $X$. Its $E_2$-page is
\begin{equation}
\label{general_Adams_E2}
	E_2^{s,t} = \Ext_\cA^{s,t}(\wH^*(X; \Z/2), \Z/2) \Longrightarrow \pi_{t-s}(X)_2^\wedge.
\end{equation}
Here $\cA$ is the $2$-primary Steenrod algebra.
\begin{rem}
The usual bigrading convention for Adams spectral sequences places $E_r^{s,t}$ at coordinates $(t-s, s)$. We follow
this convention. The \term{topological degree} of an element at coordinates $(t-s, s)$ in an Adams spectral
sequence refers to $t-s$, and $s$ is called its \term{filtration}.
\end{rem}
There is a general change-of-rings theorem, where if $\mathcal B$ is a graded Hopf algebra, $\mathcal
C\subset\mathcal B$ is a graded Hopf subalgebra, and $M$ and $N$ are graded $\mathcal B$-modules, then there is a
natural isomorphism
\begin{equation}
\label{change_of_rings}
	\Ext_{\mathcal B}^{s,t}(\mathcal B\otimes_{\mathcal C} M, N)\overset\cong\longrightarrow \Ext_{\mathcal
	C}^{s,t}(M, N).
\end{equation}
When $X = \ko\wedge Y$ or $\ku\wedge Y$, this greatly simplifies the $E_2$-page of~\eqref{general_Adams_E2}. Inside
the mod $2$ Steenrod algebra $\cA$, define the subalgebras $\cA(1)\coloneqq\ang{\Sq^1, \Sq^2}$ and $\cE(1)\coloneqq
\ang{Q_0, Q_1}$;\footnote{These generators are given in two different bases of $\cA$; the relations between them
are $Q_0 = \Sq^1$ and $Q_1 = \Sq^1\Sq^2 + \Sq^2\Sq^1$.} then, Stong~\cite{Sto63} showed $\widetilde
H^*(\ko;\Z/2)\cong\cA\otimes_{\cA(1)}\Z/2$ and Adams~\cite{Ada61} showed $\widetilde
H^*(\ku;\Z/2)\cong\cA\otimes_{\cE(1)}\Z/2$. Both $\cA(1)$ and $\cE(1)$ are Hopf subalgebras of $\cA$
so~\eqref{change_of_rings} says we need only consider
\begin{subequations}
\begin{align}
	E_2^{s,t} &= \Ext_{\cA(1)}^{s,t}(\wH^*(X;\Z/2), \Z/2) \Longrightarrow \widetilde\ko_{t-s}(X)_2^\wedge\\
	E_2^{s,t} &= \Ext_{\cE(1)}^{s,t}(\wH^*(X;\Z/2), \Z/2) \Longrightarrow \widetilde\ku_{t-s}(X)_2^\wedge.
	\label{E1_adams_general}
\end{align}
\end{subequations}
This line of reasoning, first used by Davis~\cite{Dav74}, is by now a standard trick in algebraic topology. For
further reading, we recommend the paper of Beaudry-Campbell~\cite{BC18}, who go into detail about how to define and
calculate these Ext groups and work several examples over $\cA(1)$. There are fewer worked examples
of~\eqref{E1_adams_general} in the literature; see Bruner-Greenlees~\cite{BG03}, Nguyen~\cite{Ngu09},
Francis~\cite[\S 5]{Fra11} and Al-Boshmki~\cite{AB16} for closely related calculations.

Our notation is standard in the $\cA(1)$-case, but since examples for $\cE(1)$ are sparser, we record here a few
notational conventions for working with $\cE(1)$-modules and this spectral sequence. When we draw $\cE(1)$-modules,
we will use solid straight lines to denote $Q_0$-actions and dashed curved lines to denote $Q_1$-actions.
Therefore, for example, $\cE(1)$ as a module over itself looks like this.
\begin{equation}
\begin{gathered}
	\begin{tikzpicture}[scale=0.5]
	\Eone{0}{0}{}
	\end{tikzpicture}
\end{gathered}
\end{equation}
For any $\cE(1)$-module $M$, $H^{*,*}(\cE(1))\coloneqq \Ext_{\cE(1)}^{*,*}(\Z/2, \Z/2)$ acts on
$\Ext_{\cE(1)}^{s,t}(M, \Z/2)$, analogously to the case of $\cA(1)$-modules; if $M = \tH^*(X;\Z/2)$, then
just as over $\cA(1)$, tracking this action through the Adams spectral sequence provides information about the
action of $\ku_*$ on $\widetilde\ku_*(X)$. Differentials are equivariant for this action, just like for the Adams
spectral sequence over $\cA(1)$. Since $\cE(1)$ is an exterior algebra, Koszul duality provides an isomorphism of
bigraded algebras
\begin{equation}
\label{E1Z2Ext}
	H^{*,*}(\cE(1))\cong\Z/2[h_0, v_1],
\end{equation}
where $\abs{h_0} = (1, 1)$ and $\abs{v_1} = (1, 3)$~\cite[Example 4.5.6]{BC18}. We will denote an $h_0$-action by a
vertical line, and a $v_1$-action by a lighter diagonal line. Like for $\ko$, $h_0$ lifts to multiplication by $2$;
$v_1$ lifts to the action of the Bott element $\beta\in \ku_2$~\cite[\S 2.1]{BG03}.

We will often write $\Ext_{\cA(1)}(M)$ for $\Ext_{\cA(1)}^{s,t}(M, \Z/2)$, and similarly for $\cE(1)$; when it is
clear which subalgebra we are working over, we will just write $\Ext(M)$.

\subsubsection{The Atiyah-Hirzebruch spectral sequence}
The (homological) Atiyah-Hirzebruch spectral sequence~\cite{AH61} for oriented bordism has signature
\begin{equation}
\label{MSO_AH}
	E^2_{p,q} = \widetilde H_p(X; \Omega^\SO_*) \Longrightarrow \Omega_{p+q}^\SO(X).
\end{equation}
In general, using the Atiyah-Hirzebruch spectral sequence can feel different depending on application-specific
details, so we point the reference-minded reader to García-Etxebarria-Montero~\cite[\S 2.2.2, \S 3]{GEM19} for an
introduction and some examples which may be helpful.

%

\subsection{Adams filtration 0 phases are secretly bosonic}
\label{s_secretly_bosonic}
In \cref{change_symmetry}, we defined a map from phase homology with symmetry type $\SO$ to phase homology with
symmetry types $\Spin$ or $\Spinc$ and interpreted it as regarding bosonic SPT phases as fermionic SPT phases in a
trivial way. Physicists studying fermionic SPT phases are often interested in the cokernel of this map, which is
thought of as the group of intrinsically fermionic SPT phases. Because bosonic crystalline phases are relatively
well-understood, e.g.\ in the work of Hermele, Huang, Song, and their collaborators~\cite{HSHH17, SHFH17, HH18,
SHQFH19, SFQ20, SXH20} and via the bosonic crystalline equivalence principle of Thorngren-Else~\cite{TE18}, we are
most interested in intrinsically fermionic SPT phases.

The structure of the Adams spectral sequence allows us to identify the image of this bosonic-to-fermionic map on
phase homology with little extra work. For more about the Adams spectral sequence, see~\S\ref{s_sseq_summary}; for
now, we need only that phase homology groups, reinterpreted through \cref{FCEP} as groups of invertible field
theories, are computed as homotopy groups of spectra, and that the homotopy groups of any spectrum $M$ come with a
canonical filtration called the \term{(mod $2$) Adams filtration}
\begin{equation}
\label{Adams_filtration}
	\pi_nM = F^0_n\supseteq F^1_n\supseteq F^2_n\supseteq\dotsb
\end{equation}
For more information, see~\cite[\S 4.7]{BC18}. This has two properties which are important for us.
\begin{enumerate}
	\item\label{Adams_knows_Adams} The Adams spectral sequence computes the Adams filtration: after $2$-completing,
	the associated graded of~\eqref{Adams_filtration} is the $E_\infty$-page of the Adams spectral sequence, in
	that $E_\infty^{s,t} = \mathrm{gr}_s\pi_{t-s}M$.
	\item\label{Adams_knows_whats_boring} If $M = \MTH$ is a Thom spectrum whose homotopy groups compute bordism
	groups, then in some cases of interest, notably the ones studied in this paper, elements of the associated
	graded in degree $0$ correspond to the $2$-primary part of the group of deformation classes of invertible TFTs
	which depend on something weaker than an $H$-structure, such as a spin IFT which is defined by evaluating an
	oriented IFT on spin manifolds.
\end{enumerate}
This means we can identify which invertible TFTs really use the $H$-structure, and which do not.

Now a little more detail. We do not need to say much more about~\eqref{Adams_knows_Adams}: we depict Adams spectral
sequences on a grid with coordinates $(t-s, s)$, such as in \cref{a1moddih2mod4}, right, so $F^0_n/F^1_n$ is found
in the $E_\infty$-page at coordinate $(n, 0)$.

For~\eqref{Adams_knows_whats_boring}, we make a simplifying assumption: that for the specific degree $n$ we are
investigating, $\pi_n\MTH$ is $2$-torsion. This assumption holds in all cases where we want to study the Adams
filtration in this article, but if you want to relax it, see \cref{what_if_not_2torsion}. The assumption implies
that up to extension questions on the $E_\infty$-page, the mod $2$ Adams spectral sequence fully determines
$\pi_n\MTH$,\footnote{Some extension questions can be addressed using the $H^{*,*}(\cA(1))$-action on the
$E_\infty$-page, but there are also \term{hidden extensions} which are harder to address. None of the calculations
we make in this article manifest hidden non-split extensions; one example where they do occur is $H =
\Spin\times_{\Z/2}\Z/8$~\cite{DDHM}.} and that the natural map
\begin{equation}
	(\pi_n(\MTH))^\vee\coloneqq \Hom(\pi_n(\MTH), \C^\times)\longrightarrow [\MTH, \Sigma^{n+1}I_\Z]
\end{equation}
is an isomorphism.

To pass from bordism groups to isomorphism class of invertible field theories, we must take character duals
$A\mapsto A^\vee\coloneqq \Hom(A, \C^\times)$. This is a good thing, actually: a degree-$0$ element of
$\mathrm{gr}_\bullet\pi_n(\MTH)$ does not usually uniquely lift to an element of $\pi_n\MTH$: the ambiguity is
$F^1_n$. But in $(\pi_n(\MTH))^\vee$, we get a subgroup: the surjection
\begin{subequations}
\begin{equation}
	\pi_n(\MTH)\longtwoheadrightarrow \pi_n(\MTH)/F^1_n \cong \mathrm{gr}_0\pi_n(\MTH)
\end{equation}
passes under character duality to an inclusion
\begin{equation}
	(\mathrm{gr}_0\pi_n(\MTH))^\vee \hooklongrightarrow (\pi_n(\MTH))^\vee.
\end{equation}
\end{subequations}
Therefore, in a mild abuse of notation, we refer to this subgroup of $(\pi_n(\MTH))^\vee$, identified with a
subgroup of the group isomorphism classes of invertible TFTs with $H$-structure, as the group of \term{Adams
filtration $0$ invertible TFTs with $H$-structure}.

It is a theorem~\cite[\S 8.4]{FH19b} that this subgroup consists of theories closely related to classical
Dijkgraaf-Witten theories~\cite[\S 1]{FQ93}.\footnote{These theories are not quite the same thing as classical
Dijkgraaf-Witten theories, which are TFTs of oriented manifolds with a principal $G$-bundle, and which use
$\R/\Z$-valued cohomology, rather than $\Z/2$-valued cohomology. Unoriented generalizations of classical
Dijkgraaf-Witten theory are studied in more detail in work the author~\cite[\S 3.1]{Deb20}, You~\cite{You20},
Kim~\cite[\S 6]{Kim18}, and Gagnon-Ririe and Young~\cite{GRY24}.} Isomorphism classes of these invertible TFTs are
determined by their partition functions~\cite[\S 5.3]{FH16}, so we specify these theories by their partition
functions, which are bordism invariants $\Omega_n^H\to\C^\times$.

For the Adams spectral sequence, $E_2^{0,n} = \Ext^{0,n}_{\cA}(\wH^*(\MTH;\Z/2); \Z/2)$ is canonically identified
with
\begin{equation}
	\Hom_{\cA}(\wH^*(\MTH;\Z/2), \Sigma^n \Z/2),
\end{equation}
which is a subspace of
\begin{equation}
	\Hom_{\Ab}(\wH^n(\MTH;\Z/2), \Z/2)\cong (\wH^n(\MTH;\Z/2))^\vee.
\end{equation}
The fourth quadrant of the Adams spectral sequence is empty, so $E_\infty^{0,n}$ is a subspace of $E_2^{0,n}$. Take
the sequence of maps
\begin{subequations}
\begin{equation}
	\mathrm{gr}_0\pi_n(\MTH) = E_\infty^{0,n} \hooklongrightarrow E_2^{0,n} \hooklongrightarrow (\wH^n(\MTH;
	\Z/2))^\vee
\end{equation}
and apply character duality:
\begin{equation}
	(\mathrm{gr}_0\pi_n(\MTH))^\vee \longtwoheadleftarrow (E_2^{0,n})^\vee \longtwoheadleftarrow \wH^n(\MTH;\Z/2).
\end{equation}
Now compose with the Thom isomorphism to obtain
\begin{equation}
	\zeta\colon H^n(BH;\Z/2) \longtwoheadrightarrow (\mathrm{gr}_0\pi_n(\MTH))^\vee.
\end{equation}
\end{subequations}
That is, a degree-$n$ mod $2$ cohomology class of $BH$ determines an isomorphism class of Adams filtration $0$
invertible TFTs, and all Adams filtration $0$ invertible TFTs arise in this way. The map need not be injective,
e.g.\ by the Wu formula when $H = \O$.\footnote{See~\cite{BP64, BP65, Wil73, Pap78} for some discussion of the
kernel for various $H$.}

Tracing this through Thom's collapse map tells us that given a cohomology class $\theta\in H^n(BH;\Z/2)$, the
partition function $\zeta(\theta)$ is the bordism invariant which takes a closed $n$-manifold with $H$-structure
$(M, f\colon M\to BH)$ and returns
\begin{equation}
\label{zeta_part_fn}
	\zeta(\theta)(M, f) = (-1)^{\ang{f^*\theta, [M]}}.
\end{equation}
That is, use the $H$-structure to pull $\theta$ back to $M$, then evaluate it on the mod $2$ fundamental class.
This construction uses some aspects of the $H$-structure on $M$, but in the cases relevant to this paper, it is
insensitive to the difference between $\Spin$ and $\O$, which is believed to pass to the physicists' distinction
between fermionic and bosonic phases.
\begin{lem}
If $H = \Spin\times_{\mu_2}\wG$ or $H = \Spin^c\times_{\mu_2} \wG$, where $\wG$ is in \cref{crystalline_data}, and
$H' \coloneqq \O\times G$, then the map $H\to H'$ of tangential structures induces a surjective map
$H^*(BH';\Z/2)\to H^*(BH;\Z/2)$, and therefore the partition functions~\eqref{zeta_part_fn} of the Adams filtration
$0$ theories only depend on the underlying $H'$-structure of an $H$-manifold.
\end{lem}
\begin{proof}
First, the $\Spin$ case. We established a shearing equivalence $\MTH\cong \MTSpin\wedge X$, where $X$ is a Thom
spectrum of a rank-zero virtual vector bundle over $BG$, and this equivalence fits into a homotopy commutative
diagram
\begin{subequations}
\begin{equation}
\begin{gathered}
\begin{tikzcd}
	\MTH & {\MTSpin\wedge X} \\
	{\MTO\wedge (BG)_+} & {\MTO\wedge X.}
	\arrow["\simeq", from=1-1, to=1-2]
	\arrow[from=1-1, to=2-1]
	\arrow[from=1-2, to=2-2]
	\arrow[from=2-1, to=2-2]
\end{tikzcd}
\end{gathered}
\end{equation}
Apply mod $2$ cohomology and invoke the Thom isomorphism to obtain a commutative diagram
\begin{equation}
\begin{gathered}
\begin{tikzcd}
	{H^*(BH;\Z/2)} & {H^*(B\Spin\times BG;\Z/2)} \\
	{H^*(B\O\times BG;\Z/2)} & {H^*(B\O\times         BG;\Z/2)}
	\arrow["\cong"', from=1-2, to=1-1]
	\arrow[from=2-1, to=1-1]
	\arrow["\xi"', from=2-2, to=1-2]
	\arrow["\id"', from=2-2, to=2-1]
\end{tikzcd}
\end{gathered}
\end{equation}
The map $H^*(B\O;\Z/2)\to H^*(B\Spin;\Z/2)$ is surjective, so the Künneth formula implies $\xi$ is too, so the
left-hand arrow $H^*(BH';\Z/2)\to H^*(BH;\Z/2)$ is as well.

For $H = \Spinc$, the proof is the same -- $B\Spinc$ has an additional characteristic class $c_1\in
H^2(B\Spinc;\Z)$, but its mod $2$ reduction is $w_2$, so $\xi$ is still surjective. See Harada-Kono~\cite{HK86} for
a complete account of $H^*(B\Spin_n^c;\Z/2)$.
\end{subequations}
\end{proof}





\begin{rem}
\label{what_if_not_2torsion}
In all cases that one might reasonably encounter, the bordism group $\pi_n(\mathit{MTH})$ is finitely generated, so
we can ask what happens if it contains $p$-torsion for an odd prime $p$ or free summands. For a $p$-torsion
summand, the story is very similar: one instead uses the mod $p$ Adams filtration on $\pi_nM_p^\wedge$, which is
detected by the $\Z/p$-Adams spectral sequence. This has almost the same signature as the $\Z/2$-Adams spectral
sequence we use in this paper, except that $\Z/2$ is replaced with $\Z/p$ and the Steenrod algebra is over $\Z/p$
instead of $\Z/2$.  Because the mod $p$ Thom isomorphism requires an orientation, the story is a little more
nuanced for tangential structures which do not induce an orientation.

For free summands in $\pi_nM$, there is no analogous story. The invertible field theories in question are not
topological; see Freed~\cite[Lecture 9]{Fre19}, Grady-Pavlov~\cite{GP21}, and Grady~\cite{Gra23}. 
The Adams filtration does not tell the whole story. For example, consider 3d invertible spin
field theories, classified by
\begin{equation}
	[\MTSpin, \Sigma^4I_\Z]\overset{\cong}{\longrightarrow} \Hom(\Omega_4^\Spin, \Z)\cong\Z,
\end{equation}
generated by the map $\vp$ sending a spin $4$-manifold to its signature
divided by $16$~\cite{Roh52}. As the signature does not depend on the spin structure, $16\vp$ generates
$\Hom(\Omega_4^\SO, \Z)$,\footnote{This follows from the fact that the signature defines an isomorphism
$\sigma\colon \Omega_4^\SO\to\Z$, which follows from the fact that $\CP^2$, with signature $1$, generates
$\Omega_4^\SO$~\cite[Remarque following Corollaire IV.18]{ThomThesis}.} and therefore the image of the forgetful
map $[\MTSO, \Sigma^4I_\Z]\to[\MTSpin, \Sigma^4I_\Z]$ is identified with the subgroup $16\Z$. That is,
a 3d spin invertible
field theory only depends on the underlying orientation iff it is $q$ times a generator, where $16\mid q$. So for
free summands in the abelian group of isomorphism classes of invertible field theories, the Adams filtration
approach does not work, and one must use other methods.
\end{rem}

\subsection{A few utility lemmas}
	\label{s_util_lem}
\begin{defn}
\label{locsysnot}
Let $A$ be an abelian group, $X$ be a connected space, and $\alpha\in H^1(X;\Z/2)$. Then $A_\alpha$ denotes the
local system on $X$ given by the $\Z[\pi_1(X)]$-module with underlying abelian group $A$ and in which
$g\in\pi_1(X)$ acts on $A$ by $(-1)^{\alpha(g)}$, where we interpret $\alpha$ as a map $\pi_1(X)\to\Z/2$ under the
identification $H^1(X;\Z/2)\cong\Hom(\pi_1(X), \Z/2)$.
\end{defn}
Usually $\alpha$ will be the first Stiefel-Whitney class of a vector bundle, as in the following lemma.
\begin{prop}[{Čadek~\cite[Theorem 1]{Cad99}}]
\label{twistedz2}
Let $\sigma\to B\Z/2$ denote the tautological line bundle.
\begin{enumerate}
	\item $H^k(B\Z/2;\Z_{w_1(\sigma)})$ is isomorphic to $\Z/2$ in odd degrees and $0$ in even degrees.
	\item\label{Zn_odd} If $n$ is odd, $H^k(B\Z/2; (\Z/n)_{w_1(\sigma)}) \cong 0$ for all $k$.
	\item\label{Zn_even} If $n$ is even, $H^k(B\Z/2; (\Z/n)_{w_1(\sigma)}) \cong \Z/2$ for all $k$.
\end{enumerate}
\end{prop}
Čadek discusses the case of $\Z$ coefficients; for parts~\eqref{Zn_odd} and~\eqref{Zn_even}, use the universal
coefficient theorem.
\begin{thm}[Thom isomorphism]
\label{thom_iso}
Let $A$ be an abelian group and $V\to X$ be a rank-zero virtual vector bundle. Then there is a natural isomorphism
\begin{equation}
        U\colon H^k(X; A_{w_1(V)})\overset\cong\longrightarrow \tH^*(X^V; A).
\end{equation}
\end{thm}
So when $V$ is oriented or $A = \Z/2$, we can use untwisted cohomology on the left. In this case, we will denote
elements of $\tH^*(X^V;A)$ by $Ux$, where $x\in H^k(X;A)$. When $A = \Z/2$, one can ask whether the Thom
isomorphism intertwines the action of the Steenrod algebra. It does not, but the difference is benign and easy
to calculate.
\begin{prop}[{Thom~\cite[Théorème II.2]{Tho52}}]
\label{wu_f}
Let $V\to X$ be a rank-zero virtual vector bundle and $x\in H^*(X;\Z/2)$. In $\tH^*(X^V;\Z/2)$, $\Sq(Ux) =
Uw(V)\Sq(x)$.
\end{prop}
Here $\Sq = (1 + \Sq^1 + \Sq^2 + \dots)$ is the total Steenrod square and $w(V) = 1 + w_1(V) + w_2(V) + \dots$ is
the total Stiefel-Whitney class.

There are also Thom
isomorphisms for certain generalized homology theories; we will make use of the following three.
\begin{prop}
\label{tautological_Thom}
Let $V\to X$ be a rank-zero virtual vector bundle and $H\in\set{\SO, \Spin, \Spinc}$. If $V$ has an $H$-structure,
there is a natural isomorphism $\Omega_*^H(X)\overset\cong\to\tOmega_*^H(X^V)$.
\end{prop}

We will repeatedly use the following theorem to show some differentials and extensions are trivial in the Adams
spectral sequence.
\begin{thm}[Margolis~\cite{Mar74}]
\label{margolis}
Let $\cB$ be a sub-Hopf algebra of the Steenrod algebra and $Y$ be a spectrum with $\widetilde
H^*(Y;\Z/2)\cong\cA\otimes_\cB\Z/2$ (so that the change-of-rings trick works for computing $2$-completed
$Y$-homology). For any spectrum $X$, there is a splitting
\begin{equation}
	Y\wedge X\simeq F\vee \overline X,
\end{equation}
where $F$ is an Eilenberg-Mac Lane spectrum for a graded $\Z/2$-vector space and $\tH^*(\overline X;\Z/2)$
has no free summands as an $\cA$-module.
\end{thm}
The upshot is that in the Adams spectral sequence for computing $\pi_*(Y\wedge X)_2^\wedge$, the piece of the
$E_2$-page coming from free summands of $\tH^*(X;\Z/2)$ as a $\cB$-module do not emit or receive
nontrivial differentials, and do not participate in nontrivial extensions.
\begin{lem}
\label{torsion_k_theory}
Let $G$ be a finite group and $E\to BG$ be a rank-zero virtual vector bundle.
\begin{enumerate}
	\item If $4\mid n$, $\widetilde\ko_n(BG^E)\otimes\Q \cong H_0(BG;\Q_{w_1(E)})$; if $4\nmid n$,
	$\widetilde\ko_n(BG^E)$ is torsion.
	\item The same is true for $\widetilde\ku_n(BG^E)$, except divisibility by $4$ is replaced by divisibility by
	$2$.
\end{enumerate}
\end{lem}
\begin{proof}
Atiyah-Hirzebruch~\cite{AH61} proved that the Chern character defines an equivalence
\begin{equation}
	\mathit{ch}\colon \ku\wedge H\Q\overset\simeq\longrightarrow \bigvee_{k\ge 0} \Sigma^{2k}H\Q.
\end{equation}
The Thom isomorphism theorem establishes that $\tH_*(BG^E;\Q)\cong H_*(BG;\Q_{w_1(E)})$, and since $G$ is
finite, this vanishes above degree zero by Maschke's theorem.

The proof for $\ko$-theory is the same, except first using the complexification map $c\colon \ko\to\ku$:
\begin{equation}
	\mathit{ch}\circ c\colon \ko\wedge H\Q\overset\simeq\longrightarrow \bigvee_{k\ge 0} \Sigma^{4k}H\Q.
	\qedhere
\end{equation}
\end{proof}
Choosing $E$ to be the trivial bundle shows the conclusions also hold for the torsion in $\widetilde\ko_*(BG)$ and
$\widetilde\ku_*(BG)$.

\begin{lem}[Adem-Milgram]
\label{ademmilgram}
Fix a prime $p$, and let $H$ be a subgroup of a finite group $G$ with $[G:H]$ coprime to $p$ and $P$ be a Sylow
$p$-subgroup of $H$. Assume $P$ is abelian and that $N_H(P)/P = N_G(P)/P$; then the restriction map
$\rho_{H,G}\colon H^*(BG;\Z/p)\to H^*(BH;\Z/p)$ is an isomorphism.
\end{lem}
\begin{proof}
This is a slight strengthening of theorems of Swan~\cite{Swa60} and Adem-Milgram~\cite[Theorems II.6.6 and
II.6.8]{AM04}, who prove that if $K$ is a finite group with abelian $p$-Sylow subgroup $P$, then the restriction
map $H^*(BK;\Z/p)\to H^*(BP;\Z/p)^{N_K(P)}$ is an isomorphism. In our setting, the data of $P$ and $N(P)/P$ are
identical for $G$ and $H$, so both restriction maps $r_{P,G}\colon H^*(BG;\Z/p)\to H^*(BP; \Z/p)^N$ and
$r_{P,H}\colon H^*(BH;\Z/p)\to H^*(BP; \Z/p)^N$ are isomorphisms. Since $r_{P,G} = r_{P,H}\circ\rho_{G,H}$, we are
done.
\end{proof}

\begin{lem}[{Bock-to-$\Sq^1$ lemma~\cite[\S 30.4]{BH59}}]
\label{bock_to_sq1}
Let $\beta\colon H^k(\bl;\Z/2)\to H^{k+1}(\bl;\Z)$ denote the integral Bockstein. Then $\beta(x)\bmod 2 =
\Sq^1(x)$.
\end{lem}
In the mixed unoriented case, \cref{shear_D_thm,shear_A_thm} ask us to study Thom spectra for determinants of
representations. We use the following lemma to simplify them.
\begin{lem}
\label{det_splitting}
Let $\lambda\colon G\to\O_d$ be a faithful representation whose image contains a reflection and $V_\lambda\to BG$
be the associated vector bundle. Then the splitting of the surjection
\begin{equation}
	\xymatrix@1{G\ar[r]^\lambda & \O_d\ar[r]^-{\pi_0} & \Z/2}
\end{equation}
lifts to a splitting of the Thom spectrum $(BG)^{\Det(V_\lambda)-1}$ as
\begin{equation}
	(BG)^{\Det(V_\lambda)-1} \overset\simeq\longrightarrow (B\Z/2)^{\sigma-1} \vee M,
\end{equation}
and the inclusion $\tH^*(M;\Z/2)\inj \tH^*((BG)^{\Det(V_\lambda)-1};\Z/2)$ is injective with image a complementary
vector space to the subspace spanned by $\set{\oU w_1(V_\lambda)^k\mid k\ge 0}$.
\end{lem}
\begin{proof}
Let $g\in G$ be an element sent to $\lambda$ by a reflection. Then $g^2 = 1$, so the maps $\ang g\inj G\surj \Z/2$
compose to an isomorphism. Upon taking Thom spectra, these can be identified with maps $(B\Z/2)^{\sigma-1}\to
(BG)^{\Det(V_\lambda)-1}\to (B\Z/2)^{\sigma-1}$ composing to (a map homotopy equivalent to) the identity, which
splits off $(B\Z/2)^{\sigma-1}$. The image of the map
$\tH^*((B\Z/2)^{\sigma-1};\Z/2)\to\tH^*((BG)^{\Det(V_\lambda)-1};\Z/2)$ is spanned by $\set{\oU
w_1(V_\lambda)^k\mid k\ge 0}$, and the image of $\tH^*(M;\Z/2)$ is a complementary subspace.
\end{proof}
\begin{lem}
\label{spinc_odd_primes}
Let $X$ be a space or connective spectrum. Then there is an isomorphism
\begin{equation}
	\Omega_n^{\Spin^c}(X)\otimes \Z[1/2] \overset\cong\longrightarrow \bigoplus_{k=0}^{\lfloor n/2 \rfloor}
	\Omega_{n-2k}^\SO(X)\otimes\Z[1/2].
\end{equation}
\end{lem}
\begin{proof}
It suffices to establish an equivalence of spectra
\begin{equation}
\label{spinc_spectrum_odd_primes}
	\MTSpin^c[1/2] \overset\simeq\longrightarrow \bigvee_{k\ge 0} \Sigma^{2k}\MTSO[1/2],
\end{equation}
as the lemma then follows by taking homology of $X$ with respect to both sides. As discussed in
\S\ref{s_sseq_summary}, there is an equivalence $\MTSpin^c[1/2]\simeq \MTSO[1/2]\wedge (B\T)_+$; then~\cite[Lemma
7.24]{MathSmith} shows
\begin{equation}
	\MTSO\wedge (B\T)_+\simeq \bigvee_{k\ge 0}\Sigma^{2k}\MTSO,
\end{equation}
proving~\eqref{spinc_spectrum_odd_primes}.
\end{proof}

\section{Examples: rotations and reflections}
\label{rot_refl}
	\subsection{Warmup: reflections}
		\label{reflection}
The simplest example of the fermionic crystalline equivalence principle occurs when the spatial symmetry is $\Z/2$
acting by a reflection. This symmetry can mix with $\mu_2\subset\Spin_d$, and there are two cases. The
following principle is well-established in physics literature; see Shiozaki-Shapourian-Ryu~\cite{SSR17b} and
Song-Huang-Fu-Hermele~\cite[\S VII]{SHFH17}.
\begin{itemize}
	\item If $\Z/2$ and $\mu_2$ do not mix (often written that the reflection squares to $1$), then the
	classification matches the classification of \pinp invertible field theories.
	\item Conversely, if $\Z/2$ and $\mu_2$ do mix (often written that the reflection squares to $(-1)^F$),
	the classification matches that of \pinm invertible field theories.
\end{itemize}
Condensed-matter theorists also study theories with time-reversal symmetry. Though this is also an antiunitary
symmetry that can mix with $\mu_2$, the classification in terms of pin structures is opposite that of
reflections: when time-reversal symmetry does not mix with fermion parity, we get \pinm, and when it does mix, we
get \pinp. This is also well-established in physics, and is discussed by
Kapustin-Thorngren-Turzillo-Wang~\cite{KTTW15}, Freed-Hopkins~\cite{FH16}, and others.

The difference between these two correspondences is a first hint that the fermionic crystalline equivalence
principle must be more complicated than the bosonic version; this point is raised by Thorngren-Else~\cite[\S
V.A]{TE18} and Cheng-Wang~\cite[\S II.C]{CW18}.
\begin{table}[h!]
\begin{tabular}{c c c c}
\toprule
$d$ & Class D, spinless & Class D, spin-$1/2$ & Class A\\
& \S\ref{unmixed_reflection} & \S\ref{mixed_reflection} & \S\ref{spinc_refl}\\
\midrule
$1$ & $\Z/2$ & $\Z/8$ & $\Z/4$\\
$2$ & $\Z/2$ & $0$ & $0$\\
$3$ & $\Z/16$ & $0$ & $\Z/8\oplus\Z/2$\\
$4$ & $0$ & $0$ & $0$\\
\bottomrule
\end{tabular}
\caption{$\Z/2$-equivariant phase homology groups for the cases in which $\Z/2$ acts by a reflection. As discussed
in \S\ref{reflection}, these arise as the homotopy groups of the Anderson duals of $\MTPin^+$, $\MTPin^-$, and
$\MTPin^c$. For this group action, the spinless and spin-$1/2$ classifications in class A coincide.}
\label{reflections_table}
\end{table}

\subsubsection{Class D, spinless}
\label{unmixed_reflection}
When the reflection does not mix with the internal symmetry group, our ansatz is exactly that of
Freed-Hopkins. In this setting, $\Z/2$ acts on $\R^d$ as $(d-1)+\sigma$, where $k$ denotes the rank-$k$ trivial
representation and $\sigma$ denotes the sign representation. Let $f_0^D$ denote the equivariant local system of
symmetry types for the class D spinless case.
Arguing as in~\cite[(3.6)]{FH19}, in space dimension $d$ we see that
\begin{equation}
\label{spinreflnomix}
	\Ph_0^{\Z/2}(\R^d; f_0^D)\cong [\MTSpin\wedge (B\Z/2)^{1-\sigma}, \Sigma^{d+2}I_\Z].
\end{equation}
Using~\eqref{pinpsplitting}, $\MTSpin\wedge (B\Z/2)^{1-\sigma}\simeq\MTPin^+$, identifying these phase homology
groups as homotopy groups of the Anderson dual of $\MTPin^+$, as expected. Finally, to obtain the specific
groups in \cref{reflections_table}, we use the preexisting calculations of \pinp bordism from~\cite{Gia73a,
KT90Pinp, KT90}.
\subsubsection{Class D, spin-$1/2$}
\label{mixed_reflection}
Again $\Z/2$ acts by $d-1+\sigma$, and this time, reflection mixes with fermion parity. Let $f_{1/2}^D$ denote the
equivariant local system of symmetry types for this case. The associated bundle to the $\Z/2$-representation given
by reflection is not \pinm, so by \cref{shear_D_thm},
\begin{equation}
	\Ph_0^{\Z/2}(\R^d; f_{1/2}^D)\cong [\MTSpin\wedge (B\Z/2)^{\Det(\sigma)-1}, \Sigma^{d+2}I_\Z].
\end{equation}
Because $\sigma$ is a line bundle, $\Det(\sigma) = \sigma$. Using~\eqref{pinmsplitting}, $\MTSpin\wedge
(B\Z/2)^{\sigma-1}\simeq\MTPin^-$, so these phase homology groups are identified with homotopy groups of the Anderson
dual of $\MTPin^-$ as predicted. These bordism groups are calculated in~\cite{ABP69, KT90}.
\subsubsection{Class A}
\label{spinc_refl}
For \spinc phases (those of Altland-Zirnbauer class A), the spinless and spin-$1/2$ classifications coincide:
$V_\lambda$ is \pinc, so \cref{shear_A_thm} tells us to consider $\MTSpinc\wedge (B\Z/2)^{1-\sigma}$ in both cases,
and by~\eqref{pincsplitting}, this spectrum is equivalent to $\MTPin^c$. 

Bahri-Gilkey~\cite{BG87a, BG87b} compute \pinc bordism groups,\footnote{In low degrees, Beaudry-Campbell~\cite[\S
5.6]{BC18} compute low-degree \pinc bordism groups using the Adams spectral sequence over $\cA(1)$, using that
$\MTPinc\simeq\MTSpin\wedge \Sigma^{-2}\MU_1\wedge \Sigma^{-1}\MO_1$. One can also compute using the Adams spectral
sequence over $\cE(1)$, as in \S\ref{spinc_dihedral}; we found this to be a fun and useful exercise for getting
comfortable with this variation of the Adams spectral sequence.} giving us the phase homology groups in
\cref{reflections_table}.
\subsubsection{Comparison with prior work}
\label{refl_compare}
Reflection-equivariant fermionic phases have been studied by many teams of researchers with many methods. Their
results agree with each other, and with us.
\begin{description}
	\item[Class D, spinless] These phases, especially the $\Z/16$ in $d = 3$, are studied by
	Song-Huang-Fu-Hermele~\cite[\S V.A]{SHFH17}, Hsieh-Cho-Ryu~\cite[\S IV]{HCR16},
	Shiozaki-Shapourian-Ryu~\cite[\S II.B, \S II.D]{SSR17b}, Guo-Ohmori-Putrov-Wan-Wang~\cite[\S 10.7]{GOPWW18},
	Mao-Wang~\cite{MW20}, Ning-Mao-Li-Wang~\cite{NMLW21},
	Manjunath-Calvera-Barkeshli~\cite[Table III]{MCB23},
	and Zhang-Ning-Qi-Gu~\cite[Table I, \S S-3.2]{ZNQG25}.
	\item[Class D, spin-1/2] Song-Huang-Fu-Hermele~\cite[\S V.B]{SHFH17}, Shapourian-Shiozaki-Ryu~\cite{SSR17a,
	SSR17b}, Guo-Ohmori-Putrov-Wan-Wang~\cite[\S 10.7]{GOPWW18},
	Bultinck-Williamson-Haegeman-Verstraete~\cite[\S IX]{BWHV17},
	Manjunath-Calvera-Barkeshli~\cite[Table III]{MCB23},
	and Zhang-Ning-Qi-Gu~\cite[Table I, \S S-3.2]{ZNQG25}.
	\item[Class A] These phases have been studied by Isobe-Fu~\cite{IF15}, Hong-Fu~\cite{HF17},
	Shapourian-Shiozaki-Ryu~\cite{SSR17a, SSR17b}, Song-Huang-Fu-Hermele~\cite[\S 4]{SHFH17},
	Shiozaki-Shapourian-Gomi-Ryu~\cite[\S V]{SSGR18}, Ning-Mao-Li-Wang~\cite{NMLW21},
	Lee-Shiozaki-Hsieh~\cite{LSH24, LSH24a}, and Zhang-Ning-Qi-Gu~\cite[Table II, \S S-3.2]{ZNQG25}.
\end{description}
In addition, Zhang-Ning~\cite{ZN21} study a $d = 2$ class D spinless reflection-symmetric SPT in the context of their
crystalline-equivalent bulk-boundary correspondence.
%
%

	\subsection{Inversions}
		\label{s_inv}
\term{Inversion symmetry} is the $\Z/2$-symmetry on $\R^d$ acting by $(x_1,\dotsc,x_d)\mapsto (-x_1,
\dotsc,-x_d)$. This offers another relatively simple example of the FCEP, but with a new feature in the spin-$1/2$
case: the classes in $H^2(B\Z/2;\Z/2)$ specified by the extension $1\to \Z/2\to \wG\to \Z/2\to 1$ and by $w_2(\lambda)
+ w_1(\lambda)^2$ are not always equal. This does not change very much, as we explain in \S\ref{spinful_inversion}
below.

\begin{table}[h!]
\begin{tabular}{c c c c}
\toprule
$d$ & Class D, spinless & Class D, spin-$1/2$ & Class A\\
& \S\ref{spinless_inversion} & \S\ref{spinful_inversion} & \S\ref{spinc_inversion}\\
\midrule
$1$ & $\Z/2$ & $\Z/8$ & $\Z/4$\\
$2$ & $\Z$ & $\Z\oplus\Z/8$ & $\Z^2\oplus \Z/4$\\
$3$ & $0$ & $\Z/16$ & $\Z/8\oplus\Z/2$\\
$4$ & $0$ & $\Z\oplus\Z/16$ & $\Z^2\oplus \Z/8\oplus\Z/2$\\
\bottomrule
\end{tabular}
\caption{$\Z/2$-equivariant phase homology groups for the cases where $\Z/2$ acts as inversion. The symmetry type
whose Thom spectrum determines these groups depends on $d$; see the referenced sections for which symmetry types
appear.}
\label{inversions_table}
\end{table}

\subsubsection{Class D, spinless case}
\label{spinless_inversion}
First, the case for which inversion symmetry and fermion parity do not mix. The $\Z/2$-action on $\R^d$ is a direct
sum of $d$ copies of the sign representation $\sigma$, so as a $\Z/2$-space, $\R^d$ is denoted $d\sigma$. This case
is covered by Freed-Hopkins~\cite[Example 3.5]{FH19}, and the phase homology groups are
\begin{equation}
	[\MTSpin\wedge (B\Z/2)^{d-d\sigma}, \Sigma^{d+2}I_\Z].
\end{equation}
The spectra $\MTSpin\wedge (B\Z/2)^{d-d\sigma}$ are periodic in $d$.
\begin{lem}
\label{4foldinv}
If $d' - d$ is divisible by $4$, $\MTSpin\wedge (B\Z/2)^{d(1-\sigma)}\simeq\MTSpin\wedge (B\Z/2)^{d'(1-\sigma)}$.
\end{lem}
\begin{proof}
This is an instance of \cref{relative_Thom}, using that spin structures satisfy the 2-out-of-3 property and that,
since $4\sigma$ is spin, so is $(d' - d)(1-\sigma)$.
\end{proof}
Thus we have only to determine $\MTSpin\wedge (B\Z/2)^{d(1-\sigma)}$ for small $d$.
\begin{itemize}
	\item When $d = 0$, we get $\MTSpin\wedge (B\Z/2)_+$.
	\item When $d = 1$,~\eqref{pinpsplitting} tells us $\MTSpin\wedge (B\Z/2)^{1-\sigma}\simeq\MTPin^+$.
	\item For $d = 2$, we have $\MTSpin\wedge (B\Z/2)^{2-2\sigma}$.\footnote{Campbell~\cite[\S 7.8]{Cam17} shows
	this spectrum is equivalent to $\mathit{MT}(\Spin\times_{\Z/2}\Z/4)$. Bordism for this symmetry type, called
	spin-$\Z/4$ bordism or spin\textsuperscript{$c/2$} bordism, is used in several places in recent mathematical
	physics literature, including~\cite{Cam17, Hsi18, FH19, GEM19, TY19, DL20b, GOPWW18, HKT19, WW19a, Wan20,
	MV21}.}
	\item When $d = -1$,~\eqref{pinmsplitting} gives $\MTSpin\wedge (B\Z/2)^{\sigma-1}\simeq\MTPin^-$.
\end{itemize}
The low-degree homotopy groups of these spectra that we need are computed by Giambalvo~\cite{Gia73a} and
Kirby-Taylor~\cite{KT90Pinp, KT90} (the \pinp case); Anderson-Brown-Peterson~\cite{ABP69} and
Kirby-Taylor~\cite{KT90} (the \pinm case); Giambalvo~\cite{Gia73} (the case $d = 2$); and
Mahowald-Milgram~\cite{MM76} (the $\mathrm{spin}\times\Z/2$ case). Thus we obtain the phase homology groups for the
spinless class D case in \cref{inversions_table}.


\subsubsection{Class D, spin-$1/2$ case}
\label{spinful_inversion}
Now we consider the case where the inversion symmetry and $\mu_2\subset\Pin_d^-$ mix as specified by the nontrivial
extension $1\to\mu_2\to\Z/4\to\Z/2\to 1$. This is not classified by $w_2 + w_1^2$ of the associated bundle to the
spatial representation: in the language of \S\ref{ferm_crys}, $\lambda\not\cong\xi$. Instead, this extension is
classified by $w_2(\sigma) + w_1(\sigma)^2$, and $\sigma$ is not \pinm, so if $f_{1/2}^D$ denotes the class D
spin-$1/2$ equivariant local system of symmetry types on $\R^d$, \cref{shear_D_thm} computes $\Ph_*^{\Z/2}(\R^d;
f_{1/2}^D)$ using the Thom spectrum of the virtual bundle
\begin{equation}
	-V \boxplus (\sigma + \sigma - d\sigma) \cong -V \boxplus (d-2)(1-\sigma).
\end{equation}
Thus
\begin{equation}
	\Ph_0^{\Z/2}(\R^d; f_{1/2}^D)\cong [\MTSpin\wedge (B\Z/2)^{(d-2)(1-\sigma)}, \Sigma^{d+2}I_\Z],
\end{equation}
and \cref{4foldinv} says the domain is again $4$-periodic, but differently from the spinless case.
\begin{itemize}
	\item When $d = 0$, we have $\MTSpin\wedge (B\Z/2)^{2-2\sigma}$.
	\item When $d = 1$, we have $\MTSpin\wedge (B\Z/2)^{\sigma-1}\simeq\MTPin^-$.
	\item When $d = 2$, we have $\MTSpin\wedge (B\Z/2)_+.$
	\item When $d = -1$, we have $\MTSpin\wedge (B\Z/2)^{1-\sigma}\simeq\MTPin^+$.
\end{itemize}
In the degrees we need, these bordism groups are computed in the same references we gave above
in~\S\ref{spinless_inversion}, and the relevant phase homology groups appear in \cref{inversions_table}.
\begin{rem}
This fourfold periodicity in the tangential structure appears in a few other contexts in mathematical physics, such
as recent work of
Kapustin-Thorngren-Turzillo-Wang~\cite[\S 8]{KTTW15},
Tachikawa-Yonekura~\cite[\S 3]{TY19},
Hason-Komargodski-Thorngren~\cite[\S 4.4]{HKT19},
Córdova-Ohmori-Shao-Yan~\cite{COSY19},
and
Wan-Wang-Zheng~\cite[\S 6.7]{WWZ19}, as well as in earlier work in the topology literature by Ekholm~\cite{Ekh98}
and Hambleton-Su~\cite[\S 4.C]{HS13}. See~\cite[Example 7.8]{MathSmith} for a review.
\end{rem}
\subsubsection{Class A}
\label{spinc_inversion}
In class A, whether with spinless or spin-$1/2$ fermions, the FCEP predicts by way of \cref{shear_A_thm}
that an inversion symmetry in dimension $d$ leads us to study $\MTSpinc\wedge (B\Z/2)^{d-d\sigma}$. For
any vector bundle $V\to X$, $V\oplus V\cong V\otimes\underline\C$, and complex vector bundles are \spinc, so by
\cref{relative_Thom}, we can remove factors of $2-2\sigma$ from $d - d\sigma$ without changing the Thom spectrum,
so we want to study $\MTSpinc\wedge (B\Z/2)_+$ when $d$ is even and $\MTSpinc\wedge (B\Z/2)^{1-\sigma}\simeq
\MTPinc$ when $d$ is odd.

We discussed \pinc bordism in \S\ref{spinc_refl}. Bahri-Gilkey~\cite{BG87a, BG87b} also compute
$\Omega_*^\Spinc(B\Z/2)$: they establish that the \term{Smith homomorphism}
$\tOmega_n^\Spinc(B\Z/2)\to\Omega_{n-1}^\Pinc$, which sends a \spinc manifold $M$ and principal
$\Z/2$-bundle $P\to M$ to the induced \pinc structure on a smooth submanifold representative of the Poincaré dual of
$w_1(P)\in H^1(M;\Z/2)$, is an isomorphism for all $n$; thus we get the groups in \cref{inversions_table} by
applying the universal property~\eqref{IZproperty} of $I_\Z$ to either $\Omega_*^{\Pinc}$ or
$\Omega_*^{\Spinc}\oplus\Omega_{*-1}^{\Pinc}$, depending on dimension.
\subsubsection{Comparison with prior work}
\label{inversion_compare}
Inversion-symmetric SPT phases are pretty well-studied, even in the fermionic case, and our phase homology
calculations reproduce classifications of the following inversion-symmetric phases in the literature.
\begin{description}
	\item[Class D, spinless] These phases are studied by Shiozaki-Xiong-Gomi~\cite[\S V.B]{SXG18},
	Cheng-Wang~\cite[\S III]{CW18}, and Zhang-Ning-Qi-Gu~\cite[Table I, \S III.A]{ZNQG25}.
	\item[Class D, spin-1/2] These phases are studied by You-Xu~\cite[\S III]{YX14},
	Shiozaki-Shapourian-Ryu~\cite{SSR17a, SSR17b}, Cheng-Wang~\cite[\S III]{CW18}, and Shiozaki-Xiong-Gomi~\cite[\S
	V.A]{SXG18}.
	\item[Class A] These phases are studied by You-Xu~\cite[\S IV.A.3]{YX14}, Shiozaki-Shapourian-Ryu~\cite[\S
	V.B]{SSR17b}, Song-Huang-Fu-Hermele~\cite[\S IV]{SHFH17}, Lee-Shiozaki-Hsieh~\cite[\S 5.2]{LSH24}, and
	Zhang-Ning-Qi-Gu~\cite[Table II, \S IV.A]{ZNQG25}.
	Shiozaki-Shapourian-Ryu also study the phases corresponding to the $\Z/2^{k+2}$ summand in $[\MTPinc,
	\Sigma^{2k+3}I_\Z]$ in arbitrary odd dimensions.\footnote{The presence of this summand follows from the
	existence of a $\Z/2^{k+2}$ summand in $\Omega_{2k+2}^{\Pinc}$, which is proven by Bahri-Gilkey~\cite{BG87b}.}
\end{description}
\begin{rem}
Guo-Ohmori-Putrov-Wan-Wang~\cite[\S 10.8]{GOPWW18} also study inversion-symmetric fermionic phases
from a bordism-theoretic perspective, in both the spinless and spin-$1/2$ cases. Their results disagree with
ours, and with the rest of the literature, because they use different symmetry types to model inversion-equivariant
fermionic phases. Likewise, Zhang-Ning-Qi-Gu~\cite[Table I, \S III.A]{ZNQG25} also study class D spin-1/2
inversion-symmetric phases, and obtain a different classification from ours: they choose a different way for $\Z/2$
to mix with fermion parity. They also suggest (\textit{ibid.}, \S I.B) that the same thing occurs for
rotoreflection symmetry, which we have not studied in this paper (though see~\cite[\S 7]{DYY25a}).
\end{rem}


%
%

	\subsection{Rotations}
		\label{s_rotations}
We turn to the case of phases equivariant for the cyclic group $C_n$ acting by rotation on a plane. These phases
have been studied by several groups of authors, and our results are consistent with prior work; see
\S\ref{cyclic_compare} for more information.

Let $\lambda\colon C_n\to\SO_2$ denote this representation and $V_\lambda\to BC_n$ be the associated vector bundle.
One can directly check that $C_n\to\SO_2$ lifts across $\Spin_2\to\SO_2$ iff $n$ is odd.

\begin{table}[h!]
\begin{tabular}{c c c c c}
\toprule
	 && Class D, spinless & Class D, spin-$1/2$ & Class A\\
$d$ &$n$ & \S\ref{unmixed_rotation} & \S\ref{mixed_rotation} & \S\ref{spinc_rot}\\
\midrule
$2$ & $0\bmod 4$ & \tblref{unmixedcyc}{$\Z\oplus\Z/(n/2)$} & \tblref{cyclic0mod4}{$\Z\oplus\Z/2n\oplus\Z/2$} &
		\tblref{spinccyclicbordism}{$\Z^2\oplus\Z/2n\oplus\Z/(n/2)$} \\
	& $2\bmod 4$ & \tblref{unmixedcyc}{$\Z\oplus\Z/(n/2)$} & \tblref{2mod4_cyc_spin_bordism}{$\Z\oplus \Z/4n$} &
		\tblref{spinccyclicbordism}{$\Z^2\oplus\Z/2n\oplus\Z/(n/2)$} \\
	& $1,3\bmod 4$ & \tblref{unmixedcyc}{$\Z\oplus\Z/n$} & \tblref{odd_cyc_spin_bordism}{$\Z\oplus\Z/n$} &
		\tblref{spinccyclicbordism}{$\Z^2\oplus\Z/n\oplus\Z/n$} \\
\addlinespace
$3$ & $0\bmod 4$ & \tblref{unmixedcyc}{$0$} & \tblref{cyclic0mod4}{$0$} &
		\tblref{spinccyclicbordism}{$0$} \\
	& $2\bmod 4$ & \tblref{unmixedcyc}{$0$} & \tblref{2mod4_cyc_spin_bordism}{$0$} &
		\tblref{spinccyclicbordism}{$0$} \\
	& $1,3\bmod 4$ & \tblref{unmixedcyc}{$0$} & \tblref{odd_cyc_spin_bordism}{$0$} &
		\tblref{spinccyclicbordism}{$0$} \\
\bottomrule
\end{tabular}
\caption{$C_n$-equivariant phase homology groups for the cases in which $C_n$ acts by rotations.
For the spinless class D case, these
are classified by $[\MTSpin\wedge (BC_n)^{2-V_\lambda}, \Sigma^{d+1}I_\Z]$; for spin-$1/2$ class D, by
$[\MTSpin\wedge (BC_n)_+, \Sigma^{d+1}I_\Z]$; and for class A, both spinless and spin-$1/2$, by $[\MTSpinc\wedge
(BC_n)_+, \Sigma^{d+1}I_\Z]$.}
\label{spin_table}
\end{table}

%
\subsubsection{Class D, spinless case}
\label{unmixed_rotation}
In this case, $C_n$ does not mix with $\mu_2\subset\Spin$, and \cref{shear_D_thm} reduces
\cref{equivariant_ansatz} to the computation of $[\MTSpin\wedge (BC_n)^{2-V_\lambda}, \Sigma^{d+2}I_\Z]$ if $n$ is
even, or $[\MTSpin\wedge (BC_n)_+]$, if $n$ is odd.
\begin{lem}
\label{oriented_cyc_bordism}
$\Omega_3^\SO(BC_n)\cong\Z/n$, $\Omega_4^\SO(BC_n)\cong\Z$, and $\Omega_5^\SO(BC_n)$ is torsion.
\end{lem}
\begin{proof}
Compute with the Atiyah-Hirzebruch spectral sequence for oriented bordism; it collapses for $p+q\le 4$, and the
$5$-line of the $E^2$-page is torsion, implying $\Omega_5^\SO(BC_n)$ is torsion.
\end{proof}
\begin{cor}[{Bruner-Greenlees~\cite[Example 7.3.2, \S 12.2.D]{BG10},
García-Etxebarria and Montero~\cite[\S C.2]{GEM19}}]
\label{odd_cyc_spin_bordism}
For $n$ odd, $\Omega_3^\Spin(BC_n)\cong\Z/n$, $\Omega_4^\Spin(BC_n)\cong\Z$, and $\Omega_5^\Spin(BC_n)$ is torsion.
\end{cor}
\begin{proof}
Because $n$ is odd, $BC_n$ is stably trivial at $2$, and $\MTSpin\to\MTSO$ is an equivalence away from $2$.
\end{proof}

\begin{thm}
\label{unmixedcyc}
If $n$ is even,
$\tOmega_3^\Spin((BC_n)^{2-V_\lambda})\cong\Z/(n/2)$, $\tOmega_4^\Spin((BC_n))^{2-V_\lambda})\cong\Z$, and
$\tOmega_5^\Spin((BC_n)^{2-V_\lambda})$ is torsion.
\end{thm}
\begin{proof}
The computation breaks into $2$-primary and odd-primary pieces. The forgetful map $\Omega_*^\Spin\to\Omega_*^\SO$
is an odd-primary isomorphism, and because $2-V_\lambda$ is orientable, there is a Thom isomorphism
$\widetilde\Omega_*^\SO((BC_n)^{2-V_\lambda})\cong\Omega_*^\SO(BC_n)$. Thus, \cref{oriented_cyc_bordism} takes care
of the odd-primary part.

Write $n = 2^\ell m$, where $m$ is odd. Then the map $BC_{2^\ell}\to BC_n$ is a stable $2$-primary equivalence,
because it induces an isomorphism on mod $2$ cohomology, so for the $2$-primary piece it suffices to understand the
case $n = 2^\ell$. Campbell~\cite[Theorem 1.8]{Cam17} studies $\Omega_d^\Spin((BC_{2^\ell})^{2-V_\lambda})$,
obtaining $\Z/2^{\ell-1}$ when $d = 3$, $\Z$ when $d = 4$, and torsion when $d = 5$, which suffices.\footnote{There
are a few other computations of $\widetilde \Omega_*^\Spin((BC_{2^\ell})^{2-V_\lambda})$ in low degrees by other
methods. For $\ell = 1$, see Giambalvo~\cite{Gia73a}, García-Etxebarria and Montero~\cite[(C.21)]{GEM19}, and
Freed-Hopkins~\cite[\S 5]{FH19}.  For $\ell > 1$, see Botvinnik-Gilkey~\cite[\S 5]{BG97} and
Davighi-Lohitsiri~\cite[\S A.4]{DL20b}; Botvinnik-Gilkey only report the orders of the bordism groups, but their
computations show that the groups we need are cyclic.  Be aware that Campbell and Davighi-Lohitsiri consider a
different vector bundle than $2-V_\lambda$, though their calculations apply to this case.}
\end{proof}

\subsubsection{Class D, spin-$1/2$ case}
\label{mixed_rotation}
\Cref{shear_D_thm} asks us to compute $[\MTSpin\wedge (BC_n)_+, \Sigma^{d+2}I_\Z]$,
which~\eqref{IZproperty} tells us in terms of $\Omega_*^\Spin(BC_n)$. For $n$ odd, we already saw this in
\cref{odd_cyc_spin_bordism}.
\begin{prop}
\label{2mod4_cyc_spin_bordism}
Let $n\equiv 2\bmod 4$. Then $\Omega_3^\Spin(BC_n)\cong\Z/4n$, $\Omega_4^\Spin(BC_n)\cong\Z$, and
$\Omega_5^\Spin(BC_n)$ is torsion.
\end{prop}
\begin{proof}
Inclusion $BC_2\to BC_n$ is a $2$-local equivalence, so the fact that the $2$-torsion is $\Z/8$ in degree $3$ and
vanishes in degree $4$ follows as soon as we know that for $\Omega_*^\Spin(BC_2)$. This was originally done by
Mahowald-Milgram~\cite{MM76} but has been computed in a few other places, including Mahowald~\cite[Lemma
7.3]{Mah82}, Bruner-Greenlees~\cite[Example 7.3.1]{BG10}, Siegemeyer~\cite[Theorem 2.1.5]{Sie13}, and
García-Etxebarria and Montero~\cite[(C.18)]{GEM19}. What remains is odd-primary information, which is equivalent to
the odd-primary part of oriented bordism, which we computed in \cref{oriented_cyc_bordism}.
\end{proof}

\begin{prop}
\label{cyclic0mod4}
For $n\equiv 0\bmod 4$, $\Omega_3^\Spin(BC_n)\cong\Z/2\oplus\Z/2n$, $\Omega_4^\Spin(BC_n)\cong\Z$, and
$\Omega_5^\Spin(BC_n)$ is torsion.
\end{prop}
\begin{proof}
Write $n = 2^\ell m$, where $m$ is odd. As in the proof of \cref{unmixedcyc}, the $2$-primary part of the answer is
detected by $BC_{2^\ell}\to BC_n$, and the odd-primary part of the answer is detected by oriented bordism.
Davighi-Lohitsiri~\cite[\S A.3]{DL20b} compute $\Omega_k^\Spin(BC_{2^\ell})$ for $k\le 6$, giving the $2$-primary
summand, and for the odd-primary part we use \cref{oriented_cyc_bordism}.
\end{proof}
Botvinnik-Gilkey-Stolz~\cite[Theorem 2.4]{BGS97},
Bruner-Greenlees~\cite[Example 7.3.3]{BG10},
Siegemeyer~\cite[\S 2.2]{Sie13},
Bárcenas-García-Hernández-Reinauer~\cite[Theorem 3.17]{BGR24},
and Braeger-Debray-Dierigl-Heckman-Montero~\cite[\S 13]{BDDHM25}
do special cases of this computation, by a variety of methods.
\subsubsection{Class A}
\label{spinc_rot}
The representation of $C_n$ on $\R^2$ by rotations is unitary (under the standard identification $\R^2 = \C$),
hence \spinc, so in both the spinless and spin-$1/2$ cases, we consider $\Omega_*^{\Spin^c}(BC_n)$: in the spinless
case, we have a Thom isomorphism $\tOmega_*^{\Spin^c}((BC_n)^{2-V_\lambda})\overset\cong\to
\Omega_*^{\Spin^c}(BC_n)$ (\cref{tautological_Thom}), and in the spin-$1/2$ case, $\Det(V_\lambda)$ is trivial, so
\cref{shear_A_thm} also gives us the \spinc bordism of $BC_n$.
\begin{thm}
\label{spinccyclicbordism}
The first few \spinc bordism groups of $BC_n$ are
\begin{alignat*}{2}
	\Omega_0^\Spinc(BC_{2k}) &\cong \Z\qquad\qquad\qquad & \Omega_0^\Spinc(BC_{2k+1}) &\cong \Z\\
	\Omega_1^\Spinc(BC_{2k}) &\cong \Z/2k & \Omega_1^\Spinc(BC_{2k+1}) &\cong \Z/(2k+1)\\
	\Omega_2^\Spinc(BC_{2k}) &\cong \Z & \Omega_2^\Spinc(BC_{2k+1}) &\cong \Z\\
	\Omega_3^\Spinc(BC_{2k}) &\cong \Z/4k\oplus\Z/k & \Omega_3^\Spinc(BC_{2k+1}) &\cong (\Z/(2k+1))^{\oplus 2}\\
	\Omega_4^\Spinc(BC_{2k}) &\cong \Z^2 & \Omega_4^\Spinc(BC_{2k+1}) &\cong \Z^2,
\end{alignat*}
and $\Omega_5^\Spinc(BC_n)$ is torsion for all $n$.
\end{thm}
\begin{proof}
Write $n = 2^\ell\cdot m$, where $m$ is odd. It suffices to compute the $2$-primary piece and
$\Omega_*^\Spinc(BC_n)\otimes\Z[1/2]$; \cref{spinc_odd_primes,oriented_cyc_bordism} tell us the latter, so we focus
on the $2$-local computation. The inclusion $C_{2^\ell}\to C_n$ is stably a $2$-primary equivalence, so
for the $2$-primary piece it suffices to determine $\Omega_*^\Spinc(BC_{2^\ell})$. Bahri-Gilkey~\cite[Theorem
1]{BG87b} compute these groups; when $\ell = 0$ they are $\Omega_*^\Spinc(\pt)$, which begins $\Z$, $0$,
$\Z$, $0$, $\Z^2$, $0$; and when $\ell\ne 0$ we have the same free summands as when $\ell = 0$, but additional
torsion summands: $\Omega_1^\Spinc(BC_{2^\ell}) \cong \Z/2^\ell$, and $\Omega_3^\Spinc(BC_{2^\ell})\cong
\Z/2^{\ell-1}\oplus\Z/2^{\ell+1}$.
%
\end{proof}

\subsubsection{Comparison with prior work}
\label{cyclic_compare}
Rotation-equivariant phases in class D have been studied by several groups, including
Shiozaki-Shapourian-Ryu~\cite[\S IV.C]{SSR17b},
Guo-Ohmori-Putrov-Wan-Wang~\cite[\S 10.9]{GOPWW18}, and
Freed-Hopkins~\cite[\S 5]{FH19}, who all restrict to the case $n = 2$;
Manjunath-Calvera-Barkeshli~\cite[Table III]{MCB23} and
Zhang-Ning-Qi-Gu~\cite[Table I, \S\S S-3.1, S-3.7, S-3.14, S-3.19]{ZNQG25}, who restrict to $d = 3$;
and most comprehensively by
Cheng-Wang~\cite[\S IV, \S V]{CW18} and
Cheng-Wang-Yang~\cite{CWY24},
who consider arbitrary $n$ and both the spinless and spin-$1/2$ cases in $d = 2,3$.
Zhang-Ning~\cite{ZN21} study a $d = 2$ class D spinless $\Z/2$-rotation-symmetric SPT in the context of their
crystalline-equivalent bulk-boundary correspondence.

Freed-Hopkins begin from the same ansatz as us so agreement is no surprise. In the remaining cases, there is almost
complete agreement: all classifications compute the same torsion summands, but they all miss the free summand in $d
= 2$. This is not a discrepancy, however: many authors restrict to considering phases whose low-energy effective
theories are expected to be topological field theories, which in the ansatz of
Freed-Hopkins~\cite[\FHrefllink{page.20}{\S\S 5.3--5.4}]{FH16} amounts to considering the torsion subgroup of the
classification using $I_\Z\MTH$. The non-topological theories corresponding to the free summand have been discussed
in a few references, including Freed~\cite[Lecture 9]{Fre19} and Wan-Wang~\cite[\S 7.1]{WW19a}; they correspond to
non-topological, invertible field theories by a theorem of Grady~\cite{Gra23}.

Rotation-equivariant phases in class A are studied by
Shiozaki-Shapourian-Ryu~\cite[\S IV.D]{SSR17b},
Shiozaki-Xiong-Gomi~\cite[\S V.C.1]{SXG18},
Lu-Vishwanath-Khalaf~\cite{LVK19},
Lee-Shiozaki-Hsieh~\cite[\S 4.2]{LSH24},
Herzog-Arbeitman-Bernevig-Song~\cite[\S B.2]{HABS24},
and Zhang-Ning-Qi-Gu~\cite[Table II, \S\S S-3.1, S-3.7, S-3.14, S-3.19]{ZNQG25}.
All of these classifications agree with ours on torsion; some do not consider the free summand as before, and
Shiozaki-Xiong-Gomi's computation completely matches ours.
Again, the free summand corresponds to non-topological invertible field theories.

	\subsection{Rotations and reflections}
		\label{s_dihedral}
In this section, we compute the phase homology groups corresponding to phases on $\R^d$ equivariant for the
$D_{2n}$-action of rotations and reflections in a given plane. Zhang-Wang-Yang-Qi-Gu~\cite{ZWYQG19},
Herzog-Arbeitman-Bernevig-Song~\cite[\S B.3]{HABS24},
and
Zhang-Ning-Qi-Gu~\cite{ZNQG25} also study these phases; we compare our results to theirs in
\S\ref{dihedral_discrepancy}. In addition, Zhang-Ning~\cite{ZN21} study a $d = 2$ class D spinless $D_8$-symmetric
SPT in the context of their crystalline-equivalent bulk-boundary correspondence.
\begin{table}[h!]
\begin{tabular}{c r c c c c c}
\toprule
&& Class D, spinless & Class D, spin-$1/2$ & Class A, spinless & Class A, spin-$1/2$\\
$d$ & $n\phantom{\bmod 4}$ & \S\ref{unmixed_dihedral} & \S\ref{mixed_dihedral} & \S\ref{spinc_dihedral} &
	\S\ref{s_spinc_mixed_dih}\\
\midrule
$2$ & $0\bmod 4$ & \tblref{dih_spinless_0mod4_thm}{$(\Z/2)^{\oplus 2}$} &
\tblref{spinless_D0mod4_thm}{$(\Z/2)^{\oplus 2}$} & \tblref{dih_spinc_0mod4}{$\Z/2n$} &
\tblref{bord_gps_spinc_dih_mixed_0mod4}{$\Z/(n/2)\oplus (\Z/2)^{\oplus 2}$}\\
	& $2\bmod 4$ & \tblref{unmixed_dih_2mod4}{$\Z/2$} & \tblref{mixed_dih_2mod4}{$(\Z/2)^{\oplus 2}$} &
	\tblref{dih_spinc_2mod4}{$\Z/2n$} & \tblref{bord_gps_spinc_dih_mixed_2mod4}{$\Z/n\oplus\Z/2$}\\
	& $1,3\bmod 4$ & \tblref{unmixed_dih_odd}{$\Z/2$} & \tblref{mixed_dih_odd}{$0$} & \tblref{dih_spinc_odd}{$\Z/n$} &
	\tblref{spinc_mixed_odd_thm}{$\Z/n$}\\
\addlinespace
$3$ & $0\bmod 4$ & \tblref{dih_spinless_0mod4_thm}{$(\Z/2)^{\oplus 4}$} & \tblref{spinless_D0mod4_thm}{$0$} &
\tblref{dih_spinc_0mod4}{$(\Z/2)^{\oplus 4}$} & \tblref{bord_gps_spinc_dih_mixed_0mod4}{$\Z/8\oplus \Z/4\oplus
\Z/2$}\\
	& $2\bmod 4$ & \tblref{unmixed_dih_2mod4}{$(\Z/2)^{\oplus 3}$} & \tblref{mixed_dih_2mod4}{$0$} &
	\tblref{dih_spinc_2mod4}{$(\Z/2)^{\oplus 4}$} & \tblref{bord_gps_spinc_dih_mixed_2mod4}{$\Z/8\oplus
	\Z/4\oplus\Z/2$}\\
	& $1,3\bmod 4$ & \tblref{unmixed_dih_odd}{$\Z/16$} & \tblref{mixed_dih_odd}{$0$} &
	\tblref{dih_spinc_odd}{$\Z/8\oplus\Z/2$} & \tblref{spinc_mixed_odd_thm}{$\Z/8\oplus\Z/2$}\\
\bottomrule
\end{tabular}
\caption{$D_{2n}$-equivariant phase homology groups, where $D_{2n}$ acts through rotations and reflections. These
come from computations of $\tOmega_*^\Spin(X_n)$ and $\tOmega_*^{\Spinc}(X_n)$, where $X_n$ is one of
$(BD_{2n})^{2-V_\lambda}$ or $(BD_{2n})^{\Det(V_\lambda)-1}$. See \S\ref{s_dihedral} for details and proofs.}
\label{dihedral_table}
\end{table}

Let $\lambda$ be the standard real $2$-dimensional representation of $D_{2n}$ and $V_\lambda\to BD_{2n}$ be the
associated vector bundle.
%
%
%
Let $s$ be a reflection in $D_{2n}$ and $r$ a rotation through the angle $2\pi/n$. Then, define $x,y\in
H^1(BD_{2n};\Z/2) = \Hom(D_{2n},\Z/2)$ by
\begin{subequations}
\begin{align}
	x(s^\ell r^m) &\coloneqq \ell\bmod 2\\
	y(s^\ell r^m) &\coloneqq m\bmod 2.
\end{align}
\end{subequations}
In the representation $\lambda$, $s^\ell r^m\in D_{2n}$ acts by an orientation-reversing endomorphism iff $\ell$ is
odd, so $w_1(V_\lambda) = x$.
\begin{prop}[{Munkholm~\cite[\S 5]{Mun69}, Snaith~\cite[Theorem 4.6]{Sna13}}]
\label{dihmod2coh}
\hfill
\begin{enumerate}
	\item If $n$ is odd, $H^*(BD_{2n};\Z/2)\cong\Z/2[x]$.
	\item If $n\equiv 0\bmod 4$, $H^*(BD_{2n};\Z/2)\cong\Z/2[x,y,w]/(xy + y^2)$, where $\abs w = 2$ and $w =
	w_2(V_\lambda)$.
	\item If $n\equiv 2\bmod 4$, $H^*(BD_{2n};\Z/2)\cong\Z/2[x, y]$.
\end{enumerate}
\end{prop}
\begin{lem}
\label{w2D2mod4}
For $n\equiv 2\bmod 4$, $w_2(V_\lambda) = xy + y^2$.
\end{lem}
\begin{proof}
Since $s, r^{n/2}\in D_{2n}$ commute, there is a map $j\colon\Z/2\times\Z/2\to D_{2n}$ sending $(1,0)\mapsto s$ and
$(0,1)\mapsto r^{n/2}$. The pullback map $j^*\colon H^*(BD_{2n};\Z/2)\to H^*(B\Z/2\times B\Z/2;\Z/2)$ sends $x$ and
$y$ to linearly independent elements of $H^1(B\Z/2\times B\Z/2;\Z/2)$: one way to see this is to identify the
pullback map with the map $\Hom(D_{2n}, \Z/2)\to\Hom(\Z/2\times\Z/2, \Z/2)$ given by precomposing with $j$. Thus
$j^*$ is an isomorphism on $H^1(\bl;\Z/2)$. For both $BD_{2n}$ and $B\Z/2\times B\Z/2$, the mod $2$ cohomology ring
is the free symmetric algebra on $H^1(\bl;\Z/2)$, so $j^*$ is an isomorphism of cohomology rings.

Thus we can compute $w_2(V_\lambda)$ by regarding $\lambda$ as a $\Z/2\times\Z/2$ representation. Let $\ell_1
\subset \lambda$ be the fixed locus of $s$, which is a subspace, and $\ell_2$ be its orthogonal complement. Then
$\lambda = \ell_1\oplus\ell_2$ as $(\Z/2\times\Z/2)$-representations. Both $s$ and $r^{n/2}$ act nontrivially on
$\ell_2$; on $\ell_1$, $s$ acts trivially and $r^{n/2}$ acts nontrivially. Thus $w(\ell_1) = 1 + j^*(y)$,
$w(\ell_2) = 1 + j^*(x) + j^*(y)$, and
\begin{equation}
	w_2(j^*V_\lambda) = w_2(\ell_1) + w_1(\ell_1)w_1(\ell_2) + w_2(\ell_2) = j^*(y(x+y)).
	\qedhere
\end{equation}
\end{proof}
\begin{lem}
\label{odd_dihedral_equiv}
Suppose $n$ is odd and $i\colon\Z/2\inj D_{2n}$ is the inclusion of $\ang s$. Let $V\to BD_{2n}$ be a virtual vector
bundle such that $w_1(V)$, as an element $\Hom(D_{2n}, \Z/2)$, is nonzero on $s$. Then, the induced map of Thom
spectra $\widehat\imath\colon (B\Z/2)^{i^*V}\to (BD_{2n})^{V}$ is a $2$-primary homotopy equivalence.
\end{lem}
\begin{proof}
By the homology Whitehead theorem, it suffices to show $\widehat\imath$ induces an isomorphism on mod $2$
cohomology. The Thom isomorphism rewords our question to be about the map $H^*(BD_{2n};\Z/2)\to
H^*(B\Z/2;\Z/2)$, and \cref{dihmod2coh} tells us that both $H^*(B\Z/2;\Z/2)$ and $H^*(BD_{2n};\Z/2)$ are abstractly
isomorphic to $\Z/2[x]$ with $\abs x = 1$; we will show $i^*x_{BD_{2n}} = x_{B\Z/2}$, implying $i^*$ is a ring
isomorphism. Since $x$ is the only nonzero degree-one element and $V$ and $i^*V$ are both unorientable, $x =
w_1(V)$ and $i^*x = w_1(i^*V)\ne 0$.
%
\end{proof}
We will need the next calculations to determine the odd-primary torsion subgroups of the phase homology groups
we calculate. Recall that $x\in H^1(BD_{2n};\Z/2)$ is equal to $w_1(V_\lambda)$.
\begin{lem}[{Handel~\cite[Theorems 5.8, 5.9]{Han93}}]
\label{twisted_handel}
\begin{equation}
	H_k(BD_{2n}; (\Z[1/2])_x)\cong \begin{cases}
		\Z/n, &k\equiv 1\bmod 4\\
		0, &\text{\rm otherwise.}
	\end{cases}
\end{equation}
\end{lem}
Handel calculates $H^*(BD_{2n};\Z_x)$; use the universal coefficient theorem to switch to $\Z[1/2]$-homology.
\begin{prop}
\label{twisted_dihedral_SO}
\label{twisted_dihedral_SO_U}
Suppose $V\to BD_{2n}$ is a rank-zero virtual vector bundle with $w_1(V) = x$.
\begin{enumerate}
	\item The odd-torsion subgroup of $\tOmega_k^\Spin((BD_{2n})^V)$ is isomorphic to the odd-torsion subgroup of
	$\Z/n$ for $k = 1$, and vanishes for $k = 0, 2, 3$, and $4$.
	\item The odd-torsion subgroup of $\tOmega_k^\Spinc((BD_{2n})^V)$ is isomorphic to the odd-torsion subgroup of
	$\Z/n$ for $k = 1$ and $3$, and vanishes for $k = 0, 2$, and $4$.
\end{enumerate}
\end{prop}
\begin{proof}
For the first part, apply the Atiyah-Hirzebruch spectral sequence for the completion of spin bordism at primes
other than $2$. Since $w_1(V) = x$, the Thom isomorphism identifies $\wH_*((BD_{2n})^V)\cong H_*(BD_{2n};\Z_x)$,
and by \cref{twisted_handel} we know these groups away from $2$. The only nonzero entry in the $E^2$-page of total
degree less than $5$ is $E^2_{1,0}\cong\Z/n$, so the spectral sequence collapses in the desired range and we
conclude. For the second half of the proposition, use the first half and \cref{spinc_odd_primes}.
\end{proof}
\subsubsection{Class D, spinless case}
\label{unmixed_dihedral}
Since we are in the spin-$0$ case, \cref{shear_D_thm} tells us to compute
$\tOmega_*^\Spin((BD_{2n})^{2-V_\lambda})$.
\begin{prop}
\label{unmixed_dih_odd}
For $n$ odd, the first few spin bordism groups of $(BD_{2n})^{2-V_\lambda}$ are
\begin{align*}
	\tOmega_0^\Spin((BD_{2n})^{2-V_\lambda}) &\cong \Z/2\\
	\tOmega_1^\Spin((BD_{2n})^{2-V_\lambda}) &\cong \Z/n\\
	\tOmega_2^\Spin((BD_{2n})^{2-V_\lambda}) &\cong \Z/2\\
	\tOmega_3^\Spin((BD_{2n})^{2-V_\lambda}) &\cong \Z/2\\
	\tOmega_4^\Spin((BD_{2n})^{2-V_\lambda}) &\cong \Z/16,
\end{align*}
and $\tOmega_5^\Spin((BD_{2n})^{2-V_\lambda})$ is torsion.
\end{prop}
\begin{proof}
To compute the $2$-torsion subgroups of these bordism groups, apply \cref{odd_dihedral_equiv} with $2-V_\lambda$ to
get a $2$-primary stable equivalence $(BD_{2n})^{2-V_\lambda}\simeq (B\Z/2)^{1-\sigma}$, then~\eqref{pinpsplitting}
to get $\tOmega_*^\Spin((B\Z/2)^{1-\sigma})\cong\Omega_*^{\Pin^+}$. Low-degree \pinp bordism groups are calculated
in~\cite{Gia73a, KT90Pinp, KT90}. For the odd-torsion subgroups, use \cref{twisted_dihedral_SO}.
\end{proof}
Now we turn to the case where $n\equiv 2\bmod 4$.
\begin{thm}
\label{unmixed_dih_2mod4}
When $n\equiv 2\bmod 4$, the first few spin bordism groups of $(BD_{2n})^{2-V_\lambda}$ are
\begin{align*}
	\tOmega_0^\Spin((BD_{2n})^{2-V_\lambda}) &\cong \Z/2\\
	\tOmega_1^\Spin((BD_{2n})^{2-V_\lambda}) &\cong \Z/n\\
	\tOmega_2^\Spin((BD_{2n})^{2-V_\lambda}) &\cong \Z/2\\
	\tOmega_3^\Spin((BD_{2n})^{2-V_\lambda}) &\cong \Z/2\\
	\tOmega_4^\Spin((BD_{2n})^{2-V_\lambda}) &\cong (\Z/2)^{\oplus 3},
\end{align*}
and $\tOmega_5^{\Spin}((BD_{2n})^{2-V_\lambda})$ is torsion.
\end{thm}
As usual, this gives the $n\equiv 2\bmod 4$ entries in \cref{dihedral_table}.
\begin{proof}
We will use the Adams spectral sequence at the prime $2$ to compute $\tOmega_d^\Spin((BD_{2n})^{2-V_\lambda})$ for $d\le 7$. This only
sees $2$-primary information, but we already calculated the odd-torsion subgroup in \cref{twisted_dihedral_SO}.
Recall that $w_1(V_\lambda) = x$ and (from \cref{w2D2mod4}) $w_2(V_\lambda) = xy + y^2$; thus $w_1(2 - V_\lambda) =
x$ and $w_2(2-V_\lambda) = x^2+xy+y^2$. \Cref{wu_f} then tells us the Steenrod squares in $\tH^*((BD_{2n})^{2-V_\lambda};\Z/2)$, e.g.\
$\Sq^1(\oU) = \oU x$ and $\Sq^2(\oU) = \oU (x^2+xy+y^2)$. Continuing in this vein determines the $\cA(1)$-module
structure on $\tH^*((BD_{2n})^{2-V_\lambda};\Z/2)$ in low degrees, as shown in \cref{a1moddih2mod4}, left.
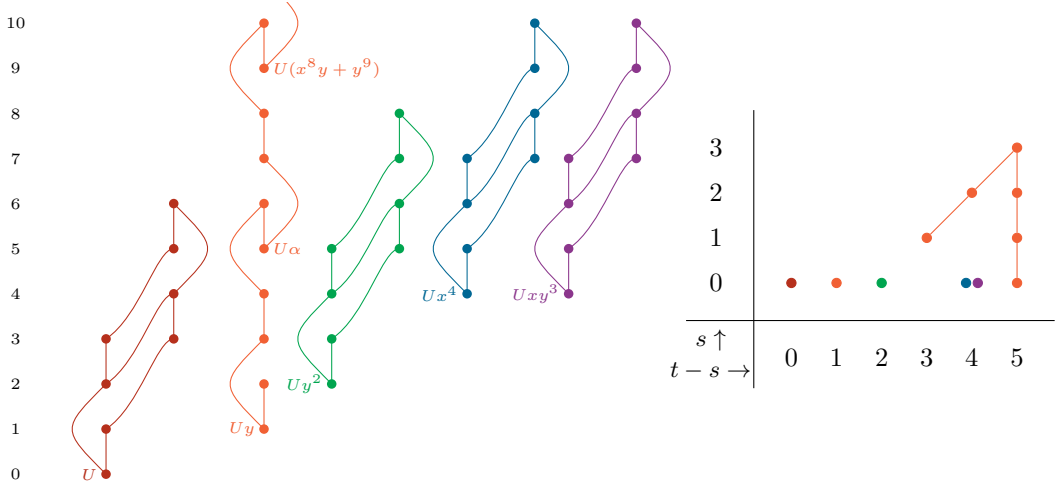
\begin{figure}[h!]
\begin{subfigure}[c]{0.525\textwidth}
\begin{tikzpicture}[scale=0.6, every node/.style = {font=\tiny}]
	\foreach \y in {0, ..., 10} {
		\node at (-1.5, \y) {$\y$};
	}
	\begin{scope}[BrickRed]
		\Aone{0}{0}{$\oU$}
	\end{scope}

	\begin{scope}[RedOrange]
	\tikzpt{3.5}{1}{$\oU y$}{};
	\foreach \y in {2, ..., 10} {
		\tikzpt{3.5}{\y}{}{};
	}
	\node[right] at (3.5, 5) {$\oU\alpha$};
	\node[right] at (3.5, 9) {$\oU(x^8y + y^9)$};
	\foreach \y in {1, 3, ..., 9} {
		\sqone(3.5, \y);
	}
	\sqtwoL(3.5, 1);
	\sqtwoL(3.5, 4);
	\sqtwoR(3.5, 5, );
	\sqtwoL(3.5, 8);
	\begin{scope}
		\clip (3, 8.5) rectangle (5, 10.45);
		\sqtwoR(3.5, 9, );
	\end{scope}
	\end{scope}

	\begin{scope}[Green]
		\Aone{5}{2}{$\oU y^2$}
	\end{scope}

	\begin{scope}[MidnightBlue]
		\Aone{8}{4}{$\oU x^4$}
	\end{scope}

	\begin{scope}[Fuchsia]
		\Aone{10.25}{4}{$\oU xy^3$}
	\end{scope}

%
\end{tikzpicture}
\end{subfigure}\qquad
\begin{subfigure}[c]{0.35\textwidth}
\begin{sseqdata}[name=AdamsD2mod4, classes=fill, xrange={0}{5}, yrange={0}{3}, scale=0.6,
x label = {$\displaystyle{s\uparrow \atop t-s\rightarrow}$},
x label style = {font = \small, xshift = -17ex, yshift=3ex}]
\class[BrickRed](0, 0)
\begin{scope}[RedOrange]
	\class(1, 0)
	\class(3, 1)
	\class(4, 2)
	\class(5, 3)
	\foreach \y in {2, ..., 0} {
		\class(5, \y)
		\structline(5, \y)(5, \y+1)
	}
	\structline(3, 1)(4, 2)
	\structline(4, 2)(5, 3)
\end{scope}
\class[Green](2, 0)
\class[MidnightBlue](4, 0)
\class[Fuchsia](4, 0)
\end{sseqdata}
\label{adamsD2mod4}
\printpage[name=AdamsD2mod4, page=2]
\end{subfigure}
\caption{Left: the $\cA(1)$-module structure on $\tH^*((BD_{2n})^{2-V_\lambda};\Z/2)$ in low degrees, when $n\equiv
2\bmod 4$. Here $\alpha\coloneqq x^4y+y^5$. The submodule pictured here contains all elements of degree at most
$5$. Right: the $E_2$-page of the corresponding Adams spectral sequence computing $\ko$-theory.}
\label{a1moddih2mod4}
\end{figure}
We obtain a splitting as $\cA(1)$-modules:
\begin{equation}
\label{dihedral2mod4summands}
	\tH^*((BD_{2n})^{2-V_\lambda};\Z/2)\cong \textcolor{BrickRed}{\cA(1)}\oplus
		\textcolor{RedOrange}{\Sigma R_0}\oplus
		\textcolor{Green}{\Sigma^2\cA(1)}\oplus
		\textcolor{MidnightBlue}{\Sigma^4\cA(1)}\oplus
		\textcolor{Fuchsia}{\Sigma^4\cA(1)}\oplus P.
\end{equation}
The $\cA(1)$-module $\textcolor{RedOrange}{R_0}$ is defined to be $\wH^*((B\Z/2)^{1-\sigma};\Z/2)$; the copy
appearing here is the indecomposable summand containing $\oU y$. The submodule $P$ contains no elements of degree
below $6$, so is irrelevant for our low-degree computations; we need to determine $\Ext(M)$ for the remaining
summands. For $\Sigma^k\cA(1)$, there is a single $\Z/2$ summand in topological degree $k$ and filtration $0$, and
for $\textcolor{RedOrange}{\Sigma R_0}$, see~\cite[\S 2]{GMM68} or~\cite[Figure 24]{BC18}. Putting these together,
we display the $E_2$-page of this Adams spectral sequence in \cref{a1moddih2mod4}, right. In this range, a
combination of $h_1$-equivariance and Margolis' theorem (\cref{margolis}) forces all differentials to vanish, and
Margolis' theorem implies there are no hidden extensions, so we are done.
\end{proof}
\begin{rem}
The module $\textcolor{RedOrange}{R_0}$ is an infinite version of the \term{Milgram modules} $Q_{0,n}$ and
$Q_{1,n}$~\cite{Mil75, DGM81}; see~\cite[Figure 1]{BL21} for a picture.
\end{rem}

Finally, consider the case that $n\equiv 0\bmod 4$.
\begin{thm}
\label{dih_spinless_0mod4_thm}
Let $n\equiv 0\bmod 4$.
\begin{align*}
	\tOmega_0^\Spin((BD_{2n})^{2-V_\lambda}) &\cong \Z/2\\
	\tOmega_1^\Spin((BD_{2n})^{2-V_\lambda}) &\cong \Z/n\\
	\tOmega_2^\Spin((BD_{2n})^{2-V_\lambda}) &\cong \Z/2\\
	\tOmega_3^\Spin((BD_{2n})^{2-V_\lambda}) &\cong (\Z/2)^{\oplus 2}\\
	\tOmega_4^\Spin((BD_{2n})^{2-V_\lambda}) &\cong (\Z/2)^{\oplus 4},
\end{align*}
and $\tOmega_5^\Spin((BD_{2n})^{2-V_\lambda})$ is torsion.
\end{thm}
\begin{proof}[Proof of \cref{dih_spinless_0mod4_thm}]
First, by \cref{twisted_dihedral_SO}, the only odd-primary torsion in $\tOmega_k^\Spin((BD_{2n})^{2-V_\lambda})$ for $k\le 4$ is in
degree $1$. Draw the Atiyah-Hirzebruch spectral sequence
\begin{equation}
	E^2_{p,q} = \widetilde H_p((BD_{2n})^{2-V_\lambda}; \Omega_q^\Spin) \Longrightarrow \tOmega_{p+q}^\Spin((BD_{2n})^{2-V_\lambda}).
\end{equation}
After applying the Thom isomorphism, this needs as input $H_*(BD_{2n};\Z_x)$ and $H_*(BD_{2n};\Z/2)$. The former
can be determined using Handel's calculation~\cite[Theorem 5.8]{Han93} of $H^*(BD_{2n}; \Z_x)$, and the latter can
be determined from \cref{dihmod2coh}; in both cases use the universal coefficient theorem to pass from homology to
cohomology. We obtain $E^2_{1,0}\cong\Z/n$ and $E^2_{0,1}\cong\Z/2$, so there are three options for
$\tOmega_1^\Spin((BD_{2n})^{2-V_\lambda})$: $\Z/n$, $\Z/n\oplus\Z/2$, or $\Z/2n$. We will address this ambiguity later.

Using \cref{dihmod2coh}, $w_1(2-V_\lambda) = x$ and $w_2(2-V_\lambda) = w + x^2$. Hence (\cref{wu_f}) $\Sq^1(\oU) =
\oU x$ and $\Sq^2(\oU) = \oU(w+x^2)$. We also need the Steenrod squares of $x$, $y$, and $w$. For degree reasons,
$\Sq(x) = x+x^2$ and $\Sq(y) = y+y^2$.
\begin{lem}[{\cite[\S 4.1]{Mal13}}]
\label{0mod4_w_lem}
$\Sq(w) = w + wx + w^2$.
\end{lem}
These and the Cartan formula determine the $\cA(1)$-module structure on $\tH^*((BD_{2n})^{2-V_\lambda};\Z/2)$, thanks to \cref{wu_f}.
In \cref{a1moddih0mod4}, left, we display this structure in low degrees.
\begin{figure}[h!]
\begin{subfigure}[c]{0.53\textwidth}
\begin{tikzpicture}[scale=0.6, every node/.style = {font=\tiny}]
	\foreach \y in {0, ..., 10} {
		\node at (-1.5, \y) {$\y$};
	}

	\begin{scope}[BrickRed]
		\Aone{0}{0}{$\oU$}
	\end{scope}

	\draw[gray, dashed, thick] (4, 1) -- (4, 2);

	\begin{scope}[RedOrange]
		\tikzpt{2.5}{2}{$\oU y^2$}{};
		\tikzpt{2.5}{3}{}{};
		\tikzptR{4}{1}{$\oU y$}{};
		\tikzpt{4}{3}{}{};
		\tikzpt{4}{4}{}{};
		\tikzpt{4}{5}{}{};
		\tikzpt{4}{6}{}{};
		
		\sqone(2.5, 2);
		\sqone(4, 3);
		\sqone(4, 5);
		\sqtwoL(4, 1);
		\sqtwoR(4, 4, );
		\sqtwoCR(2.5, 2);
		\sqtwoCR(2.5, 3);
	\end{scope}

	\tikzptR{4}{2}{$\oU w$}{Goldenrod!67!black};

	\begin{scope}[Green]
		\Joker{6}{4}{$\oU w^2$}
	\end{scope}

	\begin{scope}[PineGreen]
		\Aone{8}{4}{$\oU x^4$}
	\end{scope}

	\begin{scope}[MidnightBlue]
		\Aone{10.25}{4}{$\oU y^4$}
	\end{scope}

	\begin{scope}[Fuchsia]
		\SpanishQnMark{13}{5}{$\oU w^2y$}
	\end{scope}

%
\end{tikzpicture}
\end{subfigure}
\qquad
\begin{subfigure}[c]{0.4\textwidth}
\begin{sseqdata}[name=AdamsD0mod4, classes=fill, xrange={0}{5}, yrange={0}{4}, scale=0.7, Adams grading,
>=stealth,
x label = {$\displaystyle{s\uparrow \atop t-s\rightarrow}$},
x label style = {font = \small, xshift = -19ex, yshift=3.7ex}]
\class[BrickRed](0,0)
\begin{scope}[RedOrange]
	\class(1,0)\AdamsTower{}
	\class(2, 0)
	\class(3, 1)
	\structline(2, 0)(3, 1)
	\class(5, 2)\AdamsTower{}
\end{scope}
\begin{scope}[white]
	\class(2, 1)\AdamsTower{}
	\class(5, 0)
	\class(5, 1)
	\class(6, 1)
	\class(6, 2)
\end{scope}
\begin{scope}[Goldenrod!67!black]
	\class(2, 0)\AdamsTower{}
	\class(3, 1)
	\class(4, 2)
	\structline(2, 0, 2)(3, 1, 2)
	\structline(3, 1, 2)(4, 2)
	\class(6, 3)\AdamsTower{}
\end{scope}
\class[Green](4, 0)
\class[Green](6, 1)\AdamsTower{Green}
\class[PineGreen](4, 0)
\class[MidnightBlue](4, 0)
\class[Fuchsia](5, 0)\AdamsTower{Fuchsia}
\class[Fuchsia!67!white](6, 0)
\class[Fuchsia!75!black](6, 0)
\d[gray]2(2, 0, -1)
\d[gray]2(5, 0, -1)
\end{sseqdata}
\printpage[name=AdamsD0mod4, page=2]
\end{subfigure}
\caption{Left: the low-degree mod $2$ cohomology of $(BD_{2n})^{2-V_\lambda}$ over $\cA(1)$, $n\equiv 0\bmod 4$.
This summand contains all elements in degrees $5$ and below. The dashed line indicates that the $\Z/2^r$ Bockstein
maps $\oU y$ to $\oU w$, which we need in \cref{first_dih_differential}. Right: the $E_2$-page of the Adams
spectral sequence computing $\widetilde{\ko}_*((BD_{2n})^{2-V_\lambda})_2^\wedge$. See
\cref{first_dih_differential} for how to address the differential in topological degree $2$ and
\cref{D0mod4_nodiff} to show the differential in topological degree $5$ vanishes.}
\label{a1moddih0mod4}
\end{figure}
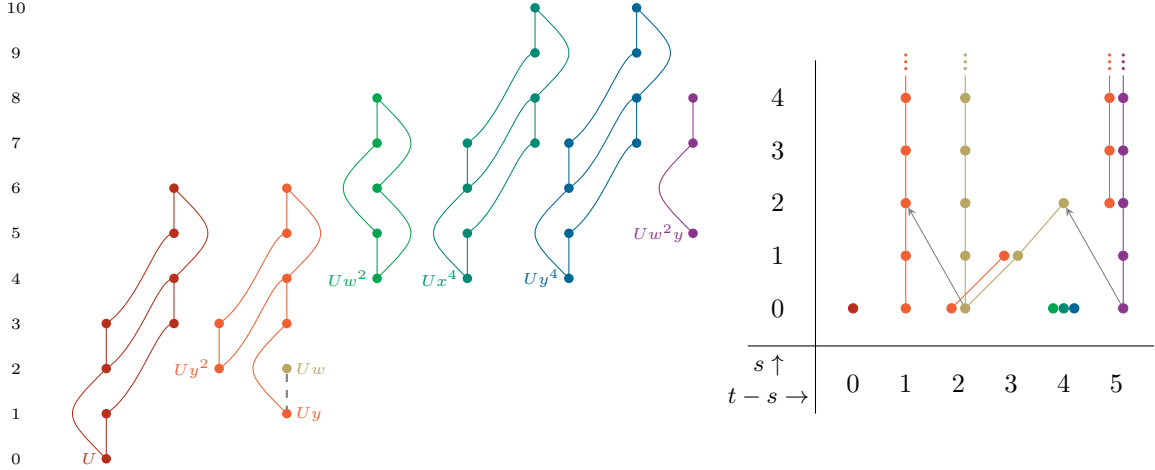

In particular,
\begin{equation}
\label{dihedral0mod4summands}
	\tH^*((BD_{2n})^{2-V_\lambda};\Z/2)\cong \textcolor{BrickRed}{\cA(1)}\oplus
		\textcolor{RedOrange}{\Sigma R_2}\oplus
		\textcolor{Goldenrod!67!black}{\Sigma^2 \Z/2}\oplus
		\textcolor{Green}{\Sigma^4 J}\oplus
		\textcolor{PineGreen}{\Sigma^4 \cA(1)}\oplus
		\textcolor{MidnightBlue}{\Sigma^4\cA(1)}\oplus
		\textcolor{Fuchsia}{\Sigma^5\uQ}\oplus P,
\end{equation}
where $P$ is $5$-connected, and we define $\textcolor{RedOrange}{R_2}$, $\textcolor{Green}{J}$, and
$\textcolor{Fuchsia}{\uQ}$ as follows. First, $\textcolor{RedOrange}{R_2}$ is defined to be the kernel of the
augmentation map $\cA(1)\to\Z/2$;\footnote{The $\cA(1)$-module $R_2$ is named the \term{Elephant} in~\cite{BM23}.}
the indecomposable summand in~\eqref{dihedral0mod4summands} isomorphic to $\textcolor{RedOrange}{\Sigma R_2}$ is
generated by $\oU y$ and $\oU y^2$. The \term{Joker} is the $\cA(1)$-module
$\textcolor{Green}{J}\coloneqq\cA(1)/(\Sq^3)$; here it is generated by $\oU w^2$. Finally,
$\textcolor{Fuchsia}{\uQ}\coloneqq\cA(1)/(\Sq^1, \Sq^2\Sq^3)$ and is  called the \term{upside-down question mark};
here it is generated by $\oU w^2y$. For each of these summands $M$ in~\eqref{dihedral0mod4summands},
$\Ext_{\cA(1)}^{s,t}(M,\Z/2)$ is known in the degrees relevant to us -- except for $P$, which is too high-degree to
affect our calculations anyways.
\begin{itemize}
	\item For $\Sigma^k\cA(1)$ there is a single $\Z/2$ in bidegree $s = 0$, $t = k$.
	\item Adams-Priddy~\cite[\S 3]{AP76} first calculated Ext of $\textcolor{RedOrange}{R_2}$,
	$\textcolor{Green}{J}$, and $\textcolor{Fuchsia}{\uQ}$; see~\cite[Figure 29]{BC18} for a picture.
	\item Liulevicius~\cite[Theorem 3]{Liu62} first calculated the Ext groups of
	$\textcolor{Goldenrod!67!black}{\Z/2}$; for a picture, see~\cite[Figure 20]{BC18}.
\end{itemize}
Put these together to obtain the $E_2$-page as in \cref{a1moddih0mod4}, right. \Cref{torsion_k_theory} tells us the
$E_\infty$-page is torsion, so there must be nonzero differentials in the range shown, though not necessarily the
$d_2$s pictured.

The first nonzero differential is a $d_r$ from the $2$-line to the $1$-line; by $h_0$-equivariance, it kills the
entire yellow tower in the $2$-line. Since a $d_r$ differential decreases $t-s$ by $1$ and increases $s$ by $r$, on
the $E_{r+1}$-page, the $2$-line contains only the first $r$ summands of the orange tower, and the $3$-line
contains only the orange $\textcolor{RedOrange}{\Z/2}$ summand in degree $s = 0$. There can be no further
differentials to or from the $1$- or $2$-lines, so we obtain $\Z/2^r$ in degree $1$ and $\Z/2$ in degree $2$.
\begin{lem}
\label{first_dih_differential}
$2^r$ is the largest power of $2$ dividing $n$, i.e.\ $\tOmega_1^\Spin((BD_{2n})^{2-V_\lambda})\cong\Z/n$.
\end{lem}
\begin{proof}
The May-Milgram theorem~\cite{MM81} identifies Adams spectral sequence differentials between towers with Bockstein
spectral sequence differentials. What it means here is that the lemma statement is equivalent to the
statement that the Bockstein map $\beta\colon H^1(\bl;\Z/2^r)\to H^2(\bl;\Z/2)$ associated to the short exact
sequence
\begin{equation}
	\shortexact{\Z/2}{\Z/2^{r+1}}{\Z/2^r}{}
\end{equation}
carries a preimage of $\oU y$ to $\oU w$. Both of these classes are in the image of the pullback map induced by
$(B\Z/n)^{2-V_\lambda}\to(BD_{2n})^{2-V_\lambda}$, and the Bockstein is natural with respect to the Thom
isomorphism, so we just have to check this in the cohomology of $B\Z/n$, where it is true~\cite{Cam17, DL20b}.
\end{proof}

The next differential that might be nonzero, and which is the only possibly nonzero differential to or from an
element of degree $3$ or $4$, is $d_2\colon E_2^{0,5}\to E_2^{2,6}$. If this $d_2 = 0$, there is also an extension
problem in degree $t-s = 4$ of the form
\begin{equation}
\label{D0mod4xtn}
	\shortexact{\textcolor{Goldenrod!67!black}{\Z/2}}{\tOmega_4^\Spin((BD_{2n})^{2-V_\lambda})}{\textcolor{Green}{\Z/2}
	\oplus \textcolor{PineGreen}{\Z/2} \oplus \textcolor{MidnightBlue}{\Z/2}}.
\end{equation}
\begin{lem}
\label{D0mod4_nodiff}
This $d_2$ vanishes, and the extension~\eqref{D0mod4xtn} splits.
\end{lem}
\begin{proof}
We will prove this by mapping to a simpler Adams spectral sequence that has already been studied, as depicted in
\cref{sseq_comparison_dihedral}.

Because $V_\lambda$ is the pullback of the tautological bundle $V_2\to B\O_2$ along $B\lambda\colon BD_{2n}\to
B\O_2$, we obtain a map of Thom spectra $f\colon (BD_{2n})^{2-V_\lambda}\to (B\O_2)^{2-V_2}$; the codomain is often
denoted $\Sigma^2\MTO_2$. Under $f$, our $\oU w\in \tH^2((BD_{2n})^{2-V_\lambda};\Z/2)$ is the pullback of $\oU
w_2\in\tH^2(\Sigma^2\MTO_2)$.

The spin bordism of $\Sigma^2\MTO_2$ is identified with the bordism theory of the group $\Pin^{\tilde c+}\coloneqq
(\Pin^+ \ltimes \Spin_2)/\mu_2$. Invertible field theories for this tangential structure are believed to
correspond to invertible topological phases of Altland-Zirnbauer type AII~\cite[\FHrefllink{page.77}{(9.25)},
\FHrefllink{page.95}{(10.2)}]{FH16}.\footnote{For further discussion, see also Metlitski~\cite{Met15} and
Seiberg-Witten~\cite[\S A.4]{SW16}.}

Several authors study the Adams spectral sequence for $\Omega_*^{\Pin^{\tilde c+}} \cong
\tOmega_*^\Spin(\Sigma^2\MTO_2)$ in low degrees, including Freed-Hopkins~\cite[\FHrefllink{page.97}{Figure 5}, case
$s = -2$]{FH16}, Campbell~\cite[Example 6.10]{Cam17}, and Wan-Wang-Zheng~\cite[\S 6.2.3]{WWZ19}. Their work shows
$\oU w_2\in\tH^2(\Sigma^2\MTO_2;\Z/2)$ generates a $\textcolor{Dandelion}{\Sigma^2\Z/2}$ summand as an
$\cA(1)$-submodule of $\tH^*(\Sigma^2\MTO_2;\Z/2)$, and therefore $f^*$ restricts to an isomorphism from
that $\textcolor{Dandelion}{\Sigma^2\Z/2}$ summand to our $\textcolor{Goldenrod!67!black}{\Sigma^2\Z/2}$ summand generated by
$\oU w$. This means the submodule of the $E_2$-page for $\tOmega_*^\Spin((BD_{2n})^{2-V_\lambda})$ coming from
$\textcolor{Goldenrod!67!black}{\Sigma^2\Z/2}$ maps isomorphically onto the submodule of the $E_2$-page for
$\tOmega_*^\Spin(\Sigma^2\MTO_2)$ coming from the $\textcolor{Dandelion}{\Sigma^2\Z/2}$ generated by $\oU w_2$ --- and
crucially, in that spectral sequence, $E_2^{0,5} \cong 0$. See the commutative diagram of pink arrows in
\cref{sseq_comparison_dihedral}. Thus the image of our $d_2$ under $f$ vanishes, and the map between
these spectral sequences on $E_2^{2,6}$s (the targets of these $d_2$s) is an isomorphism, so our $d_2$ also
vanishes.
\begin{figure}[h!]
\includegraphics[width=0.65\textwidth]{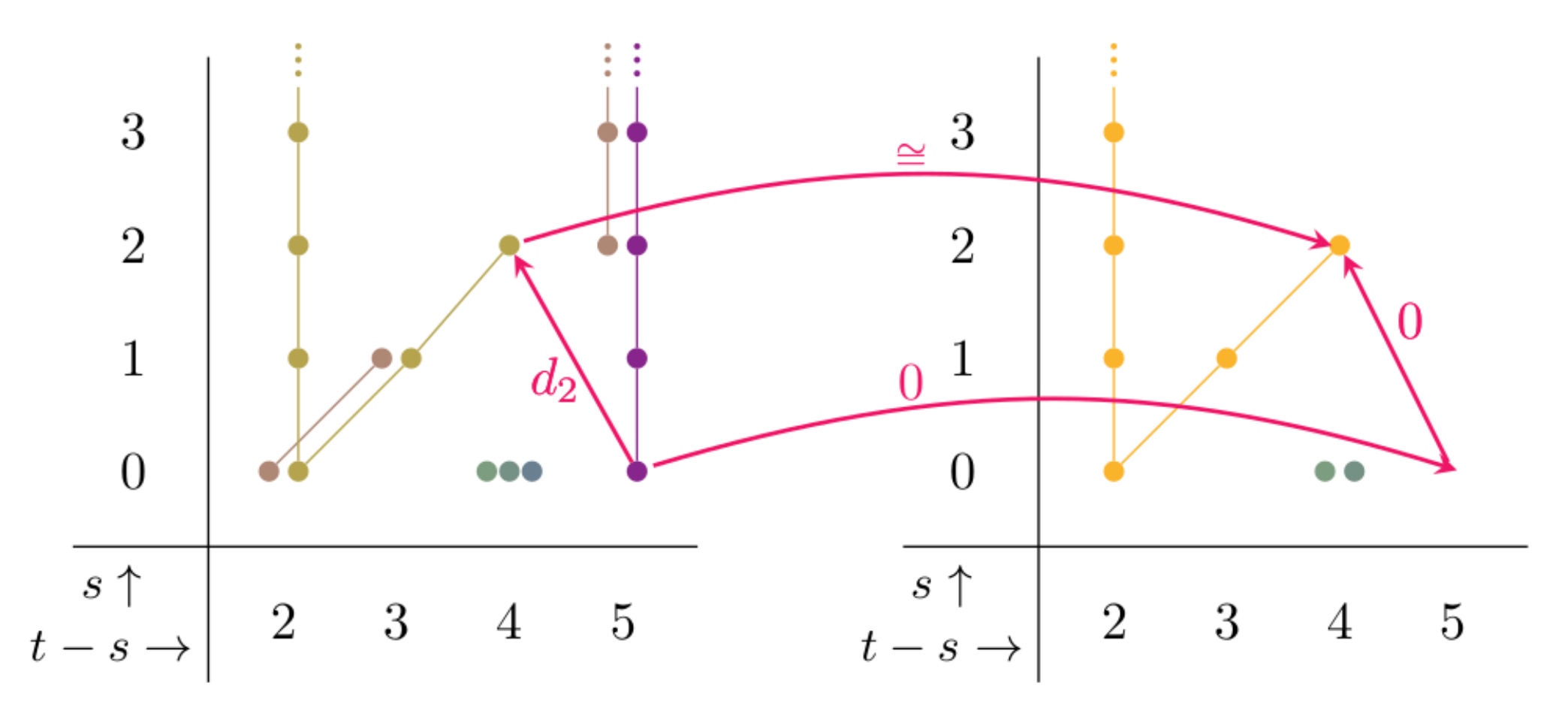}
\caption{The map $(BD_{2n})^{2-V_\lambda}\to \Sigma^2\MTO_2$ induces a map between the Adams spectral sequences computing their
$\ko$-theory groups. We use this in \cref{D0mod4_nodiff} to show the pictured $d_2$ vanishes, as the square of pink
arrows in the above figure is commutative. The right-hand side of this figure, which displays
$\Ext(\wH^*(\Sigma^2\MTO_2;\Z/2))$, is adapted from Campbell~\protect\cite[Figure 6.9]{Cam17}.}
\label{sseq_comparison_dihedral}
\end{figure}

Now suppose~\eqref{D0mod4xtn} does not split; then, there are elements $x, y\in \tOmega_4^\Spin((BD_{2n})^{2-V_\lambda})$ such that $x
= 2y$ and the image of $x$ image in the $E_\infty$-page of the Adams spectral sequence is the nonzero element of
$E_\infty^{2,6}\cong \textcolor{Goldenrod!67!black}{\Z/2}$. Then $f$ maps this
$\textcolor{Goldenrod!67!black}{\Z/2}$ isomorphically onto a $\Z/2$ in the $E_\infty$-page for $\Sigma^2\MTO_2$, so
$f_*(x)\ne 0$. But $\Omega_4^{\Pin^{\tilde c+}}\cong (\Z/2)^{\oplus 3}$~\cite[\FHrefllink{page.91}{Theorem
9.87}]{FH16}, so no matter where $y$ maps to, $2y = x\mapsto 0$, which is a problem.
\end{proof}
We have thus determined $\tOmega_d^\Spin((BD_{2n})^{2-V_\lambda})_2^\wedge$ for $d = 3,4$, so we are done.
\end{proof}

%
%
%
\subsubsection{Class D, spin-$1/2$ case}
\label{mixed_dihedral}
\begin{lem}
\label{dih_mixed_not_pinm}
$V_\lambda$ is not \pinm.
\end{lem}
\begin{proof}
For $n$ even, this follows by pulling back along $BC_n\to BD_{2n}$: we saw in \S\ref{s_rotations} that the pullback
is not spin, so $V_\lambda$ cannot be \pinm. For $n$ odd, pull back along the map $B\Z/2\to BD_{2n}$ induced by the
inclusion of a reflection; the pullback is not \pinm, so neither is $V_\lambda$.
\end{proof}
Therefore by \cref{shear_D_thm}, we consider $(BD_{2n})^{\Det(V_\lambda)-1}$.
\begin{prop}
\label{mixed_dih_odd}
For $n$ odd, the first few spin bordism groups of $(BD_{2n})^{\Det(V_\lambda)-1}$ are
\begin{align*}
	\tOmega_0^\Spin((BD_{2n})^{\Det(V_\lambda)-1}) &\cong \Z/2\\
	\tOmega_1^\Spin((BD_{2n})^{\Det(V_\lambda)-1}) &\cong \Z/2n\\
	\tOmega_2^\Spin((BD_{2n})^{\Det(V_\lambda)-1}) &\cong \Z/8\\
	\tOmega_3^\Spin((BD_{2n})^{\Det(V_\lambda)-1}) &\cong 0\\
	\tOmega_4^\Spin((BD_{2n})^{\Det(V_\lambda)-1}) &\cong 0,
\end{align*}
and $\tOmega_5^\Spin((BD_{2n})^{\Det(V_\lambda)-1})$ is torsion.
\end{prop}
\begin{proof}
To compute the $2$-torsion subgroups of these bordism groups, apply \cref{odd_dihedral_equiv} with
$\Det(V_\lambda)-1$ get a $2$-primary stable equivalence $(BD_{2n})^{\Det(V_\lambda)-1}\simeq (B\Z/2)^{\sigma-1}$,
then~\eqref{pinmsplitting} to get $\tOmega_*^\Spin((B\Z/2)^{1-\sigma})\cong\Omega_*^{\Pin^-}$. Low-degree \pinm
bordism groups are calculated in~\cite{ABP69, KT90}. For the odd-torsion subgroups, use \cref{twisted_dihedral_SO}.
\end{proof}
\begin{thm}
\label{mixed_dih_2mod4}
When $n\equiv 2\bmod 4$, the first few bordism groups of $(BD_{2n})^{\Det(V_\lambda)-1}$ are
\begin{align*}
	\tOmega_0^\Spin((BD_{2n})^{\Det(V_\lambda)-1}) &\cong \Z/2\\
	\tOmega_1^\Spin((BD_{2n})^{\Det(V_\lambda)-1}) &\cong \Z/n\oplus\Z/2\\
	\tOmega_2^\Spin((BD_{2n})^{\Det(V_\lambda)-1}) &\cong \Z/8\oplus\Z/4\\
	\tOmega_3^\Spin((BD_{2n})^{\Det(V_\lambda)-1}) &\cong \Z/2\oplus\Z/2\\
	\tOmega_4^\Spin((BD_{2n})^{\Det(V_\lambda)-1}) &\cong 0,
\end{align*}
and $\tOmega_5^\Spin((BD_{2n})^{\Det(V_\lambda)-1})$ is torsion.
\end{thm}
\begin{proof}
We establish a $2$-primary isomorphism $\tOmega_*^\Spin((BD_{2n})^{\Det(V_\lambda)-1})\cong \Omega_*^{\Pin^-}(B\Z/2)$, so the free and
$2$-torsion part of the spin bordism groups of $(BD_{2n})^{\Det(V_\lambda)-1}$ are isomorphic to the \pinm bordism groups of $B\Z/2$.
Once we finish this, we use work of Guo-Ohmori-Putrov-Wan-Wang~\cite[\S 7.2.1]{GOPWW18} computing
$\Omega_k^{\Pin^-}(B\Z/2)$ in degrees $5$ and below to get the $2$-primary part; for the odd-primary torsion, we
use \cref{twisted_dihedral_SO} as usual.
\begin{lem}
The inclusion $i\colon \Z/2\times\Z/2\inj D_{2n}$ given by a reflection and a half-turn induces a $2$-primary
equivalence of Thom spectra $(B(\Z/2\times\Z/2))^{i^*\Det(V_\lambda)-1}\overset\simeq\to
(BD_{2n})^{\Det(V_\lambda)-1}$.
\end{lem}
\begin{proof}
The map $Bi\colon B(\Z/2\times\Z/2)\to BD_{2n}$ induces an equivalence on mod $2$ cohomology, and therefore by the
Thom isomorphism theorem also induces an equivalence on the mod $2$ cohomology of the Thom spectra in question.
This suffices by the stable Whitehead theorem.
%
\end{proof}
The stable bundle $i^*\Det(V_\lambda)\to B(\Z/2\times\Z/2)$ splits as an exterior direct sum
$\sigma\boxplus\underline 0$, where $\sigma\to B\Z/2$ is the tautological line bundle. Therefore the Thom spectrum
also splits: $(B(\Z/2\times\Z/2))^{i^*\Det(V_\lambda) - 1}\simeq (B\Z/2)^{\sigma-1}\wedge (B\Z/2)_+$. Therefore
by~\eqref{pinmsplitting},
\begin{equation}
	\tOmega_*^{\Spin}((BD_{2n})^{\Det(V_\lambda)-1}) \cong \tOmega_*^\Spin(B\Z/2)^{\sigma-1}\wedge (B\Z/2)_+)\cong
	\Omega^{\Pin^-}(B\Z/2).
	\qedhere
\end{equation}
\end{proof}

Finally, let $n\equiv 0\bmod 4$.
%
%
Recall $H^*(BD_{2n};\Z/2)\cong\Z/2[x, y, w]/(xy+y^2)$ with $\abs x
= \abs y = 1$ and $\abs w = 2$, so $\Sq(x) = x+x^2$ and $\Sq(y) = y+y^2$, and from \cref{0mod4_w_lem}, $\Sq(w) = w
+ wx + w^2$.  The Stiefel-Whitney classes of $\Det(V_\lambda)$ tell us by way of \cref{wu_f} that if $\oU$ is the
Thom class, $\Sq^1(\oU) = \oU x$ and $\Sq^2(\oU) = 0$ in the cohomology of $(BD_{2n})^{\Det(V_\lambda)-1}$.
\begin{thm}
\label{spinless_D0mod4_thm}
For $n\equiv 0\bmod 4$, the first few bordism groups of $(BD_{2n})^{\Det(V_\lambda)-1}$ are
\begin{align*}
	\tOmega_0^\Spin((BD_{2n})^{\Det(V_\lambda)-1}) &\cong \Z/2\\
	\tOmega_1^\Spin((BD_{2n})^{\Det(V_\lambda)-1}) &\cong \Z/n\oplus\Z/2\\
	\tOmega_2^\Spin((BD_{2n})^{\Det(V_\lambda)-1}) &\cong \Z/8\oplus\Z/4\\
	\tOmega_3^\Spin((BD_{2n})^{\Det(V_\lambda)-1}) &\cong \Z/2\oplus\Z/2\\
	\tOmega_4^\Spin((BD_{2n})^{\Det(V_\lambda)-1}) &\cong 0,
\end{align*}
and $\tOmega_5^\Spin((BD_{2n})^{\Det(V_\lambda)-1})$ is torsion.
\end{thm}
\begin{proof}
First, by \cref{twisted_dihedral_SO}, the only odd-primary torsion in $\tOmega_k^\Spin((BD_{2n})^{\Det(V_\lambda)-1})$ for $k\le 4$ is in
degree $1$. Draw the Atiyah-Hirzebruch spectral sequence
\begin{equation}
	E^2_{p,q} = \widetilde H_p((BD_{2n})^{\Det(V_\lambda)-1}; \Omega_q^\Spin) \Longrightarrow \tOmega_{p+q}^\Spin(X).
\end{equation}
After applying the Thom isomorphism, this needs as input $H_*(BD_{2n};\Z_x)$ and $H_*(BD_{2n};\Z/2)$. The former
can be determined using Handel's calculation~\cite[Theorem 5.8]{Han93} of $H^*(BD_{2n}; \Z_x)$, and the latter can
be determined from \cref{dihmod2coh}; in both cases use the universal coefficient theorem to pass from homology to
cohomology. Since $E^2_{1,0}\cong\Z/n$ and $E^2_{0,1}\cong\Z/2$, there are three options for
$\tOmega_1^\Spin((BD_{2n})^{\Det(V_\lambda)-1})$: $\Z/n$, $\Z/n\oplus\Z/2$, or $\Z/2n$. We will learn which one is correct in our analysis of
the $2$-primary part below.

For the $2$-primary part, we use the Adams spectral sequence as usual. By \cref{det_splitting}, a choice of a
reflection in $D_{2n}$ induces a splitting
\begin{equation}
	(BD_{2n})^{\Det(V_\lambda)-1} \overset\simeq\longrightarrow (B\Z/2)^{\sigma-1} \vee M_n,
\end{equation}
such that the map $\wH^*(M_n;\Z/2)\to \wH^*((BD_{2n})^{\Det(V_\lambda)-1};\Z/2)$ is injective with image complementary to the subspace
spanned by $\set{Ux^i\mid i\ge 0}$. This splits $\widetilde E_*((BD_{2n})^{\Det(V_\lambda)-1})$ for any generalized homology theory $E$; we
want spin bordism, and we focus on $\tOmega_*^\Spin(M_n)$, adding in the summands arising from
$\tOmega_*^\Spin((B\Z/2)^{\sigma-1})\cong\Omega_*^{\Pin^-}$ at the end. The $\cA(1)$-module structure on
$\wH^*(M_n;\Z/2)$ is determined by its image in $\wH^*((BD_{2n})^{\Det(V_\lambda)-1};\Z/2)$, which we know by \cref{wu_f}.
Using this, we draw this $\cA(1)$-module structure in
\cref{spinful0mod4a1}, left.

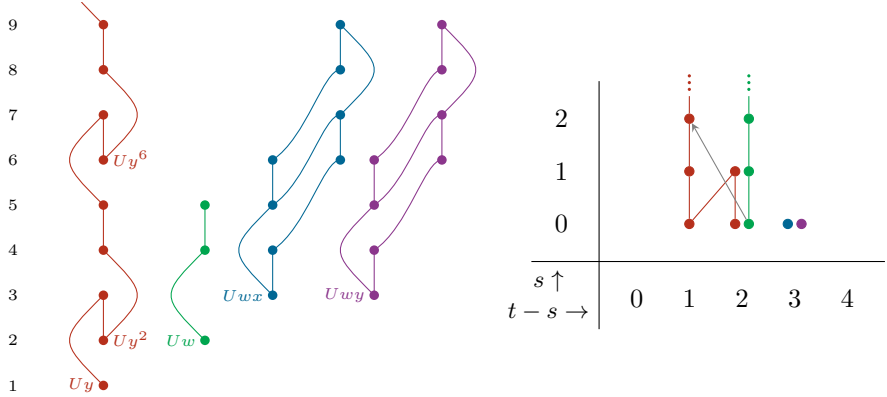
\begin{figure}[h!]
\begin{subfigure}[c]{0.35\textwidth}
\begin{tikzpicture}[scale=0.6, every node/.style = {font=\tiny}]
	\foreach \y in {1, ..., 9} {
		\node at (-2, \y) {$\y$};
	}
	
	\begin{scope}[BrickRed]
		\foreach \y in {3, 4, 5, 7, 8, 9}  {
			\tikzpt{0}{\y}{}{};
		}
		\tikzpt{0}{1}{$\oU y$}{};
		\tikzptR{0}{2}{$\oU y^2$}{};
		\tikzptR{0}{6}{$\oU y^6$}{};
		\foreach \y in {2, 4, ..., 8} {
			\sqone(0, \y);
		}
		\foreach \y in {1, 5} {
			\sqtwoL(0, \y);
		}
		\begin{scope}
			\clip (-0.5, 8.5) rectangle (0.5, 9.5);
			\sqtwoL(0, 9);
		\end{scope}
		\foreach \y in {2, 6} {
			\sqtwoR(0, \y, );
		}
	\end{scope}
	\begin{scope}[Green]
		\SpanishQnMark{2.25}{2}{$\oU w$};
	\end{scope}
	\begin{scope}[MidnightBlue]
		\Aone{3.75}{3}{$\oU wx$}{};
	\end{scope}
	\begin{scope}[Fuchsia]
		\Aone{6}{3}{$\oU wy$}{};
	\end{scope}
\end{tikzpicture}
\end{subfigure}
\qquad
\begin{subfigure}[c]{0.4\textwidth}
\begin{sseqdata}[name=AdamsD0mod4spinful, classes=fill, xrange={0}{4}, yrange={0}{2}, scale=0.7, Adams grading,
	>=stealth, x label = {$\displaystyle{s\uparrow \atop t-s\rightarrow}$},
	x label style = {font = \small, xshift = -17ex, yshift=3.7ex}]
	\begin{scope}[BrickRed]
		\class(1, 0)\AdamsTower{}
		\class(2, 0)
		\class(2, 1)\structline
		\structline(1, 0)(2, 1)

		\class(5, 2)\AdamsTower{}
		\class(6, 3)
		\structline(5, 2)(6, 3)
	\end{scope}
	\begin{scope}[white] 
		\class(2, 2)\AdamsTower{}
		\class(5, 0)
		\class(5, 1)
	\end{scope}
	\begin{scope}[Green]
		\class(2, 0)\AdamsTower{}
	\end{scope}
	\class[MidnightBlue](3, 0)
	\class[Fuchsia](3, 0)
	\begin{scope}[Fuchsia]
		\class(5, 0)\AdamsTower{}
		\class(6, 1)
		\structline(5, 0, -1)(6, 1)
	\end{scope}
	\d[gray]2(2, 0, -1)
\end{sseqdata}
\printpage[name=AdamsD0mod4spinful, page=2]
\end{subfigure}
\caption{Left: the $\cA(1)$-module structure on $\tH^*(M_n;\Z/2)$ in low degrees. The pictured summand contains all
elements in degrees $4$ and below. Right: the Ext of this module, which is the $E_2$-page of the Adams spectral
sequence converging to $\widetilde{\ko}_*(M_n)$. See the proof of \cref{spinless_D0mod4_thm} for more information.}
\label{spinful0mod4a1}
\end{figure}

As $\cA(1)$-modules,
\begin{equation}
\label{a1_str_mixed_dih_0mid4}
	\wH^*(M_n;\Z/2)\cong \textcolor{BrickRed}{\Sigma R_1} \oplus
		\textcolor{Green}{\Sigma^2\uQ} \oplus
		\textcolor{MidnightBlue}{\Sigma^3\cA(1)} \oplus
		\textcolor{Fuchsia}{\Sigma^3\cA(1)} \oplus P,
\end{equation}
where $P$ is $4$-connected; we will see below that the $4$-line is empty, so there are no nonzero differentials
from $\Ext(P)$ to anything we care about. Here $\textcolor{BrickRed}{\Sigma
R_1}$ is the indecomposable summand containing $\oU y$. For the $\Sigma^k\cA(1)$ summands, we know the Ext; for
$\textcolor{BrickRed}{\Sigma R_1}$, see~\cite[Figure 26]{BC18}, and for $\textcolor{Green}{\Sigma^2\uQ}$,
see~\cite[Figure 29]{BC18}. Assembling these, we display the $E_2$-page for $t-s\le 5$ in \cref{spinful0mod4a1},
right. \Cref{torsion_k_theory} implies $\tOmega_5^\Spin((BD_{2n})^{\Det(V_\lambda)-1})$ is torsion, as claimed, and that there must be a
differential $d_r$ from the infinite tower in topological degree $2$ to the infinite tower in topological degree
$1$, though it might not be the $d_2$ pictured.\footnote{In fact, $r$ is the largest number such that $2^r\mid n$.
Like in the proof of \cref{first_dih_differential}, one can deduce this using the Bockstein from $\oU y$ to $\oU w$
and the May-Milgram theorem.} Margolis' theorem and $h_0$-equivariance rule out any other nonzero differentials to
or from elements with $t-s\le 4$. Therefore in this range, $E_{r+1} = E_\infty$. The infinite tower in topological
degree $2$ is killed by the differential, as are all but $r$ of the $\textcolor{BrickRed}{\Z/2}$ summands of the
infinite tower in topological degree $1$. The first few $2$-completed spin bordism groups of $M_n$ are therefore
$\Z/2^r$ in degree $1$, $\Z/4$ in degree $2$, $\Z/2\oplus\Z/2$ in degree $3$, and $0$ in degrees $0$ and $4$.

Finally, we add in the \pinm bordism summands as computed in~\cite{ABP69, KT90}: a $\Z/2$ in degrees $0$ and $1$, a
$\Z/8$ in degree $2$, and $0$ otherwise. In particular, since the $2$-torsion subgroup of $\tOmega_1^\Spin((BD_{2n})^{\Det(V_\lambda)-1})$ is
of the form $\Z/2\oplus\Z/2^r$, $\tOmega_1^\Spin((BD_{2n})^{\Det(V_\lambda)-1})\cong \Z/n\oplus\Z/2$.
\end{proof}

\subsubsection{Class A, spinless case}
\label{spinc_dihedral}
In this case, \cref{shear_A_thm} asks us to compute the \spinc bordism of $(BD_{2n})^{2-V_\lambda}$.
%
\begin{thm}
\label{dih_spinc_odd}
\label{spinc_mixed_odd_thm}
For $n$ odd, the first few spin bordism groups of $(BD_{2n})^{2-V_\lambda}$ are
\begin{align*}
	\tOmega_0^\Spinc((BD_{2n})^{2-V_\lambda}) &\cong \Z/2\\
	\tOmega_1^\Spinc((BD_{2n})^{2-V_\lambda}) &\cong \Z/n\\
	\tOmega_2^\Spinc((BD_{2n})^{2-V_\lambda}) &\cong \Z/4\\
	\tOmega_3^\Spinc((BD_{2n})^{2-V_\lambda}) &\cong \Z/n\\
	\tOmega_4^\Spinc((BD_{2n})^{2-V_\lambda}) &\cong \Z/8\oplus\Z/2,
\end{align*}
and $\tOmega_5^\Spinc((BD_{2n})^{2-V_\lambda})$ is torsion.
\end{thm}
\begin{proof}
To compute the $2$-torsion subgroups of these bordism groups, apply \cref{odd_dihedral_equiv} with $2-V_\lambda$ to
get a $2$-primary stable equivalence $(BD_{2n})^{2-V_\lambda}\simeq (B\Z/2)^{1-\sigma}$, then~\eqref{pincsplitting}
to get $\tOmega_*^{\Spinc}((B\Z/2)^{1-\sigma})\cong\Omega_*^{\Pin^c}$. The \pinc bordism groups we need are
calculated by Bahri-Gilkey~\cite{BG87a, BG87b}. For the odd-torsion subgroups, use \cref{twisted_dihedral_SO_U}.
\end{proof}
\begin{thm}
\label{dih_spinc_2mod4}
Let $n\equiv 2\bmod 4$; then the low-degree \spinc bordism of $(BD_{2n})^{2-V_\lambda}$ is
\begin{align*}
	\tOmega_0^\Spinc((BD_{2n})^{2-V_\lambda}) &\cong \Z/2\\
	\tOmega_1^\Spinc((BD_{2n})^{2-V_\lambda}) &\cong \Z/n\\
	\tOmega_2^\Spinc((BD_{2n})^{2-V_\lambda}) &\cong (\Z/2)^{\oplus 2}\\
	\tOmega_3^\Spinc((BD_{2n})^{2-V_\lambda}) &\cong \Z/2n\\
	\tOmega_4^\Spinc((BD_{2n})^{2-V_\lambda}) &\cong (\Z/2)^{\oplus 4},
\end{align*}
and $\tOmega_5^\Spinc((BD_{2n})^{2-V_\lambda})$ is torsion.
\end{thm}
\begin{proof}
First, \cref{twisted_dihedral_SO_U} computes the odd-torsion subgroups: a $\Z/n$ in degrees $1$ and $3$, and
nothing else below degree $5$.

To compute the $2$-primary information we use the Adams spectral sequence over $\cE(1)$, which converges to
$\widetilde{\ku}_*((BD_{2n})^{2-V_\lambda})$, together with Anderson-Brown-Peterson's isomorphism
$\tOmega_n^{\Spinc}((BD_{2n})^{2-V_\lambda})\overset\cong\to \widetilde{\ku}_n((BD_{2n})^{2-V_\lambda})\oplus\widetilde{\ku}_{n-4}((BD_{2n})^{2-V_\lambda})$ for $n\le
7$~\cite{ABP67}.

The $\cA(1)$-module structure on $\tH^*((BD_{2n})^{2-V_\lambda};\Z/2)$ that we calculated in~\eqref{dihedral2mod4summands} and
displayed in \cref{a1moddih2mod4}, left, determines the $\cE(1)$-module structure: as $\cE(1)$-modules,
$\cA(1)\cong\cE(1)\oplus \Sigma^2\cE(1)$~\cite[\S 4.7.2]{Bay94}. Therefore
\begin{equation}
	\tH^*((BD_{2n})^{2-V_\lambda};\Z/2) \cong\textcolor{BrickRed!80!black}{\cE(1)} \oplus
		\textcolor{RedOrange}{\Sigma R_0} \oplus
		\textcolor{BrickRed!80!white}{\Sigma^2\cE(1)} \oplus
		\textcolor{Green!80!black}{\Sigma^2\cE(1)} \oplus
		\textcolor{Green!80!white}{\Sigma^4\cE(1)}\oplus
		\textcolor{MidnightBlue!80!black}{\Sigma^4\cE(1)} \oplus
		\textcolor{Fuchsia!80!black}{\Sigma^4\cE(1)} \oplus P,
\end{equation}
where $P$ is $5$-connected; we draw a picture of this $\cE(1)$-module in \cref{e1_d2n_2mod4}, left.
\begin{figure}[h!]
\begin{subfigure}[c]{0.55\textwidth}
\begin{tikzpicture}[scale=0.6, every node/.style = {font=\tiny}] 
\foreach \y in {0, ..., 8} {
	\node at (-1.5, \y) {$\y$};
}
\begin{scope}[BrickRed!80!black]
	\Eone{0}{0}{$\oU$}{}
\end{scope}
\begin{scope}[BrickRed!80!white]
	\Eone{4}{2}{$\oU x^2$}
\end{scope}
\begin{scope}[RedOrange]
	\tikzpt{2}{1}{$\oU y$}{};
	\foreach \y in {2, ..., 8} {
		\tikzpt{2}{\y}{}{};
	}
	\foreach \y in {1, 3, ..., 7} {
		\sqone(2, \y);
	}
	\begin{scope}
		\clip (1, 0.8) rectangle (3, 8.2);
		\foreach \y in {1, 5} {

			\qoneR(2, \y);
			\qoneL(2, \y+2);
		}
	\end{scope}
\end{scope}
\begin{scope}[Green!80!black]
	\Eone{6}{2}{$\oU y^2$}{}
\end{scope}
\begin{scope}[Green!80!white]
	\Eone{8.5}{4}{$\oU x^2y^2$}
\end{scope}
\begin{scope}[MidnightBlue!80!black]
	\Eone{10.5}{4}{$\oU x^4$}
\end{scope}
\begin{scope}[Fuchsia!80!black]
	\Eone{12.5}{4}{$\oU xy^3$}
\end{scope}
\end{tikzpicture}
\end{subfigure}
\begin{subfigure}[c]{0.4\textwidth}
	\begin{sseqdata}[name=D2mod4E1, classes=fill, scale=0.7, xrange={0}{5}, yrange={0}{3},
	x label = {$\displaystyle{s\uparrow \atop t-s\rightarrow}$},
	x label style = {font = \small, xshift = -19.75ex, yshift=3.7ex}]
	\class[BrickRed!80!black](0, 0)
	\class[BrickRed!80!white](2, 0)
	\begin{scope}[RedOrange]
		\class(1, 0)
		\foreach \x [count=\ymax] in {3, 5, 7, 9} {
			\class(\x, 0)
			\foreach \y in {1, ..., \ymax} {
				\class(\x, \y)\structline
				\structline[RedOrange!40!white](\x, \y)(\x - 2, \y - 1)
			}
		}
	\end{scope}
	\class[Green!80!black](2, 0)
	\class[Green!80!white](4, 0)
	\class[MidnightBlue!80!black](4, 0)
	\class[MidnightBlue!80!white](6, 0)
	\class[Fuchsia!80!black](4, 0)
	\class[Fuchsia!80!white](6, 0)
	\class[MidnightBlue](6, 0)
	\class[Fuchsia](6, 0)
	\end{sseqdata}
	\printpage[name=D2mod4E1, page=2]
\end{subfigure}
\caption{Left: the $\cE(1)$-module structure on $\tH^*((BD_{2n})^{2-V_\lambda};\Z/2)$, $n\equiv 2\bmod 4$, in low
degrees. The pictured submodule contains all elements in degrees $5$ and below. Right: the Adams $E_2$-page
computing $\widetilde{\ku}_*((BD_{2n})^{2-V_\lambda})$.}
\label{e1_d2n_2mod4}
\end{figure}
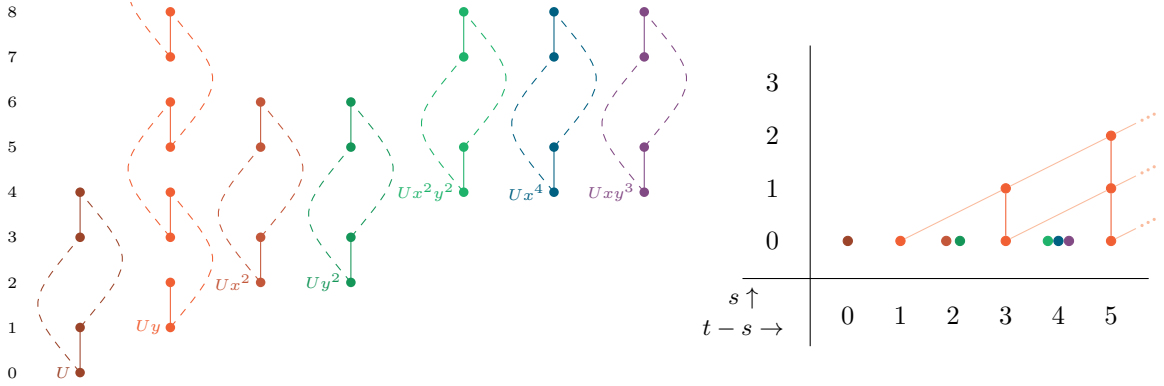

Next Ext. For $\Sigma^k\cE(1)$, there is a unique $\Z/2$ summand, in degree $s = 0$, $t = k$; for
$\textcolor{RedOrange}{\Sigma R_0}$, we must work a little harder.
\begin{prop}
\label{the_ext_of_H}
There is an isomorphism of $H^{*,*}(\cE(1))$-modules
\begin{equation}
	\Ext_{\cE(1)}^{s,t}(\textcolor{RedOrange}{R_0}, \Z/2) \cong \Z/2[a_0, a_1, a_2,\dotsc]/(h_0a_0 = 0, v_1a_i =
	h_0a_{i+1}: i\ge 0),
\end{equation}
with $a_i\in\Ext^{0,2i}$.
\end{prop}
We draw this $\cE(1)$-module in \cref{simpler_ext_H}, right. Dhankhar-Field-Nigam-Quigley-Yang~\cite[Figure
2]{DFNQY25} give an independent proof of the $\Z/2[h_0]$-module structure on $\Ext_{\cE(1)}(R_0)$.
\begin{proof}
Our proof uses as input $\Ext_{\cE(1)}(N_1)$, where $N_1$ is defined to be the $\cA$-module $\Sigma^{-1}\widetilde
H^*(\RP^2;\Z/2)$, with two $\Z/2$ summands connected by a $\Sq^1$; this in turn defines its $\cA(1)$- and
$\cE(1)$-module structures. Davis-Mahowald~\cite[\S 2]{DM81} calculate $\Ext_{\cE(1)}(N_1)$ as a graded vector
space but we also need its $H^{*,*}(\cE(1))$-module structure.

Let $\ang{Q_1}\subset\cE(1)$ denote the subalgebra generated by $Q_1$, which is a two-dimensional vector space over
$\Z/2$. As $\cE(1)$-modules, $N_1\cong\cE(1)\otimes_{\ang{Q_1}} \Z/2$, so by the change-of-rings
theorem~\eqref{change_of_rings}, there are isomorphisms of $H^{*,*}(\cE(1))$-modules
\begin{equation}
\label{N1extcalc}
	\Ext_{\cE(1)}(N_1)\cong\Ext_{\ang{Q_1}}(\Z/2)\cong\Z/2[v_1],
\end{equation}
with $v_1\in\Ext_{\cE(1)}^{1,3}(N_1, \Z/2)$. The rightmost isomorphism in~\eqref{N1extcalc} uses Koszul
duality~\cite[Remark 4.5.4]{BC18}, which applies because $\ang{Q_1}$ is an exterior algebra.

Now for $\textcolor{RedOrange}{R_0}$, we use the extension of $\cE(1)$-modules
\begin{equation}
\label{recursive_H_SES}
	\shortexact{\textcolor{RubineRed}{\Sigma^2 R_0}}{\textcolor{RedOrange}{R_0}}{\textcolor{Periwinkle}{N_1}},
\end{equation}
drawn in \cref{simpler_ext_H}, left.
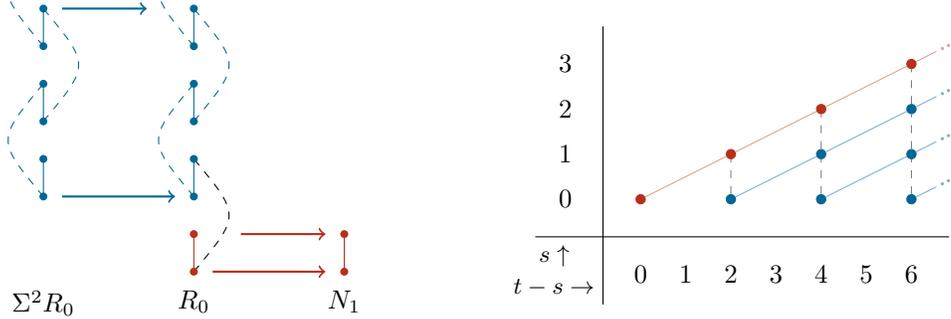
\begin{figure}[h!]
\centering
\begin{subfigure}[c]{0.4\textwidth}
\begin{tikzpicture}[scale=0.5]
\qoneR(0, 0);
\begin{scope}[RubineRed] 
\foreach \x in {-4, 0} {
	\foreach \y in {2, ..., 7} {
		\tikzpt{\x}{\y}{}{};
	}
	\foreach \y in {2, 4, 6} {
		\sqone(\x, \y);
	}
	\qoneL(\x, 2);
	\qoneR(\x, 4);
	\begin{scope}
		\clip (\x - 1, 5.9) rectangle (\x + 1, 7.25);
		\qoneL(\x, 6);
	\end{scope}
}
\draw[thick, ->](-3.5, 2) -- (-0.5, 2);
\draw[thick, ->](-3.5, 7) -- (-1.25, 7);
\end{scope}
\begin{scope}[Periwinkle] 
\foreach \x in {0, 4} {
	\tikzpt{\x}{0}{}{};
	\tikzpt{\x}{1}{}{};
	\sqone(\x, 0);

}
\draw[thick, ->](0.5, 0) -- (3.5, 0);
\draw[thick, ->](1.25, 1) -- (3.5, 1);
\end{scope}
\node[below] at (-4, -0.25) {$\Sigma^2 R_0$};
\node[below] at (0, -0.25) {$R_0$};
\node[below] at (4, -0.25) {$N_1$};
\end{tikzpicture}
\end{subfigure}
\begin{subfigure}[c]{0.4\textwidth}
\begin{sseqdata}[name=ext_H_recurse, classes=fill, xrange={0}{6}, yrange={0}{3}, scale=0.6,
x label = {$\displaystyle{s\uparrow \atop t-s\rightarrow}$},
class labels = { below = 0.07em, font=\small },
x label style = {font = \small, xshift = -19.5ex, yshift=3ex}]
	\begin{scope}[Periwinkle] 
		\class(0, 0)
		\class(2, 1)
			\structline[Periwinkle!40!white](0, 0)(2, 1)
		\class(4, 2)
			\structline[Periwinkle!40!white](2, 1)(4, 2)
		\class(6, 3)
			\structline[Periwinkle!40!white](4, 2)(6, 3)
		\class(8, 4)
			\structline[Periwinkle!40!white](6, 3)(8, 4)
		\classoptions["a_0"](0, 0)
	\end{scope}
	\begin{scope}[RubineRed] 
		\class(2, 0)
		\class(4, 0)
		\class(4, 1)\structline[gray, dashed]
		\structline[RubineRed!40!white](2, 0)(4, 1)
		
		\class(6, 0)
		\class(6, 1)\structline[gray, dashed]
		\class(6, 2)\structline[gray, dashed]
		\structline[RubineRed!40!white](4, 0)(6, 1)
		\structline[RubineRed!40!white](4, 1)(6, 2)
		
		\class(8, 1)
		\class(8, 2)
		\class(8, 3)
		\structline[RubineRed!40!white](6, 0)(8, 1)
		\structline[RubineRed!40!white](6, 1)(8, 2)
		\structline[RubineRed!40!white](6, 2)(8, 3)
		\classoptions["a_1"](2, 0)
		\classoptions["a_2"](4, 0)
		\classoptions["a_3"](6, 0)
	\end{scope}
	\structline[gray, dashed](2, 0)(2, 1)
	\structline[gray, dashed](4, 1)(4, 2)
	\structline[gray, dashed](6, 2)(6, 3)
\end{sseqdata}
\printpage[name=ext_H_recurse, page=2]
\end{subfigure}
\caption{Left: the extension~\eqref{recursive_H_SES}. Right: the long exact sequence it induces of Ext groups. See
the proof of \cref{the_ext_of_H} for why the long exact sequence looks like this; the key feature is that there are
no elements in odd topological degree, so all boundary maps vanish. The dashed lines are $h_0$-extensions which are
not implied by the long exact sequence, but are shown in the proof of \cref{the_ext_of_H}.}
\label{simpler_ext_H}
\end{figure}

At first, all we know is $\Ext(N_1)$. Because this lives solely in even topological degrees, and $\Sigma^2 R_0$ is
$2$-connected, the long exact sequence diagram is empty in topological degree $1$, so the boundary map
\begin{equation}
	\delta\colon \Ext_{\cE(1)}^{s,s+1}(\textcolor{RedOrange}{R_0}, \Z/2)\to\Ext_{\cE(1)}^{s,s}(N_1, \Z/2)
\end{equation}
vanishes, which tells us the line $t-s = 0$ in $\Ext(\textcolor{RedOrange}{R_0})$ consists of a single $\Z/2$ summand
in filtration $0$. Therefore the line $t-s=2$ in the long exact sequence diagram consists of two $\Z/2$
summands: one in filtration $1$ coming from $N_1$, and one in filtration $0$ coming from $\Sigma^2 R_0$. Since
the $1$-line of the diagram is empty and $\Ext(N_1)$ is concentrated in even degrees, the $3$-line of the diagram
is empty, so there are no differentials to the $2$-line. Continuing in this way produces \cref{simpler_ext_H},
right.

Finally, acting by $h_0\in H^{*,*}(\cE(1))$ defines an isomorphism
\begin{equation}
	\Ext_{\cE(1)}^{0,2}(\textcolor{RedOrange}{R_0}, \Z/2)\to\Ext_{\cE(1)}^{1, 3}(\textcolor{RedOrange}{R_0}, \Z/2).
\end{equation}
This can be checked directly from the definition: begin with the unique nontrivial map
$\textcolor{RedOrange}{R_0}\to\Sigma^2\Z/2$ and act on it by an extension representing $h_0$ (namely the extension
$0\to\Sigma \Z/2\to N_1\to\Z/2\to 0$); the result is a nontrivial extension.
\end{proof}
With $\Ext(\textcolor{RedOrange}{\Sigma R_0})$ in hand, we return to our goal of computing $\widetilde\ku_*((BD_{2n})^{2-V_\lambda})$.
We draw the $E_2$-page of the Adams spectral sequence in~\eqref{e1_d2n_2mod4}, right. Margolis' theorem
(\cref{margolis}) forces all differentials in this range to vanish, except possible differentials with target the
$7$-line, and there can be no hidden extensions in the range depicted. Thus for $n = 2k < 7$,
$\widetilde\ku_*((BD_{2n})^{2-V_\lambda})\cong (\Z/2)^{\oplus k+1}$ and for $n = 2k+1 < 8$, $\widetilde\ku_*((BD_{2n})^{2-V_\lambda})\cong\Z/2^{k+1}$; we
finish with the fact that $\Omega_k^{\Spinc}(\bl)\cong\ku_k(\bl)\oplus\ku_{k-4}(\bl)$ for connective spectra and
$k\le 7$, so we can read off the \spinc bordism groups from the $\ku$-homology groups.
\end{proof}

%
\begin{thm}
\label{dih_spinc_0mod4}
If $n\equiv 0\bmod 4$, the first few \spinc bordism groups of $(BD_{2n})^{2-V_\lambda}$ are
\begin{align*}
	\tOmega_0^\Spinc((BD_{2n})^{2-V_\lambda}) &\cong \Z/2\\
	\tOmega_1^\Spinc((BD_{2n})^{2-V_\lambda}) &\cong \Z/n\\
	\tOmega_2^\Spinc((BD_{2n})^{2-V_\lambda}) &\cong (\Z/2)^{\oplus 2}\\
	\tOmega_3^\Spinc((BD_{2n})^{2-V_\lambda}) &\cong \Z/2n\\
	\tOmega_4^\Spinc((BD_{2n})^{2-V_\lambda}) &\cong (\Z/2)^{\oplus 4},
\end{align*}
and $\tOmega_5^\Spinc((BD_{2n})^{2-V_\lambda})$ is torsion.
\end{thm}
\begin{proof}[Proof of \cref{dih_spinc_0mod4}]
By \cref{twisted_dihedral_SO_U}, the odd-primary torsion is isomorphic to the odd-primary torsion of $\Z/n$ in
degrees $1$ and $3$ and vanishes in degrees $0$, $2$, and $4$.

At $2$, we use the Adams spectral sequence. We described the $\cA(1)$-module structure on $\tH^*((BD_{2n})^{2-V_\lambda};\Z/2)$
in~\eqref{dihedral0mod4summands} and draw it in \cref{a1moddih0mod4}; this determines the $\cE(1)$-module
structure, with isomorphisms of $\cE(1)$-modules $\cA(1)\cong\cE(1)\oplus\Sigma^2\cE(1)$ and $R_2\cong
\uQ\oplus\Sigma\cE(1)$~\cite[\S 4.7.2]{Bay94} and $J\cong \cE(1)\oplus\Sigma^2\Z/2$, which one can check directly.
Hence as $\cE(1)$-modules,
\begin{equation}
	\tH^*((BD_{2n})^{2-V_\lambda} ;\Z/2) \cong \textcolor{BrickRed!80!black}{\cE(1)} \oplus
		\textcolor{RedOrange!80!black}{\Sigma\uQ}
		\oplus \textcolor{BrickRed!80!white}{\Sigma^2\cE(1)} \oplus
		\textcolor{RedOrange!80!white}{\Sigma^2\cE(1)} \oplus
		\textcolor{Goldenrod!67!black}{\Sigma^2\Z/2} \oplus
		\textcolor{Green}{\Sigma^4\cE(1)} \oplus
		\textcolor{PineGreen}{\Sigma^4\cE(1)} \oplus
		\textcolor{MidnightBlue}{\Sigma^4\cE(1)} \oplus
		\textcolor{Fuchsia}{\Sigma^5\uQ} \oplus P,
\end{equation}
where $P$ is $5$-connected. We draw this $\cE(1)$-module in \cref{E1D0mod4picture}, left.
\begin{figure}[h!]
\makebox[\textwidth][c]{
\begin{subfigure}[c]{0.64\textwidth}
\begin{tikzpicture}[scale=0.6, every node/.style = {font=\tiny}]
\foreach \y in {0, ..., 8} {
	\node at (-1.5, \y) {$\y$};
}
\begin{scope}[BrickRed!80!black]
	\Eone{0}{0}{$\oU$}{}
\end{scope}
\begin{scope}[BrickRed!80!white]
	\Eone{4.57}{2}{$\oU x^2$}{}
\end{scope}
\draw[dashed, thick, gray] (2.5, 1) -- (2.5, 2);
\begin{scope}[RedOrange!80!black]
	\EoneQnMark{2.5}{1}{$\oU y$};
\end{scope}
\begin{scope}[RedOrange!80!white]
	\Eone{6.5}{2}{$\oU y^2$}
\end{scope}
	\tikzptR{2.5}{2}{$\oU w$}{Goldenrod!67!black};
\begin{scope}[Green]
	\Eone{8.5}{4}{$\oU w^2$}
\end{scope}
\begin{scope}[PineGreen]
	\Eone{10.5}{4}{$\oU x^4$}
\end{scope}
\begin{scope}[MidnightBlue]
	\Eone{12.5}{4}{$\oU y^4$}
\end{scope}
\begin{scope}[Fuchsia]
	\EoneQnMark{14.75}{5}{$\oU w^2y$}
\end{scope}
\end{tikzpicture}
\end{subfigure}
\begin{subfigure}[c]{0.4\textwidth}
\begin{sseqdata}[name=D0mod4E1, classes=fill, xrange={0}{5}, yrange={0}{3}, scale=0.7, Adams grading, >=stealth,
x label = {$\displaystyle{s\uparrow \atop t-s\rightarrow}$},
x label style = {font = \small, xshift = -19ex, yshift=3.7ex}]
\class[BrickRed!80!black](0, 0)
\class[BrickRed!80!white](2, 0)
\class[RedOrange!80!white](2, 0)
\begin{scope}[RedOrange!80!black]
	\class(1, 0)\AdamsTower{} \class(1, 5)
	\class(3, 0)\AdamsTower{} \class(3, 5)
	\class(5, 1)\AdamsTower{}
	\class(7, 2)\AdamsTower{}
\end{scope}
\begin{scope}[white]
	\class(2, 1)\AdamsTower{}
	\class(2, 1)\AdamsTower{}
	\class(5, 0)
\end{scope}
\structline[RedOrange!40!white](1, 2)(3, 3)
\begin{scope}[Goldenrod!67!black]
	\class(2, 0)\AdamsTower{}
	\class(4, 1)\AdamsTower{}
	\class(6, 2)\AdamsTower{}
	\class(8, 3)\class(8, 4)
	\structline[Goldenrod](2, 0, -1)(4, 1)
\end{scope}
\d[gray]2(2, 0, -1)
\d[gray]2(4, 1, -1)
\class[Green](4, 0)
\begin{scope}[Green!80!black]
	\class(6, 0)\AdamsTower{}
\end{scope}
\class[PineGreen](4, 0)
\class[PineGreen!80!white](6, 0)
\class[MidnightBlue](4, 0)
\class[MidnightBlue!80!white](6, 0)
\begin{scope}[Fuchsia]
	\class(5, 0)\AdamsTower{}
	\class(7, 0)\AdamsTower{}
	\class(9, 1)\class(9, 2)\class(9, 3)\class(9, 4)
\end{scope}
\class[Fuchsia!67!white](6, 0)
\class[Fuchsia!75!black](6, 0)
\end{sseqdata}
\printpage[name=D0mod4E1, page=2]
\end{subfigure}
}
\caption{Left: the $\cE(1)$-module structure on $\tH^*((BD_{2n})^{2-V_\lambda};\Z/2)$, $n\equiv 0\bmod 4$, in low
degrees. The pictured submodule contains all elements in degrees $5$ and below. The gray dashed line indicates that the
$\Z/2^r$ Bockstein maps a preimage of $\oU y$ to $\oU w$, which we use in the proof of \cref{dih_spinc_0mod4}.
Right: the $E_2$-page for the Adams spectral sequence computing $\widetilde{\ku}_*((BD_{2n})^{2-V_\lambda})$. The
two pictured differentials are related by a $v_1$-action.}
\label{E1D0mod4picture}
\end{figure}

We saw $\Ext(\Z/2)$ in~\eqref{E1Z2Ext}, and Adams-Priddy~\cite[\S 3]{AP76} show
\begin{equation}
\label{ExtN2}
	\Ext_{\cE(1)}^{s,t}(\uQ, \Z/2)\cong \Ext_{\cE(1)}^{s+1, t+1}(\Z/2, \Z/2),
\end{equation}
with the isomorphism intertwining the $H^{*,*}(\cE(1))$-actions.
We can therefore draw the $E_2$-page of the Adams spectral sequence in \cref{E1D0mod4picture}, right. We hide most
$v_1$-actions to declutter the diagram. 

The first differential that could be nonzero is from the $2$-line to the $1$-line; as differentials are
$h_0$-equivariant, if a $d_r$ differential is nonzero on one summand in the tower on the $2$-line, then it is
nonzero on the entire tower, so we refer to differentials between towers. The May-Milgram theorem~\cite{MM81}
characterizes differentials between towers: there is a $d_r$ differential between those two towers iff the
Bockstein $\beta\colon H^1(\bl;\Z/2^r)\to H^2(\bl;\Z/2)$ carries a preimage of $\oU y$ to $\oU w$.  The Thom
isomorphism is natural with respect to this Bockstein, so it suffices to know whether $\beta(y) = w$ in
$H^2(BD_{2n};\Z/2)$, and we saw this in the proof of \cref{first_dih_differential}, where $r$ is the largest number
such that $2^r\mid n$. This means that the $2$-torsion in $\tOmega_1^\Spinc((BD_{2n})^{2-V_\lambda})$ is isomorphic to that of $\Z/n$,
so along with our odd-torsion computation we see that $\tOmega_1^\Spinc((BD_{2n})^{2-V_\lambda})\cong\Z/n$.

The other differential we need to resolve in range goes from the tower in the $4$-line to the tower in the
$3$-line. Action by $v_1\in\ku_2$ carries the tower in the $2$-line to the tower in the $4$-line, and the tower in
the $1$-line to the tower in the $3$-line, and differentials are $v_1$-equivariant, so there is also a $d_r$
differential between these towers. As seen in \cref{E1D0mod4picture}, right, on the $E_\infty$-page there are $r+1$
$\textcolor{RedOrange!80!black}{\Z/2}$ summands on the $3$-line, all connected, so together with our odd-torsion
computation we see that $\tOmega_3^\Spinc((BD_{2n})^{2-V_\lambda})\cong\Z/2n$.

There can be no other nonzero differentials in range, and Margolis' theorem precludes any hidden extensions, so we
are done.
\end{proof}

\subsubsection{Class A, spin-$1/2$ case}
	\label{s_spinc_mixed_dih}
\begin{lem}
\label{dihedral_when_pinc}
$V_\lambda$ is \pinc iff $n$ is odd.
\end{lem}
\begin{proof}
For $n$ odd, we saw that inclusion of a reflection defines a map $B\Z/2\to BD_{2n}$ which is an isomorphism on mod
$2$ cohomology. Therefore we can compute Stiefel-Whitney classes of $V_\lambda$ by pulling back to $B\Z/2$, and we
saw that the pullback bundle is stably equivalent to a line bundle, so $w_2 = 0$.

For $n$ even, recall that $V_\lambda$ is \pinc iff $\beta(w_2(V_\lambda)) = 0$, where $\beta\colon H^k(\bl;\Z/2)\to
H^{k+1}(\bl;\Z)$ is the integral Bockstein. \Cref{bock_to_sq1} means it suffices to show $\Sq^1(w_2(V_\lambda))\ne
0$. In the notation of \cref{dihmod2coh}, for $n\equiv 2\bmod 4$, $w_2(V_\lambda) = xy + y^2$, and $\Sq^1(xy + y^2)
= x^2y + xy^2\ne 0$. For $n\equiv 0\bmod 4$, $w_2(V_\lambda) = w$, and by \cref{0mod4_w_lem}, $\Sq^1(w)\ne 0$.
\end{proof}
Therefore for $n$ odd, we consider $(BD_{2n})^{2-V_\lambda}$. We computed
$\Omega_k^\Spinc((BD_{2n})^{2-V_\lambda})$ for $k\le 4$ in \cref{dih_spinc_odd}.

For $n$ even, \cref{shear_A_thm} directs us to the \spinc bordism of $(BD_{2n})^{\Det(V_\lambda)-1}$.
\begin{thm}
\label{bord_gps_spinc_dih_mixed_2mod4}
If $n\equiv 2\bmod 4$,  the first few \spinc bordism groups of $(BD_{2n})^{\Det(V_\lambda)-1}$ are
\begin{align*}
	\tOmega_0^{\Spinc}((BD_{2n})^{\Det(V_\lambda)-1}) &\cong \Z/2\\
	\tOmega_1^{\Spinc}((BD_{2n})^{\Det(V_\lambda)-1}) &\cong \Z/n\\
	\tOmega_2^{\Spinc}((BD_{2n})^{\Det(V_\lambda)-1}) &\cong \Z/4\oplus\Z/2\\
	\tOmega_3^{\Spinc}((BD_{2n})^{\Det(V_\lambda)-1}) &\cong \Z/n\oplus\Z/2\\
	\tOmega_4^{\Spinc}((BD_{2n})^{\Det(V_\lambda)-1}) &\cong \Z/8\oplus\Z/4\oplus\Z/2.
\end{align*}
Because \cref{torsion_k_theory} implies $\tOmega_5^\Spinc((BD_{2n})^{\Det(V_\lambda)-1})$ is torsion, the phase homology groups for this
symmetry type are $\Z/n\oplus\Z/2$ for $d = 2$ and $\Z/8\oplus\Z/4\oplus\Z/2$ for $d = 3$.
\end{thm}
The $2$-local isomorphism $\tOmega_*^\Spin((BD_{2n})^{\Det(V_\lambda)-1})\cong\Omega_*^{\Pin^-}(B\Z/2)$ we used in
\cref{mixed_dih_2mod4} implies a $2$-local isomorphism $\tOmega_*^{\Spinc}((BD_{2n})^{\Det(V_\lambda)-1})\cong
\Omega_*^{\Pinc}(B\Z/2)$, so when $n = 2$, these are also the \pinc bordism groups of $B\Z/2$. This may be of
independent interest.
\begin{proof}
We can read the odd-primary torsion off of~\cref{twisted_dihedral_SO_U}, thus for the rest of the proof we
implicitly localize at $2$.
Recall from the proof of \cref{mixed_dih_2mod4} that $(BD_{2n})^{\Det(V_\lambda)-1}\simeq (B\Z/2)^{\sigma-1}\wedge
(B\Z/2)_+$. There is a well-known homotopy equivalence $S_\sigma\colon \Sigma^\infty B\Z/2\simeq (B\Z/2)^\sigma$
(see for example~\cite[Lemma 2.6.5]{Koc96}), essentially reducing the problem to the computation of
$\ku_*(B\Z/2\times B\Z/2)$. In a little more detail, combining $S_\sigma$ with the stable splitting $\Sigma^\infty
X_+\simeq \mathbb S \vee \Sigma^\infty X$, we have
\begin{equation}
\label{k4_smith}
	(B\Z/2)^{\sigma-1}\wedge (B\Z/2)_+ \simeq \Sigma^{-1}\Sigma^\infty B\Z/2 \vee
	\Sigma^{-1}\Sigma^\infty (B\Z/2\wedge B\Z/2).
\end{equation}
Ossa~\cite[Proposition 3]{Oss89} shows an equivalence
\begin{equation}
\label{ossa}
    \ku\wedge B\Z/2\wedge B\Z/2 \simeq (\ku\wedge \Sigma^2 B\Z/2) \vee \Sigma^2 H(\Z/2[u, v]),
\end{equation}
where the third term refers to a generalized Eilenberg-Mac Lane spectrum on the graded abelian group $\Z/2[u,
v]$, with $\abs u = \abs v = 2$.\footnote{Ossa's splitting~\eqref{ossa} or its analogue on homotopy groups has also
been proven in several other ways: see Johnson-Wilson~\cite{JW97}, Bruner~\cite[Corollary 3.3]{Bru99},
Bruner-Greenlees~\cite[Example 4.11.2]{BG03}, Powell~\cite{Pow14}, and Bruner-Mira-Stanley-Snaith~\cite[Theorem
2.12]{BMSS15}.} Combining~\eqref{k4_smith} and~\eqref{ossa}, we have
\begin{equation}
	\ku\wedge (B\Z/2)^{\sigma-1}\wedge (B\Z/2)_+ \simeq (\ku\wedge \Sigma^{-1}B\Z/2) \vee (\ku\wedge\Sigma B\Z/2)
	\vee \Sigma H(\Z/2[u, v]).
\end{equation}
To finish the proof, take the groups $\widetilde{\ku}_*(B\Z/2)$, computed by Hashimoto~\cite[Theorem 3.1]{Has83},
and shift them down by $1$, resp.\ up by $1$ for the first two summands, and append a $\Z/2$ summand in degree $1$
and a $(\Z/2)^{\oplus 2}$ summand in degree $3$ for $\Sigma H(\Z/2[u, v])$.
\end{proof}

\begin{thm}
\label{bord_gps_spinc_dih_mixed_0mod4}
If $n\equiv 0\bmod 4$,  the first few \spinc bordism groups of $(BD_{2n})^{\Det(V_\lambda)-1}$ are
\begin{align*}
	\tOmega_0^{\Spinc}((BD_{2n})^{\Det(V_\lambda)-1}) &\cong \Z/2\\
	\tOmega_1^{\Spinc}((BD_{2n})^{\Det(V_\lambda)-1}) &\cong \Z/n\\
	\tOmega_2^{\Spinc}((BD_{2n})^{\Det(V_\lambda)-1}) &\cong \Z/4\oplus \Z/2\\
	\tOmega_3^{\Spinc}((BD_{2n})^{\Det(V_\lambda)-1}) &\cong \Z/(n/2) \oplus (\Z/2)^{\oplus 2}\\
	\tOmega_4^{\Spinc}((BD_{2n})^{\Det(V_\lambda)-1}) &\cong \Z/8\oplus\Z/4\oplus\Z/2.
\end{align*}
Because \cref{torsion_k_theory} implies $\tOmega_5^\Spinc((BD_{2n})^{\Det(V_\lambda)-1})$ is torsion, the phase homology groups for this
symmetry type are $\Z/(n/2)\oplus (\Z/2)^{\oplus 2}$ for $d = 2$ and $\Z/8\oplus\Z/4\oplus\Z/2$ for $d = 3$.
\end{thm}
\begin{proof}
We closely follow the proof of \cref{spinless_D0mod4_thm}. For odd-primary torsion, use
\cref{twisted_dihedral_SO_U} to see that the odd-primary torsion in the range we care about is isomorphic to the
odd torsion in $\Z/n$ in degrees $1$ and $3$, and is $0$ in degrees $0$, $2$, and $4$.

On to the prime $2$. In \cref{spinless_D0mod4_thm}, we established a splitting $(BD_{2n})^{\Det(V_\lambda)-1}\simeq (B\Z/2)^{\sigma-1}\vee
M_n$, allowing us to focus solely on $\tOmega_*^\Spin(M_n)$:
$\tOmega_*^{\Spinc}((B\Z/2)^{\sigma-1})\cong\Omega_*^{\Pinc}$~\eqref{pincsplitting}, and we know \pinc bordism
groups thanks to Bahri-Gilkey~\cite{BG87a, BG87b}. In~\eqref{a1_str_mixed_dih_0mid4}, we determined the
$\cA(1)$-module structure on $\tH^*(M_n;\Z/2)$ in low degrees, and the isomorphisms of $\cE(1)$-modules $R_1\cong
\Z/2\oplus\Sigma R_0$ and $\cA(1)\cong\cE(1)\oplus \Sigma^2\cE(1)$ mean that as $\cE(1)$-modules,
\begin{equation}
\label{spinc_mixed_dih_0mod4_e1}
	\tH^*(M_n;\Z/2) \cong \textcolor{BrickRed!80!black}{\Sigma\Z/2} \oplus
		\textcolor{BrickRed!80!white}{\Sigma^2 R_0} \oplus
		\textcolor{Green}{\Sigma^2\uQ} \oplus
		\textcolor{MidnightBlue}{\Sigma^3\cE(1)} \oplus
		\textcolor{Fuchsia}{\Sigma^3\cE(1)} \oplus P,
\end{equation}
where $P$ is $4$-connected. A priori, $\Ext(P)$ could have nonzero differentials to elements of the $4$-line, but
we will see that this does not happen without needing to compute $\Ext(P)$.
In \cref{dih_0mod4_spinc_mixed_figures}, left, we draw~\eqref{spinc_mixed_dih_0mod4_e1}. To determine the
$E_2$-page of the Adams spectral sequence, see~\eqref{E1Z2Ext} for $\Ext(\Z/2)$, \cref{the_ext_of_H} for
$\Ext(\textcolor{BrickRed!80!white}{R_0})$, and~\eqref{ExtN2} for $\Ext(\textcolor{Green}{\uQ})$. We draw the
$E_2$-page of the Adams spectral sequence for $\widetilde{\ku}_*(M_n)$, as in \cref{dih_0mod4_spinc_mixed_figures},
right --- though for legibility, most $v_1$-actions are hidden. \Cref{torsion_k_theory} implies there must be
differentials in this range, though not necessarily the $d_2$s pictured.
\begin{figure}[h!]
\begin{subfigure}[c]{0.3\textwidth}
\begin{tikzpicture}[scale=0.6, every node/.style = {font=\tiny}]
	\foreach \y in {1, ..., 7} {
		\node at (-2, \y) {$\y$};
	}
	\draw[thick, dashed, gray] (1.5, 1) -- (1.5, 2);
	\tikzpt{1.5}{1}{$\oU y$}{BrickRed!80!black};
	\begin{scope}[BrickRed!80!white]
		\tikzpt{0}{2}{$\oU y^2$}{};
		\foreach \y in {3, ..., 7} {
			\tikzpt{0}{\y}{}{};
		}
		\foreach \y in {2, 4, 6} {
			\sqone(0, \y);
		}
		\qoneL(0, 2);
		\qoneR(0, 4);
		\begin{scope}
			\clip (-1, 5.5) rectangle (0.5, 7.5);
			\qoneL(0, 6);
		\end{scope}
	\end{scope}
	\begin{scope}[Green]
		\EoneQnMark{1.5}{2}{$\oU w$};
	\end{scope}
	\begin{scope}[MidnightBlue]
		\Eone{3}{3}{$\oU wx$};
	\end{scope}
	\begin{scope}[Fuchsia]
		\Eone{5}{3}{$\oU wy$};
	\end{scope}
\end{tikzpicture}
\end{subfigure}
\qquad
\begin{subfigure}[c]{0.4\textwidth}
\begin{sseqdata}[name=mixedclassAdih0mod4, classes=fill, xrange={0}{4}, yrange={0}{3}, scale=0.7,
        x label = {$\displaystyle{s\uparrow \atop t-s\rightarrow}$},
        x label style = {font = \small, xshift = -16.25ex, yshift=3.7ex}, >=stealth, Adams grading]
\begin{scope}[BrickRed!80!black]
	\class(1, 0)\AdamsTower{}
	\class(3, 1)\AdamsTower{}
	\class(5, 2)\AdamsTower{}
\end{scope}
\begin{scope}[BrickRed!80!white]
	\class(2, 0)
	\class(4, 0)
	\class(4, 1)\structline
\end{scope}
\begin{scope}[White]
	\class(2, 1)\AdamsTower{}
	\class(4, 2)\AdamsTower{}
	\class(5, 1)\class(5,1)\AdamsTower{}
\end{scope}
\structline[BrickRed!60!white](1, 2)(3, 3)
\begin{scope}[Green]
	\class(2, 0)\AdamsTower{}
	\class(4, 0)\AdamsTower{}
	\structline[Green!40!white](2, 0, -1)(4, 1, -1)
\end{scope}
\class[MidnightBlue](3, 0)
\class[Fuchsia](3, 0)
\class[MidnightBlue!80!white](5, 0)
\class[Fuchsia!80!white](5, 0)
\begin{scope}[Fuchsia]
	\class(5, 0)\AdamsTower{}
\end{scope}
\d[gray]2(2, 0, -1)
\d[gray]2(4, 0, -1)
\d[gray]2(4, 1, -1)
\end{sseqdata}
\printpage[name=mixedclassAdih0mod4, page=2]
\end{subfigure}
\caption{Left: the $\cE(1)$-module structure on $\tH^*(M_n;\Z/2)$ in low degrees. The pictured submodule contains
all elements in degrees $4$ and below. The dashed line indicates a $\Z/2^r$ Bockstein, which we use to resolve a
differential. Right: Ext of this submodule, which is the $E_2$-page of the Adams spectral
sequence computing $\widetilde{\ku}_*(M_n)$ for $t-s\le 4$. Most $v_1$-actions are hidden for readability. See the
proof of \cref{bord_gps_spinc_dih_mixed_0mod4} for more information.}
\label{dih_0mod4_spinc_mixed_figures}
\end{figure}

For $\tOmega^{\Spin^c}_1(M_n)$ to be torsion, there must be a differential $d_r$ from the $2$-line to the $1$-line;
then, $\tOmega^{\Spin^c}_1(M_n)\cong \Z/2^r$, and since $\Omega_1^{\Pin^c} \cong 0$, $\tOmega^{\Spin^c}_1(M_n)\cong
\Z/2^r$ as well. Differentials between towers, such as this $d_r$, are characterized by the May-Milgram
theorem~\cite{MM81}, and just as in the proof of \cref{dih_spinc_0mod4}, we conclude $r$ is the largest natural
number such that $2^r\mid n$. Combining this with our odd-torsion calculation, $\tOmega_1^\Spinc(M_n)\cong\Z/n$.

Continuing in increasing topological degree, this $d_r$ kills the entire orange tower in the $2$-line, and we infer
$\tOmega^{\Spin^c}_2(M_n)\cong\Z/2$. The green and blue summands in the $3$-line survive and split off by Margolis'
theorem. $v_1$-equivariance of differentials implies that $d_r\colon E_2^{s,4+s}\to E_2^{s+2, s+3}$ is nonzero,
and again maps the orange tower to the dark red tower, leaving a single $\textcolor{BrickRed!80!black}{\Z/2}$
summand in $E_3^{1,4}$. There can be no further differentials to the $3$-line, so $\tOmega^{\Spinc}_3(M_n)_2^\wedge\cong
\Z/2^{r-1}\oplus (\Z/2)^{\oplus 2}$. Our odd-primary calculation then tells us that $\tOmega_3^\Spinc(M_n)\cong
\Z/(n/2)\oplus (\Z/2)^{\oplus 2}$. Finally, the orange tower in the $4$-line is killed by the $d_r$ we most
recently discussed, and the two light red $\textcolor{BrickRed!80!white}{\Z/2}$ summands in the $4$-line cannot
emit or receive differentials. Thus as promised $\Ext(P)$ does not have nonzero differentials to the $4$-line, so
we conclude by adding the \pinc bordism summands back in.
\end{proof}

\subsubsection{Comparison with other literature}
\label{dihedral_discrepancy}
At the time this paper appeared on arXiv, interacting fermionic phases equivariant for a dihedral group $D_{2n}$
acting by rotations and reflections had only been studied by Zhang-Wang-Yang-Qi-Gu~\cite{ZWYQG19}, who considered both spinless and spin-$1/2$ phases in
dimension $2+1$ for all $n$, and in Altland-Zirnbauer class D. They also study systems without a spatial symmetry,
using the extended supercohomology classification of Wang-Gu~\cite{WG18a, WG18b} to classify these phases and
discuss the FCEP for dihedral groups. We find complete agreement with their results.\footnote{At the time of
writing, the arXiv version of \cite{ZWYQG19} contains an error that was fixed by a correction to the published
version. The agreement stated above is with the corrected published version.}

More recently, 
Herzog-Arbeitman-Bernevig-Song~\cite[\S B.2]{HABS24} studied 2d class A interacting fermionic phases with spinless $D_{2n}$
symmetry, and
Zhang-Ning-Qi-Gu~\cite[Tables I and II, \S\S S-3.5, S-3.11, S-3.17, S-3.23]{ZNQG25} studied
interacting fermionic phases with this $D_{2n}$ symmetry in classes D and A, with both spinless and spin-$1/2$
symmetry, in dimension $3+1$. Both teams restricted to $n = 2,3,4,6$. Their results also completely agree with ours.
\begin{rem}
\label{super_coh}
The phases we classify are realized by the extended supercohomology classifications of Wang-Gu~\cite{WG18a, WG18b}
and Kapustin-Thorngren~\cite{KT17}.\footnote{These classifications concern phases with an internal $D_{2n}$
symmetry, but the fermionic crystalline equivalence principle allows us to pass back and forth.}
Gaiotto-Johnson-Freyd~\cite[\S\S 5.4--5.6]{GJF19} determine that the extended supercohomology classification à
la~\cite{KT17, WG18a} is the cohomology of $(BD_{2n})^{2-V_\lambda}$ or $(BD_{2n})^{\Det(V_\lambda)-1}$ with
respect to a spectrum they call $\mathrm{fGP}_{\le 2}^\times$, which is equivalent to the $(-3)$-connected cover of
$I_\Z\MTSpin$. Wang-Gu's refinement in~\cite{WG18b} corresponds instead to the spectrum $\mathrm{fGP}^\times$,
equivalent to the $(-7)$-connected cover of $I_\Z\MTSpin$.\footnote{The reader may at this point wonder why our
classification is a generalized \emph{homology} theory, while these extended supercohomology classifications are
generalized \emph{cohomology} theories. This is a subtle point. The passage between homology and cohomology occurs
because in these dimensions, we may approximate $\MTSpin$ by $\KO$ due to Anderson-Brown-Peterson's~\cite{ABP67}
study of the connectivity of the Atiyah-Bott-Shapiro map~\cite{ABS}, then use that $\KO$ is shifted Anderson
self-dual~\cite{And69, FMS07a, HS14, Ric16, GM17, HLN20} to pass between $I_\Z\KO$-homology and
$\Sigma^4\KO$-cohomology.  See Freed-Hopkins~\cite[\S 5.1]{FH19} for further discussion.}

The connective covering maps induce comparison maps from the classifications of fermionic phases using extended
supercohomology to the classification of fermionic phases under our ansatz. For $\mathrm{fGP}^\times$, the map is
sufficiently connected as to be an isomorphism between the classifications of $(d+1)$-dimensional phases for all
$d\le 5$. For $\mathrm{fGP}_{\le 2}^\times$, the map is not always an isomorphism even for $d = 2$: the cokernel
when computing supercohomology of $X$ is $\tH^0(X;\Z)$, and this is nonzero e.g.\ for $X = (BC_n)^{2-V_\lambda}$
from~\S\ref{s_rotations}. But for dihedral groups, $\tH^0((BD_{2n})^\xi;\Z)$ vanishes whenever $\xi\to BD_{2n}$ is
a rank-$0$ unorientable virtual vector bundle, so in this case the comparison map is an isomorphism.
\end{rem}

	\subsection{$D_{2n}$ acting by rotations}
\label{para_symm}
The dihedral group $D_{2n}$ can act on $\R^3$ in an orientation-preserving manner, where $C_n\subset D_{2n}$ acts
by rotations in a plane and a preimage of the generator of $D_{2n}/C_n\cong\Z/2$ acts by a rotation perpendicular
to that plane. Said differently, this point group is defined by a representation $\lambda\colon D_{2n}\to\SO_3$
which decomposes as $\rho\oplus\sigma$, where $\rho$ is the standard two-dimensional representation by rotations
and reflections, and $\sigma\colon D_{2n}\to\O_1$ is the sign representation, which is the determinant of $\rho$.
Confusingly, this point group is sometimes called ``three-dimensional dihedral symmetry;'' in this convention, the
three-dimensional action by $\rho\oplus\underline\R$ is called \term{pyramidal symmetry}.

At the time this paper first appeared on arXiv, interacting fermionic phases for this $D_{2n}$ symmetry had not
appeared in the literature. Since then, Zhang-Ning-Qi-Gu~\cite[Tables I and II, \S\S S-3.4, S-3.10,
S-3.16, S-3.22]{ZNQG25} computed the groups of these phases for both spinless and spin-$1/2$ fermions, in classes D
and A, for $n = 2,3,4,6$; their methods are quite different, and their results completely match ours.
\begin{table}[h!]
\begin{tabular}{r c c c c}
\toprule
& Class D, spinless & Class D, spin-$1/2$ & Class A, spinless & Class A, spin-$1/2$\\
$n$ & \S\ref{para_D_spinless} & \S\ref{para_D_spinful} & \S\ref{para_A_spinless} & \S\ref{para_A_spinful}\\
\midrule
$0\bmod 4$ & $\tblref{dihedral_twspin_bordism_0mod4}{\Z/}2$ & $\tblref{D2n_spin_bord_even}{(\Z/2)^{\oplus 2}}$ &
	$\tblref{unmixed_A_para_0mod4}{0}$ & $\tblref{spinc_dihedral_bordism}{(\Z/2)^{\oplus 2}}$\\
$2\bmod 4$ & $\tblref{dihedral_twspin_bordism_2mod4}{0}$ & $\tblref{D2n_spin_bord_even}{(\Z/2)^{\oplus 2}}$ &
	$\tblref{unmixed_A_para_2mod4}{0}$ & $\tblref{2mod4_spinc_bord_dihedral}{(\Z/2)^{\oplus 3}}$\\
$1,3\bmod 4$ & $\tblref{odd_3d_dih_D_spinless}{0}$ & $\tblref{D2n_spin_bord_odd}{0}$ &
$\tblref{odd_dihedral_spinc_bordism}{0}$ & $\tblref{para_odd_dih_spinful}{0}$\\
\bottomrule
\end{tabular}
\caption{$D_{2n}$-equivariant phase homology groups, where $D_{2n}$ acts faithfully on $\R^3$ by rotations. These
arise as homotopy groups of Anderson duals of $\MTSpin\wedge X_n$ and $\MTSpinc\wedge X_n$, where $X_n$ is one of
$(BD_{2n})^{3-V_\lambda}$ or $(BD_{2n})_+$. See \S\ref{para_symm} for details and proofs.}
\label{para_n_gonal_table}
\end{table}

For any representation $\phi\colon D_{2n}\to\O_d$, let $V_\phi\to BD_{2n}$ denote the associated vector bundle.
\begin{lem}\hfill
\label{3d_dih_not_pinc}
\begin{enumerate}
	\item\label{spinc_not_spin} If $n$ is odd, $V_\lambda$ is \pinc but not \pinm.
	\item\label{nt_pinc} If $n$ is even, $V_\lambda$ is not \pinc.
\end{enumerate}
\end{lem}
\begin{proof}
For~\eqref{nt_pinc}, we show that if $\beta$ is the integral Bockstein, $\beta w_2(V_\lambda)\ne 0$. By
\cref{bock_to_sq1}, it suffices to show $\Sq^1(w_2(V_\lambda))\ne 0$. For $n\equiv 2\bmod 4$,
\begin{subequations}
\begin{equation}
\label{w2_2mod4_3d}
	w_2(V_\lambda) = w_2(V_\rho) + w_1(V_\rho)w_1(V_\sigma) + w_2(V_\sigma) = x^2 + xy + y^2,
\end{equation}
and $\Sq^1(x^2 + xy + y^2) = x^2y + xy^2$. For $n\equiv 0\bmod 4$,
\begin{equation}
\label{w2_0mod4_3d}
	w_2(V_\lambda) = w_2(V_\rho) + w_1(V_\rho)w_1(V_\sigma) + w_2(V_\sigma) = w + x^2,
\end{equation}
\end{subequations}
and $\Sq^1(w + x^2) = wx$, so in neither case is $V_\lambda$ \pinc.

Now~\eqref{spinc_not_spin}. Choose $i\colon \Z/2\inj D_{2n}$ given by a splitting of $D_{2n}\surj
D_{2n}/C_n\cong\Z/2$; restricting to $\Z/2$ along $i$, $\lambda$ decomposes as $2\sigma \oplus\R$. Therefore
$i^*V_\lambda\to B\Z/2$ is \spinc but not spin: $w_2(2V_\sigma) = w_1(V_\sigma)^2 = x^2$, and for any vector bundle
$V$, $V\oplus V$ admits a complex structure, hence a \spinc structure. In particular, $\beta(w_2(i^*V_\lambda))\ne
0$. The maps $\Z/2\inj D_{2n}\surj\Z/2$ compose to the identity, so the induced maps on cohomology also compose to
the identity.  Therefore $\beta(w_2(V_\lambda))\ne 0$ too.
\end{proof}

These propositions are the analogues of \cref{twisted_dihedral_SO,twisted_dihedral_SO_U}, helping us calculate
odd-primary torsion in phase homology groups.
\begin{lem}[{Handel~\cite[Theorems 5.2, 5.3]{Han93}}]
\begin{equation}
	\tH_k(BD_{2n};\Z[1/2]) \cong \begin{cases}
		\Z/n, &k\equiv 3\bmod 4\\
		0, &\text{\rm otherwise.}
	\end{cases}
\end{equation}
\end{lem}
As usual, Handel computes $H^*(BD_{2n};\Z)$, and it is up to us to change to homology with $\Z[1/2]$ coefficients.
\begin{prop}
\label{untwisted_dihedral_SO}
Suppose $V$ is a rank-zero oriented virtual vector bundle.
\begin{enumerate}
	\item\label{untw_Spin} The odd-torsion subgroup of $\tOmega_k^\Spin((BD_{2n})^V)$ is isomorphic to the
	odd-torsion subgroup of $\Z/n$ when $k = 3$ and vanishes for all other $k\le 6$.
	\item\label{untw_Spinc} The odd-torsion subgroup of $\tOmega_k^\Spinc((BD_{2n})^V)$ is isomorphic to the
	odd-torsion subgroup of $\Z/n$ when $k = 3$ and $k = 5$ and vanishes for all other $k\le 6$.
\end{enumerate}
\end{prop}
\begin{proof}
It suffices to work at odd primes. There is an odd-primary equivalence $\MTSpin\to\MTSO$
and, since $V$ is oriented, there is a Thom isomorphism $\MTSO\wedge (BD_{2n})_+\overset\simeq\to
\MTSO\wedge (BD_{2n})^V$. Therefore for~\eqref{untw_Spin} it suffices to study $\tOmega_*^\SO(BD_{2n})$.
Using
\cref{untwisted_dihedral_SO} for input, as well as the Künneth formula to determine $H_*(BD_{2n}\wedge
B\T)_p^\wedge$, one sees that the Atiyah-Hirzebruch spectral sequence computing these bordism groups collapses
for degree reasons in total degree $6$ and below. Then \cref{spinc_odd_primes} takes care of~\eqref{untw_Spinc}.
\end{proof}
\subsubsection{Class D, spinless case}
\label{para_D_spinless}
Let $f_0^D$ denote the equivariant local system of symmetry types for this case. \Cref{shear_D_thm} tells us that
to compute $\Ph_*^{D_{2n}}(\R^3, f_0^D)$, we should study the spin bordism of $X_n\coloneqq (BD_n)^{3-V_\lambda}$.
\begin{prop}
\label{odd_3d_dih_D_spinless}
Suppose $n$ is odd. Then
\begin{align*}
	\tOmega_0^\Spin(X_n) &\cong \Z\\
	\tOmega_1^\Spin(X_n) &\cong \Z/4\\
	\tOmega_2^\Spin(X_n) &\cong 0\\
	\tOmega_3^\Spin(X_n) &\cong \Z/n\\
	\tOmega_4^\Spin(X_n) &\cong \Z\\
	\tOmega_5^\Spin(X_n) &\cong \Z/16\\
	\tOmega_6^\Spin(X_n) &\cong 0,
\end{align*}
and therefore $\Ph_0^{D_{2n}}(\R^3, f_0^D) \cong 0$.
\end{prop}
\begin{proof}
\Cref{untwisted_dihedral_SO} shows that $\tOmega_k^\Spin(X_n)$ lacks odd-primary torsion for $k = 4,5$, so it
suffices to work at $2$. The inclusion $\Z/2\inj D_{2n}$ induces an isomorphism $H^*(BD_{2n};\Z/2)\to
H^*(B\Z/2;\Z/2)$, as we saw in the proof of \cref{odd_dihedral_equiv}, hence by naturality of the Thom isomorphism
gives an isomorphism
\begin{equation}
\label{XnOddDih}
	\tH^*(X_n;\Z/2)\overset\cong\longrightarrow \tH^*((B\Z/2)^{3-V_\lambda|_{B\Z/2}};\Z/2).
\end{equation}
Restricted to $\Z/2$, $\lambda\cong 2\sigma\oplus \R$, so by the stable Whitehead theorem,~\eqref{XnOddDih} gives a
stable $2$-primary equivalence $X_n\simeq (B\Z/2)^{2-2\sigma}$, or to $(B\Z/2)^{2\sigma-2}$ by \cref{relative_Thom}
(since for any vector bundle $V$, $V^{\oplus 4}$ admits a spin structure by the Whitney sum formula).
Giambalvo~\cite[\S 3]{Gia73} computes
$\tOmega_k^\Spin((B\Z/2)^{2\sigma-2})$, obtaining the free and $2$-torsion summands we claim in the theorem
statement.
\end{proof}
\begin{prop}[{Pedrotti~\cite[Theorem 8.0.8]{Ped17}}]
\label{dihedral_twspin_bordism_2mod4}
For $n\equiv 2\bmod 4$, $\tOmega_4^\Spin(X_n)\cong\Z$, and by \cref{torsion_k_theory} $\tOmega_5^\Spin(X_n)$ is
torsion. Therefore $\Ph_0^{D_{2n}}(\R^3, f_0^D)$ vanishes.
\end{prop}
\begin{rem}
Pedrotti reports this computation in terms of $w_1$ and $w_2$ of $3-V_\lambda$, rather than $\lambda$ itself, so we
should check that our characteristic classes agree with his: we want $w_1(3-V_\lambda) = 0$ and $w_2(3-V_\lambda) =
x^2 + xy + y^2$. Indeed $\Im(\lambda)\subset\SO_3$, so $V_\lambda$ is orientable, and from~\eqref{w2_2mod4_3d} that
$w_2(V_\lambda) = x^2 + xy + y^2$. Since $w_1(V_\lambda) = 0$, these are also $w_1$ and $w_2$ of $3-V_\lambda$, as
desired.
\end{rem}
\begin{prop}[{Pedrotti~\cite[Theorem 9.0.14]{Ped17}}]
\label{dihedral_twspin_bordism_0mod4}
For $n\equiv 0\bmod 4$, $\tOmega_4^\Spin(X_n)\cong\Z\oplus\Z/2$, and by \cref{torsion_k_theory}
$\tOmega_5^\Spin(X_n)$ is torsion. Therefore $\Ph_0^{D_{2n}}(\R^3, f_0^D)\cong\Z/2$.
\end{prop}
Pedrotti takes as input $w_1(3-V_\lambda) = 0$ and $w_2(3-V_\lambda) = w+x^2$, which agrees with the classes of
$V_\lambda$ (e.g.~\eqref{w2_0mod4_3d}). Beware that what we call $x$ he calls $y$, and vice versa!
\subsubsection{Class D, spin-$1/2$ case}
\label{para_D_spinful}
Let $f_{1/2}^D$ denote the equivariant local system of symmetry types for this case.
\Cref{3d_dih_not_pinc,shear_D_thm} tell us that to compute $\Ph_*^{D_{2n}}(\R^3, f_{1/2}^D)$, we should study the
spin bordism of $(BD_n)^{\Det(V_\lambda)-1}$. Since $V_\lambda$ is orientable, this is isomorphic to
$\Omega_4^\Spin(BD_{2n})$.
\begin{prop}[{Davighi-Gripaios-Lohitsiri~\cite[\S A.5]{DGL22}}]
\label{D2n_spin_bord_odd}
Suppose $n$ is odd. Then $\Omega_4^\Spin(BD_{2n})\cong\Z$ and $\Omega_5^\Spin(BD_{2n})\cong 0$.
\end{prop}
Thus
$\Ph_0^{D_{2n}}(\R^3, f_{1/2}^D)\cong 0$.
\begin{prop}[{Pedrotti~\cite[Theorems 8.0.4, 9.0.3]{Ped17}, Davighi-Gripaios-Lohitsiri~\cite[\S A.5]{DGL22}}]
\label{D2n_spin_bord_even}
For $n$ even, $\Omega_4^\Spin(BD_{2n})\cong\Z\oplus (\Z/2)^{\oplus 2}$.
\end{prop}
Pedrotti~\cite[Theorems 8.0.4 and 9.0.3]{Ped17} shows $\Omega_4^\Spin(BD_{2n})\cong\Z\oplus H_4(BD_{2n};\Z)$, and
the latter is computed by Handel~\cite[Theorem 5.2]{Han93}.
Bruner-Greenlees~\cite[Corollary 8.5.9]{BG10} also compute $\Omega_4^\Spin(BD_{2n})$ when $n$ is a power of $2$.

$\Omega_5^\Spin(BD_{2n})$ is torsion~\cite[(A.30)]{DGL22}, so $\Ph_0^D(\R^3, f_{1/2}^D)\cong
\Z/2\oplus\Z/2$.
\subsubsection{Class A, spinless case}
\label{para_A_spinless}
Let $f_0^A$ denote the equivariant local system of symmetry types in the spinless type A case. By
\cref{3d_dih_not_pinc}, we should compute $\tOmega_*^\Spinc(X_n)$, where $X_n\coloneqq (BD_{2n})^{3-V_\lambda}$.

When $n$ is odd, $V_\lambda$ is \spinc, so there is a Thom isomorphism $\MTSpin^c\wedge X_n\simeq\MTSpin^c\wedge
(BD_{2n})_+$.
\begin{thm}
\label{odd_dihedral_spinc_bordism}
Suppose $n$ is odd. Then
\begin{align*}
	\Omega_0^\Spinc(BD_{2n}) &\cong \Z\\
	\Omega_1^\Spinc(BD_{2n}) &\cong \Z/2\\
	\Omega_2^\Spinc(BD_{2n}) &\cong \Z\\
	\Omega_3^\Spinc(BD_{2n}) &\cong \Z/4n\\
	\Omega_4^\Spinc(BD_{2n}) &\cong \Z^2\\
	\Omega_5^\Spinc(BD_{2n}) &\cong \Z/8n\oplus\Z/2\\
	\Omega_6^\Spinc(BD_{2n}) &\cong \Z^2.
\end{align*}
Therefore $\Ph_0^{D_{2n}}(\R^3, f_0^A)\cong 0$.
\end{thm}
\begin{proof}
\Cref{untwisted_dihedral_SO} accounts for the odd-primary torsion, so we just have to work at $2$. The map
$\Z/2\inj D_{2n}$ induced by a choice of reflection defines an isomorphism on mod $2$ cohomology, therefore by the
stable Whitehead theorem is a $2$-local stable equivalence. Therefore it defines an isomorphism
$\Omega_*^\Spinc(B\Z/2)_2^\wedge\to\Omega_*^\Spinc(BD_{2n})_2^\wedge$, and the \spinc bordism of $B\Z/2$ is
computed by Bahri-Gilkey~\cite{BG87a, BG87b}.
\end{proof}
\begin{thm}
\label{unmixed_A_para_2mod4}
Suppose $n\equiv 2\bmod 4$. Then the first few \spinc bordism groups of $X_n$ are
\begin{align*}
	\tOmega_0^\Spinc(X_n) &\cong \Z\\
	\tOmega_1^\Spinc(X_n) &\cong (\Z/2)^{\oplus 2}\\
	\tOmega_2^\Spinc(X_n) &\cong \Z\\
	\tOmega_3^\Spinc(X_n) &\cong \Z/n\oplus (\Z/2)^{\oplus 2}\\
	\tOmega_4^\Spinc(X_n) &\cong \Z^2,
\end{align*}
and $\tOmega_5^\Spinc(X_n)$ is torsion.
\end{thm}
\begin{proof}
The odd-torsion subgroups can be read off of~\eqref{untwisted_dihedral_SO}. For the $2$-primary part, we use the
Adams spectral sequence over $\cE(1)$. Letting $\oU$ denote the Thom class, we saw $w_1(V_\lambda) = 0$, so
$\Sq^1(\oU) = 0$, and~\eqref{w2_2mod4_3d} $w_2(V_\lambda) = x^2 + xy + y^2$, so $\Sq^2(\oU) = \oU(x^2+xy+y^2)$.
Using this and the Cartan formula, we have an $\cE(1)$-module isomorphism
\begin{equation}
	\tH^*(X_n;\Z/2) \cong \textcolor{BrickRed}{\uQ} \oplus
		\textcolor{RedOrange}{\Sigma\cE(1)} \oplus
		\textcolor{Goldenrod!67!black}{\Sigma\cE(1)} \oplus
		\textcolor{Green}{\Sigma^3\cE(1)} \oplus
		\textcolor{MidnightBlue}{\Sigma^3\cE(1)} \oplus
		\textcolor{Fuchsia}{\Sigma^3\cE(1)} \oplus P,
\end{equation}
where $P$ is $4$-connected. We draw this in \cref{2mod4_3d_dih_spin_unmixed}, left. A priori $\Ext(P)$ could have
nonzero differentials to the $4$-line and therefore affect our computation, but we will see that this cannot happen
without needing to determine $\Ext(P)$.

\begin{figure}[h!]
\begin{subfigure}[c]{0.4\textwidth}
\begin{tikzpicture}[scale=0.6, every node/.style = {font=\tiny}]
	\foreach \y in {0, ..., 7} {
		\node at (-2, \y) {$\y$};
	}

	\begin{scope}[BrickRed]
		\EoneQnMark{0}{0}{$\oU$};
	\end{scope}
	\begin{scope}[RedOrange]
		\Eone{1.5}{1}{$\oU x$};
	\end{scope}
	\begin{scope}[Goldenrod!67!black]
		\Eone{3.5}{1}{$\oU y$};
	\end{scope}
	\begin{scope}[Green]
		\Eone{5.5}{3}{$\oU x^3$};
	\end{scope}
	\begin{scope}[MidnightBlue]
		\Eone{7.5}{3}{$\oU y^3$}{};
	\end{scope}
	\begin{scope}[Fuchsia]
		\Eone{9.5}{3}{$\oU x^3y$}{};
	\end{scope}
\end{tikzpicture}
\end{subfigure}
\qquad\qquad
\begin{subfigure}[c]{0.4\textwidth}
\begin{sseqdata}[name=spintwdih2mod4, classes=fill, xrange={0}{4}, yrange={0}{2}, scale=0.7,
	x label = {$\displaystyle{s\uparrow \atop t-s\rightarrow}$},
	x label style = {font = \small, xshift = -16.25ex, yshift=4ex}, >=stealth, Adams grading]
\begin{scope}[BrickRed]
	\class(0, 0)\AdamsTower{}
	\class(2, 0)\AdamsTower{}
	\class(4, 1)\AdamsTower{}
	\class(6, 2)\class(6, 3)\class(6, 4)
	\begin{scope}[BrickRed!40!white]
		\structline(0, 0)(2, 1)
		\structline(2, 0)(4, 1)
		\foreach \y in {1, 2} {
			\structline(0, \y)(2, \y+1)
			\structline(2, \y)(4, \y+1)
			\structline(4, \y)(6, \y+1)
		}
	\end{scope}
\end{scope}
\class[RedOrange](1, 0)
\class[Goldenrod!75!black](1, 0)
\class[Green](3, 0)
\class[MidnightBlue](3, 0)
\class[Fuchsia](3, 0)
\end{sseqdata}
\printpage[name=spintwdih2mod4, page=2]
\end{subfigure}

\caption{Left: the $\cE(1)$-module structure on $\tH^*(X_n;\Z/2)$ when $n\equiv 2\bmod 4$. The pictured summand
contains all elements in degrees $4$ and below. Right: the $E_2$-page of the corresponding Adams spectral sequence
computing $\widetilde{\ku}_*(X_n)_2^\wedge$.}
\label{2mod4_3d_dih_spin_unmixed}
\end{figure}

We saw $\Ext_{\cE(1)}(\textcolor{BrickRed}{\uQ})$ in~\eqref{ExtN2}, so we can draw the $E_2$-page of the
Adams spectral sequence in \cref{2mod4_3d_dih_spin_unmixed}, right. $h_0$-equivariance rules out nonzero
differentials in degrees $3$ and below, but a priori there could be a nonzero differential from the $5$-line to
then $4$-line. To rule this out, use \cref{torsion_k_theory} to see that $\widetilde{\ku}_4(X_n)$ has one free
summand. Therefore there cannot be any nonzero differentials to the $4$-line: $h_0$-equivariance would mean that if
there were such a differential, it would kill all but finitely many summands in the $4$-line of the $E_2$-page,
preventing $\widetilde{\ku}_4(X_n)$ from having a free part.
\end{proof}

\begin{thm}
\label{unmixed_A_para_0mod4}
When $n\equiv 0\bmod 4$, the first few \spinc bordism groups of $X_n$ are
\begin{align*}
	\tOmega_0^\Spinc(X_n) &\cong \Z\\
	\tOmega_1^\Spinc(X_n) &\cong (\Z/2)^{\oplus 2}\\
	\tOmega_2^\Spinc(X_n) &\cong \Z\\
	\tOmega_3^\Spinc(X_n) &\cong \Z/n\oplus (\Z/2)^{\oplus 2}\\
	\tOmega_4^\Spinc(X_n) &\cong \Z^2,
\end{align*}
and $\tOmega_5^\Spinc(X_n)$ is torsion. Therefore $\Ph_0^{D_{2n}}(\R^3, f_0^A)\cong 0$.
\end{thm}
\begin{proof}
The odd-torsion subgroups are calculated in \cref{untwisted_dihedral_SO}. For the $2$-torsion, we use the Adams
spectral sequence over $\cE(1)$. Recall that $w_1(V_\lambda) = 0$ and (from~\eqref{w2_0mod4_3d}) $w_2(V_\lambda) =
w + x^2$, so $w_1(3-V_\lambda) = 0$ and $w_2(3-V_\lambda) = w + x^2$. Thus in $\tH^*(X_n;\Z/2)$, $\Sq^1(\oU) = 0$
and $\Sq^2(\oU) = \oU(w + x^2)$. Using this and the Cartan formula, we can compute the $\cE(1)$-action on
$\tH^*(X_n;\Z/2)$, and find that
\begin{equation}
	\tH^*(X_n;\Z/2) \cong \textcolor{BrickRed}{\uQ} \oplus
		\textcolor{RedOrange}{\Sigma\cE(1)} \oplus
		\textcolor{Goldenrod!67!black}{\Sigma\cE(1)} \oplus
		\textcolor{Green}{\Sigma^3\cE(1)} \oplus
		\textcolor{PineGreen}{\Sigma^3\cE(1)} \oplus
		\textcolor{MidnightBlue}{\Sigma^3\Z/2} \oplus
		\textcolor{Fuchsia}{\Sigma^4\uQ} \oplus P,
\end{equation}
where $P$ is $4$-connected. We draw this in \cref{0mod4_3d_dih_fig}, left. We will see in a moment that $\Ext(P)$
has no nonzero differentials to elements in degree $4$ and below, which means we can ignore it in our computations.
We saw $\Ext_{\cE(1)}(\textcolor{BrickRed}{\uQ})$ in~\eqref{ExtN2}, so we can draw the $E_2$-page of the
Adams spectral sequence in \cref{2mod4_3d_dih_spin_unmixed}, right.
\begin{figure}[h!]
\begin{subfigure}[c]{0.4\textwidth}
\begin{tikzpicture}[scale=0.6, every node/.style = {font=\tiny}]
	\foreach \y in {0, ..., 7} {
		\node at (-2, \y) {$\y$};
	}

	\begin{scope}[BrickRed]
		\EoneQnMark{0}{0}{$\oU$};
	\end{scope}
	\begin{scope}[RedOrange]
		\Eone{1.5}{1}{$\oU x$};
	\end{scope}
	\begin{scope}[Goldenrod!67!black]
		\Eone{3.5}{1}{$\oU y$};
	\end{scope}
	\begin{scope}[Green]
		\Eone{5.5}{3}{$\oU x^3$};
	\end{scope}
	\begin{scope}[PineGreen]
		\Eone{7.5}{3}{$\oU y^3$};
	\end{scope}
	\draw[thick, gray, dashed] (9.75, 3) -- (9.75, 4);
	\tikzpt{9.75}{3}{$\oU wy$}{MidnightBlue};
	\begin{scope}[Fuchsia]
		\EoneQnMark{9.75}{4}{$\oU w^2$}{};
	\end{scope}
\end{tikzpicture}
\end{subfigure}
\qquad\qquad
\begin{subfigure}[c]{0.4\textwidth}
\begin{sseqdata}[name=spintwdih0mod4, classes=fill, xrange={0}{4}, yrange={0}{2}, scale=0.7,
	x label = {$\displaystyle{s\uparrow \atop t-s\rightarrow}$},
	x label style = {font = \small, xshift = -16.25ex, yshift=4ex}, >=stealth, Adams grading]
\begin{scope}[BrickRed]
	\class(0, 0)\AdamsTower{}
	\class(2, 0)\AdamsTower{}
	\class(4, 1)\AdamsTower{}
	\class(6, 2)\class(6, 3)\class(6, 4)
\end{scope}
\class[RedOrange](1, 0)
\class[Goldenrod!75!black](1, 0)
\class[Green](3, 0)
\class[PineGreen](3, 0)
\begin{scope}[white]
	\class(3, 1)\class(3, 1)
	\class(3, 2)\class(3, 2)
	\class(3, 3)\class(3, 3)
	\class(4, 0)
\end{scope}
\begin{scope}[MidnightBlue]
	\class(3, 0)\AdamsTower{}
	\class(5, 1)\class(5, 2)\class(5, 3)
\end{scope}
\begin{scope}[Fuchsia]
	\class(4, 0)\AdamsTower{}
\end{scope}
\d[gray]2(4, 0, -1)(3, 2, -1)
\end{sseqdata}
\printpage[name=spintwdih0mod4, page=2]
\end{subfigure}
\caption{Left: the $\cE(1)$-module structure on $\tH^*(X_n;\Z/2)$ when $n\equiv 0\bmod 4$. The pictured summand
contains all elements in degrees $4$ and below. The gray dashed line indicates a $\Z/2^r$ Bockstein, where $r$ is
the largest number for which $2^r\mid n$; this is not part of the $\cE(1)$-module structure, but we use it in
\cref{unmixed_A_para_0mod4} to resolve a differential. Right: the $E_2$-page of the corresponding Adams spectral
sequence computing $\widetilde{\ku}_*(X_n)_2^\wedge$; $v_1$-actions are hidden for legibility. We will see in
\cref{unmixed_A_para_0mod4} that there is a $d_r$ from the purple tower in the $4$-line to the $3$-line, though it
is not always the $d_2$ pictured.}
\label{0mod4_3d_dih_fig}
\end{figure}
Margolis' theorem (\cref{margolis}) implies the only possible nonzero differentials from an element of topological
degree $4$ or below are the differentials from a tower in the $4$-line to the blue tower in the $3$-line, and
\cref{torsion_k_theory} implies $\widetilde{ku}_4(X_n)$ has free rank $1$, so for some $r$ this differential $d_r$
is nonzero. Moreover, its source must be the purple tower: the red tower is in the image of $v_1\colon
E_2^{s,s+2}\to E_2^{s+1, s+5}$, so if $d_r(x) = y$ for any element $x$ of the red tower in degree $4$, then $y$ is
also in the image of $v_1$, but the blue tower is not in this image. Therefore we know that $d_r$ kills the entire
purple tower in degree $4$, and the red tower survives to the $E_\infty$-page: the red tower supports no nonzero
differentials to the $3$-line, and if there were a differential from the $5$-line to the red tower, $h_0$-linearity
guarantees it would kill all but finitely many summands of the red tower, contradicting \cref{torsion_k_theory}.

It remains only to determine the value of $r$. In $H^*(BD_{2n};\Z/2)$, the $\Z/2^k$ Bockstein carries (a preimage
of) $wy$ to $w^2$, where $k$ is the largest number such that $2^k\mid n$. This can be checked by, e.g., pulling
back to $BC_n$, where this Bockstein is discussed by~\cite{Cam17, DL20b}. The Thom isomorphism theorem implies the
$\Z/2^k$ Bockstein sends (a preimage of) $\oU wy$ to $\oU w^2$, and therefore by the May-Milgram
theorem~\cite{MM81}, $r = k$.
\end{proof}
\subsubsection{Class A, spin-$1/2$ case}
\label{para_A_spinful}
Let $f_{1/2}^A$ denote the equivariant local system of symmetry types in the spin-$1/2$ type A case. In this case
the ansatz tells us to study $\Omega_*^\Spinc(BD_{2n})$.
\begin{prop}
\label{para_odd_dih_spinful}
For $n$ odd, $\Ph_0^{D_{2n}}(\R^3, f_{1/2}^A) = 0$.
\end{prop}
\begin{proof}
This follows from our computation of $\Omega_k^\Spinc(BD_{2n})$ in \cref{odd_dihedral_spinc_bordism}.
\end{proof}
\begin{thm}
\label{2mod4_spinc_bord_dihedral}
Suppose $n\equiv 2\bmod 4$. Then
\begin{align*}
	\Omega_0^\Spinc(BD_{2n}) &\cong \Z\\
	\Omega_1^\Spinc(BD_{2n}) &\cong (\Z/2)^{\oplus 2}\\
	\Omega_2^\Spinc(BD_{2n}) &\cong \Z\oplus\Z/2\\
	\Omega_3^\Spinc(BD_{2n}) &\cong \Z/n \oplus (\Z/4)^{\oplus 2}\oplus (\Z/2)^{\oplus 2}\\
	\Omega_4^\Spinc(BD_{2n}) &\cong \Z^2\oplus (\Z/2)^{\oplus 3}\\
	\Omega_5^\Spinc(BD_{2n}) &\cong \Z/n \oplus (\Z/8)^{\oplus 2}\oplus\Z/4\oplus (\Z/2)^{\oplus 5}\\
	\Omega_6^\Spinc(BD_{2n}) &\cong \Z^2\oplus (\Z/2)^{\oplus 6}.
\end{align*}
\end{thm}
\begin{proof}
We calculated the odd-primary torsion in these bordism groups in \cref{untwisted_dihedral_SO}; now the $2$-primary
part. The inclusion $\Z/2\times\Z/2\to D_{2n}$ given by a reflection and a rotation by $\pi$ induces an isomorphism
on mod $2$ cohomology, so by the stable Whitehead theorem, $BD_{2n}\to B(\Z/2\times\Z/2)$ is an equivalence after
stabilizing and $2$-completing. Therefore $\Omega_*^\Spinc(BD_{2n})_2^\wedge
\overset\cong\to\Omega_*^\Spinc(B(\Z/2\times\Z/2))_2^\wedge$; since $\MTSpinc\to\ku\vee\Sigma^4\ku$ is an
isomorphism in degrees $8$ and below, it suffices to know $\ku_*(B(\Z/2\times\Z/2))$.

Recall from~\eqref{ossa} Ossa's splitting~\cite[Proposition 3]{Oss89} of $\ku\wedge B\Z/2\wedge B\Z/2$; combining
this with the stable splitting
\begin{equation}
	\Sigma^\infty(B\Z/2\times B\Z/2)_+ \simeq \Sph\vee \Sigma^\infty B\Z/2\vee \Sigma^\infty B\Z/2\vee
	\Sigma^\infty (B\Z/2\wedge B\Z/2),
\end{equation}
we see that $\ku_*(B(\Z/2\times\Z/2))$ can be assembled from the following pieces.
\begin{enumerate}
	\item $\ku_*(\pt)$, which contributes $\Z$ in even degrees and $0$ in odd degrees.
	\item Two copies of $\widetilde{\ku}_*(B\Z/2)$. Hashimoto~\cite[Theorem 3.1]{Has83} shows each copy vanishes in
	even degrees and is isomorphic to $\Z/2^{k+1}$ in odd degree $2k+1$.
	\item $\widetilde{\ku}_*(\Sigma^2 B\Z/2)$. Hashimoto (\textit{ibid}.) shows this vanishes in even degrees and
	is isomorphic to $\Z/2^k$ in odd degree $2k+1$.
	\item $\pi_*(\Sigma^2H\Z/2[u, v])$, which contributes $0$ in degrees $0$ and $1$ and $(\Z/2)^{\oplus(k-1)}$ in
	degrees $k\ge 2$.
\end{enumerate}
Putting this together and adding in the odd-primary torsion, we obtain $\ku_k(BD_{2n})$ for $k\le 6$; using the
Anderson-Brown-Peterson isomorphism $\Omega_k^\Spinc(BD_{2n})\cong \ku_k(BD_{2n})\oplus\ku_{k-4}(BD_{2n})$, valid
for $k < 8$, we obtain the bordism groups in the theorem statement.
\end{proof}
\begin{thm}
\label{spinc_dihedral_bordism}
Suppose $n\equiv 0\bmod 4$. Then 
\begin{align*}
	\Omega_0^\Spinc(BD_{2n}) &\cong \Z\\
	\Omega_1^\Spinc(BD_{2n}) &\cong (\Z/2)^{\oplus 2}\\
	\Omega_2^\Spinc(BD_{2n}) &\cong \Z\oplus\Z/2\\
	\Omega_3^\Spinc(BD_{2n}) &\cong \Z/n\oplus (\Z/4)^{\oplus 2}\\
	\Omega_4^\Spinc(BD_{2n}) &\cong \Z^2 \oplus (\Z/2)^{\oplus 2},
\end{align*}
and $\Omega_5^\Spinc(BD_{2n})$ is torsion. Therefore $\Ph_0^{D_{2n}}(\R^3, f_{1/2}^A)\cong (\Z/2)^{\oplus 2}$.
\end{thm}
For $n = 4$ this is due to Bruner-Greenlees~\cite[Theorem 3.5.1]{BG03}.
\begin{proof}
We focus on $\ku\wedge BD_{2n}$; as usual, we append $\ku_*(\pt)$ to pass from reduced to unreduced $\ku$-homology,
assemble $2$-primary \spinc bordism from $\ku_*(\bl)\oplus\ku_{*-4}(\bl)$, and use \cref{untwisted_dihedral_SO}
for the odd-primary torsion. Mitchell-Priddy~\cite[Theorem A]{MP84} show that for $n\equiv 0\bmod 4$, there is a
prime $q$ of the form $\pm 1\bmod 8$ and a $2$-local homotopy equivalence\footnote{Mitchell-Priddy exhibit this
when $n$ is in addition a power of $2$; for the general case, let $n = 2^\ell m$ for $m$ odd; then the inclusion
$D_{2^\ell}\inj D_n$ induces an isomorphism on mod $2$ cohomology, hence a $2$-local stable equivalence
$\Sigma^\infty (BD_{2^\ell})_+\simeq_{(2)} \Sigma^\infty (BD_{2n})_+$.}
\begin{equation}
	\Sigma^\infty BD_{2n}\simeq_{(2)} \Sigma^\infty B\PSL_2(\F_q) \vee L(2)\vee L(2)\vee \Sigma^\infty B\Z/2 \vee
	\Sigma^\infty B\Z/2
\end{equation}
for a certain spectrum $L(2)$ defined in~\cite{MP83}. Thus, to prove the theorem, we will calculate
$\widetilde{\ku}_*(B\PSL_2(\F_q))$ and $\ku_*(L(2))$ in low degrees in \cref{ku_L2,ku_PSL}, then sum these groups
with two copies of $\widetilde{\ku}_*(B\Z/2)$, computed by Hashimoto~\cite[Theorem 3.1]{Has83}.
\begin{lem}
\label{ku_L2}
There are natural numbers $\alpha_i$ such that there is a homotopy equivalence
\begin{equation}
	\ku\wedge L(2)\simeq\bigvee_{\alpha_i: i\ge 1} \Sigma^{\alpha_i} H\Z/2,
\end{equation}
where $\alpha_1 = 4$, $\alpha_2 = 6$, and for $i\ge 3$, $\alpha_i\ge 8$.
\end{lem}
\begin{proof}
Bayen~\cite[Proposition 3.4.4, \S 3.5.3]{Bay94} shows that as an $\cA(1)$-module, $H^*(L(2);\Z/2)$ is isomorphic to
a direct sum of shifts of $\cA(1)$ beginning with a $\Sigma^4\cA(1)$ and no other $\Sigma^k\cA(1)$ summands for $k
\le 7$. Since $\cA(1)\cong\cE(1)\oplus \Sigma^2\cE(1)$ as $\cE(1)$-modules, we learn that as an $\cE(1)$-module,
$H^*(L(2);\Z/2)$ is free on generators in degrees at least $8$, together with generators in degrees $4$ and $6$.
Margolis' theorem (\cref{margolis}) then implies the lemma statement.
\end{proof}
\begin{lem}
\label{ku_PSL}
Let $q$ be a prime of the form $\pm 1\bmod 8$. Then there is an $r \ge 2$ depending on $q$ such that, modulo
odd-primary torsion, $\widetilde{\ku}_2(B\PSL_2(\F_q))\cong\Z/2$, $\widetilde{\ku}_3(B\PSL_2(\F_q))\cong \Z/2^r$,
and $\widetilde{\ku}_k(B\PSL_2(\F_q)) = 0$ for $k = 0,1,4$.
\end{lem}
\begin{proof}
Bayen~\cite[Proposition 3.4.2]{Bay94} establishes an $\cA(1)$-module isomorphism $\widetilde
H^*(B\PSL_2(\F_q);\Z/2)\cong \Sigma^2 J\oplus \Sigma^3\uQ\oplus P$ for a $5$-connected $\cA(1)$-module $P$. Since
$J\cong \cE(1)\oplus\Sigma^2\Z/2$ as $\cE(1)$-modules, as we discussed in the proof of \cref{dih_spinc_0mod4}, we
therefore have an $\cE(1)$-module isomorphism
\begin{equation}
	\widetilde H^*(B\PSL_2(\F_q);\Z/2) \cong \textcolor{Green}{\Sigma^2\cE(1)} \oplus
		\textcolor{MidnightBlue}{\Sigma^3 \uQ} \oplus
		\textcolor{Fuchsia}{\Sigma^4\Z/2} \oplus P.
\end{equation}
We draw this in \cref{spinc_dihedral_figure}, left. Using~\eqref{ExtN2} and~\eqref{E1Z2Ext}
for $\Ext_{\cE(1)}(\textcolor{MidnightBlue}{\uQ})$ and $\Ext_{\cE(1)}(\textcolor{Fuchsia}{\Z/2})$, respectively, we
draw the Adams $E_2$-page in \cref{spinc_dihedral_figure}, right. This Adams spectral sequence does not collapse a
priori, and indeed by \cref{torsion_k_theory}, there must be a nonzero $d_r$ differential from the $h_0$-tower in
topological degree $4$ to the $h_0$-tower in topological degree $3$ for some $r$. After that differential, there
can be no further differentials or extension problems in the range we consider, so we have finished the proof.
\end{proof}

\begin{figure}[h!]
\begin{subfigure}[c]{0.25\textwidth}
\begin{tikzpicture}[scale=0.6, every node/.style = {font=\tiny}]
	\foreach \y in {2, ..., 6} {
		\node at (-1.5, \y) {$\y$};
	}
	\begin{scope}[Green]
		\Eone{0}{2}{$\nu_2$};
	\end{scope}
	\begin{scope}[MidnightBlue]
		\EoneQnMark{2.5}{3}{$\overline\nu_3$};
	\end{scope}
	\tikzptR{2.5}{4}{$\nu_2^2$}{Fuchsia};
\end{tikzpicture}
\end{subfigure}
\begin{subfigure}[c]{0.35\textwidth}
\begin{sseqdata}[name=spincdih0mod4, classes=fill, xrange={0}{4}, yrange={0}{3}, scale=0.7,
	x label = {$\displaystyle{s\uparrow \atop t-s\rightarrow}$},
	x label style = {font = \small, xshift = -16.25ex, yshift=4ex}, >=stealth, Adams grading]
	\class[Green](2, 0)
	\begin{scope}[MidnightBlue]
		\class(3, 0)\AdamsTower{}
		\class(5, 0)\AdamsTower{}
	\end{scope}
	\begin{scope}[Fuchsia]
		\class(4, 0)\AdamsTower{}
	\end{scope}
	\d[gray]2(4, 0, 1)(3, 2, -1)
\end{sseqdata}
\printpage[name=spincdih0mod4, page=2]
\end{subfigure}
\caption{Left: the $\cE(1)$-module structure on $\tH^*(B\PSL_2(\F_q);\Z/2)$ in low degrees, where $q$ is a prime of
the form $\pm 1\bmod 8$. Names of cohomology classes are as in Mitchell-Priddy~\cite[Theorem 2.6]{MP84}.
Right: the $E_2$-page of the Adams spectral sequence computing $\widetilde{\ku}(B\PSL_2(\F_q))_2^\wedge$.
$v_1$-actions are hidden for readability. We will see in \cref{ku_PSL} that there is a nonzero
differential from the $4$-line to the $3$-line, though it is not necessarily the $d_2$ pictured.}
\label{spinc_dihedral_figure}
\end{figure}

Thus we know all the $\ku$-homology groups of the summands of $BD_{2n}$ in the range we need, so we can write down
the \spinc bordism groups in the theorem statement.
\end{proof}

\section{Examples: tetrahedral, octahedral, and icosahedral symmetries}
\label{3d_pt}
We now consider the sporadic point groups in three dimensions: chiral and full tetrahedral symmetry
(\S\ref{s_chiral_tet}, resp.\ \S\ref{s_full_tet}), pyritohedral
symmetry (\S\ref{pyrit}), chiral and full octahedral symmetry (\S\ref{s_chiral_oct}, resp.\ \S\ref{s_full_oct}),
and chiral and full icosahedral symmetry (\S\ref{s_chiral_ico}, resp.\ \S\ref{s_full_ico}). At the time this paper
first appeared on arXiv, interacting fermionic phases with these symmetry groups had not appeared in the
literature, so our calculations offered predictions in physics. Since then, Zhang-Ning-Qi-Gu~\cite[Tables I and II,
\S\S III and IV]{ZNQG25} computed the groups of these phases (except for the icosahedral symmetries) using
different methods, finding nearly complete agreement. The few discrepancies turned out to be calculational errors
in the first version of this paper; in the current version, which fixes these errors, there is complete agreement
with~\cite{ZNQG25}. We thank Zheng-Cheng Gu, Yang Qi, and Weicheng Ye for helpful discussions on this matter.

\begin{table}[h!]
\begin{tabular}{l c c c c c}
\toprule
Point group & Ref. & Class D, spinless & Class D, spin-$1/2$ & Class A, spinless & Class A, spin-$1/2$\\
\midrule
Chiral tet.\ ($A_4$, $T$) & \S\ref{s_chiral_tet} & $0$ & $0$ & $0$ & $0$\\
Pyrit.\ ($A_4\times\Z/2$, $T_h$) & \S\ref{pyrit} & $(\Z/2)^{\oplus 3}$ & $\Z/2$ &
	$\Z/4\oplus (\Z/2)^{\oplus 3}$ & \tblref{mixed_pyr_classA_thm}{$\Z/8\oplus (\Z/2)^{\oplus 3}$}\\
Full tet.\ ($S_4$, $T_d$) & \S\ref{s_full_tet} & $(\Z/2)^{\oplus 3}$ & $0$ &
$(\Z/2)^{\oplus 4}$ & $\Z/8\oplus (\Z/2)^{\oplus 2}$\\
\addlinespace
Chiral oct.\ ($S_4$, $O$) & \S\ref{s_chiral_oct} & $\Z/2$ & $\Z/2$ & $0$ & $\Z/2$ \\
Full oct.\ ($S_4\times\Z/2$, $O_h$) & \S\ref{s_full_oct} & $(\Z/2)^{\oplus 5}$ & $(\Z/2)^{\oplus 2}$ &
$\Z/4\oplus (\Z/2)^{\oplus 4}$ & $\Z/8\oplus \Z/4\oplus (\Z/2)^{\oplus 4}$\\
\addlinespace
Chiral icos.\ ($A_5$, $I$) & \S\ref{s_chiral_ico} & $0$ & $0$ & $0$ & $0$\\
Full icos.\ ($A_5\times\Z/2$, $I_h$) & \S\ref{s_full_ico} & $(\Z/2)^{\oplus 3}$ & $\Z/2$ &
$\Z/4\oplus (\Z/2)^{\oplus 3}$ & $\Z/8\oplus (\Z/2)^{\oplus 3}$\\
\bottomrule
\end{tabular}
\caption{Phase homology groups in dimension $3+1$ equivariant with respect to various tetrahedral, octahedral, and
icosahedral symmetries and the ways they can mix with fermion parity. See the referenced sections for how the
fermionic crystalline equivalence principle associates this data with symmetry types for invertible TFTs.}
\label{TOI_table}
\end{table}

	\subsection{Chiral tetrahedral symmetry}
		\label{tetrahedral}
		\label{s_chiral_tet}
We compute phase homology groups equivariant for a chiral tetrahedral symmetry $\lambda\colon A_4\to\SO_3$.
We will
show that our ansatz implies there are no nontrivial phases with either spinless or spin-$1/2$ fermions in both
class D and class A.
As usual, $V_\lambda\to BA_4$ denotes the
vector bundle associated to $\lambda$.

\begin{prop}
\label{a4z2}
$H^*(BA_4;\Z/2)\cong\Z/2[u,v,w]/(u^3 + v^2 + w^2 + vw)$, where $\abs u = 2$ and $\abs v = \abs w = 3$.
$\Sq(u) = u + v + w + u^2$, $\Sq(v) = v+u^2 + uw + v^2$, and $\Sq(w) = w + u^2 + uv + w^2$.
\end{prop}
Except for the Steenrod operations, this result can be found in several places, such as~\cite{King}
and~\cite[Theorem III.1.3]{AM04}, so we will be brief.
\begin{proof}[Proof sketch]
Use the Lyndon-Hochschild-Serre spectral sequence~\cite{Lyn48, Ser50, HS53} for the short exact sequence $1\to
\Z/2\times\Z/2\to A_4\to\Z/3\to 1$; the mod 2 cohomology of $\Z/3$ is trivial, so the spectral sequence collapses,
and
\begin{equation}
\label{A4Klein}
	H^*(BA_4;\Z/2)\cong H^0(B\Z/3; H^*(B\Z/2\times B\Z/2;\Z/2)) = H^*(B\Z/2\times B\Z/2;\Z/2)^{\Z/3}.
\end{equation}
We can choose this $\Z/3$-action to be such that a generator of $\Z/3$ acts on $\Z/2\times\Z/2 =
\set{1,\alpha,\beta,\alpha+\beta}$ by $\alpha\mapsto \alpha+\beta$, $\beta\mapsto\alpha$, and
$\alpha+\beta\mapsto\beta$. In a mild abuse of notation, we identify $\Z/2\times\Z/2$ with $H^1(B\Z/2\times
B\Z/2;\Z/2) \cong \Hom(\Z/2\times\Z/2, \Z/2)$: these are dual $\Z/2$-vector spaces, and we have a basis for one,
which we identify with the dual basis vectors of the other. Thus $H^*(B\Z/2\times B\Z/2;\Z/2)\cong\Z/2[\alpha,
\beta]$.

The unique nonzero degree-$2$ cohomology class fixed by $\Z/3$ is $u\coloneqq \alpha^2 + \alpha\beta + \beta^2$,
and two linearly independent degree-$3$ classes fixed by $\Z/3$ are $v\coloneqq \alpha^3 + \alpha^2\beta + \beta^3$
and $w\coloneqq \alpha^3 + \alpha\beta^2 + \beta^3$, whence the relation.

For the Steenrod squares, the identification in~\eqref{A4Klein} of $H^*(BA_4;\Z/2)$ as a subalgebra of
$H^*(B\Z/2\times B\Z/2;\Z/2)$ is the pullback map for $B\Z/2\times B\Z/2\to BA_4$, hence $\cA$-equivariant, so we
can compute $\Sq(u)$ in $H^*(B\Z/2\times B\Z/2;\Z/2)$; the computation follows from $\Sq(\alpha) = \alpha+\alpha^2$
and $\Sq(\beta) = \beta + \beta^2$.
\end{proof}
\begin{lem}
\label{SWalt}
$w_1(V_\lambda) = 0$ and $w_2(V_\lambda) = u$.
\end{lem}
\begin{proof}
Since $V_\lambda$ is orientable, $w_1(V_\lambda) = 0$, and since $V_\lambda$ is not spin, $w_2(V_\lambda)\ne 0$.
Since $H^2(BA_4;\Z/2)\cong\Z/2\cdot u$, $w_2(V_\lambda) = u$.
\end{proof}
One way to see that this representation is not spin is to look at the \term{binary tetrahedral group}
$\mathit{2T}$, defined to be the preimage of $A_4\subset\SO_3$ under the double cover $\Spin_3\surj\SO_3$. If
$V_\lambda$ were spin, $\mathit{2T}$ would be a split extension of $A_4$ by $\mu_2$, but it is not split.
%
%
\subsubsection{Class D, spinless case}
If $A_4$ does not mix with the symmetry type, our ansatz reduces to that of Freed-Hopkins, which reduces the
computation of these $A_4$-equivariant phase homology groups to the computation of $[\MTSpin\wedge
(BA_4)^{3-V_\lambda}, \Sigma^5I_\Z]$.
\begin{thm}
\label{spin_bord_X}
The first few spin bordism groups of $X\coloneqq (BA_4)^{3-V_\lambda}$ are
\begin{align*}
	\tOmega_0^\Spin(X) &\cong\Z\\
	\tOmega_1^\Spin(X) &\cong\Z/3\\
	\tOmega_2^\Spin(X) &\cong 0\\
	\tOmega_3^\Spin(X) &\cong\Z/6\\
	\tOmega_4^\Spin(X) &\cong\Z\\
	\tOmega_5^\Spin(X) &\cong \Z/18\oplus\Z/2\\
	\tOmega_6^\Spin(X) &\cong\Z/2\\
	\tOmega_7^\Spin(X) &\cong \Z/9.
\end{align*}
Thus if $f_0^D$ denotes the $A_4$-equivariant local system of symmetry types for this case, $\Ph_0^{A_4}(\R^3,
f_0^D) = 0$.
\end{thm}
\begin{proof}
At the prime $2$, we use the Adams spectral sequence; if $p$ is an odd prime, the map
$\tOmega_*^\Spin(X)\to\tOmega_*^\SO(X)$ is an isomorphism on $p$-torsion, and we will determine the $p$-torsion
part of $\tOmega_*^\SO(X)$.

First, the $2$-primary piece.
Letting $\oU$ denote the mod $2$ Thom class as usual, $\Sq^1(\oU) = 0$ and $\Sq^2(\oU) = \oU u$. This and
\cref{a4z2} allow us to determine the $\cA(1)$-module structure on $\tH^*(X;\Z/2)$ in low degrees, as depicted in
\cref{a1moda4}, left.
\begin{figure}[h!]
\begin{subfigure}[c]{0.4\textwidth}
\begin{tikzpicture}[scale=0.6, every node/.style = {font=\tiny}]
  	\foreach \y in {0, ..., 11} {
		\node at (-2, \y) {$\y$};
	}
	\begin{scope}[MidnightBlue]
		\SpanishQnMark{0}{0}{$\oU$}
	\end{scope}

	\begin{scope}[Green]
		\Aone{2}{3}{$\oU w$}{}
	\end{scope}

	\begin{scope}[Fuchsia]
		\Aone{5.5}{5}{$\oU uv$}{}
	\end{scope}
\end{tikzpicture}
\end{subfigure}
\qquad\quad
\begin{subfigure}[c]{0.4\textwidth}
\begin{sseqdata}[name=AdamsA4, classes = fill, scale=0.5, xrange={0}{7}, yrange={0}{5},
x label = {$\displaystyle{s\uparrow \atop t-s\rightarrow}$},
x label style = {font = \small, xshift = -19ex, yshift=2ex}]
\class[MidnightBlue](0,0)\AdamsTower{MidnightBlue}
\class[Green](3, 0)
\class[MidnightBlue](4, 1)\AdamsTower{MidnightBlue}

\class[Fuchsia](5, 0)
\class[MidnightBlue](5, 2)
\class[MidnightBlue](6, 3)
\structline[MidnightBlue](4, 1)(5, 2)
\structline[MidnightBlue](5, 2)(6,3)
\end{sseqdata}
\printpage[name=AdamsA4, page=2]
\end{subfigure}
\caption{Left: the $\cA(1)$-module structure on $\tH^*((BA_4)^{3-V_\lambda};\Z/2)$ in low degrees. This submodule
contains all elements of degree at most $8$. Right: the $E_2$-page of the Adams spectral sequence calculating
$\widetilde\ko_*((BA_4)^{3-V_\lambda})$, given by $\Ext_{\cA(1)}^{s,t}(\tH^*((BA_4)^{3-V_\lambda};\Z/2), \Z/2)$.}
\label{a1moda4}
\end{figure}

Hence as $\cA(1)$-modules,
\begin{equation}
\label{A4cohsplitting}
	\tH^*(X;\Z/2)\cong \textcolor{MidnightBlue}{\uQ}
		\oplus \textcolor{Green}{\Sigma^3\cA(1)}
		\oplus \textcolor{Fuchsia}{\Sigma^5\cA(1)} \oplus P,
\end{equation}
where $P$ is $8$-connected.
Because we only care about degrees $6$ and below, $P$ is irrelevant
for us, and for the remaining summands in~\eqref{A4cohsplitting}, $\Ext_{\cA(1)}^{s,t}(\bl, \Z/2)$ has already been
computed. For $\Sigma^k\cA(1)$, there's a single $\Z/2$ with $s = 0$, $t = k$; for $\uQ$, see~\cite[Figure
29]{BC18}. We put this together and display the $E_2$-page for our spectral sequence in \cref{a1moda4}, right. A
combination of $h_0$-equivariance and Margolis' theorem (\cref{margolis}) rules out nontrivial differentials and
hidden extensions.
Therefore the $2$-primary part of $\tOmega_k^\Spin(X)$ has a single free summand each in degrees $0$ and $4$, is
$0$ in degrees $1$ and $2$, is $\Z/2$ in degrees $3$ and $6$, and is $\Z/2\oplus\Z/2$ in degree $5$.

For the odd-primary part, we use the fact that $\Omega_*^\Spin\to\Omega_*^\SO$ is an equivalence after inverting
$2$. Moreover, because $\lambda$ factors through $\SO_3$, $V_\lambda\to BA_4$ is orientable, so there is a Thom
isomorphism $\tOmega_*^\SO(X)\overset\cong\to\Omega_*^\SO(BA_4)$. Hence we just need the odd-primary part of
$\Omega_*^\SO(BA_4)$, which is isomorphic to the odd-primary part of $\Omega_*^\Spin(BA_4)$. In the degrees we care
about, this is isomorphic to $\ko_*(BA_4)$, and Bruner-Greenlees~\cite[\S 7.7.E]{BG10} show that the odd-primary
torsion in $\ko_*(BA_4)$ below degree $6$ consists of $\Z/3$ summands in degrees $1$ and $3$ and $\Z/9$ summands in
degrees $5$ and $7$.
\end{proof}
\subsubsection{Class D, spin-$1/2$ case}
In this case, the symmetries mix as specified by the group extension giving the binary tetrahedral group. 
\begin{thm}
\label{chi_tet_spinful}
The $A_4$-equivariant phase homology group for the class D, spin-$1/2$ symmetry type in 3d is trivial.
\end{thm}
\begin{proof}
Let $f_{1/2}^D$ denote the local system on $\R^3$ assigned to this symmetry type.  Since $V_\lambda$ is not \pinm
(if it were, it would be \pinm and orientable, hence spin), \cref{shear_D_thm} says $\Ki_0^{A_4}(\R^3;
f_{1/2}^D)\cong [\MTSpin\wedge (BA_4)_+, \Sigma^5I_\Z]$. Bruner-Greenlees~\cite[\S 7.7.E]{BG10} show
$\ko_4(BA_4)\cong\Z$ and $\ko_5(BA_4)$ is torsion, so this phase homology group vanishes.
\end{proof}
This bordism group is also studied by Davighi-Gripaios-Lohitsiri~\cite[\S A.2]{DGL22}.
\subsubsection{Class A}
\label{chiral_tet_A_unmixed}
\begin{lem}
\label{tet_not_pinc}
$V_\lambda\to BA_4$ is not \pinc.
\end{lem}
\begin{proof}
If $\beta\colon H^2(\bl;\Z/2)\to H^3(\bl;\Z)$ denotes the integral Bockstein, we want to show $\beta w_2(V_\lambda)
\ne 0$. By \cref{bock_to_sq1}, it suffices to show $\Sq^1(w_2(V_\lambda))\ne 0$.
\Cref{SWalt} gives $w_2(V_\lambda) = b$, and $\Sq^1b = ab + c$.
\end{proof}
Therefore \cref{shear_A_thm} computes $\Ph_0^{A_4}(\R^3; f_{1/2}^A)$ in terms of the
\spinc bordism of $(BA_4)^{\Det(V_\lambda)-1}$. Since $V_\lambda$ is orientable, this is isomorphic to the \spinc
bordism of $BA_4$. For $\Ph_0^{A_4}(\R^3; f_0^A)$, we use $(BA_4)^{3 - V_\lambda}$, as usual.
\begin{thm}
\label{spinc_chi_tet_thm}
The low-degree \spinc bordism groups of $(BA_4)^{3-V_\lambda}$ and $BA_4$ are
\begin{alignat*}{2}
	\tOmega_0^\Spinc((BA_4)^{3-V_\lambda}) &\cong \Z \qquad\qquad & \Omega_0^\Spinc(BA_4) &\cong \Z\\
	\tOmega_1^\Spinc((BA_4)^{3-V_\lambda}) &\cong \Z/3 \qquad\qquad & \Omega_1^\Spinc(BA_4) &\cong \Z/3\\
	\tOmega_2^\Spinc((BA_4)^{3-V_\lambda}) &\cong \Z \qquad\qquad & \Omega_2^\Spinc(BA_4) &\cong \Z\oplus\Z/2\\
	\tOmega_3^\Spinc((BA_4)^{3-V_\lambda}) &\cong \Z/6\oplus\Z/3 \qquad\qquad & \Omega_3^\Spinc(BA_4) &\cong
	\Z/6\oplus\Z/3\\
	\tOmega_4^\Spinc((BA_4)^{3-V_\lambda}) &\cong \Z^2 \qquad\qquad & \Omega_4^\Spinc(BA_4) &\cong \Z^2,
\end{alignat*}
and in both cases, $\Omega_5^\Spinc$ is torsion. Hence both $\Ph_0^{A_4}(\R^3; f_0^A)$ and $\Ph_0^{A_4}(\R^3;
f_{1/2}^A)$ vanish.
\end{thm}
\begin{proof}
\Cref{spinc_odd_primes} reduces the calculation of the odd-primary torsion to the respective oriented bordism
computations; as discussed in the proof of \cref{spin_bord_X}, a Thom isomorphism implies that for
both $X = (BA_4)^{3-V_\lambda}$ and $X = BA_4$, $\Omega_*^\SO(X)\cong\Omega_*^\SO(BA_4)$, and the odd-primary
torsion in those groups is isomorphic to the odd-primary torsion in the bordism groups in the statement of
\cref{spin_bord_X}. Thus we focus on $p = 2$.

As usual, we compute $\ku$-homology and compute $\Omega_k^{\Spin^c}(\bl)$ as $\ku_k(\bl)\oplus\ku_{k-4}(\bl)$,
valid for spaces and connective spectra when $k\le 7$.
Then we use the Adams spectral sequence over $\cE(1)$ to compute $2$-completed $\ku$-homology.

First we run the Adams spectral sequence for $(BA_4)^{3-V_\lambda}$. Use the $\cA(1)$-module structure on $\tH^*((BA_4)^{3-V_\lambda};\Z/2)$
from~\eqref{A4cohsplitting} (drawn in \cref{a1moda4}, left) to compute that the $\cE(1)$-module structure is
\begin{equation}
	\tH^*((BA_4)^{3-V_\lambda};\Z/2)\cong \textcolor{MidnightBlue}{\uQ} \oplus
		\textcolor{Green!75!black}{\Sigma^3\cE(1)} \oplus
		\textcolor{Green!75!white}{\Sigma^5\cE(1)} \oplus
		\textcolor{Fuchsia}{\Sigma^5\cE(1)}
		\oplus P,
\end{equation}
where $P$ is $6$-connected. We draw this in \cref{E1modunmixed_chiral_tet}, left. We computed
$\Ext_{\cE(1)}^{s,t}(\textcolor{MidnightBlue}{\uQ}, \Z/2)$ in \eqref{ExtN2}, and $P$ is too high-degree to be
relevant to us, so the $E_2$-page of the Adams spectral sequence for $\widetilde\ku_*((BA_4)^{3-V_\lambda})$ is given in
\cref{E1modunmixed_chiral_tet}, right. Margolis' theorem (\cref{margolis}) implies this spectral sequence collapses
and there are no extension problems, so we conclude.

\begin{figure}[h!]
\begin{subfigure}[c]{0.35\textwidth}
\begin{tikzpicture}[scale=0.6, every node/.style = {font=\tiny}]
\foreach \y in {0, ..., 9} {
	\node at (-1.5, \y) {$\y$};
}
\begin{scope}[MidnightBlue]
	\EoneQnMark{0}{0}{$\oU$}
\end{scope}
\begin{scope}[Green!75!black]
	\Eone{1.5}{3}{$\oU w$}
\end{scope}
\begin{scope}[Green!75!white]
	\Eone{3.75}{5}{$\oU uw$}
\end{scope}
\begin{scope}[Fuchsia]
	\Eone{5.75}{5}{$\oU uv$}
\end{scope}
\end{tikzpicture}
\end{subfigure}
\begin{subfigure}[c]{0.45\textwidth}
\begin{sseqdata}[name=spincunmixedA4, classes=fill, scale=0.6, xrange={0}{6}, yrange={0}{3},
x label = {$\displaystyle{s\uparrow \atop t-s\rightarrow}$},
x label style = {font = \small, xshift = -19ex, yshift=3ex}]
\begin{scope}[MidnightBlue]
	\class(0, 0)\AdamsTower{}
	\class(2, 0)\AdamsTower{}
	\class(4, 1)\AdamsTower{}
	\class(6, 2)\AdamsTower{}
	\foreach \x in {0, 2} {
		\foreach \y in {0, ..., 3} {
			\structline[MidnightBlue!40!white](\x, \y)(\x + 2, \y + 1)
		}
	}
	\foreach \y in {1, ..., 3} {
		\structline[MidnightBlue!40!white](4, \y)(6, \y + 1)
	}
	\class(8, 3)\class(8, 4)
	\structline[MidnightBlue!40!white](6, 2)(8, 3)
	\structline[MidnightBlue!40!white](6, 3)(8, 4)

\end{scope}
\class[Green!75!black](3, 0)
\class[Green!75!white](5, 0)
\class[Fuchsia](5, 0)
\end{sseqdata}
\printpage[name=spincunmixedA4, page=2]
\end{subfigure}
\caption{Left: the $\cE(1)$-module structure on $\tH^*((BA_4)^{3-V_\lambda};\Z/2)$ in low degrees. The picture
includes all elements in degrees $6$ and below. Right: $\Ext_{\cE(1)}^{s,t}(\tH^*((BA_4)^{3-V_\lambda};\Z/2),
\Z/2)$, the $E_2$-page of the Adams spectral sequence for $\widetilde\ku_*((BA_4)^{3-V_\lambda})$.}
\label{E1modunmixed_chiral_tet}
\end{figure}

On to $BA_4$. As before, $\ku_*(BA_4)$ splits as $\ku_*(\pt)\oplus\widetilde\ku_*(BA_4)$, and we focus on the
latter.
\begin{lem}[{Bruner-Greenlees~\cite[\S 5.2]{BG03}}]
\label{BG_alternating}
There is a $2$-local homotopy equivalence
\begin{equation}
	\ku\wedge BA_4\simeq_{(2)} \ku\wedge\Sigma^2 B\Z/2 \vee H\mathcal M,
\end{equation}
where $H\mathcal M$ denotes the generalized Eilenberg-Mac Lane spectrum on a graded abelian group $\mathcal
M\cong\Z/2\cdot\set{a_2, a_6, \dotsc}$ with $\abs{a_2} = 2$, $\abs{a_6} = 6$, and all other basis elements in
degrees at least $8$.
\end{lem}
Bruner-Greenlees provide the proof strategy but do not list the specific degrees of the first few summands, so we
extract them from the details of the proof.
\begin{proof}
Mitchell-Priddy~\cite[Theorem A, (0.1)]{MP84} prove a $2$-local stable splitting
\begin{equation}
\label{stable_klein_four}
	\Sigma^\infty (B\Z/2\wedge B\Z/2)\simeq_{(2)} \Sigma^\infty BA_4 \vee L(2)\vee L(2),
\end{equation}
where $L(2)$ is the same spectrum as in the proof of \cref{spinc_dihedral_bordism}. Smash~\eqref{stable_klein_four}
with $\ku$, then use \cref{ku_L2} to simplify $\ku\wedge L(2)$:
\begin{equation}
\label{ku_klein_four}
	\ku\wedge B\Z/2\wedge B\Z/2\simeq_{(2)} \ku\wedge BA_4\vee \Sigma^4 H\Z/2\vee \Sigma^4 H\Z/2\vee \Sigma^6
	H\Z/2\vee \Sigma^6 H\Z/2\vee\dotsb,
\end{equation}
where the ``$\dotsb$'' refers to additional $\Sigma^k H\Z/2$ summands with $k\ge 8$. Now
compare~\eqref{ku_klein_four} with the splitting~\eqref{ossa} arising from Ossa's theorem~\cite[Proposition
3]{Oss89}:
\begin{equation}
\label{ossa_compare}
	\ku\wedge \Sigma^2 B\Z/2 \vee \Sigma^2 H(\Z/2[u, v])\simeq_{(2)} \ku\wedge BA_4\vee \Sigma^4 H\Z/2\vee \Sigma^4
	H\Z/2\vee \Sigma^6 H\Z/2\vee \Sigma^6 H\Z/2\vee\dotsb,
\end{equation}
where $\abs u = \abs v = 2$, so
\begin{equation}
\label{expl_expa}
	\Sigma^2 H(\Z/2[u,v])\simeq \Sigma^2 H\Z/2\vee (\Sigma^4 H\Z/2)^{\vee 2}\vee (\Sigma^6 H\Z/2)^{\vee 3}\vee
	\dotsb.
\end{equation}
We may cancel $\Sigma^k H\Z/2$ summands out from both sides of~\eqref{ossa_compare}, using the explicit
expansion~\eqref{expl_expa}; doing so, we obtain the lemma statement.
\end{proof}
Thus the computation of $\widetilde{\ku}_*(BA_4)$ in the degrees we need reduces to the computation of
$\widetilde{\ku}_*(B\Z/2)$, which was done by Hashimoto~\cite[Theorem 3.1]{Has83}, so we are all set.
\end{proof}

	\subsection{Pyritohedral symmetry}
		\label{pyrit}
\term{Pyritohedral symmetry} is the action of $G\coloneqq A_4\times\Z/2$ on $\R^3$ in which $A_4$ acts as the
orientation-preserving symmetries of a tetrahedron and $\Z/2$ acts through inversion; let $\lambda$ denote this
representation and $V_\lambda\to BG$ be the associated vector bundle. Because $G$ splits as a direct product, it is
easier to analyze than full tetrahedral symmetry (i.e.\ chiral tetrahedral symmetry together with a reflection), as
we will see in this and the next section.


\subsubsection{Spinless case}
By the twisted Künneth formula, $H^*((BG)^{3-V_\lambda})$ is $2$-torsion; therefore
$\tOmega_*^\Spin((BG)^{3-V_\lambda})$ also lacks odd-primary torsion.
so we just have to work with the Adams spectral sequence at $p = 2$. In the rest of this section, all cohomology is
with $\Z/2$ coefficients unless otherwise stated.
\begin{prop}
\label{spinless_pyrit_bordism}
The first several spin bordism groups of $(BG)^{3-V_\lambda}$ are
\begin{align*}
	\tOmega_0^\Spin((BG)^{3-V_\lambda}) &\cong \Z/2\\
	\tOmega_1^\Spin((BG)^{3-V_\lambda}) &\cong 0\\
	\tOmega_2^\Spin((BG)^{3-V_\lambda}) &\cong \Z/2\\
	\tOmega_3^\Spin((BG)^{3-V_\lambda}) &\cong \Z/2\\
	\tOmega_4^\Spin((BG)^{3-V_\lambda}) &\cong (\Z/2)^{\oplus 3}\\
	\tOmega_5^\Spin((BG)^{3-V_\lambda}) &\cong (\Z/2)^{\oplus 3}\\
	\tOmega_6^\Spin((BG)^{3-V_\lambda}) &\cong \Z/16\oplus (\Z/2)^{\oplus 2}\\
	\tOmega_7^\Spin((BG)^{3-V_\lambda}) &\cong (\Z/2)^{\oplus 2}.
\end{align*}
\end{prop}
\begin{proof}
We employ a trick to reduce the amount of direct computations. We will replace $(3-V_\lambda)\to BG$ with a virtual vector
bundle $E\to BG$ with the same first two Stiefel-Whitney classes, but which splits as an exterior sum over $BA_4$
and $B\Z/2$. The Thom spectrum $(BG)^{E}$ has two nice properties: its spin bordism coincides with that of
$(BG)^{3-V_\lambda}$, but $(BG)^E$ splits as a smash product of Thom spectra over $BA_4$ and $B\Z/2$, simplifying
the calculation of the $E_2$-page of the Adams spectral sequence.

The Künneth formula and \cref{a4z2} together imply
\begin{equation}
	H^*(BG) \cong \Z/2[x, u, v, w]/(u^3+v^2+w^2+vw),
\end{equation}
where $\abs x = 1$, $\abs u = 2$, and $\abs v = \abs w = 3$, and that $\Sq(x) = x + x^2$ and the Steenrod squares
of $u$, $v$, and $w$ are as in \cref{a4z2}.
\begin{lem}
\label{SWpyrit}
The first two Stiefel-Whitney classes of $V$ are $w_1(V_\lambda) = x$ and $w_2(V_\lambda) = u+x^2$.
\end{lem}
\begin{proof}
Since this representation contains orientation-reversing symmetries, $w_1(V_\lambda)$ must be nonzero, so is $x$.
For $w_2$, we saw in \cref{SWalt} that when one restricts to $A_4\subset A_4\times\Z/2$, one has
$w_2(V_\lambda|_{BA_4}) = u$; when one restricts to $\Z/2$, this is $3$ copies of the sign representation, hence has
$w_2(V_\lambda|_{B\Z/2}) = x^2$.
\end{proof}
Let $E\to BG$ be the virtual vector bundle
\begin{equation}
	E \coloneqq 4 - (V_\lambda|_{BA_4} \boxplus {-\sigma)},
\end{equation}
where $\sigma\to B\Z/2$ is the tautological line bundle. The Whitney sum formula implies for $i = 1,2$, $w_i(E)
= w_i(3-V_\lambda)$. That is, $(3-V_\lambda) - E$ is spin, so by the relative Thom
isomorphism (\cref{relative_Thom}),
\begin{equation}
	\tOmega_*^\Spin((BG)^{3-V_\lambda}) \cong \tOmega_*^\Spin((BG)^E).
\end{equation}
%
Because $E\to BG$ is an external sum,
\begin{equation}
	(BG)^E \simeq (BA_4)^{3-V_\lambda}\wedge (B\Z/2)^{\sigma-1}.
\end{equation}
We know the $\cA(1)$-module structures on the low-degree cohomology of both summands, and the Künneth formula tells
us to tensor them together (over $\Z/2$) to determine the $\cA(1)$-module structure on $\wH^*((BG)^E)$.

In~\eqref{A4cohsplitting}, we computed the $\cA(1)$-module structure on $\wH^*((BA_4)^{3-V_\lambda})$ in low
degrees, and split off two $\Sigma^k\cA(1)$ summands. Margolis' theorem (\cref{margolis}) promotes that to a
splitting of spectra
\begin{equation}
	\ko\wedge (BA_4)^{3-V_\lambda} \simeq \textcolor{Green}{\Sigma^3 H\Z/2}\vee \textcolor{Fuchsia}{\Sigma^5 H\Z/2}
	\vee \textcolor{MidnightBlue}{Y},
\end{equation}
such that as an $\cA$-module,
\begin{equation}
	\wH^*(\textcolor{MidnightBlue}{Y}) \cong \cA\otimes_{\cA(1)}(\textcolor{MidnightBlue}{\uQ} \oplus P),
\end{equation}
where $P$ is $7$-connected. When we smash $(B\Z/2)^{\sigma-1}$ back in, each $\Sigma^k H\Z/2\wedge
(B\Z/2)^{\sigma-1}$ contributes a summand of $\wH_{n-k}((B\Z/2)^{\sigma-1})$ to
$\widetilde\ko_n((BG)^{E})$, i.e.\ a $\Z/2$-summand in each degree $\ell\ge k$. The upshot for $\cA(1)$-modules
is
\begin{equation}
\label{A1_BZ2}
	\Sigma^k\cA(1)\otimes_{\Z/2} \wH^*((B\Z/2)^{\sigma-1}) \cong \bigoplus_{\ell\ge k}\Sigma^\ell \Z/2.
\end{equation}
Margolis' theorem lifts this to split off corresponding $\Sigma^\ell H\Z/2$ summands. Therefore there is a spectrum
$Y'$ such that
\begin{equation}
\label{pyrit_BG_splitting}
	\widetilde\ko_n((BG)^E) \cong \pi_n(Y') \oplus
	\textcolor{Green}{\wH_{n-3}((B\Z/2)^{\sigma-1})} \oplus
	\textcolor{Fuchsia}{\wH_{n-5}((B\Z/2)^{\sigma-1})}
\end{equation}
and as $\cA$-modules,
\begin{equation}
\label{pyrit_part2}
	\wH^*(Y')\cong \cA\otimes_{\cA(1)} (\textcolor{MidnightBlue}{\uQ}\oplus P) \otimes_{\Z/2}
	\wH^*((B\Z/2)^{\sigma-1}).
\end{equation}
The change-of-rings theorem~\eqref{change_of_rings} thus applies to the $E_2$-page of the Adams spectral sequence
calculating $\pi_*(Y')$, yielding
\begin{equation}
\label{reduced_SS_pyrit}
	E_2^{s, t} \cong \Ext_{\cA(1)}^{s,t}((\textcolor{MidnightBlue}{\uQ}\oplus P)\otimes_{\Z/2}
	\wH^*((B\Z/2)^{\sigma-1}), \Z/2).
\end{equation}
We will work with this spectral sequence, adding in the summands corresponding to $\textcolor{Green}{\Sigma^3
H\Z/2}$ and $\textcolor{Fuchsia}{\Sigma^5 H\Z/2}$ afterwards.
%

Our first order of business is to compute the tensor product in~\eqref{reduced_SS_pyrit}. The $\cA(1)$-module
structure on $\wH^*((B\Z/2)^{\sigma-1})$ can be found in~\cite[Figure 4]{BC18}.
\begin{lem}
\label{pyritohedral_tensor_products}
There is an isomorphism of $\cA(1)$-modules $\uQ\otimes_{\Z/2} \tH^*((B\Z/2)^{\sigma-1}) \cong \cA(1)\oplus
\Sigma^2 R_0\oplus \Sigma^4\cA(1)\oplus P$, where $P$ is $7$-connected.
\end{lem}
\begin{proof}
Compute directly, by hand or by computer.
\end{proof}
\label{pyrit_unmixed_ss}

By~\eqref{pyrit_BG_splitting} and~\eqref{pyrit_part2}, we can work with~\eqref{reduced_SS_pyrit}, then add in the
$\Z/2$ summands coming from the $\Sigma^k H\Z/2$ summands at the end. \cref{pyritohedral_tensor_products} tells us
the $E_2$-page of~\eqref{reduced_SS_pyrit} is
\begin{equation}
\label{E2_unmixed_pyrit}
	E_2^{s,t} \cong \Ext(\textcolor{BrickRed}{\cA(1)} \oplus \textcolor{RedOrange}{\Sigma^2 R_0} \oplus
	\textcolor{MidnightBlue}{\Sigma^4\cA(1)} \oplus P).
\end{equation}
Since $P$ is $7$-connected, its Ext is concentrated in degrees irrelevant to us, and we ignore it.
$\Ext(\textcolor{RedOrange}{\Sigma^2 R_0})$ is computed in the degrees we need by Beaudry-Campbell~\cite[Figures
23, 24]{BC18}; using this, the $E_2$-page of~\eqref{reduced_SS_pyrit} is
\begin{sseqdata}[name=pyritspinless, classes=fill, scale=0.5, xrange={0}{7}, yrange={0}{3},
x label = {$\displaystyle{s\uparrow \atop t-s\rightarrow}$},
x label style = {font = \small, xshift = -19.5ex, yshift=2ex}]
	\class[BrickRed](0, 0)
	\begin{scope}[RedOrange]
		\class(2, 0)
		\class(4, 1)
		\class(5, 2) \structline(4, 1)(5, 2)
		\class(6, 0)
		\class(6, 1)\structline
		\class(6, 2)\structline
		\class(6, 3)\structline
		\structline(5, 2)(6, 3)
	\end{scope}
	\class[MidnightBlue](4, 0)
\end{sseqdata}
\begin{equation}
\begin{gathered}
	\printpage[name=pyritspinless, page=2]
\end{gathered}
\end{equation}
Margolis' theorem and $h_1$-equivariance of differentials immediately imply there are no nontrivial differentials
or extension problems below degree $8$, so we conclude.
\end{proof}
\subsubsection{Class D, spin-$1/2$ case}
\label{pyrit_mixed_ss}
\label{D_mixed_py}
\Cref{shear_D_thm} computes the relevant phase homology groups for this case
in terms of the spin bordism of $(BA_4\times B\Z/2)^{\Det(V_\lambda) - 1}$. The isomorphism
$\Det(V_\lambda)\cong 0\boxplus \sigma$ provides an isomorphism $(BA_4\times B\Z/2)^{\Det(V_\lambda)-1}\simeq
(BA_4)_+\wedge (B\Z/2)^{\sigma-1}$; \eqref{pinmsplitting}
thus implies the spin bordism of this spectrum computes the \pinm bordism
of $BA_4$, which could be independently interesting.
\begin{thm}
\label{spinful_pyrit_bordism}
The first few spin bordism groups of $(BA_4\times B\Z/2)^{\Det(V_\lambda)-1}$ are
\begin{align*}
	\tOmega_0^\Spin((BA_4\times B\Z/2)^{\Det(V_\lambda)-1}) &\cong \Z/2\\
	\tOmega_1^\Spin((BA_4\times B\Z/2)^{\Det(V_\lambda)-1}) &\cong \Z/2\\
	\tOmega_2^\Spin((BA_4\times B\Z/2)^{\Det(V_\lambda)-1}) &\cong \Z/8\oplus\Z/2\\
	\tOmega_3^\Spin((BA_4\times B\Z/2)^{\Det(V_\lambda)-1}) &\cong \Z/2\oplus\Z/2\\
	\tOmega_4^\Spin((BA_4\times B\Z/2)^{\Det(V_\lambda)-1}) &\cong \Z/2.
\end{align*}
Since $\tOmega_5^\Spin((BA_4\times B\Z/2)^{\Det(V_\lambda)-1})$ is torsion by \cref{torsion_k_theory}, $\Ph_0^{A_4\times\Z/2}(\R^3; f^D_{1/2})\cong\Z/2$.
\end{thm}
\begin{proof}
By the twisted Künneth formula, $\tH^*((BA_4\times B\Z/2)^{\Det(V_\lambda)-1})$ has no odd-primary torsion, and therefore neither does
$\tOmega_*^\Spin((BA_4\times B\Z/2)^{\Det(V_\lambda)-1})$, so it suffices to work at the prime $2$, which we do.

Use \cref{det_splitting} to split $(BA_4\times B\Z/2)^{\Det(V_\lambda)-1}\simeq (B\Z/2)^{\sigma-1}\vee M$, where the map $\tH^*(M;\Z/2)\to\tH^*((BA_4\times B\Z/2)^{\Det(V_\lambda)-1};\Z/2)$
is injective with image a complimentary subspace to $\Z/2\cdot \set{\oU x^k\mid k\ge 0}$.

As usual, $w_1(\Det(V_\lambda)-1) = w_1(V_\lambda) = x$ and $w_2(\Det(V_\lambda)-1) = 0$. We also need to know the
$\cA$-action on $H^*(BG;\Z/2)$; the Künneth formula determines this using as input the $\cA$-action on
$H^*(BA_4;\Z/2)$, which we computed in \cref{a4z2}, and the $\cA$-action on $H^*(B\Z/2;\Z/2)$, which is standard.
Using this, we can determine the $\cA(1)$-module structure on $\tH^*(M;\Z/2)$. We obtain an
isomorphism of $\cA(1)$-modules
\begin{equation}
\label{D_mixed_py_mod}
	\wH^*(M;\Z/2) \cong \textcolor{BrickRed}{\Sigma^2R_3} \oplus
		\textcolor{Green}{\Sigma^3\cA(1)} \oplus
		\textcolor{MidnightBlue}{\Sigma^3\cA(1)} \oplus
		\textcolor{Fuchsia}{\Sigma^4\cA(1)} \oplus P,
\end{equation}
where $P$ is $4$-connected. We will see in \cref{pyrit_mixed_drawn}, right, that for $t-s \le 4$, $E_2^{s,t}$ is
concentrated in Adams filtration $0$; this and the $4$-connectedness of $P$ imply its contribution to the
$E_2$-page cannot affect the spectral sequence in degrees $t-s \le 4$, which is all we need. We draw these
summands, except for $P$, in \cref{pyrit_mixed_drawn}, left.
\begin{figure}[h!]
\begin{subfigure}[c]{0.4\textwidth}
\begin{tikzpicture}[scale=0.6, every node/.style = {font=\tiny}]
	\foreach \y in {2, ..., 10} {
		\node at (-2, \y) {$\y$};
	}
	\begin{scope}[BrickRed]
        \tikzpt{0}{2}{$\oU u$}{};
        \tikzpt{0}{3}{}{};
        \tikzpt{0}{4}{}{};
        \tikzpt{0}{5}{}{};
        \tikzpt{0}{7}{}{};
        \tikzpt{1.5}{5}{}{};
		\tikzpt{1.5}{6}{}{};
        \sqone(0, 2);
        \sqone(0, 4);
        \sqone(1.5, 5);
        \sqtwoL(0, 2);
        \sqtwoCR(0, 3);
        \sqtwoCR(0, 4);
        \sqtwoL(0, 5);
		
		\tikzptR{0}{6}{$\oU \alpha$}{};
		\tikzpt{0}{8}{}{};
		\tikzpt{0}{9}{}{};
		\tikzptR{0}{10}{$\oU (u^2x^6 + v^2x^3 + w^2x^3)$}{};
		\tikzpt{0}{11}{}{};

		\sqone(0, 6);
		\sqone(0, 8);
		\sqtwoR(0, 6, );
		\begin{scope}
			\clip (-1, 8.5) rectangle (1, 11.5);
			\sqtwoL(0, 9);
			\sqone(0, 10);
			\sqtwoR(0, 10, );
		\end{scope}
	\end{scope}
	\begin{scope}[Green]
		\Aone{2.625}{3}{$\oU ux$}
	\end{scope}
	\begin{scope}[MidnightBlue]
		\Aone{5}{3}{$\oU v$}
	\end{scope}
	\begin{scope}[Fuchsia]
		\Aone{7.5}{4}{$\oU ux^2$}
	\end{scope}
\end{tikzpicture}
\end{subfigure}
\qquad\qquad
\begin{subfigure}[c]{0.4\textwidth}
\begin{sseqdata}[name=pyritmixed, classes=fill, xrange={0}{4}, yrange={0}{2}, scale=0.6,
	x label = {$\displaystyle{s\uparrow \atop t-s\rightarrow}$},
	x label style = {font = \small, xshift = -15.5ex, yshift=2.8ex}]
	\class[BrickRed](2, 0)
	\class[Green](3, 0)
	\class[MidnightBlue](3, 0)
	\class[Fuchsia](4, 0)
\end{sseqdata}
\printpage[name=pyritmixed, page=2]
\end{subfigure}
\caption{Left: the $\cA(1)$-module structure on $\tH^*(M;\Z/2)$ in low degrees. This picture includes all summands
in degrees $4$ and below. Here $\alpha\coloneqq u^2x^2 + v^2 + w^2$. Right: the $E_2$-page of the corresponding
Adams spectral sequence.}
\label{pyrit_mixed_drawn}
\end{figure}

Yu~\cite[Theorem 3.1]{Yu95} calculates $\Ext(\textcolor{BrickRed}{R_3})$ in the range we need (see also
Beaudry-Campbell~\cite[Figures 32,
33]{BC18}), and we can draw the $E_2$-page of the
Adams spectral sequence in \cref{pyrit_mixed_drawn}, right. This collapses, so we add in the \pinm bordism summands
we need from~\cite{ABP69, KT90} to obtain the groups in the theorem.
\end{proof}
\begin{rem}
Finite truncations of $\textcolor{BrickRed}{R_3}$ are the duals of the Milgram modules $Q_{3,n}$~\cite{Mil75,
DGM81}. See~\cite[Figure 1]{BL21} for a picture.
\end{rem}

\subsubsection{Class A, spinless case}
\label{s_spinc_pyr}
We saw in \cref{shear_A_thm} that
$\Ph_0^{A_4\times\Z/2}(\R^3; f_0^A)$ is determined by $\tOmega_*^\Spinc((BA_4\times B\Z/2)^{3-V_\lambda})$.
\begin{thm}
\label{spinc_pyrit_bord}
The first few \spinc bordism groups of $(BA_4\times B\Z/2)^{3-V_\lambda}$ are
\begin{align*}
	\tOmega_0^\Spinc((BA_4\times B\Z/2)^{3-V_\lambda}) &\cong \Z/2\\
	\tOmega_1^\Spinc((BA_4\times B\Z/2)^{3-V_\lambda}) &\cong 0\\
	\tOmega_2^\Spinc((BA_4\times B\Z/2)^{3-V_\lambda}) &\cong (\Z/2)^{\oplus 2}\\
	\tOmega_3^\Spinc((BA_4\times B\Z/2)^{3-V_\lambda}) &\cong \Z/2\\
	\tOmega_4^\Spinc((BA_4\times B\Z/2)^{3-V_\lambda}) &\cong \Z/4\oplus(\Z/2)^{\oplus 3}\\
	\tOmega_5^\Spinc((BA_4\times B\Z/2)^{3-V_\lambda}) &\cong (\Z/2)^{\oplus 3}\\
	\tOmega_6^\Spinc((BA_4\times B\Z/2)^{3-V_\lambda}) &\cong \Z/8\oplus (\Z/2)^{\oplus 6}\\
	\tOmega_7^\Spinc((BA_4\times B\Z/2)^{3-V_\lambda}) &\cong (\Z/2)^{\oplus 5},
\end{align*}
so $\Ph_0^{A_4\times\Z/2}(\R^3; f^A_0)\cong \Z/4\oplus(\Z/2)^{\oplus 3}$.
\end{thm}
\begin{proof}
The twisted Thom isomorphism and twisted Künneth formula imply $\tH^*((BA_4\times B\Z/2)^{3-V_\lambda};\Z)$ is $2$-torsion.  Therefore for any odd
prime $p$, the mod $p$ Whitehead theorem~\cite[Chapitre III, Théorème 3]{Ser53} implies $\tOmega_*^\Spinc((BA_4\times B\Z/2)^{3-V_\lambda})$ also
has no $p$-torsion. This leaves only $p = 2$, for which we use the Adams spectral sequence over $\cE(1)$.

We determined the $\cA(1)$-module structure on $\tH^*((BG)^{3-V_\lambda})$ as given in~\eqref{E2_unmixed_pyrit}, together
with an $\Sigma^\ell\cA(1)$ for $\ell = 3,4$, and two $\Sigma^\ell\cA(1)$ summands for $\ell \ge 5$. This
determines the $\cE(1)$-module structure: as $\cE(1)$-modules, $\cA(1)\cong\cE(1)\oplus\Sigma^2\cE(1)$, so
\begin{equation}
	\tH^*((BA_4\times B\Z/2)^{3-V_\lambda};\Z/2) \cong \textcolor{MidnightBlue}{\Sigma^2 R_0} \oplus
	\textcolor{BrickRed}{V'\otimes_{\Z/2}\cE(1)} \oplus P,
\end{equation}
where $V'$ is a graded $\Z/2$-vector space with a basis of homogeneous elements in degrees $0$, $2$, $3$, $4$, $4$,
$5$, $5$, $5$, $6$, $6$, $6$, $6$, $7$, $7$, $7$, and $7$, and $P$ is $7$-connected. Therefore the $E_2$-page is
\begin{sseqdata}[name=pyritspinc, classes=fill, scale=0.7, xrange={0}{7}, yrange={0}{3},
x label = {$\displaystyle{s\uparrow \atop t-s\rightarrow}$},
x label style = {font = \small, xshift = -23.5ex, yshift=4ex}]
\begin{scope}[BrickRed]
	\class(0, 0)
	\class(2, 0)
	\class(3, 0)
	\class(4, 0)
	\class(5, 0)
	\class(5, 0)
	\class(5, 0)
	\class(6, 0)
	\class(6, 0)
	\class(6, 0)
	\class(6, 0)
	\class(7, 0)
	\class(7, 0)
	\class(7, 0)
	\class(7, 0)
\end{scope}
\begin{scope}[MidnightBlue]
	\class(2, 0)
	\class(4, 0)
	\class(4, 1)\structline
	\structline[MidnightBlue!40!white](2, 0, -1)(4, 1)
	\class(6, 0)
	\class(6, 1)\structline
	\structline[MidnightBlue!40!white](4, 0, -1)(6, 1)
	\class(6, 2)\structline
	\structline[MidnightBlue!40!white](4, 1)(6, 2)
	\foreach \y in {1, 2, 3} {
		\class(8, \y)
		\structline[MidnightBlue!40!white](8, \y)(6, \y - 1, -1)
	}
\end{scope}
\class[BrickRed](4, 0)
\end{sseqdata}
\begin{equation}
\begin{gathered}
	\printpage[name=pyritspinc, page=2]
\end{gathered}
\end{equation}
By Margolis' theorem, there are no nontrivial differentials or extension problems in this range.
\end{proof}

\subsubsection{Class A, spin-$1/2$ case}
To compute the phase homology groups for this case,
\cref{shear_A_thm} asks us to investigate the \spinc bordism of $(BA_4\times B\Z/2)^{\Det(V_\lambda)-1}\simeq
(BA_4\times B\Z/2)^{0\boxplus \sigma - 1}$; we know $V_\lambda$ is not \pinc because we saw in \cref{tet_not_pinc}
that the pullback of $V_\lambda$ along $BA_4\to BA_4\times B\Z/2$ is not \pinc.
\begin{thm}
\label{mixed_pyr_classA_thm}
The first few \spinc bordism groups of $(BA_4\times B\Z/2)^{\Det(V_\lambda)-1}$ are
\begin{align*}
	\tOmega_0^{\Spin^c}((BA_4\times B\Z/2)^{\Det(V_\lambda)-1}) &\cong \Z/2\\
	\tOmega_1^{\Spin^c}((BA_4\times B\Z/2)^{\Det(V_\lambda)-1}) &\cong 0\\
	\tOmega_2^{\Spin^c}((BA_4\times B\Z/2)^{\Det(V_\lambda)-1}) &\cong \Z/4\oplus\Z/2\\
	\tOmega_3^{\Spin^c}((BA_4\times B\Z/2)^{\Det(V_\lambda)-1}) &\cong (\Z/2)^{\oplus 2}\\
	\tOmega_4^{\Spin^c}((BA_4\times B\Z/2)^{\Det(V_\lambda)-1}) &\cong \Z/8\oplus (\Z/2)^{\oplus 3}\
\end{align*}
By \cref{torsion_k_theory}, $\tOmega_5^{\Spinc}((BA_4\times B\Z/2)^{\Det(V_\lambda)-1})$ is torsion, so $\Ph_0^{A_4\times\Z/2}(\R^3; f_{1/2}^A)\cong
\Z/8\oplus (\Z/2)^{\oplus 3}$.
\end{thm}
\begin{proof}
After the usual reduction from \spinc bordism to $\ku$ homology, we want to understand the homotopy groups of
$\ku\wedge (BA_4)_+\wedge (B\Z/2)^{\sigma-1}$, which we will do by constructing a $2$-local stable splitting. First
use the equivalence $(B\Z/2)^{\sigma-1}\simeq \Sigma^{-1}\Sigma^\infty B\Z/2$~\cite[Lemma 2.6.5]{Koc96} to reduce
to $\Sigma^{-1}\ku\wedge (BA_4)_+\wedge B\Z/2$, then split $\Sigma^\infty (BA_4)_+\simeq \Sigma^\infty BA_4\vee
\mathbb S$, so that we have a $\Sigma^{-1}\ku\wedge B\Z/2$ summand, whose homotopy groups are
$\widetilde{\ku}_{*+1}(B\Z/2)$, computed by Hashimoto~\cite[Theorem 3.1]{Has83}, together with
$\Sigma^{-1}\ku\wedge BA_4\wedge B\Z/2$. Take the result of \cref{BG_alternating}, apply $\Sigma^{-1}$, and smash
with $B\Z/2$ to obtain a $2$-local equivalence
\begin{equation}
\label{A4_Z2_split}
	\Sigma^{-1}\ku\wedge BA_4\wedge B\Z/2\simeq_{(2)} \ku\wedge \Sigma B\Z/2\wedge B\Z/2\vee \Sigma H\Z/2\wedge
	B\Z/2\vee \Sigma^5 H\Z/2\vee B\Z/2\vee \dotsb,
\end{equation}
where ``$\dotsb$'' refers to additional $H\Z/2\wedge B\Z/2$ summands in degrees $7$ and above. For any space $X$,
there is a homotopy equivalence betwee $H\Z/2\wedge X$ and the generalized Eilenberg-Mac Lane spectrum on the
graded abelian group $H^*(X;\Z/2)$, simplifying the $H\Z/2\wedge B\Z/2$ summands in~\eqref{A4_Z2_split}. Combine
this with Ossa's theorem~\cite[Proposition 3]{Oss89} as in \cref{ossa} to simplify~\eqref{A4_Z2_split} into
\begin{equation}
	\Sigma^{-1}\ku\wedge BA_4\wedge B\Z/2\simeq_{(2)} \ku\wedge \Sigma^3 B\Z/2\vee H\mathcal M,
\end{equation}
where $\mathcal M$ is a generalized Eilenberg-Mac Lane spectrum on the graded abelian group $\Z/2\cdot\set{a_2,
a_3, b_3, a_4, a_5, b_5, c_5, a_6, b_6,\dotsc}$, where $\abs{a_i} = \abs{b_i} = \abs{c_i} = i$ and all unlisted
generators have degree at least $7$. Use Hashimoto~\cite[Theorem 3.1]{Has83} for $\widetilde{\ku}_*(\Sigma^3
B\Z/2)$, and for $H\mathcal M$, we obtain a $\Z/2$ summand in degree $k$ for each basis element of degree $k$.
Using the reductions noted above, this finishes the proof of the theorem.
\end{proof}

	\subsection{Full tetrahedral symmetry}
		\label{s_full_tet}
The full group of symmetries of the tetrahedron, including reflections, is the symmetric group $S_4$, acting via
the representation $\lambda\colon S_4\to\O_3$, which is isomorphic to the quotient of the four-dimensional real
permutation representation by the fixed line $\R\cdot (1, 1, 1, 1)$.
\begin{prop}[{\cite[\S 2.3]{Ngu09}}]
\label{mod2S4coh}
$H^*(BS_4;\Z/2)\cong\Z/2[a,b,c]/(ac)$, with $\abs a = 1$, $\abs b = 2$, and $\abs c = 3$. The Steenrod squares of
the generators are $\Sq(a) = a+a^2$, $\Sq(b) = b + ab + c + b^2$, and $\Sq(c) = c + bc + c^2$.\footnote{The ring
structure on $H^*(BS_4;\Z/2)$ was known earlier, due to Cardenas~\cite{Car65}; see~\cite[Example VI.1.13]{AM04}.}
\end{prop}
Let $V_\lambda\to BS_4$ denote the associated vector bundle to $\lambda$.
\begin{prop}[{Milgram~\cite[Theorem 2.4.1]{Mil72}}]
\label{S4Steenrod}
$w_1(V_\lambda) = a$, $w_2(V_\lambda) = b$, and $w_3(V_\lambda) = c$.
\end{prop}
Milgram uses the four-dimensional permutation representation $W$ instead of $V_\lambda$, but $W\cong
V_\lambda\oplus \underline\R$ as $S_4$-representations.
We need the next calculation to determine the odd-primary torsion subgroups of the phase homology groups that we
calculate.
\begin{lem}
\label{S4_odd_twisted_coh}
Suppose $V\to BS_4$ is a rank-zero virtual vector bundle with $w_1(V) = a$. Then
the inclusion $i\colon S_3\inj S_4$ defines an isomorphism
\begin{equation}
	\tH_*((BS_3)^{i^*V}) \otimes\Z[1/2]\to \tH_*((BS_4)^V) \otimes\Z[1/2].
\end{equation}
\end{lem}
\begin{proof}
The commutative diagram of short exact sequences
\begin{equation}
\label{SES_diag}
\begin{gathered}
	\xymatrix{
		1\ar[r] & A_3\ar[r]\ar@{^(->}[d] & S_3\ar[r]\ar@{^(->}[d]^i & \Z/2\ar@{=}[d]\ar[r] & 1\\
		1\ar[r] & A_4\ar[r] & S_4\ar[r] & \Z/2\ar[r] & 1
	}
\end{gathered}
\end{equation}
induces a map between their Lyndon-Hochschild-Serre spectral sequences with signatures
\begin{equation}
	E^2_{p,q} = H_p(B\Z/2; \underline{H_q(BA_k; \Z[1/2])}\otimes (\Z[1/2])_x) \Longrightarrow H_{p+q}(BS_k;
	(\Z[1/2])_{w_1(V)}),
\end{equation}
where $\underline{H_q(BA_k; \Z[1/2])}$ means the local system on $B\Z/2$ induced by the action of $\Z/2$ on $A_k$
as specified by the extension $1\to A_k\to S_k\to \Z/2\to 1$, and $x$ is the generator of $H^1(B\Z/2;\Z/2)$.

We claim the map on these spectral sequences is an isomorphism on $E^2$-pages. It is a standard theorem that if $H$
is the $p$-Sylow subgroup of a finite group $G$, and $p$ is invertible in a commutative ring $R$, then the pullback
map $H^*(BG; R)\to H^*(BH; R)$ is an isomorphism (e.g.\ the transfer provides an inverse). Thus
the map
$H_*(BA_4;\Z[1/2])\to H_*(BA_3;\Z[1/2])$ is an isomorphism, and this isomorphism intertwines the $\Z/2$-actions on
$\underline{H_*(BA_k; \Z[1/2])}\otimes (\Z[1/2])_x$, because~\eqref{SES_diag} commutes. Therefore it induces an
isomorphism on all $E^r$-pages, hence also on what these spectral sequences converge to.
\end{proof}
The top row in~\eqref{SES_diag} can be identified with $1\to \Z/3\to D_6\to\Z/2\to 1$, so by the same lines of
reasoning as in \cref{twisted_dihedral_SO,twisted_dihedral_SO_U} we deduce
\begin{subequations}
\begin{align}
\label{twisted_symm_SO}
	\tOmega_k^\Spin((BS_4)^V) \otimes\Z[1/2] &\cong \begin{cases}
		\Z/3, &k = 1\\
		0, &k = 0, 2, 3, 4
	\end{cases}\\
\label{twisted_symm_SO_U}
	\tOmega_k^\Spinc((BS_4)^V)\otimes\Z[1/2] &\cong \begin{cases}
		\Z/3, &k=1,3\\
		0, &k=0, 2, 4.
	\end{cases}
\end{align}
\end{subequations}

\subsubsection{Class D, spinless case}
\label{chiral_tet_spinless} 
As usual in the spinless case for unorientable representations, \cref{shear_D_thm} asks us to consider
$\tOmega_*^\Spin((BS_4)^{3-V_\lambda})$.
\begin{thm}
\label{ctsthm}
The first few spin bordism groups of $(BS_4)^{3-V_\lambda}$ are
\begin{align*}
    \tOmega_0^\Spin((BS_4)^{3-V_\lambda}) &\cong \Z/2\\
    \tOmega_1^\Spin((BS_4)^{3-V_\lambda}) &\cong \Z/3\\
    \tOmega_2^\Spin((BS_4)^{3-V_\lambda}) &\cong \Z/2\\
    \tOmega_3^\Spin((BS_4)^{3-V_\lambda}) &\cong \Z/2\\
    \tOmega_4^\Spin((BS_4)^{3-V_\lambda}) &\cong (\Z/2)^{\oplus 3},
\end{align*}
and $\tOmega_5^\Spin((BS_4)^{3-V_\lambda})$ is torsion. 
\end{thm}
This is the most involved computation in this paper, due to a tricky Adams differential. We will thus prove
\cref{ctsthm} in a few steps. First, the odd-primary information is in~\eqref{twisted_symm_SO}. Then, for the
$2$-primary part, we extract as much information as we can from the Adams spectral sequence in \cref{ctsthm_part1},
reducing the computation to the question of whether $\eta\in\Omega_1^\Spin$ acts nontrivially as a map
$\tOmega_3^\Spin((BS_4)^{3-V_\lambda})\to\tOmega_4^\Spin((BS_4)^{3-V_\lambda})$. Finally, in \cref{ctsthm_part2},
we show that $\eta$ acts trivially, which will finish the proof of \cref{ctsthm}.
\begin{prop}
\label{ctsthm_part1}\hfill
\begin{enumerate}
    \item 
 	\Cref{ctsthm} is true for $\tOmega_k^\Spin((BS_4)^{3-V_\lambda})$ for $k\in\set{0,1,2,3,5}$.
    \item Exactly one of the following two cases occurs:
    \begin{itemize}
        \item $\tOmega_4^\Spin((BS_4)^{3-V_\lambda})\cong \Z/4\oplus (\Z/2)^{\oplus 2}$ and
        $\eta\colon \tOmega_3^\Spin((BS_4)^{3-V_\lambda})\to \tOmega_4^\Spin((BS_4)^{3-V_\lambda})$ is nonzero, or
        \item $\tOmega_4^\Spin((BS_4)^{3-V_\lambda})\cong(\Z/2)^{\oplus 3}$ and $\eta\colon
        \tOmega_3^\Spin((BS_4)^{3-V_\lambda})\to \tOmega_4^\Spin((BS_4)^{3-V_\lambda})$ is the zero map.
    \end{itemize}
\end{enumerate}
\end{prop}
\begin{proof}
For odd-primary information, see~\eqref{twisted_symm_SO}. For $2$-primary information, we will again use the Adams
spectral sequence over $\cA(1)$. Our first task is to write down $\tH^*((BS_4)^{3-V_\lambda};\Z/2)$ as an $\cA(1)$-module in low
degrees, using \cref{S4Steenrod} to deduce $w_1(3-V_\lambda) = a$ and $w_2(3-V_\lambda) = a^2+b$. We describe this
$\cA(1)$-module structure in low degrees in \cref{spinlesssymmetrica1}, left.
\begin{figure}[h!]
\begin{subfigure}[c]{0.54\textwidth}
\begin{tikzpicture}[scale=0.6, every node/.style = {font=\tiny}]
    \foreach \y in {0, ..., 12} {
        \node at (-1.5, \y) {$\y$};
    }
    \begin{scope}[BrickRed]
        \Aone{0}{0}{$\oU$}
    \end{scope}

    \begin{scope}[RedOrange]
        \tikzpt{2.25}{2}{$\oU b$}{};
        \tikzpt{2.25}{3}{}{};
        \sqone(2.25, 2);
    \end{scope}

    \begin{scope}[Green]
        \Aone{3.5}{4}{$\oU a^4$};
    \end{scope}

    \begin{scope}[MidnightBlue]
        \tikzpt{6}{4}{$\oU b^2$}{};
        \tikzpt{6}{5}{}{};
        \tikzpt{6}{7}{}{};
        \tikzpt{6}{8}{}{};
        \tikzpt{7.5}{6}{}{};
        \tikzpt{7.5}{7}{}{};
        \tikzpt{9}{5}{$\oU bc$}{};
        \tikzpt{9}{6}{}{};
        \tikzpt{9}{8}{}{};
        \tikzpt{9}{9}{}{};
        \sqone(6, 4);
        \sqtwoCR(6, 4);
        \sqtwoL(6, 5);
        \sqone(6, 7);
        \sqtwoCR(7.5, 7);
        \sqtwoCL(7.5, 6);
        \sqone(7.5, 6);
        \sqone(9, 5);
        \sqtwoCL(9, 5);
        \sqtwoR(9, 6, );
        \sqone(9, 8);
    \end{scope}

    \begin{scope}[Fuchsia]
        \Aone{11}{6}{$\oU a^2b^2$}
    \end{scope}
\end{tikzpicture}
\end{subfigure}
\begin{subfigure}[c]{0.45\textwidth}
\begin{sseqdata}[name=spinlessS4, classes=fill, xrange={0}{7}, yrange={0}{3}, scale=0.6, Adams grading, >=stealth,
x label = {$\displaystyle{s\uparrow \atop t-s\rightarrow}$},
x label style = {font = \small, xshift = -21ex, yshift=3ex}]
\class[BrickRed](0, 0)
\begin{scope}[RedOrange]
    \class(2, 0)
    \class(3, 1)
    \class(4, 2)
    \structline(2, 0)(3, 1)
    \structline(3, 1)(4, 2)
    \class(6, 3)
    \class(4, 1)
    \structline(4, 1)(4, 2)
    \class(5, 2)
    \structline(4, 1)(5, 2)
    \structline(5, 2)(6, 3)
\end{scope}
\class[Green](4, 0)
\begin{scope}[MidnightBlue]
    \class(4, 0)
    \class(5, 0)
    \class(6, 1)
    \structline(5, 0)(6, 1)
\end{scope}
\class[Fuchsia](6, 0)
\d[gray]2(5, 0)
\end{sseqdata}
\label{spinless_s4_sseq}
\printpage[name=spinlessS4, page=2]
\end{subfigure}
\caption{Left: the $\cA(1)$-module structure on $\tH^*((BS_4)^{3-V_\lambda};\Z/2)$ in low degrees. This submodule
contains all elements of degree at most $7$. Right: the $E_2$-page of the Adams spectral sequence computing
$\tOmega^{\Spin}_*((BS_4)^{3-V_\lambda})$.}
\label{spinlesssymmetrica1}
\end{figure}

Let $\textcolor{MidnightBlue}{\Sigma^4 N_2}$ denote the submodule generated by $\oU b^2$ and $\oU bc$, which is a
nontrivial extension of $J$ by $\Sigma J$.\footnote{We propose calling $\textcolor{MidnightBlue}{N_2}$ the
\term{butterfly}; it also appears in~\cite[\S 5]{Wil73}, \cite[Lemma 8.3]{Pea14}, \cite[\S 6]{Pow15},
and~\cite[Figure 16]{WWZ19}.}
Then there is an isomorphism
\begin{equation}
\label{symspinlcoh}
    \tH^*((BS_4)^{3-V_\lambda};\Z/2)\cong \textcolor{BrickRed}{\cA(1)}\oplus
        \textcolor{RedOrange}{\Sigma^2 N_1}\oplus
        \textcolor{Green}{\Sigma^4\cA(1)}\oplus
        \textcolor{MidnightBlue}{\Sigma^4 N_2}\oplus
        \textcolor{Fuchsia}{\Sigma^6\cA(1)} \oplus P,
\end{equation}
The indecomposable summand isomorphic to $\textcolor{RedOrange}{\Sigma^2 N_1}$ is generated by $\oU b$, and $P$ has
no elements in degrees below $8$, and therefore is irrelevant for our low-degree computations. As before, we know
what a $\Sigma^k\cA(1)$ summand contributes to the $E_2$-page. Wilson~\cite[Figure 4]{Wil73} computes
$\Ext(\textcolor{RedOrange}{N_1})$ as a bigraded vector space, and the actions of $h_0$ and $h_1$ are stated
in~\cite[\S 3]{BB96} and~\cite[Figure 15]{WWZ19}: all $h_0$- and $h_1$-actions that could be nonzero when looking
at bidegrees are in fact nonzero.\footnote{One way to compute these $h_0$- and $h_1$-actions would be to use the long exact sequences in
Ext associated to the two short exact sequences
\begin{subequations}
\begin{gather}
    \shortexact{\Sigma\Z/2}{\textcolor{RedOrange}{N_1}}{\Z/2}{}\\
    \shortexact{\textcolor{RedOrange}{\Sigma^2 N_1}}{\uQ}{\Z/2},
\end{gather}
\end{subequations}
together with the fact that the boundary maps in the long exact sequences commute with the
$H^{*,*}(\cA(1))$-action. Wilson~\cite[Figure 3]{Wil73} actually computes the first of these two long exact
sequences but does not report
the $h_0$- or $h_1$-actions.}

Alternatively, one can compute $\Ext(\textcolor{RedOrange}{N_1})$, by using
a well-known explicit ($12$-shifted) $4$-periodic minimal resolution\footnote{After some practice with
$\cA(1)$-modules, writing this minimal resolution down is straightforward, if a little tedious; we found it a
helpful exercise when learning this material and the interested reader might too. Though this minimal resolution is
certainly known, it is not explicitly written in many places; the resolution will not be televised.}
    \begin{equation}
    \label{N1minres}
        \xymatrix@C=0.4cm{
            \textcolor{RedOrange}{N_1} & \cA(1)\ar[l]_-{f_0} & \Sigma^2\cA(1)\oplus\Sigma^3\cA(1)\ar[l]_-{f_1} &
            \Sigma^4\cA(1)\oplus\Sigma^5\cA(1)\ar[l]_-{f_2} & \Sigma^7\cA(1)\ar[l]_-{f_3} &
            \Sigma^{12}\cA(1)\ar[l]_-{f_4} & \Sigma^{14}\cA(1)\oplus\Sigma^{15}\cA(1)\ar[l]_-{\Sigma^{12}f_1} &
            \dots\ar[l]_-{\Sigma^{12}f_2}
        }
    \end{equation}
The dimension of $\Ext_{\cA(1)}^{s,t}(\textcolor{RedOrange}{N_1}, \Z/2)$ is the number of summands of
$\Sigma^t\cA(1)$ in the $s^{\mathrm{th}}$ module in the extension.
 This (shifted up by $2$ for $\textcolor{RedOrange}{\Sigma^2 N_1}$) gives the orange
summands in~\cref{spinlesssymmetrica1}, right.

Wilson~\cite[Figure 6]{Wil73} computes $\Ext(\textcolor{MidnightBlue}{\Sigma^4 N_2})$ as a bigraded vector space,
but we also want the $H^{*,*}(\cA(1))$-module structure. For this, we use a convenient shortcut: the
kernel of the map $f_2$
in~\eqref{N1minres} is isomorphic to $\textcolor{MidnightBlue}{\Sigma^4 N_2}$. Thus, the sequence~\eqref{N1minres}
except for the first two terms forms a minimal resolution for $\textcolor{MidnightBlue}{\Sigma^4 N_2}$, so for
every $s,t\ge 0$, there is an isomorphism
\begin{equation}
    \Ext_{\cA(1)}^{s,t}(\textcolor{MidnightBlue}{N_2}, \Z/2) \cong \Ext_{\cA(1)}^{s+2,
    t+4}(\textcolor{RedOrange}{N_1}, \Z/2)
\end{equation}
equivariant for the $H^{*,*}(\cA(1))$-actions on both sides. This gives us the blue summands
in~\cref{spinlesssymmetrica1}, right.
Now we can draw the $E_2$-page for the Adams spectral sequence for $\tOmega_*^\Spin((BS_4)^{3-V_\lambda})$, and do so in
\cref{spinlesssymmetrica1}, right.

Margolis' theorem and $h_i$-equivariance of differentials imply there is a single differential in this range that
could be nonzero, namely the pictured $d_2\colon E_2^{0,5}\to E_2^{2,6}$. Either this differential vanishes, or it
doesn't; we will show these two cases correspond to the two cases in the proposition statement.

First suppose this differential vanishes. There are no further nonzero differentials in the range we care about,
but we must address two extension questions in degree $4$:
\begin{subequations}
\begin{gather}
    \label{deg4_first_xtn}
    \shortexact{\textcolor{RedOrange}{\Z/2}}{A}{\textcolor{Green}{\Z/2}\oplus
    \textcolor{MidnightBlue}{\Z/2}}{}\\
    \label{deg4_second_xtn}
    \shortexact{\textcolor{RedOrange}{\Z/2}}{\tOmega_4^\Spin((BS_4)^{3-V_\lambda})}{A}{}
\end{gather}
\end{subequations}

For~\eqref{deg4_first_xtn},
assume the sequence does not split; then, $\tOmega^{\Spin}_4((BS_4)^{3-V_\lambda})$ has an element $x$ such that
$2x\ne 0$ and if $y$ is the image of $2x$ in the $E_\infty$-page, then $h_1y\ne 0$. This fact lifts to a nonzero
action by $\eta\in\ko_1$ carrying $2x$ to some element $z\in\tOmega^{\Spin}_{5}((BS_4)^{3-V_\lambda})$ such that $z
= 2\eta x$ and $z \ne 0$, but $2\eta = 0$, causing a contradiction.

Because~\eqref{deg4_first_xtn} splits and $(h_0\cdot)\colon E_\infty^{1,5}\to E_\infty^{2,6}$ is an isomorphism,
all possible extensions in~\eqref{deg4_second_xtn} give $\tOmega^{\Spin}_4((BS_4)^{3-V_\lambda})\cong\Z/4\oplus
(\Z/2)^{\oplus 2}$.
%

The last thing we need to prove in the first case of the proposition statement is that $\eta$ acts nontrivially
from topological degree $3$ to degree $4$. This follows from the nontrivial $h_1$-action
$E_\infty^{1,4}\to E_\infty^{2,6}$.

In the second case, we assume $d_2\colon E_2^{5,0}\to E_2^{4,6}$ is nonzero. Then the only interesting extension
question in topological degree $4$ on the $E_\infty$-page is~\eqref{deg4_first_xtn}, which splits for the same
reason as in the previous case. Thus if $d_2\ne 0$, $\tOmega_4^{\Spin}((BS_4)^{3-V_\lambda})\cong (\Z/2)^{\oplus
3}$. The last thing to verify is that $\eta$ is the zero map from topological degree $3$ to degree $4$. The unique
nonzero class $x$ in degree $3$ corresponds to a class in the $E_\infty$-page in filtration $1$, so $\eta x$ must
have image in the $E_\infty$-page with filtration at least $2$, since $\eta$ itself has filtration $1$. But
inspecting topological degree $4$ of the $E_\infty$-page, all nonzero elements have filtration $0$ or $1$, so $\eta
x = 0$.
\end{proof}
\begin{prop}
\label{ctsthm_part2}
$\eta\colon \tOmega_3^\Spin((BS_4)^{3-V_\lambda})\to\tOmega_3^\Spin((BS_4)^{3-V_\lambda})$ is the zero map.
\end{prop}
Combined with \cref{ctsthm_part1}, this will prove \cref{ctsthm}.
\begin{proof}
Consider the map $j$ on twisted spin bordism induced by the inclusion $A_4\inj S_4$. Since this is induced from a
map of spectra, it commutes with the action of $\eta$, giving a commutative diagram
\begin{equation}
\label{etacomm}
\begin{tikzcd}
    {\tOmega_3^\Spin((BA_4)^{3-V_\lambda|_{A_4}})} & {\tOmega_3^\Spin((BS_4)^{3-V_\lambda})} \\
    {\tOmega_4^\Spin((BA_4)^{3-V_\lambda|_{A_4}})} & {\tOmega_4^\Spin((BS_4)^{3-V_\lambda})}.
    \arrow["j", from=1-1, to=1-2]
    \arrow["\eta", from=1-1, to=2-1]
    \arrow["\eta", from=1-2, to=2-2]
    \arrow["j", from=2-1, to=2-2]
\end{tikzcd}
\end{equation}
The proposition statement is that the right-hand $\eta$ vanishes. We can compute the left-hand $\eta$ fairly
quickly: the restriction of $V_\lambda$ to $A_4$ is the representation associated to a chiral tetrahedral symmetry,
which we studied in \S\ref{s_chiral_tet}; in particular, in \cref{spin_bord_X} we computed
$\tOmega_k^\Spin((BA_4)^{3-V_\lambda|_{A_4}})$ for $k = 3,4$ to be $\Z/2$, resp.\ $\Z$. There is no nonzero
homomorphism $\Z/2\to\Z$, so the left-hand $\eta$ map in~\eqref{etacomm} is $0$. Thus it suffices to show that $j\colon
\tOmega_3^\Spin((BA_4)^{3-V_\lambda|_{A_4}}) \to \tOmega_3^\Spin((BS_4)^{3-V_\lambda})$ is surjective,
as commutativity of~\eqref{etacomm} would force the right-hand $\eta$ to vanish, and we would be done with the
proof. Thus in the rest of the proof we show $j$ is surjective.

To prove this, we will fit $j$ into a long exact sequence whose third term is tractable to calculate. This long
exact sequence is attributed to James (see, e.g., \cite[Remark 3.14]{KZ18}); see~\cite[Theorem 5.1]{MathSmith} for
a proof and a bordism-theoretic interpretation of the sequence.
\begin{thm}[James]
\label{james}
Let $V\to X$ be a virtual vector bundle and $W\to X$ be a vector bundle of rank $r$ with sphere bundle $p\colon
S(W)\to X$. Then there is a long exact sequence
\begin{equation}
\label{smith_LES}
    \dotsb\to
    \tOmega_n^\Spin(S(W)^{p^*V}) \overset{p_*}{\longrightarrow}
    \tOmega_n^\Spin(X^V) \overset{\mathrm{sm}_W}{\longrightarrow}
    \tOmega_{n-r}^\Spin(X^{V\oplus W - r}) \longrightarrow
    \tOmega_{n-1}^\Spin(S(W)^{p^*V})  \to\dotsb
\end{equation}
\end{thm}
For reasons that are still mysterious to the author, this long exact sequence seems to contain different
information than the Postnikov and Adams filtrations, making it surprisingly useful when the Atiyah-Hirzebruch and
Adams spectral sequences have difficult differentials or extension problems. See~\cite{HS13, GOPWW18, DL23,
DDHM, Deb24, DNT24, CDKPSS24, DDKLPTT23, DYY25a, DYY25b, DYY25c, JTVP25, MathSmith} for examples of other
computations using this technique.

We will apply~\eqref{smith_LES} to the case $X = BS_4$, $V = 3 - V_\lambda$, and $W = \sigma$, the real line bundle
associated to the sign homomorphism $S_4\to \Z/2\cong\O_1$. By~\cite[Lemma A.8]{DYY25a} (compare~\cite[Lemma
C.2]{DL23}), the map $p_*\colon S(\sigma)\to BS_4$ is homotopy equivalent to $j\colon BA_4\to BS_4$,
so~\eqref{smith_LES} specializes to a long exact sequence
\begin{equation}
\label{specific_smith}
\dotsb \longrightarrow
    \tOmega_3^\Spin((BA_4)^{3-V_\lambda|_{A_4}}) \overset j\longrightarrow
    \tOmega_3^\Spin((BS_4)^{3-V_\lambda}) \overset{\mathrm{sm}_\sigma}{\longrightarrow}
    \tOmega_2^\Spin((BS_4)^{2-V_\lambda + \sigma}) \longrightarrow\dotsb
\end{equation}
\textbf{In the rest of the proof, we implicitly localize at 2.} This is an exact functor, so does not affect
our argument, except that it allows us to ignore odd-primary torsion in $\tOmega_2^\Spin((BS_4)^{2-V_\lambda +
\sigma})$.\footnote{Alternatively, one can prove that there is no odd-primary torsion in this bordism group by a
small modification of the proof of \cref{SO_bord_S4}.}

Recall that we have reduced the proof of the proposition statement to showing that $j$ is surjective; by exactness,
it suffices to show $\tOmega_2^\Spin((BS_4)^{2-V_\lambda + \sigma})\cong 0$, and this is how we will finish the proof.
We do this computation with the Adams spectral sequence, so the first step is to study the $\cA(1)$-module
structure on $\tH^*((BS_4)^{2-V_\lambda + \sigma};\Z/2)$. The Whitney sum formula shows $w_1(2-V_\lambda + \sigma)
= 0$ and $w_2(2-V_\lambda + \sigma) = b$, so $\Sq^1(U) = 0$ and $\Sq^2(U) = Ub$.

The reader can check that, up to isomorphism, there is a unique $\cA(1)$-module which is a nonsplit extension of
$\uQ$ by $\Sigma J$; call this module $N_5$. This module is identical to $\uQ\oplus\Sigma J$ except that in the
former, $\Sq^2$ of the unique highest-degree nonzero element of $\uQ$ is the unique highest-degree nonzero element
of $\Sigma J$; in the split extension, $\Sq^2$ of this element vanishes. Thus, the quotients of $N_5$ and
$\uQ\oplus\Sigma J$ by the submodules of elements in degrees $5$ and above are isomorphic. This implies that if
$t-s < 4$, $\Ext^{s,t}(N_5)\cong\Ext^{s,t}(\uQ\oplus J)$ (see~\cite[\S E.2]{CDDH25}).

Using the Cartan formula and $\Sq^i(U)$ as usual, one obtains an $\cA(1)$-module isomorphism
\begin{equation}
\label{silly_Hstar}
    \tH^*((BS_4)^{2-V_\lambda + \sigma};\Z/2) \cong N_5\oplus P,
\end{equation}
where $P$ is concentrated in degrees $3$ and above, so we can ignore it. We draw a picture of the
decomposition~\eqref{silly_Hstar}, i.e.\ of the $\cA(1)$-module $N_5$, in \cref{smithfig}, left. As noted above, to
compute $\Ext(N_5)$ in topological degree $4$ and below, we may replace it with $\uQ$ and $\Sigma J$, whose Ext
groups appear in~\cite[Figure 29]{BC18}. Thus we draw the $E_2$-page of the Adams spectral sequence computing
$\tOmega_*^\Spin((BS_4)^{2-V_\lambda+\sigma})$ in degrees $2$ and below in \cref{smithfig}, right. $E_2^{s,t}\cong
0$ when $t-s = 2$, so the same is true for $E_\infty^{s,t}$, so $\tOmega_2^\Spin((BS_4)^{2-V_\lambda+\sigma})\cong
0$.
\end{proof}
\begin{figure}[h!]
\begin{subfigure}[c]{0.25\textwidth}
\begin{tikzpicture}[scale=0.6, every node/.style = {font=\tiny}]
\foreach \y in {0, ..., 5} {
        \node at (-2, \y) {$\y$};
}
\Nfive{0}{0}{$\oU$}{$\oU a$}
%
%
\end{tikzpicture}
\end{subfigure}
\begin{subfigure}[c]{0.3\textwidth}
\begin{sseqdata}[name=smiths, classes=fill, scale=0.7, xrange={0}{2}, yrange={0}{3}, Adams grading,
>=stealth,
x label = {$\displaystyle{s\uparrow \atop t-s\rightarrow}$},
x label style = {font = \small, xshift = -12ex, yshift=4ex}]
       \class(0, 0)\AdamsTower{}
       \class(1, 0)
\end{sseqdata}
    \printpage[name=smiths, page=2]
\end{subfigure}
\caption{Left: the $\cA(1)$-module structure on
$\tH^*((BS_4)^{2-V_\lambda+\sigma};\Z/2)$ in low degrees. The pictured submodule contains all elements in degrees
$2$ and below.  Right: the $E_2$-page of the corresponding Adams spectral sequence computing
$\tOmega^\Spin_*((BS_4)^{2-V_\lambda+\sigma})_2^\wedge$. We use this in the proof of \cref{ctsthm_part2}.}
\label{smithfig}
\end{figure}

\subsubsection{Class D, spin-$1/2$ case}
\label{s_full_spinful_tet}
As $V_\lambda$ is not \pinm, \cref{shear_D_thm} tells us to compute the spin bordism of
$(BS_4)^{\Det(V_\lambda)-1}$.
\begin{thm}
\label{spinful_full_tet_thm}
The first few spin bordism groups of $(BS_4)^{\Det(V_\lambda)-1}$ are
\begin{align*}
	\tOmega_0^\Spin((BS_4)^{\Det(V_\lambda)-1}) &\cong \Z/2\\
	\tOmega_1^\Spin((BS_4)^{\Det(V_\lambda)-1}) &\cong \Z/6\\
	\tOmega_2^\Spin((BS_4)^{\Det(V_\lambda)-1}) &\cong \Z/8\oplus\Z/2\\
	\tOmega_3^\Spin((BS_4)^{\Det(V_\lambda)-1}) &\cong \Z/2\\
	\tOmega_4^\Spin((BS_4)^{\Det(V_\lambda)-1}) &\cong 0,
\end{align*}
and $\tOmega_5^\Spin((BS_4)^{\Det(V_\lambda)-1})$ is torsion.
\end{thm}
\begin{proof}
Odd-primary information is computed in the range we need by~\eqref{twisted_symm_SO}. For $2$-primary information,
we use the Adams spectral sequence as usual. Recall the $\cA(1)$-module structure on
$H^*(BS_4;\Z/2)\cong\Z/2[a,b,c]/(ac)$ from \cref{mod2S4coh,S4Steenrod}. \Cref{det_splitting} shows that inclusion
of a transposition extends to a splitting
\begin{equation}
	(BS_4)^{\Det(V_\lambda)-1} \overset\simeq\longrightarrow (B\Z/2)^{\sigma-1}\vee M,
\end{equation}
and the map $\wH^*(M;\Z/2)\to\wH^*((BS_4)^{\Det(V_\lambda)-1};\Z/2)$ is injective, with image a complementary subspace to the span of
$\set{Ua^n\mid n\ge 0}$. As usual, we write down $\tH^*(M;\Z/2)$ as an $\cA(1)$-module in low degrees, using
$w_1(\Det(V_\lambda)-1) = a$ and $w_2(\Det(V_\lambda)-1) = 0$, and give the answer in \cref{spinfulsymmetrica1},
left.
\begin{figure}[h!]
\begin{subfigure}[c]{0.35\textwidth}
\begin{tikzpicture}[scale=0.6, every node/.style = {font=\tiny}]
	\foreach \y in {2, ..., 9} {
		\node at (-2, \y) {$\y$};
	}

	\begin{scope}[BrickRed]
		\AoneTruncFour{0}{2}{$\oU b$}
	\end{scope}
	
	\begin{scope}[MidnightBlue]
		\Aone{3}{3}{$\oU ab$}
	\end{scope}
\end{tikzpicture}
\end{subfigure}
\qquad
\begin{subfigure}[c]{0.35\textwidth}
\begin{sseqdata}[name=spinfulS4, classes=fill, xrange={0}{4}, yrange={0}{2}, scale=0.6,
x label = {$\displaystyle{s\uparrow \atop t-s\rightarrow}$},
x label style = {font = \small, xshift = -15.7ex, yshift=3ex}]
	\class[BrickRed](2, 0)
	\class[MidnightBlue](3, 0)
\end{sseqdata}
\printpage[name=spinfulS4, page=2]
\end{subfigure}
\caption{Left: The $\cA(1)$-module structure on $\tH^*(M;\Z/2)$ in low degrees. This submodule contains all
elements of degree at most $4$. Right: the Ext of this module, which is the beginning of the Adams spectral
sequence computing $\widetilde{\ko}_*(M)$. More information in the proof of \cref{spinful_full_tet_thm}.}
\label{spinfulsymmetrica1}
\end{figure}

Let $\textcolor{BrickRed}{\Sigma^2 N_3}$ denote the $\cA(1)$-submodule generated by $\oU b$; Baker~\cite[\S
5]{Bak18} calls it the \term{whiskered Joker}, and it also appears in~\cite{Pea14, Pow15, BDDHM25}. There is an isomorphism of $\cA(1)$-modules
\begin{equation}
\label{A1_mod_spinful_full_tet}
	\tH^*(M;\Z/2)\cong \textcolor{BrickRed}{\Sigma^2 N_3} \oplus
		\textcolor{MidnightBlue}{\Sigma^3\cA(1)} \oplus P,
\end{equation}
where $P$ contains no elements of degree less than $4$. Therefore if the $4$-line of the $E_2$-page is empty, $P$
does not enter into our calculations --- and we will see momentarily that the $4$-line is in fact empty. We know
what the $\textcolor{MidnightBlue}{\Sigma^3\cA(1)}$ summand contributes to the $E_2$-page of the Adams spectral
sequence. For $\textcolor{BrickRed}{N_3}$, we leverage what we learned from $N_1$ in \S\ref{chiral_tet_spinless}.
Specifically, the unique nonzero $\cA(1)$-module map $\cA(1)\to\textcolor{BrickRed}{N_3}$ has kernel isomorphic to
$\Sigma^5 N_1$, so a minimal resolution for $\Sigma^5 N_1$ induces a minimal resolution for
$\textcolor{BrickRed}{N_3}$ which has an additional copy of $\cA(1)$ in topological degree $0$ and filtration $0$,
and in which everything else is shifted up one in filtration, giving the red summands
in \cref{spinfulsymmetrica1}, right.

Thus the $E_2$-page for this Adams spectral sequence is as in \cref{spinfulsymmetrica1}, right. In this range, the
spectral sequence collapses. Combine this with the \pinm bordism summands from~\cite{ABP69, KT90} as usual to
obtain the groups in the theorem statement, and \cref{torsion_k_theory} finishes us off by telling us
$\tOmega_5^\Spin((BS_4)^{\Det(V_\lambda)-1})$ is torsion.
\end{proof}

\subsubsection{Class A, spinless case}
\label{s_spinc_full_tet}
\Cref{shear_A_thm} asks us to consider the \spinc bordism of $(BS_4)^{3-V_\lambda}$.
\begin{thm}
\label{spinc_full_tet_thm}
The first few \spinc bordism groups of $(BS_4)^{3-V_\lambda}$ are
\begin{align*}
	\tOmega_0^\Spinc((BS_4)^{3-V_\lambda}) &= \Z/2\\
	\tOmega_1^\Spinc((BS_4)^{3-V_\lambda}) &= \Z/3\\
	\tOmega_2^\Spinc((BS_4)^{3-V_\lambda}) &= (\Z/2)^{\oplus 2}\\
	\tOmega_3^\Spinc((BS_4)^{3-V_\lambda}) &= \Z/3\\
	\tOmega_4^\Spinc((BS_4)^{3-V_\lambda}) &= (\Z/2)^{\oplus 4},
\end{align*}
and $\tOmega_5^{\Spinc}((BS_4)^{3-V_\lambda})$ is torsion. Therefore $\Ph_0^{S_4}(\R^3; f_0^A)\cong (\Z/2)^{\oplus 4}$.
\end{thm}
\begin{proof}
We will use the Adams spectral sequence over $\cE(1)$ as usual to capture the $2$-primary information; for
odd-primary information, see~\eqref{twisted_symm_SO_U}.

We use the $\cA(1)$-module structure on $\tH^*((BS_4)^{3-V_\lambda};\Z/2)$ that we determined in~\eqref{symspinlcoh} and drew
in \cref{spinlesssymmetrica1} to determine the $\cE(1)$-module structure: as $\cE(1)$-modules,
$\cA(1)\cong\cE(1)\oplus\Sigma^2\cE(1)$, and $N_2\cong \cE(1)\oplus \Sigma\cE(1)\oplus\Sigma^2 N_1$, so as
$\cE(1)$-modules,
\begin{equation}
	\tH^*((BS_4)^{3-V_\lambda};\Z/2)\cong \textcolor{BrickRed!80!black}{\cE(1)} \oplus
		\textcolor{BrickRed!80!white}{\Sigma^2\cE(1)} \oplus
		\textcolor{RedOrange}{\Sigma^2 N_1} \oplus
		\textcolor{Green}{\Sigma^4\cE(1)} \oplus
		\textcolor{MidnightBlue!75!black}{\Sigma^4\cE(1)} \oplus
		\textcolor{MidnightBlue!75!white}{\Sigma^5\cE(1)} \oplus P,
\end{equation}
where $P$ is $5$-connected. We draw this in \cref{full_tetrahedral_spinc_a1}, left.
%
\begin{figure}[h!]
\begin{subfigure}[c]{0.45\textwidth}
\begin{tikzpicture}[scale=0.6, every node/.style = {font=\tiny}]
\foreach \y in {0, ..., 9} {
	\node at (-2, \y) {$\y$};
}
\begin{scope}[BrickRed!80!black]
	\Eone{0}{0}{$\oU$}
\end{scope}
\begin{scope}[BrickRed!80!white]
	\Eone{2}{2}{$\oU a^2$}
\end{scope}
\begin{scope}[RedOrange]
	\tikzpt{3.25}{2}{$\oU b$}{};
	\tikzpt{3.25}{3}{}{};
	\sqone(3.25, 2);
\end{scope}
\begin{scope}[Green]
	\Eone{4.5}{4}{$\oU a^4$}
\end{scope}
\begin{scope}[MidnightBlue!75!black]
	\Eone{6.5}{4}{$\oU b^2$}
\end{scope}
\begin{scope}[MidnightBlue!75!white]
	\Eone{8.75}{5}{$\oU bc$}
\end{scope}
\end{tikzpicture}
\end{subfigure}
\begin{subfigure}[c]{0.35\textwidth}
\begin{sseqdata}[name=spincfulltet, classes=fill, scale=0.6, xrange={0}{5}, yrange={0}{3},
x label = {$\displaystyle{s\uparrow \atop t-s\rightarrow}$},
x label style = {font = \small, xshift = -18ex, yshift=3ex}]
\class[BrickRed!80!black](0, 0)
\class[BrickRed!80!white](2, 0)
\begin{scope}[RedOrange]
	\class(2, 0)
	\class(4, 1)\structline[RedOrange!40!white](2, 0, -1)(4, 1)
	\class(6, 2)\structline[RedOrange!40!white](4, 1)(6, 2)
	\class(8, 3)\structline[RedOrange!40!white](6, 2)(8, 3)
\end{scope}
\class[Green](4, 0)
\class[Green!80!white](6, 0)
\class[MidnightBlue!75!black](4, 0)
\class[MidnightBlue!75!white](5, 0)
\begin{scope}[MidnightBlue!75!white]
	\class(6, 0)
	\class(8, 1)
	\structline[MidnightBlue!40!white](6, 0, -1)(8, 1)
\end{scope}
\class[Fuchsia](6, 0)
\end{sseqdata}
\printpage[name=spincfulltet, page=2]
\end{subfigure}
\caption{Left: the $\cE(1)$-module structure on $\tH^*((BS_4)^{3-V_\lambda};\Z/2)$ in low degrees. The pictured
submodule contains all elements of degree at most $5$. Right: the Ext of this module, which is the beginning of the
$E_2$-page of the Adams spectral sequence computing $\widetilde{\ku}_*((BS_4)^{3-V_\lambda})$.}
\label{full_tetrahedral_spinc_a1}
\end{figure}
Recalling $\Ext_{\cE(1)}(N_1)$ from~\eqref{N1extcalc}, the $E_2$-page of the Adams spectral sequence is in
\cref{full_tetrahedral_spinc_a1}, right.
There can be no differentials in the range drawn for degree reasons, and Margolis' theorem (\cref{margolis})
implies there are no nontrivial extensions, either, so we are done.
\end{proof}

\subsubsection{Class A, spin-$1/2$ case}
	\Cref{shear_A_thm} says that to compute the phase homology groups for this case, we should investigate the \spinc
bordism of $(BS_4)^{\Det V_\lambda - 1}$: we know $V_\lambda$ is not \pinc because its pullback along
$BA_4\to BS_4$ is not \pinc, as we established in \cref{tet_not_pinc}.
\begin{thm}
The first few \spinc bordism groups of $(BS_4)^{\Det(V_\lambda)-1}$ are
\begin{align*}
	\tOmega_0^{\Spin^c}((BS_4)^{\Det(V_\lambda)-1}) &\cong \Z/2\\
	\tOmega_1^{\Spin^c}((BS_4)^{\Det(V_\lambda)-1}) &\cong \Z/3\\
	\tOmega_2^{\Spin^c}((BS_4)^{\Det(V_\lambda)-1}) &\cong \Z/4\oplus\Z/2\\
	\tOmega_3^{\Spin^c}((BS_4)^{\Det(V_\lambda)-1}) &\cong \Z/6\\
	\tOmega_4^{\Spin^c}((BS_4)^{\Det(V_\lambda)-1}) &\cong \Z/8\oplus (\Z/2)^{\oplus 2}.
\end{align*}
By \cref{torsion_k_theory}, $\tOmega_5^{\Spinc}((BS_4)^{\Det(V_\lambda)-1})$ is torsion, so $\Ph_0^{S_4}(\R^3; f^A_{1/2})\cong \Z/8\oplus
(\Z/2)^{\oplus 2}$.
\end{thm}
\begin{proof}
See~\eqref{twisted_symm_SO_U} for the odd-primary torsion in $\tOmega_*^\Spinc((BS_4)^{\Det(V_\lambda)-1})$. For $2$-torsion, we reuse our
work from \S\ref{s_full_spinful_tet}. First, $(BS_4)^{\Det(V_\lambda)-1}\simeq (B\Z/2)^{\sigma-1}\vee M$, and we gave the low-degree
cohomology of $M$ as an $\cA(1)$-module in~\eqref{A1_mod_spinful_full_tet}, and drew it in
\cref{spinfulsymmetrica1}, left. This determines the $\cE(1)$-module structure on it, so we can calculate \spinc
bordism of $M$ using the Adams spectral sequence. For the other summand, we have
$\tOmega_*^\Spinc((B\Z/2)^{\sigma-1})\cong\Omega_*^{\Pin^c}$~\eqref{pincsplitting}, so we direct-sum in the \pinc
bordism groups computed by Bahri-Gilkey~\cite{BG87a, BG87b}.

There are isomorphisms of $\cE(1)$-modules $\cA(1)\cong\cE(1)\oplus\Sigma^2\cE(1)$ and
$N_3\cong\cE(1)\oplus \Sigma^2N_1$. Therefore as an $\cE(1)$-module,
\begin{equation}
	\tH^*(M;\Z/2) \cong \textcolor{BrickRed!80!black}{\Sigma^2\cE(1)} \oplus
	\textcolor{MidnightBlue}{\Sigma^3\cE(1)} \oplus \textcolor{BrickRed!80!white}{\Sigma^4 N_1} \oplus P,
\end{equation}
where $P$ is $4$-connected. As usual for these cases, we will see that $\Ext(\tH^*(M;\Z/2), \Z/2)$ has no nonzero
elements with $t-s = 4$ and $s > 1$, so $P$ does not affect our calculations. See \cref{mixed_full_tet_pinc_drawn},
left, for a picture of the $\cE(1)$-module structure on $\tH^*(M;\Z/2)$.
\begin{figure}[h!]
\begin{subfigure}[c]{0.3\textwidth}
\begin{tikzpicture}[scale=0.6, every node/.style = {font=\tiny}]
	\foreach \y in {2, ..., 7} {
		\node at (-2, \y) {$\y$};
	}
	\begin{scope}[BrickRed!80!black]
		\Eone{0}{2}{$\oU b$};
	\end{scope}
	\begin{scope}[MidnightBlue]
		\Eone{2.25}{3}{$\oU ab$};
	\end{scope}
	\begin{scope}[BrickRed!80!white]
		\tikzpt{4}{4}{$\oU \alpha$}{};
		\tikzpt{4}{5}{}{};
		\sqone(4, 4);
	\end{scope}
\end{tikzpicture}
\end{subfigure}
\quad
\begin{subfigure}[c]{0.3\textwidth}
\begin{sseqdata}[name=mixedfulltetclassA, classes=fill, xrange={0}{4}, yrange={0}{2}, scale=0.6,
	x label = {$\displaystyle{s\uparrow \atop t-s\rightarrow}$},
	x label style = {font = \small, xshift = -15.5ex, yshift=3ex}]

	\class[BrickRed!80!black](2, 0)
	\class[MidnightBlue](3, 0)
	\class[BrickRed!80!white](4, 0)
	\class[white](6, 1)
	\structline[BrickRed!40!white](4, 0)(6, 1)
\end{sseqdata}
	\printpage[name=mixedfulltetclassA, page=2]
\end{subfigure}
\caption{Left: the $\cE(1)$-module structure on $\tH^*(M; \Z/2)$ in low degrees; the pictured summands include all
elements in degrees $4$ and below. Here $\alpha\coloneqq a^2b + b^2$. Right: the Ext of this module, which is the
beginning of the $E_2$-page of the Adams spectral sequence computing $\widetilde\ku_*(M)$.}
\label{mixed_full_tet_pinc_drawn}
\end{figure}
We calculated $\Ext(\textcolor{BrickRed!80!white}{\Sigma^4 N_1})$ in~\eqref{N1extcalc}, so can draw the $E_2$-page
of the Adams spectral sequence in \cref{mixed_full_tet_pinc_drawn}, right. This collapses, so we add in the \pinc
bordism summands and conclude.
\end{proof}


	\subsection{Chiral octahedral symmetry}
		\label{s_chiral_oct}
Let $\lambda\colon S_4\to\O_3$ denote the representation as symmetries of an octahedron and $V_\lambda\to BS_4$
denote the associated vector bundle. Recall from \cref{mod2S4coh} the mod $2$ cohomology of $BS_4$.
\begin{lem}
\label{oct_wn}
$w_1(V_\lambda) = 0$ and $w_2(V_\lambda) = a^2 + b$.
\end{lem}
\begin{proof}
Since $\Im(\lambda)\subset\SO_3$, $w_1(V_\lambda) = 0$. We know $w_2(V_\lambda)$ restricts to $u\in H^2(BA_4;\Z/2)$
by considering tetrahedral symmetry inside octahedral symmetry and using~\cref{SWalt}, so $w_2(V_\lambda)$ could
be $a^2+b$ or $b$. The fact that $\lambda$ splits as $2\sigma\oplus\R$ when restricted to a $\Z/2$ subgroup given
by a transposition tells us $w_2(V_\lambda)$ is $a^2 + b$, not $b$.
\end{proof}
By \cref{tet_not_pinc}, the pullback of $V_\lambda$ to $BA_4$ is not \pinc, so $V_\lambda$ is not \pinc, and hence
$V_\lambda$ is also not \pinm.
\begin{lem}
\label{SO_bord_S4}
\begin{equation}
	\tOmega_k^\SO(BS_4)\otimes\Z[1/2] \cong \begin{cases}
		\Z/3, & k = 3\\
		0, &k = 0, 1, 2, 4, 5, 6.
	\end{cases}
\end{equation}
\end{lem}
\begin{proof}
Let $\ell$ be an odd prime and consider the Atiyah-Hirzebruch spectral sequence
\begin{equation}
\label{oct_s4_SO}
	E^2_{p,q} = H_p(BS_4; (\MTSO_\ell^\wedge)_q) \Longrightarrow (\MTSO_\ell^\wedge)_{p+q}(BS_4) =
	\Omega_{p+q}^\SO(BS_4)_\ell^\wedge.
\end{equation}
If $\ell\ne 3$, then $\ell\nmid\abs{S_4}$, so the $\Z_\ell$-cohomology of $BS_4$ vanishes in positive degrees
and~\eqref{oct_s4_SO} is trivial, contributing no $\ell$-torsion to $\tOmega_*^\SO(BS_4)\otimes\Z[1/2]$. For $\ell
= 3$, use Thomas' calculation of $H^*(BS_4;\Z)$ ~\cite{Tho74} and the universal coefficient theorem to show that
$H_*(BS_4;\Z_3)$ consists of $\Z_3$ in degree $0$, $\Z/3$ in degree $2$, and nothing else nonzero in degrees $5$
and below. Therefore~\eqref{oct_s4_SO} collapses, giving us the desired result.
\end{proof}

\subsubsection{Class D, spinless case}
	\Cref{shear_D_thm} computes $\Ph_k^{S_4}(\R^3; f_0^D)$ in terms of $\tOmega_*^\Spin((BS_4)^{3-V_\lambda})$, so we
study the spin bordism of $(BS_4)^{3-V_\lambda}$.
\begin{thm}
\label{oct_spinl_thm}
There is an $r\ge 2$ such that the first few spin bordism groups of $(BS_4)^{3-V_\lambda}$ are
\begin{align*}
	\tOmega_0^\Spin((BS_4)^{3-V_\lambda}) &\cong \Z\\
	\tOmega_1^\Spin((BS_4)^{3-V_\lambda}) &\cong \Z/2\\
	\tOmega_2^\Spin((BS_4)^{3-V_\lambda}) &\cong 0\\
	\tOmega_3^\Spin((BS_4)^{3-V_\lambda}) &\cong \Z/3\oplus\Z/2^r\\
	\tOmega_4^\Spin((BS_4)^{3-V_\lambda}) &\cong \Z \oplus \Z/2,
\end{align*}
and $\tOmega_5^\Spin((BS_4)^{3-V_\lambda})$ is torsion. Hence $\Ph_0^{S_4}(\R^3; f_0^D) \cong \Z/2$.
\end{thm}
The Atiyah-Hirzebruch spectral sequence allows one to show $r = 2$, so
$\tOmega_3^\Spin((BS_4)^{3-V_\lambda})\cong\Z/12$. As usual, we will not need this, so do not prove it.
\begin{proof}
For odd-primary torsion, use the fact that $\Omega_*^\Spin(\bl)\to\Omega_*^\SO(\bl)$ is an isomorphism, so it
suffices to understand $\tOmega_*^\SO((BS_4)^{3-V_\lambda})$, and that $V_\lambda\to BS_4$ is orientable, so there
is a Thom isomorphism $\Omega_k^\SO(BS_4)\to\tOmega_k^\SO((BS_4)^{3-V_\lambda})$ (\cref{tautological_Thom}); then
we can read off the odd-primary torsion from \cref{SO_bord_S4}.

%
On to the prime $2$. From \cref{mod2S4coh,S4Steenrod} we know the mod $2$ cohomology of $BS_4$ and the action of
the Steenrod algebra, and using \cref{oct_wn} we can draw $\tH^*((BS_4)^{3-V_\lambda};\Z/2)$ as an $\cA(1)$-module in low degrees,
which we do in \cref{chiral_octahedral_spinless_coh}, left.
\begin{figure}[h!]
\begin{subfigure}[c]{0.5\textwidth}
\begin{tikzpicture}[scale=0.6, every node/.style = {font=\tiny}]
\foreach \y in {0, ..., 11} {
	\node at (-1.5, \y) {$\y$};
}

\begin{scope}[BrickRed]
	\SpanishQnMark{0}{0}{$U$};
\end{scope}
\begin{scope}[RedOrange]
	\Aone{1.5}{1}{$Ua$};
\end{scope}
\begin{scope}[Goldenrod!67!black]
	\tikzpt{4.5}{3}{$Uc$}{};
\end{scope}

\begin{scope}[Green]
	\Rtwo{4.5}{4}{$Ub^2$}{$U\alpha$};
\end{scope}

\begin{scope}[MidnightBlue]
	\Aone{7.25}{5}{$Ua^5$};
\end{scope}

\begin{scope}[Fuchsia]
	\Joker{10}{5}{$Uab^2$};
\end{scope}
\end{tikzpicture}
\end{subfigure}
\begin{subfigure}[c]{0.45\textwidth}
\begin{sseqdata}[name=octahedralspinless, classes=fill, scale=0.7, xrange={0}{5}, yrange={0}{3}, Adams grading,
>=stealth,
x label = {$\displaystyle{s\uparrow \atop t-s\rightarrow}$},
x label style = {font = \small, xshift = -19ex, yshift=4ex}]
\begin{scope}[BrickRed]
	\class(0, 0)\AdamsTower{}
	\class(4, 1)\AdamsTower{}
	\class(5, 2)\structline(4, 1)(5, 2)
	\class(6, 3)\structline
\end{scope}
\class[RedOrange](1, 0)
\begin{scope}[Goldenrod!67!black]
\class(3, 0)\AdamsTower{}
\class(4, 1)\structline(3, 0)(4, 1, -1)
\class(5, 2)\structline
\end{scope}
\begin{scope}[draw=none, fill=none]
	\class(4, 0)
	\foreach \x in {0,2,3,4} {
		\class(4, \x)
	}
\end{scope}
\begin{scope}[Green]
	\class(4, 0)\AdamsTower{}
	\class(5, 0)
	\class(6, 1)\structline
\end{scope}
\class[MidnightBlue](5, 0)
\class[Fuchsia](5, 0)
\d[gray]2(4, 0, -1);
\end{sseqdata}
\label{octspinlesssseq}
	\printpage[name=octahedralspinless, page=2]
\end{subfigure}
\caption{Left: the $\cA(1)$-module structure on $\tH^*((BS_4)^{3-V_\lambda};\Z/2)$ in low degrees. The pictured
submodule contains all elements of degrees $6$ and below. Here $\alpha\coloneqq ab^2 + bc$.
Right: the $E_2$-page
of the corresponding Adams spectral sequence computing $\widetilde{\ko}_*((BS_4)^{3-V_\lambda})_2^\wedge$. We will
see there is a differential from the $4$-line to the $3$-line; it is in fact the $d_2$ depicted, though we do not
prove that.}
\label{chiral_octahedral_spinless_coh}
\end{figure}

We obtain an isomorphism of $\cA(1)$-modules
\begin{equation}
\label{spinless_chiral_octahedral_a1}
	\tH^*((BS_4)^{3-V_\lambda};\Z/2)\cong
		\textcolor{BrickRed}{\uQ} \oplus
		\textcolor{RedOrange}{\Sigma\cA(1)} \oplus
		\textcolor{Goldenrod!67!black}{\Sigma^3\Z/2} \oplus 
		\textcolor{Green}{\Sigma^4 R_2} \oplus
		\textcolor{MidnightBlue}{\Sigma^5\cA(1)} \oplus
		\textcolor{Fuchsia}{\Sigma^5 J} \oplus P,
\end{equation}
where $P$ is $6$-connected.

We draw the $E_2$-page in \cref{chiral_octahedral_spinless_coh}, right. Because differentials must be $h_0$-equivariant, they all vanish in the range pictured except possibly for those
from the $4$-line to the $3$-line, one of which is indicated in the chart. By \cref{torsion_k_theory},
$\widetilde\ko_4((BS_4)^{3-V_\lambda})\otimes\Q\cong\widetilde\ko_0((BS_4)^{3-V_\lambda})\otimes\Q$, and from \cref{chiral_octahedral_spinless_coh},
right, the latter group is isomorphic to $\Q$. Thus $\tOmega_4^\Spin((BS_4)^{3-V_\lambda})$ has exactly one free summand, so one of
the two towers in the $4$-line lives to the $E_\infty$-page, and the other admits a nonzero $d_r$ differential to
the tower in degree $3$. Thus, on the $3$-line of the $E_{r+1}$-page, there is a tower with finitely many
$\textcolor{Goldenrod!67!black}{\Z/2}$ summands, giving $\textcolor{Goldenrod!67!black}{\Z/2^r}$ in degree
$3$ as promised.\footnote{We have not determined which elements of the $4$-line the differential is nonzero on. One
way to determine this is to use that the generator of $H^{3,7}(\cA(1))\cong\Z/2$ carries the summands in the
$0$-line onto a subset of the red tower in the $4$-line. Differentials are equivariant for this action, and
differentials emerging from the $0$-line vanish, so all differentials must vanish on the red tower too.}
\end{proof}
%

\subsubsection{Class D, spin-$1/2$ case}
\Cref{shear_D_thm} computes the phase homology groups we need in terms of
$\Omega_*^\Spin(BS_4)$.
\begin{thm}
\label{s4_spin_bordism}
There is an $r\ge 2$ such that the first several spin bordism groups of $BS_4$ are
\begin{align*}
	\Omega_0^\Spin(BS_4) &\cong \Z\\
	\Omega_1^\Spin(BS_4) &\cong (\Z/2)^{\oplus 2}\\
	\Omega_2^\Spin(BS_4) &\cong (\Z/2)^{\oplus 3}\\
	\Omega_3^\Spin(BS_4) &\cong \Z/24 \oplus\Z/2^4,\\
	\Omega_4^\Spin(BS_4) &\cong \Z\oplus\Z/2\\
	\Omega_5^\Spin(BS_4) &\cong 0\\
	\Omega_6^\Spin(BS_4) &\cong \Z/2.
\end{align*}
Therefore $\Ph_0^{S_4}(\R^3; f_{1/2}^D)\cong\Z/2$.
\end{thm}
\begin{proof}
The $2$-primary part of \cref{s4_spin_bordism} was done by Bayen~\cite[Chapter 3]{Bay94}.
To determine the odd-primary torsion, use first that the forgetful map $\Omega_*^\Spin(\bl)\to\Omega_*^\SO(\bl)$ is
an isomorphism on odd-primary torsion, so we just have to determine the odd-primary torsion in
$\Omega_k^\SO(BS_4)$ for $k\le 6$, which we did in \cref{SO_bord_S4}.
%
%
\end{proof}

\subsubsection{Class A}
	As in the case of chiral tetrahedral symmetry, $V_\lambda$ does not admit a \pinc structure, since we saw in
\cref{tet_not_pinc} that its pullback along $BA_4\to BS_4$ also does not admit a \pinc structure.
\Cref{shear_A_thm} expresses $\Ph_0^{S_4}(\R^3; f_0^A)$ and $\Ph_0^{S_4}(\R^3; f_{1/2}^A)$ in
terms of the \spinc bordism of $(BS_4)^{3-V_\lambda}$ for spinless fermions and $BS_4$ for spin-$1/2$ fermions.
\begin{thm}
\label{spinc_chiral_oct_bordism}
There are integers $r,r'\ge 2$ such that the low-degree \spinc bordism groups of $(BS_4)^{3-V_\lambda}$ and $BS_4$
are
\begin{alignat*}{2}
	\tOmega_0^\Spinc((BS_4)^{3-V_\lambda}) &\cong \Z \qquad\qquad & \Omega_0^\Spinc(BS_4) &\cong\Z \\ 
	\tOmega_1^\Spinc((BS_4)^{3-V_\lambda}) &\cong \Z/2 \qquad\qquad & \Omega_1^\Spinc(BS_4) &\cong\Z/2 \\ 
	\tOmega_2^\Spinc((BS_4)^{3-V_\lambda}) &\cong \Z \qquad\qquad & \Omega_2^\Spinc(BS_4) &\cong \Z\oplus\Z/2 \\ 
	\tOmega_3^\Spinc((BS_4)^{3-V_\lambda}) &\cong \Z/6\oplus\Z/2^r \qquad\qquad & \Omega_3^\Spinc(BS_4) &\cong
	\Z/12\oplus\Z/2^{r'} \\
	\tOmega_4^\Spinc((BS_4)^{3-V_\lambda}) &\cong \Z^2 \qquad\qquad & \Omega_4^\Spinc(BS_4) &\cong\Z^2 \oplus\Z/2
	\\ 
	\tOmega_5^\Spinc((BS_4)^{3-V_\lambda}) &\cong \Z/2^{r-1}\oplus\Z/6\oplus(\Z/2)^{\oplus 3} &&\\
	\tOmega_6^\Spinc((BS_4)^{3-V_\lambda}) &\cong \Z^2. && 
\end{alignat*}
\end{thm}
One can use the Atiyah-Hirzebruch spectral sequence to show $r = r' = 2$. We do not need this, so do not go into
the details.
\begin{proof}
As usual, the calculation separates into odd-primary and $2$-primary parts. \Cref{spinc_odd_primes} reduces the
odd-primary computationd to the corresponding spin bordism computation, which we did in
\cref{oct_spinl_thm,s4_spin_bordism}, so all that remains is the $2$-primary computation.

For the spinless case, recall from \eqref{spinless_chiral_octahedral_a1} (drawn in
\cref{chiral_octahedral_spinless_coh}) the calculation of $\tH^*((BS_4)^{3-V_\lambda};\Z/2)$ as an $\cA(1)$-module.
The $\cE(1)$-module isomorphisms $\cA(1)\cong \cE(1)\oplus\Sigma^2 \cE(1)$, $R_2\cong \uQ\oplus \Sigma\cE(1)$, and
$J\cong \cE(1)\oplus\Sigma^2\Z/2$ imply that there is an isomorphism of $\cE(1)$-modules
\begin{equation}
	\tH^*((BS_4)^{3-V_\lambda};\Z/2) \cong \textcolor{BrickRed}{\uQ} \oplus
		\textcolor{RedOrange!67!black}{\Sigma\cE(1)} \oplus
		\textcolor{Goldenrod!67!black}{\Sigma^3\Z/2} \oplus
		\textcolor{RedOrange!67!white}{\Sigma^3\cE(1)} \oplus
		\textcolor{Green!80!black}{\Sigma^4 \uQ} \oplus
		\textcolor{Green!80!white}{\Sigma^5\cE(1)} \oplus
		\textcolor{MidnightBlue}{\Sigma^5\cE(1)} \oplus
		\textcolor{Fuchsia}{\Sigma^5\cE(1)} \oplus P,
\end{equation}
where $P$ is $6$-connected. We draw this in \cref{spinc_chiral_oct_unmixed}, left.
\begin{figure}[h!]
\begin{subfigure}[c]{0.58\textwidth}
\begin{tikzpicture}[scale=0.6, every node/.style = {font=\tiny}]
\foreach \y in {0, ..., 9} {
	\node at (-1.5, \y) {$\y$};
}
\begin{scope}[BrickRed]
	\EoneQnMark{0}{0}{$\oU$}
\end{scope}
\begin{scope}[RedOrange!67!black]
	\Eone{1.5}{1}{$\oU a$};
\end{scope}
\tikzptB{3}{3}{$\oU c$}{Goldenrod!67!black};
\begin{scope}[RedOrange!67!white]
	\Eone{4.5}{3}{$Ua^3$};
\end{scope}
\begin{scope}[Green!80!black]
	\EoneQnMark{7}{4}{$\oU b^2$}
\end{scope}
\begin{scope}[Green!80!white]
	\Eone{8.5}{5}{$Ubc$};
\end{scope}
\begin{scope}[MidnightBlue]
	\Eone{10.5}{5}{$\oU a^5$}
\end{scope}
\begin{scope}[Fuchsia]
	\Eone{12.5}{5}{$\oU ab^2$}
\end{scope}
\end{tikzpicture}
\end{subfigure}
\begin{subfigure}[c]{0.4\textwidth}
\begin{sseqdata}[name=spincunmixedchioct, classes=fill, scale=0.65, xrange={0}{6}, yrange={0}{3}, Adams grading,
>=stealth,x label = {$\displaystyle{s\uparrow \atop t-s\rightarrow}$},
x label style = {font = \small, xshift = -20.5ex, yshift=3ex}]
\begin{scope}[BrickRed]
	\class(0, 0)\AdamsTower{}
	\class(2, 0)\AdamsTower{}
	\class(4, 1)\AdamsTower{}
	\class(6, 2)\AdamsTower{}
	\class(8, 3)
	\class(8, 4)

\end{scope}
\class[RedOrange!67!black](1, 0)
\begin{scope}[Goldenrod!67!black]
	\class(3, 0)\AdamsTower{}
	\class(5, 1)\AdamsTower{}
	\class(7, 2)\class(7, 3)\class(7, 4)
\structline[Goldenrod!67!gray](3, 2, -1)(5, 3, -1)
\end{scope}
\begin{scope}[draw=none, fill=none]
\foreach \y in {1, ..., 4} {
	\class (3, \y)
}
\class(4, 0)
\class(6, 0)
\class(6, 1)
\end{scope}
\class[RedOrange!67!white](3, 0)
\class[Green!80!white](5, 0)
\begin{scope}[Green!80!black]
	\class(4, 0)\AdamsTower{}
	\class(6, 0)\AdamsTower{}
	\class(8, 1)\class(8, 2)\class(8, 3)\class(8, 4)
\structline[Green](4, 0, -1)(6, 1, -1);
\end{scope}
\class[MidnightBlue](5, 0)
\class[Fuchsia](5, 0)
\class(3, 5)
\class(5, 5)
\d[gray]2(4, 0, -1)
\d[gray]2(6, 0, -1)
\d[gray]2(6, 1, -1)
\end{sseqdata}
\printpage[name=spincunmixedchioct, page=2]
\end{subfigure}
\caption{Left: the $\cE(1)$-module structure on $\tH^*((BS_4)^{3-V_\lambda};\Z/2)$ in low degrees. The pictured
submodule contains all elements of degrees at most $6$. Right: the corresponding Ext, which is the $E_2$-page of
the Adams spectral sequence for $\widetilde\ku_*((BS_4)^{3-V})$. Some nonzero $v_1$-actions are hidden for
clarity.}
\label{spinc_chiral_oct_unmixed}
\end{figure}

To draw the $E_2$-page of the Adams spectral sequence, use the computations of $\Ext(\uQ)$ from~\eqref{ExtN2} and
$\Ext_{\cE(1)}(\Z/2)$ from~\eqref{E1Z2Ext} to obtain \cref{spinc_chiral_oct_unmixed}, right. For clarity, we do not
draw most $v_1$-actions. There may be differentials in this range, though we do not determine whether they are the
$d_2$s pictured.

From \cref{spinc_chiral_oct_unmixed}, right, $\widetilde\ku_0((BS_4)^{3-V_\lambda})\cong\Z$, so
\cref{torsion_k_theory} implies there is a single free summand in each even degree and the odd-degree $\ku$-groups
are torsion. Therefore, one of the towers on the $4$-line must admit a nontrivial $d_r$ differential to the tower
on the $3$-line, and in fact, $v_1$-equivariance of the differentials implies that tower on the $4$-line must be the
blue one coming from $\textcolor{Green!80!black}{\Sigma^4\uQ}$. The remaining tower must survive, so on the
$E_\infty$-page, the $3$-line has its $\textcolor{RedOrange!67!white}{\Z/2}$ summand and a
$\textcolor{Goldenrod!67!black}{\Z/2^r}$ summand coming from the red tower, and the $4$-line has a single
$\textcolor{BrickRed}{\Z}$ summand left. The results on $\widetilde\ku_5$ and $\widetilde\ku_6$ follow from $v_1$-
and $h_0$-equivariance of $d_r$.

On to the spin-$1/2$ case. We factor $\ku_*(BS_4)\cong\ku_*(\pt)\oplus\widetilde\ku_*(BS_4)$.
Mitchell-Priddy~\cite[Theorem B]{MP84} prove a $2$-local stable splitting of $BS_4$:
\begin{equation}
\label{mitchell_priddy_splitting}
   \Sigma^\infty BS_4 \simeq_{(2)} \Sigma^\infty B\Z/2 \vee \Sigma^\infty B\PSL_2(\F_7) \vee
   L(2),
\end{equation}
where $L(2)$ is as in \S\ref{para_symm}. In the degrees we need, the $2$-local $\ku$-homology of each of these
pieces is known: for $B\Z/2$, see Hashimoto~\cite[Theorem 3.1]{Has83}; for $B\PSL_2(\F_7)$, see \cref{ku_PSL}; and
for $L(2)$, see \cref{ku_L2}. Combining these summands and passing from $\ku$-homology to \spinc bordism as usual,
we arrive at the free and $2$-torsion groups in the theorem statement.
\end{proof}

	\subsection{Full octahedral symmetry}
\label{s_full_oct}
The full group of symmetries of the octahedron, including orientation-reversing ones, is isomorphic to
$G\coloneqq S_4\times\Z/2$. Let $\lambda\colon G\to\O_3$ denote the corresponding three-dimensional real
representation of $G$, and $V_\lambda\to BG$ denote the associated vector bundle. We saw in \S\ref{s_chiral_oct}
the pullback of $V_\lambda$ along $BS_4\to BG$ is not \pinc, so $V_\lambda$ is also not \pinc, and therefore is
also not \pinm.

The Künneth formula and \cref{mod2S4coh} together imply
\begin{equation}
\label{kunneth_full_oct}
	H^*(BG;\Z/2) \cong \Z/2[x, a, b, c]/(ac),
\end{equation}
where $\abs x = \abs a = 1$, $\abs b = 2$, and $\abs c = 3$.
\begin{lem}
$w_1(V_\lambda) = x$ and $w_2(V_\lambda) = b + a^2 + x^2$.
\end{lem}
\begin{proof}
For $w_1$, we know $w_1(V_\lambda)\ne 0$ because $V_\lambda$ is unorientable, but because $V_\lambda|_{BS_4}$ is
orientable, $w_1(V_\lambda)$ cannot be $a$ or $x+a$, leaving $w_1(V_\lambda) = x$.

For $w_2$, we know the pullback of $V_\lambda$ to $BS_4$ has $w_2(V_\lambda|_{S_4}) = a^2 + b$ by \cref{oct_wn}, so
$w_2(V_\lambda) = a^2 + b + c_1x^2 + c_2ax$ for some $c_1,c_2\in\Z/2$. If $i\colon B\Z/2\to BG$ is induced by the
inclusion of an inversion symmetry in $G$, then $i^*\lambda$ decomposes as a direct sum of three copies of the sign
representation, so $i^*V_\lambda\cong 3\sigma$. Therefore $i^*w_2(V_\lambda) = x^2$, so $c_1 = 1$. To show $c_2 =
0$, pull back $V_\lambda$ across the inclusion $\Z/2\times\Z/2\to\Z/2\times S_4$ which is the identity on the first
factor and the inclusion of a transposition on the second factor. If $\sigma_1$ is the unique real, one-dimensional
representation of $\Z/2\times\Z/2$ that is nontrivial on the first factor and trivial on the second factor, and
$\sigma_2$ is likewise nontrivial on the second factor but trivial on the first factor, then there is an
isomorphism of representations of $\Z/2\times\Z/2$
\begin{equation}
	V_\lambda|_{\Z/2\times\Z/2} \overset\cong\longrightarrow 2(\sigma_1\otimes\sigma_2) \oplus \sigma_1,
\end{equation}
so $w_2$ of this representation is $x^2 + a^2$, reaffirming that $c_1 = 1$ and showing that $c_2 = 0$.
\end{proof}
\subsubsection{Class D, spinless case}
\Cref{shear_D_thm} implies we should study the spin bordism of $(BG)^{3-V_\lambda}$. We will argue as we did in the
case of pyritohedral symmetry in \S\ref{pyrit}, replacing $3-V_\lambda$ with a more convenient virtual bundle whose
Adams spectral sequence is easier to calculate.
%
\begin{thm}
The first few spin bordism groups of $(BG)^{3-V_\lambda}$ are
\begin{align*}
	\tOmega_0^\Spin((BG)^{3-V_\lambda}) &\cong \Z/2\\
	\tOmega_1^\Spin((BG)^{3-V_\lambda}) &\cong \Z/2\\
	\tOmega_2^\Spin((BG)^{3-V_\lambda}) &\cong (\Z/2)^{\oplus 2}\\
	\tOmega_3^\Spin((BG)^{3-V_\lambda}) &\cong (\Z/2)^{\oplus 2}\\
	\tOmega_4^\Spin((BG)^{3-V_\lambda}) &\cong (\Z/2)^{\oplus 5},
\end{align*}
and $\tOmega_5^\Spin((BG)^{3-V_\lambda})$ is torsion.
\end{thm}
\begin{proof}
Using the twisted Künneth formula, $\tH^*((BG)^{3-V_\lambda};\Z)$ contains no odd-primary torsion,
so neither does $\tOmega_*^\Spin((BG)^{3-V_\lambda})$, so using the $2$-primary Adams spectral sequence suffices.

For the rest of this section, all homology and cohomology is understood to be with $\Z/2$ coefficients.
\begin{lem}
\label{full_oct_same_E2}
Let $E\to BG$ denote the virtual vector bundle induced from the virtual representation
\begin{equation}
	2 - (V_\lambda|_{S_4}\boxplus (-\sigma)).
\end{equation}
Then $\tOmega_*^\Spin((BG)^{3-V_\lambda})\cong\tOmega_*^\Spin((BG)^E)$.
\end{lem}
\begin{proof}
By the relative Thom isomorphism (\cref{relative_Thom}), it suffices to show $(3-V_\lambda)-E$ is spin, which we
know because $w_1(E) = x = w_1(3-V_\lambda)$ and $w_2(E) = a^2+b = w_2(3-V_\lambda)$.
\end{proof}
Because $E$ is induced from a representation which is an exterior sum, its Thom spectrum splits as
\begin{equation}
	(BG)^{E} \simeq (BS_4)^{3 - V_\lambda|_{S_4}} \wedge (B\Z/2)^{\sigma-1}.
\end{equation}
The Künneth theorem then simplifies the $E_2$-page:
\begin{equation}
	E_2^{s,t} = \Ext_{\cA(1)}^{s,t}(\tH^*((BS_4)^{3 - V_\lambda|_{S_4}}) \otimes_{\Z/2} \widetilde
	H^*((B\Z/2)^{\sigma-1}); \Z/2).
\end{equation}
Let $N_4$ denote the $\cA(1)$-module $\tH^*((B\Z/2)^{\sigma-1})$; see Campbell~\cite[Figure 6.1]{Cam17} for an
explicit description of $N_4$. We
computed $\tH^*((BS_4)^{3-V_\lambda})$ in~\eqref{spinless_chiral_octahedral_a1} (drawn in
\cref{chiral_octahedral_spinless_coh}), and now we just have to tensor each of the summands that appear in low
degrees there with $N_4$. Those summands are shifts of $\uQ$, $\cA(1)$, $\Z/2$, and $R_2$. For any $\cA(1)$-module
$M$, $\Z/2\otimes_{\Z/2} M\cong M$, and we computed $\cA(1)\otimes_{\Z/2} N_4$ in~\eqref{A1_BZ2} and
$\uQ\otimes_{\Z/2} N_4$ in \cref{pyritohedral_tensor_products}. That leaves $R_2\otimes_{\Z/2} N_4$.
\begin{lem}
\label{oct_tensor}
There is an $\cA(1)$-module isomorphism $R_2\otimes_{\Z/2} N_4\cong \cA(1)\oplus P'$ for a $0$-connected
$\cA(1)$-module $P'$.
\end{lem}
\begin{proof}
Direct computation. Specifically, if $a$ denotes the unique nonzero element of $R_2$ which is homogeneous of degree
$0$, and $U$ is the corresponding element of $N_4$, then $a\otimes U$ is the unique nonzero element of
$R_2\otimes_{\Z/2} N_4$ which is homogeneous of degree $0$. It thus suffices to show that $a\otimes U$ generates a
free summand, which will automatically split off~\cite[Remark D.7]{FH16}; to show this, it suffices to show that
$\Sq^2\Sq^2\Sq^2(a\otimes U)\ne 0$~\cite[Lemma D.6]{FH16}, which is straightforward.
\end{proof}
\begin{cor}
There is an isomorphism of $\cA(1)$-modules
\begin{equation}
	\tH^*((BG)^{3-V_\lambda}) \overset\cong\longrightarrow
	\textcolor{BrickRed}{\Sigma^2 R_0} \oplus
	\textcolor{Goldenrod!67!black}{\Sigma^3 N_4} \oplus F \oplus P,
\end{equation}
where $P$ is $4$-connected and $F$ is a free $\cA(1)$-module with a homogeneous basis of elements in degrees
$\set{\textcolor{BrickRed}{0}, \textcolor{RedOrange}{1}, \textcolor{RedOrange}{2}, \textcolor{RedOrange}{3},
\textcolor{BrickRed}{4}, \textcolor{RedOrange}{4}, \textcolor{Green}{4}}$.
\end{cor}
To calculate the $E_2$-page of the Adams spectral sequence, we look up $\Ext(\textcolor{BrickRed}{R_0})$
from~\cite[\S 2]{GMM68} or~\cite[Figures 23, 24]{BC18} and $\Ext(\textcolor{Goldenrod!67!black}{N_4})$
from~\cite[\S 2]{GMM68} or~\cite[Figure 6.3]{Cam17}, allowing us to draw the $E_2$-page in
\cref{revised_full_oct_spinless}. In this range, differentials in our spectral sequence vanish for degree reasons,
and Margolis' theorem takes care of the extension problems, finishing the proof.
\end{proof}

\begin{figure}[h!]
\begin{sseqdata}[name=fulloctfixed, classes=fill, Adams grading, scale=0.7, xrange={0}{4}, yrange={0}{2},
>=stealth, x label = {$\displaystyle{s\uparrow \atop t-s\rightarrow}$}, x label style = {font = \small, xshift =
-17ex, yshift=4ex}]
\begin{scope}[BrickRed]
	\class[fill=none](0, 0)
	\class(2, 0)
	\class[fill=none](4, 0)
	\class(4, 1)
	\class(5, 2)\structline
\end{scope}
\begin{scope}[RedOrange, fill=none]
	\class(1, 0)
	\class(2, 0)
	\class(3, 0)
	\class(4, 0)
\end{scope}
\begin{scope}[Goldenrod!67!black]
	\class(3, 0)
	\class(4, 1)\structline
	\class(5, 2)\structline
\end{scope}
\class[Green, fill=none](4, 0)
\end{sseqdata}
\printpage[name=fulloctfixed, page=2]
\caption{
The $E_2$-page of the Adams spectral sequence computing $\tOmega_*^\Spin((BG)^{3-V_\lambda})_2^\wedge$.
Unfilled dots represent the classes coming from free $\cA(1)$-module summands.}
\label{revised_full_oct_spinless}
\end{figure}

\subsubsection{Class D, spin-$1/2$ case}
Now we ask for the symmetries to mix. 
By \cref{shear_D_thm}, we consider the spin bordism of $(BS_4\times B\Z/2)^{\Det(V_\lambda)-1}$,
because $V_\lambda$ is not \pinm. The isomorphism $\Det V_\lambda\cong 0\boxplus \sigma$ provides an isomorphism
$(BS_4\times B\Z/2)^{\Det(V_\lambda)-1}\simeq (BS_4)_+\wedge (B\Z/2)^{\sigma-1}$, so~\eqref{pinmsplitting} implies the spin bordism of this spectrum
computes the \pinm bordism of $BS_4$, which could be independently interesting.
\begin{thm}
\label{spinful_full_oct_bordism}
The first few spin bordism groups of $(BS_4\times B\Z/2)^{\Det(V_\lambda)-1}$ are
\begin{align*}
	\tOmega_0^\Spin((BS_4\times B\Z/2)^{\Det(V_\lambda)-1}) &\cong \Z/2\\
	\tOmega_1^\Spin((BS_4\times B\Z/2)^{\Det(V_\lambda)-1}) &\cong (\Z/2)^{\oplus 2}\\
	\tOmega_2^\Spin((BS_4\times B\Z/2)^{\Det(V_\lambda)-1}) &\cong \Z/8\oplus\Z/4\oplus \Z/2\\
	\tOmega_3^\Spin((BS_4\times B\Z/2)^{\Det(V_\lambda)-1}) &\cong (\Z/2)^{\oplus 4}\\
	\tOmega_4^\Spin((BS_4\times B\Z/2)^{\Det(V_\lambda)-1}) &\cong (\Z/2)^{\oplus 2}.
\end{align*}
Since $\tOmega_5^\Spin((BS_4\times B\Z/2)^{\Det(V_\lambda)-1})$ is torsion by \cref{torsion_k_theory}, $\Ph_0^{S_4\times\Z/2}(\R^3;
f_{1/2}^D)\cong (\Z/2)^{\oplus 2}$.
\end{thm}
\begin{proof}
As usual, \cref{det_splitting} spits $(BS_4\times B\Z/2)^{\Det(V_\lambda)-1}$ as a sum of $(B\Z/2)^{\sigma-1}$ and another spectrum $M$, where
$\tH^*(M;\Z/2)$ is complementary in $\tH^*((BS_4\times B\Z/2)^{\Det(V_\lambda)-1};\Z/2)$ to the space spanned by $\set{\oU w_1(\lambda)^k}$. The
$(B\Z/2)^{\sigma-1}$ summand gives us \pinm bordism, and we focus on $M$.

We have $w_1(\Det(V_\lambda)-1) = w_1(V_\lambda) = x$ and $w_2(\Det V_\lambda-1) = 0$; this and the $\cA$-module
structure on $BS_4\times B\Z/2$ calculated in~\eqref{kunneth_full_oct} determine the $\cA(1)$-module structure on
$M$. We obtain an isomorphism of $\cA(1)$-modules
\begin{equation}
\label{full_oct_D_mixed}
	\wH^*(M;\Z/2) \cong \textcolor{BrickRed}{\Sigma R_5} \oplus
		\textcolor{RedOrange}{\Sigma^2 R_3} \oplus
		\textcolor{Goldenrod!75!black}{\Sigma^3\cA(1)} \oplus
		\textcolor{Green}{\Sigma^3\cA(1)} \oplus
		\textcolor{PineGreen}{\Sigma^3\cA(1)} \oplus
		\textcolor{MidnightBlue}{\Sigma^3\cA(1)} \oplus
		\textcolor{Fuchsia!67!white}{\Sigma^4\cA(1)} \oplus
		\textcolor{Fuchsia!75!black}{\Sigma^4\cA(1)} \oplus P,
	\end{equation}
where $P$ is $4$-connected. We will see momentarily that for $t-s \le 4$, $E_2^{s,t}$ is empty for $s\ge 2$; this
and the $4$-connectedness of $P$ imply its contribution to the $E_2$-page cannot affect the spectral sequence in
degrees $t-s \le 4$, which is all we need. We draw these summands, except for $P$, in \cref{full_oct_mixed_drawn}.
\begin{figure}[h!]
\begin{tikzpicture}[scale=0.6, every node/.style = {font=\tiny}]
	\foreach \y in {2, ..., 10} {
		\node at (-2, \y) {$\y$};
	}
	\begin{scope}[BrickRed]
		\tikzpt{0}{1}{$\oU a$}{};
		\tikzpt{0}{2}{}{};
		\tikzpt{0}{4}{}{};
		\tikzpt{0}{5}{}{};
		\tikzptR{1.5}{2}{$\oU ax$}{};
		\tikzpt{1.5}{3}{}{};
		\tikzpt{1.5}{4}{}{};
		\tikzpt{1.5}{5}{}{};
		\tikzpt{1.5}{6}{$\oU a^3x^3$}{};
		\tikzpt{1.5}{7}{}{};
		\tikzpt{1.5}{8}{}{};
		\tikzpt{1.5}{9}{}{};

		\sqone(0, 1);
		\sqtwoL(0, 2);
		\sqone(0, 4);
		\sqtwoCR(0, 1);
		\sqone(1.5, 2);
		\sqtwoR(1.5, 2, );
		\sqtwoCL(1.5, 3);
		\sqone(1.5, 4);
		\sqtwoR(1.5, 5, );
		\sqone(1.5, 6);
		\sqtwoL(1.5, 6);
		\sqone(1.5, 8);
		\begin{scope}
			\clip (0.5, 8.5) rectangle (2.5, 9.5);
			\sqtwoR(1.5, 9, );
		\end{scope}
	\end{scope}
	\begin{scope}[RedOrange]
		\tikzpt{3.5}{2}{$\oU b$}{};
		\tikzpt{3.5}{3}{}{};
		\tikzpt{3.5}{4}{}{};
		\tikzpt{3.5}{5}{}{};
		\tikzpt{3.5}{7}{}{};
		\tikzpt{5}{5}{}{};
		\tikzpt{5}{6}{}{};
		\sqone(3.5, 2);
		\sqone(3.5, 4);
		\sqone(5, 5);
		\sqtwoL(3.5, 2);
		\sqtwoCR(3.5, 3);
		\sqtwoCR(3.5, 4);
		\sqtwoL(3.5, 5);
				
		\tikzptR{3.5}{6}{$\oU \alpha$}{};
		\tikzpt{3.5}{8}{}{};
		\tikzpt{3.5}{9}{}{};

		\sqone(3.5, 6);
		\sqone(3.5, 8);
		\sqtwoR(3.5, 6, );
		\begin{scope}
			\clip (2.5, 8.5) rectangle (4.5, 9.5);
			\sqtwoL(3.5, 9);
			\sqone(3.5, 10);
			\sqtwoR(3.5, 10, );
		\end{scope}
	\end{scope}
	\begin{scope}[Goldenrod!75!black]
		\Aone{6}{3}{$\oU a^3$}{};
	\end{scope}
	\begin{scope}[Green]
		\Aone{8.25}{3}{$\oU ab$}{};
	\end{scope}
	\begin{scope}[PineGreen]
		\Aone{10.5}{3}{$\oU ax^2$}{};
	\end{scope}
	\begin{scope}[MidnightBlue]
		\Aone{12.75}{3}{$\oU c$}{};
	\end{scope}
	\begin{scope}[Fuchsia!67!white]
		\Aone{15}{4}{$\oU a^2b$}{};
	\end{scope}
	\begin{scope}[Fuchsia!75!black]
		\Aone{17.25}{4}{$\oU bx^2$}{};
	\end{scope}
\end{tikzpicture}
\caption{The $\cA(1)$-module structure on $\tH^*(M;\Z/2)$ in low degrees. The pictured summand contains all classes
in degrees $4$ and below. Here $\alpha\coloneqq b^2x^2 + a^2b^2 +c^2$.}
\label{full_oct_mixed_drawn}
\end{figure}

Yu~\cite[Theorem 3.1]{Yu95} calculates $\Ext(\textcolor{BrickRed}{R_5})$ and $\Ext(\textcolor{RedOrange}{R_3})$
(see also Beaudry-Campbell~\cite[Figures 32, 33, 37]{BC18}),
so we can draw the $E_2$-page of the Adams spectral sequence:
\begin{sseqdata}[name=fulloctmixed, classes=fill, xrange={0}{4}, yrange={0}{2}, scale=0.6,
	x label = {$\displaystyle{s\uparrow \atop t-s\rightarrow}$},
	x label style = {font = \small, xshift = -15.5ex, yshift=3ex}]
\begin{scope}[BrickRed]
	\class(1, 0)
	\class(2, 1)\structline
	\class(2, 0)\structline
\end{scope}
\class[white](2, 1)
\class[RedOrange](2, 0)
\class[Goldenrod!75!black](3, 0)
\class[Green](3, 0)
\class[PineGreen](3, 0)
\class[MidnightBlue](3, 0)
\class[Fuchsia!67!white](4, 0)
\class[Fuchsia!75!black](4, 0)
\end{sseqdata}
\begin{equation}
\begin{gathered}
\printpage[name=fulloctmixed, page=2]
\end{gathered}
\end{equation}
This collapses, and it and the \pinm bordism groups from the $(B\Z/2)^{\sigma-1}$ summand, which are computed
in~\cite{ABP69, KT90}, sum together to the groups in the theorem.
\end{proof}
\begin{rem}
Finite truncations of $\textcolor{BrickRed}{R_5}$ are the duals of the Milgram modules $Q_{2,n}$~\cite{Mil75,
DGM81}. See~\cite[Figure 1]{BL21} for a picture.
\end{rem}

\subsubsection{Class A, spinless case}
Let $f_0^A$ denote the local system of symmetry types for this case. We want to calculate
$\tOmega_*^\Spinc((BG)^{3-V_\lambda})$. Using the twisted Künneth formula, $\tH^*((BG)^{3-V_\lambda});\Z/2)$ is
$2$-torsion, and therefore $\tOmega_*^\Spinc((BG)^{3-V_\lambda})$ is too, so it suffices to use the $2$-primary
Adams spectral sequence.
\begin{thm}
The first few \spinc bordism groups of $(BG)^{3-V_\lambda}$ are:
\begin{align*}
	\tOmega_0^\Spinc((BG)^{3-V_\lambda}) &\cong \Z/2\\
	\tOmega_1^\Spinc((BG)^{3-V_\lambda}) &\cong \Z/2\\
	\tOmega_2^\Spinc((BG)^{3-V_\lambda}) &\cong (\Z/2)^{\oplus 3}\\
	\tOmega_3^\Spinc((BG)^{3-V_\lambda}) &\cong (\Z/2)^{\oplus 3}\\
	\tOmega_4^\Spinc((BG)^{3-V_\lambda}) &\cong \Z/4\oplus (\Z/2)^{\oplus 4},
\end{align*}
and $\tOmega_5^\Spinc((BG)^{3-V_\lambda})$ is torsion. Hence $\Ph_0^{S_4\times\Z/2}(\R^3; f_0^A)\cong \Z/4\oplus
(\Z/2)^{\oplus 4}$.
\end{thm}
\begin{proof}
The $\cA(1)$-modules $R_0$ and $N_4$ are isomorphic as $\cE(1)$-modules,\footnote{Giambalvo~\cite[\S 2]{Gia73a}
shows that if $M \coloneqq H^*(B\Spin;\Z/2)$, then the $E_2$-page of the Adams spectral sequence computing \pinp,
resp.\ \pinm bordism is isomorphic to $\Ext_{\cA(1)}(M\otimes_{\Z/2} R_0)$, resp.\ $\Ext_{\cA(1)}(M\otimes_{\Z/2}
N_4)$. Passing to $\cE(1)$ provides information on twisted \spinc structures, so the $\cE(1)$-module isomorphism
$R_0\cong N_4$ is an algebraic manifestation of the fact that the symmetry type \pinc is realized as both
$\Pin^+\times_{\mu_2}\T$ and $\Pin^-\times_{\mu_2}\T$.} and $\cA(1)\cong\cE(1)\oplus \Sigma^2\cE(1)$as
$\cE(1)$-modules, so there is an $\cE(1)$-module isomorphism
\begin{equation}
	\tH^*((BG)^{3-V_\lambda}) \cong \textcolor{BrickRed}{\Sigma^2 R_0} \oplus
	\textcolor{Goldenrod!67!black}{\Sigma^3 R_0} \oplus F_c\oplus P_c,
\end{equation}
where $P_c$ is $4$-connected and $F_c$ is a free $\cA(1)$-module with a homogeneous basis of elements in degrees
$\set{\textcolor{BrickRed}{0}, \textcolor{RedOrange}{1}, \textcolor{BrickRed}{2}, \textcolor{RedOrange}{2},
\textcolor{RedOrange}{3}, \textcolor{RedOrange}{3}, \textcolor{BrickRed}{4}, \textcolor{RedOrange}{4},
\textcolor{RedOrange}{4}, \textcolor{Green}{4}}$. Using our calculation of $\Ext_{\cE(1)}(R_0)$ from
\cref{the_ext_of_H}, we draw the Adams $E_2$-page:
\begin{sseqdata}[name=spincfulloct, classes=fill, scale=0.7, xrange={0}{4}, yrange={0}{2}, x label =
{$\displaystyle{s\uparrow \atop t-s\rightarrow}$}, x label style = {font = \small, xshift = -16.5ex, yshift=4ex}]
\class[draw=none, fill=none](4, 1)
\begin{scope}[BrickRed]
	\foreach \x in {0, 2, 4} {
		\class[fill=none](\x, 0)
	}
	\class(2, 0)
	\class(4, 1)\structline[BrickRed!40!white]
	\class(4, 0)\structline
	\class(6, 1)\structline[BrickRed!40!white]
	\class(6, 2)\structline[BrickRed!40!white](4, 1, -1)(6, 2)
\end{scope}
\begin{scope}[draw=none, fill=none]
	\foreach \x in {0, 1, 2} {
		\class(4, 1)
	}
\end{scope}
\begin{scope}[RedOrange, fill=white]
	\foreach \x in {1, 2, 3, 3, 4, 4} {
		\class(\x, 0)
	}
\end{scope}
\begin{scope}[Goldenrod!67!black]	
	\class(3, 0)
	\class(5, 1)\structline[Goldenrod!67!gray]
\end{scope}
\class[fill=none, Green](4, 0)
\end{sseqdata}
\begin{equation}
\begin{gathered}
	\printpage[name=spincfulloct, page=2]
\end{gathered}
\end{equation}
Below degree $5$, there are no nonzero differentials, because there is nothing in Adams filtration $2$ or higher.
And degree considerations rule out hidden extensions, so we are done.
\end{proof}
\subsubsection{Class A, spin-$1/2$ case}
Because $V_\lambda$ is not \pinc, \cref{shear_A_thm} tells us to compute the \spinc bordism groups of $(BS_4\times
B\Z/2)^{\Det(V_\lambda)-1}$.
\begin{thm}
The first few \spinc bordism groups of $(BS_4\times B\Z/2)^{\Det(V_\lambda)-1}$ are
\begin{align*}
	\tOmega_0^\Spinc((BS_4\times B\Z/2)^{\Det(V_\lambda)-1}) &\cong \Z/2\\
	\tOmega_1^\Spinc((BS_4\times B\Z/2)^{\Det(V_\lambda)-1}) &\cong \Z/2\\
	\tOmega_2^\Spinc((BS_4\times B\Z/2)^{\Det(V_\lambda)-1}) &\cong \Z/4\oplus (\Z/2)^{\oplus 2}\\
	\tOmega_3^\Spinc((BS_4\times B\Z/2)^{\Det(V_\lambda)-1}) &\cong (\Z/2)^{\oplus 4}\\
	\tOmega_4^\Spinc((BS_4\times B\Z/2)^{\Det(V_\lambda)-1}) &\cong \Z/8\oplus\Z/4\oplus (\Z/2)^{\oplus 4}.
\end{align*}
$\tOmega_5^\Spinc((BS_4\times B\Z/2)^{\Det(V_\lambda)-1})$ is torsion.
\end{thm}
\begin{proof}
By \cref{det_splitting}, $(BS_4\times B\Z/2)^{\Det(V_\lambda)-1}$ splits as $(B\Z/2)^{\sigma-1}\vee M$, where
$\tH^*(M;\Z/2)$ is isomorphic to a complementary subspace to the subspace $\Z/2\cdot\set{\oU x^k}$ inside
$\tH^*((BS_4\times B\Z/2)^{\Det(V_\lambda)-1};\Z/2)$. As usual, the $(B\Z/2)^{\sigma-1}$ summand contributes \pinc
bordism groups to the final answer, so we focus on $M$. The $\cA(1)$-module structure we computed
in~\eqref{full_oct_D_mixed} and drew in \cref{full_oct_mixed_drawn} tells us the $\cE(1)$-structure; here, we use
that $R_5\cong \cE(1)\oplus \Sigma R_0$~\cite[Proof of Claim 2.8]{Yu95} and $R_3\cong \cE(1)\oplus \Sigma^2 R_0$ as
$\cE(1)$-modules (see also~\cite[\S 7]{Pow14a}). Therefore, there is an $\cE(1)$-module isomorphism
\begin{equation}
\label{spinc_mixed_full_oct_a1}
	\tH^*(M;\Z/2) \cong \textcolor{BrickRed!80!black}{\Sigma\cE(1)} \oplus
		\textcolor{BrickRed!80!white}{\Sigma^2 R_0} \oplus
		\textcolor{RedOrange!80!black}{\Sigma^2\cE(1)} \oplus 
		\textcolor{Goldenrod!75!black}{\Sigma^3\cE(1)} \oplus
		\textcolor{Green}{\Sigma^3\cE(1)} \oplus
		\textcolor{MidnightBlue}{\Sigma^3\cE(1)} \oplus
		\textcolor{RedOrange!80!white}{\Sigma^4 R_0} \oplus
		\textcolor{Fuchsia!67!white}{\Sigma^4\cE(1)} \oplus
		\textcolor{Fuchsia!75!black}{\Sigma^4\cE(1)} \oplus P,
\end{equation}
where $P$ is $4$-connected. Therefore to infer anything about $\tOmega_4^\Spinc(M)$ from this spectral sequence, we
must argue that $P$ does not affect it; this will follow when we see the $t-s = 4$ line of the $E_2$-page is empty
in Adams filtration $2$ and above, so there can be no nonzero differentials from the $5$-line to the $4$-line. We
draw~\eqref{spinc_mixed_full_oct_a1} in \cref{spinc_mixed_full_oct_drawn}.
\begin{figure}[h!]
\begin{tikzpicture}[scale=0.6, every node/.style = {font=\tiny}]
	\foreach \y in {1, ..., 8} {
		\node at (-2, \y) {$\y$};
	}
	\begin{scope}[BrickRed!80!black]
		\Eone{0}{1}{$\oU a$}{};
	\end{scope}
	\begin{scope}[BrickRed!80!white]
		\tikzpt{2.25}{2}{$\oU ax$}{};
		\foreach \y in {3, ..., 9} {
			\tikzpt{2.25}{\y}{}{};
		}
		\begin{scope}
			\clip (1, 1.5) rectangle (3.5, 9.5);
			\foreach \y in {2, 4, 6, 8} {
				\sqone(2.25, \y);
			}
			\qoneL(2.25, 2);
			\qoneR(2.25, 4);
			\qoneL(2.25, 6);
			\qoneR(2.25, 8);
		\end{scope}
	\end{scope}
	\begin{scope}[RedOrange!80!black]
		\Eone{3.75}{2}{$\oU b$}{};
	\end{scope}
	\begin{scope}[Goldenrod!75!black]
		\Eone{5.9}{3}{$\oU a^3$}{};
	\end{scope}
	\begin{scope}[Green]
		\Eone{7.75}{3}{$\oU ab$}{};
	\end{scope}
	\begin{scope}[PineGreen]
		\Eone{9.75}{3}{$\oU ax^2$}{};
	\end{scope}
	\begin{scope}[MidnightBlue]
		\Eone{11.75}{3}{$\oU c$}{};
	\end{scope}
	\begin{scope}[RedOrange!80!white]
		\tikzpt{13.75}{3}{$\oU \alpha$}{};
		\foreach \y in {4, ..., 8} {
			\tikzpt{13.75}{\y}{}{};
		}
		\begin{scope}
			\clip (12.5, 2.5) rectangle (15, 8.5);
			\sqone(13.75, 3);
			\sqone(13.75, 5);
			\sqone(13.75, 7);
			\qoneL(13.75, 3);
			\qoneR(13.75, 5);
			\qoneL(13.75, 7);
		\end{scope}
	\end{scope}
	\begin{scope}[Fuchsia!67!white]
		\Eone{15.5}{4}{$\oU a^2b$}{};
	\end{scope}
	\begin{scope}[Fuchsia!75!black]
		\Eone{17.25}{4}{$\oU bx^2$}{};
	\end{scope}
\end{tikzpicture}
\caption{The $\cE(1)$-module structure on $\tH^*(M;\Z/2)$ in low degrees. Here $\alpha\coloneqq abx + b^2 + cx$.
This submodule contains all elements in degrees 4 and below.}
\label{spinc_mixed_full_oct_drawn}
\end{figure}
% ...

Recalling $\Ext(R_0)$ from \cref{the_ext_of_H}, the $E_2$-page of the Adams spectral sequence for
${\widetilde\ku}_*(M)$ is
\begin{sseqdata}[name=spincfulloctmixed, classes=fill, xrange={0}{4}, yrange={0}{2}, scale=0.6,
		x label = {$\displaystyle{s\uparrow \atop t-s\rightarrow}$},
		x label style = {font = \small, xshift = -15.5ex, yshift=3ex}]
\class[BrickRed!80!black](1, 0)
\begin{scope}[BrickRed!80!white]
	\class(2, 0)
	\class(4, 0)
	\class(4, 1)\structline
	\class[white](4, 1)
	\class(6, 1)
	\class(6, 2)
	\begin{scope}[BrickRed!45!white]
		\structline(2, 0, -1)(4, 1)
		\structline(4, 0)(6, 1)
		\structline(4, 1)(6, 2)
	\end{scope}
\end{scope}
\class[RedOrange!80!black](2, 0)
\class[Goldenrod!75!black](3, 0)
\class[Green](3, 0)
\class[PineGreen](3, 0)
\class[MidnightBlue](3, 0)
\class[RedOrange!80!white](4, 0)
	\class(6, 1)
\structline[RedOrange!35!white](4, 0, -1)(6, 1)
\class[Fuchsia!67!white](4, 0)
\class[Fuchsia!75!black](4, 0)
\end{sseqdata}
\begin{equation}
\begin{gathered}
	\printpage[name=spincfulloctmixed, page=2]
\end{gathered}
\end{equation}
In this range, the spectral sequence collapses, so we read off $\tOmega_*^{\Spinc}(M)$ and combine it with \pinc
bordism as computed in~\cite{BG87a, BG87b} to conclude.
\end{proof}

	\subsection{Chiral icosahedral symmetry}
		\label{s_chiral_ico}
Let $\lambda\colon A_5\to\SO_3$ denote the representation given by chiral icosahedral symmetry, and as usual let
$V_\lambda\to BA_5$ denote the associated vector bundle.
\begin{rem}
\label{aperiodic}
Unlike the previous symmetry groups we studied, icosahedral symmetry is incompatible with translations, and there
are no space groups whose underlying point group is either the chiral icosahedral group or the full icosahedral
group. This means one should not expect to realize any phases equivariant for these symmetry groups as a lattice
Hamiltonian system on a periodic lattice on $\R^3$. This does not rule out the possibility of interesting phases
with an icosahedral symmetry: there are examples of phases studied via lattice Hamiltonian realizations on lattices
in great generality, such as the toric code model in~\cite[\S 2.3]{Fre19}, the GDS model in~\cite{FreedmanHastings,
Deb20, FHHT19}, and the phases on aperiodic lattices studied by Huang-Wu-Liu~\cite{HWL20}. In a similar vein, it
may be possible for a Hamiltonian on an aperiodic lattice with icosahedral symmetry to model a nontrivial
crystalline SPT. See~\cite{VLPAPF19} for an example of how such an implementation might look.
\end{rem}
For icosahedral symmetry, the hard work is behind us. Let $\lambda\colon A_5\to\O_3$ denote the representation as
the orientation-preserving symmetries of the icosahedron. The restriction to $A_4\subset A_5$ corresponds to
symmetries that preserve a concentric tetrahedron. Let $V_\lambda\to BA_5$ be the associated bundle to $\lambda$.
\begin{lem}
\label{A4injA5}
The inclusion $\vp\colon A_4\inj A_5$ induces an equivalence on mod $2$ cohomology. Hence
$\vp$ induces $2$-primary equivalences $\Sigma^\infty(BA_4)_+\to\Sigma^\infty
(BA_5)_+$ and $(BA_4)^{3-\vp^*(V_\lambda)}\to (BA_5)^{3-V_\lambda}$.
\end{lem}
\begin{proof}
The first part is \cref{ademmilgram}: here $[A_5: A_4] = 5$, $P = \Z/2\times\Z/2$, and for both $A_4$ and $A_5$,
$N(P)/P\cong\Z/3$.

For the second part, the Thom isomorphism theorem tells us $\vp'\colon (BA_4)^{3-\vp^*(V_\lambda)}\to
(BA_5)^{3-V_\lambda}$ induces an isomorphism on mod $2$ cohomology. The desired $2$-primary equivalences then
follow from the mod $2$ Whitehead theorem~\cite[Chapitre III, Théorème 3]{Ser53}.
\end{proof}
We can therefore reuse the calculations we made at the prime $2$ in \S\ref{tetrahedral} to obtain the $2$-primary
parts of $\tOmega_k^\Spin((BA_5)^{3-V_\lambda})$ and $\Omega_k^\Spin(BA_5)$; the odd-primary pieces are different,
but not hard.
\begin{prop}
\label{odd_tors_a5}
The only odd-primary torsion in $H_k(BA_5)$ for $k< 7$ is contained in $H_3(BA_5)\cong\Z/30$.
\end{prop}
\begin{proof}[Proof sketch]
Since $\abs{A_5} = 60 = 2^2\cdot
3\cdot 5$, there is no $p$-primary torsion for $p > 5$, so it suffices to determine $H^k(BA_5; \Z/3)$ and
$H^k(BA_5; \Z/5)$ in low degrees. This can be done using the theorem of Adem-Milgram~\cite[Theorem II.6.8]{AM04}
mentioned above, since the Sylow $3$- and $5$-subgroups of $A_5$ are abelian.
\end{proof}
\begin{cor}
In $\tOmega_k^\Spin((BA_5)^{3-V_\lambda})$ and $\Omega_k^\Spin(BA_5)$, the only odd-primary torsion for $k<
7$ is a $\Z/15$ in degree $3$.
\end{cor}
\begin{proof}
As usual, we use the fact that $\Omega_*^\Spin\to\Omega_*^\SO$ is an isomorphism on odd-primary torsion, together
with the Thom isomorphism $\tOmega_*^\SO((BA_5)^{3-V_\lambda})\cong\Omega_*^\SO(BA_5)$, to reduce to showing the
claim for $\Omega_k^\SO(BA_5)$. For this, use the Atiyah-Hirzebruch spectral sequence
\begin{equation}
	E^2_{p,q} = H_p(BA_5;\Omega_q^\SO(\pt)) \Longrightarrow \Omega_{p+q}^\SO(BA_5).
\end{equation}
On the $E^2$-page, the only odd-primary torsion in total degree below $7$ is $\Z/15\subset E^2_{3,0} = H_3(BA_5)$.
In all differentials involving $E^r_{3,0}$, the other group is zero, so this odd-primary torsion lives to the
$E^\infty$-page.

We also must check that the free summands in total degree below $7$ do not receive differentials that produce more
odd-primary torsion. There are only two such free summands, in $E^2_{0,0}$ and $E^2_{0,4}$, and they can only
receive differentials from $2$-torsion abelian groups, so that does not happen.
\end{proof}
Now we need to combine this with the $2$-primary summands. For $(BA_5)^{3-V_\lambda}$, we need
$\Omega_*^\Spin((BA_4)^{3-\vp^*V_\lambda})$, which we computed in \cref{spin_bord_X}. For $BA_5$, we need
$\Omega_*^\Spin(BA_4)$; in the degrees we need, this is isomorphic to $\ko_*(BA_4)$, which Bruner-Greenlees compute
in~\cite[\S 7.7.E]{BG10}.
\begin{thm}
\label{spin_chiral_icos_bordism}
The low-degree spin bordism groups of $(BA_5)^{3-V}$ and $BA_5$ are
\begin{alignat*}{2}
	\tOmega_0^\Spin((BA_5)^{3-V_\lambda}) &\cong \Z\qquad & \Omega_0^\Spin(BA_5) &\cong\Z\\
	\tOmega_1^\Spin((BA_5)^{3-V_\lambda}) &\cong 0\qquad & \Omega_1^\Spin(BA_5) &\cong\Z/2\\
	\tOmega_2^\Spin((BA_5)^{3-V_\lambda}) &\cong 0\qquad & \Omega_2^\Spin(BA_5) &\cong\Z/2\oplus\Z/2\\
	\tOmega_3^\Spin((BA_5)^{3-V_\lambda}) &\cong \Z/30 \qquad & \Omega_3^\Spin(BA_5) &\cong\Z/60\\
	\tOmega_4^\Spin((BA_5)^{3-V_\lambda}) &\cong \Z \qquad & \Omega_4^\Spin(BA_5) &\cong\Z\\
	\tOmega_5^\Spin((BA_5)^{3-V_\lambda}) &\cong \Z/2\oplus\Z/2 \qquad & \Omega_5^\Spin(BA_5) &\cong 0\\
	\tOmega_6^\Spin((BA_5)^{3-V_\lambda}) &\cong \Z/2\qquad & \Omega_6^\Spin(BA_5) &\cong\Z/2.
\end{alignat*}
Hence the $0^{\mathrm{th}}$ $A_5$-equivariant phase homology groups vanish for both spinless and spin-$1/2$
fermions.
\end{thm}
Finally, class A\@. Since $V_\lambda$ is not \pinc, because its restriction to $A_4$ is not (\cref{tet_not_pinc}),
we care about $(BA_5)^{\Det(V_\lambda)-1} \cong (BA_5)_+$ in the spin-$1/2$ case, because $V_\lambda$ is orientable.
Let $f_0^A$, resp.\ $f_{1/2}^A$, denote the equivariant local systems of symmetry types for the class A spinless,
resp.\ spin-$1/2$ cases.
\begin{thm}
The low-degree \spinc bordism groups of $(BA_5)^{3-V_\lambda}$ and $BA_5$ are
\begin{alignat*}{2}
	\tOmega_0^\Spinc((BA_5)^{3-V_\lambda}) &\cong \Z\qquad & \Omega_0^\Spinc(BA_5) &\cong\Z\\
	\tOmega_1^\Spinc((BA_5)^{3-V_\lambda}) &\cong 0\qquad & \Omega_1^\Spinc(BA_5) &\cong 0\\
	\tOmega_2^\Spinc((BA_5)^{3-V_\lambda}) &\cong \Z\qquad & \Omega_2^\Spinc(BA_5) &\cong\Z\oplus \Z/2\\
	\tOmega_3^\Spinc((BA_5)^{3-V_\lambda}) &\cong \Z/30 \qquad & \Omega_3^\Spinc(BA_5) &\cong \Z/30\\
	\tOmega_4^\Spinc((BA_5)^{3-V_\lambda}) &\cong \Z^2 \qquad & \Omega_4^\Spinc(BA_5) &\cong\Z^2,
\end{alignat*}
and in both cases, $\Omega_5^\Spinc$ is torsion. Hence both $\Ph_0^{A_5}(\R^3; f^A_0)$ and $\Ph_0^{A_5}(\R^3;
f^A_{1/2})$ vanish.
\end{thm}
\begin{proof}
The calculation separates into $2$-primary and odd-primary computations; by \cref{A4injA5}, the $2$-primary pieces
are exactly as in \cref{spinc_chi_tet_thm}. For the odd-primary piece, \cref{spinc_odd_primes} reduces the question
to the corresponding question for oriented bordism, or equivalently spin bordism, which we did in
\cref{spin_chiral_icos_bordism}.
%
\end{proof}

	\subsection{Full icosahedral symmetry}
		\label{s_full_ico}
If one includes orientation-reversing symmetries of the icosahedron, the symmetry group enlarges to
$A_5\times\Z/2$, with the $\Z/2$ generated by an inversion. This symmetry group is also incompatible with
translations, so \cref{aperiodic} applies. This calculation also quickly reduces to something we already know:
restricting the representation to $A_4\times\Z/2$ yields the pyritohedral symmetry representation we studied in
\S\ref{pyrit}.
\begin{thm}
Let $\rho$ be a virtual $A_5\times\Z/2$-representation with rank zero, and let $V_\rho\to BG$ denote
the associated virtual vector bundle. Suppose that $w_1(V_\rho) = x$, where $x$ denotes the generator of
$H^1(B\Z/2;\Z/2)\subset H^1(B(A_5\times \Z/2);\Z/2)$. Then inclusion of the pyritohedral symmetry subgroup
$\vp\colon A_4\times\Z/2\inj A_5\times\Z/2$ induces a homotopy equivalence
$B(A_4\times\Z/2)^{V_\rho}\overset\simeq\to B(A_5\times\Z/2)^{V_\rho}$.
\end{thm}
\begin{proof}
By the Whitehead theorem, it suffices to establish that $\vp$ induces an isomorphism $\widetilde
H^*(B(A_5\times\Z/2)^{V_\rho};k)\to \tH^*(B(A_4\times\Z/2)^{V_\rho};k)$ for $k = \Q$ and $k = \Z/p$ for all primes
$p$.

\Cref{A4injA5} and the Künneth theorem imply that $\vp^*\colon H^*(B(A_5\times\Z/2);\Z/2)\to
H^*(B(A_4\times\Z/2);\Z/2)$ is an isomorphism. Together with the Thom isomorphism theorem, this takes care of the
case $k = \Z/2$.

Let $G$ be either of $A_4\times\Z/2$ or $A_5\times\Z/2$; the map $B\vp\colon B(A_4\times\Z/2)\to B(A_5\times\Z/2)$
allows us to think of $V_\rho$ as over $BG$ for either $G$, and make sense of the statement $w_1(V_\rho) = x$. The
Thom isomorphism implies $\tH^*((BG)^{V_\rho};\Z)\cong H^*(BG; \Z_x)$, and since $\Z_x$ arises as a pullback local
system along $BG\to B\Z/2$, the twisted Künneth formula proves $\tH^*(BG;\Z)$ is $2$-torsion. The universal
coefficient theorem then implies that when we take coefficients in $k = \Q$ or $k = \Z/p$ for $p$ odd,
$\tH^*(B(A_4\times\Z/2)^{V_\rho}; k)$ and $H^*(B(A_5\times\Z/2)^{V_\rho};k)$ vanish, so the map between them is
vacuously an isomorphism.
\end{proof}
Let $\lambda\colon A_5\times\Z/2\to\O_3$ denote the representation as the group of symmetries of an icosahedron and
$V_\lambda\to B(A_5\times\Z/2)$ denote the associated vector bundle. Then $w_1(V_\lambda) = x$.
\begin{cor}
$\vp$ induces homotopy equivalences
\begin{subequations}
\begin{gather}
	B(A_4\times\Z/2))^{3-V_\lambda} \overset\cong\longrightarrow (B(A_5\times\Z/2))^{3-V_\lambda}\\\
	(B(A_4\times\Z/2))^{\Det(V_\lambda)-1} \overset\cong\longrightarrow (B(A_5\times\Z/2))^{\Det(V_\lambda)-1}.
\end{gather}
\end{subequations}
Therefore
\begin{enumerate}
	\item \cref{spinless_pyrit_bordism} implies that $\Ph_0^{A_5\times\Z/2}(\R^3; f^D_0)\cong(\Z/2)^{\oplus 3}$;
	\item \cref{spinful_pyrit_bordism} implies that $\Ph_0^{A_5\times\Z/2}(\R^3; f^D_{1/2})\cong \Z/2$;
	\item \cref{spinc_pyrit_bord} implies that $\Ph_0^{A_5\times\Z/2}(\R^3; f^A_0)\cong \Z/4\oplus (\Z/2)^{\oplus
	3}$; and
	\item \cref{mixed_pyr_classA_thm} implies that $\Ph_0^{A_5\times\Z/2}(\R^3; f^A_{1/2})\cong \Z/8 \oplus
	(\Z/2)^{\oplus 3}$.
\end{enumerate}
\end{cor}

\section{Glide symmetry protected phases}
	\label{glide_s}
Though we have focused on point group symmetries thus far, Freed-Hopkins' ansatz~\cite{FH19} also applies to
crystallographic groups. In this section, we apply their ansatz to the group of glide symmetries; invertible phases
equivariant for this symmetry have been studied by Lu-Shi-Lu~\cite{LSL17} and Xiong-Alexandradinata~\cite{XA18},
and our results agree with theirs. In particular, Lu-Si-Lu make a conjecture classifying certain glide-symmetric
phases in all symmetry types, and we prove that their conjecture follows from Freed-Hopkins' ansatz.

The group of \term{glide symmetries} acting on $\R^d$, $d\ge 2$, is the free group on the single generator
\begin{equation}
	(x_1, x_2, \dotsc, x_d)\mapsto (x_1 + 1, -x_2, x_3, \dotsc, x_d).
\end{equation}
In previous sections, when the symmetry type is $H = \Spin$, $\Spinc$, $\Pin^\pm$, etc., the symmetry type can mix
with the group action on spacetime, corresponding physically to spinless or spin-$1/2$ fermions. Here, this cannot
happen: if $\mu_2$ denotes the kernel of the map $\Spin_n\to\SO_n$ or $\Pin_n^\pm\to\O_n$, all extensions
\begin{equation}
	\shortexact{\mu_2}{\wG}{\Z}{}
\end{equation}
split, so given one of these symmetry types, there is a unique equivariant symmetry type for this $\Z$-action with
respect to mixing with fermion parity, corresponding to the trivial local system $\underline E\to\R^d$ with value
$E\coloneqq\Map(\MTH, \Sigma^2I_\Z)$.

\begin{defn}
Recall from \cref{change_symmetry} that we defined a ``forgetful map'' $\vp\colon\Ki_*^\Z(\R^d; \underline
E)\to\Ki_*(\R^d; \underline E)$. The \term{intrinsically $\Z$-equivariant phase homology}, denoted
$\widehat{\Ki}{}_*^\Z(\R^d; \underline E)$, is the kernel of this map.
\end{defn}
This corresponds under Freed-Hopkins' ansatz to what Lu-Shi-Lu call a \term{glide SPT}: an invertible phase
equivariant for a $\Z$ glide symmetry which is trivializable when one forgets the symmetry.

Let $\TP_d(H)$ denote the abelian group of SPT phases in (spatial) dimension $d$; Freed-Hopkins' ansatz~\cite{FH16}
classifying these phases in terms of invertible field theories predicts $\TP_d(H)\cong E_{-d}$.

Lu-Shi-Lu~\cite{LSL17} studied groups of glide SPTs and conjectured a formula classifying them in terms of the
classification of ordinary SPTs. We prove the corresponding statement on phase homology groups.
\begin{thm}
\label{glidethm}
For a given symmetry type $\rho_n\colon H_n\to\O_n$, there is a natural isomorphism $\widehat{\Ki}{}_0^\Z(\R^d;
\underline E)\cong E_{-(d-1)}\otimes\Z/2$.
\end{thm}
Passing this through the ansatz, this predicts that the group of glide SPTs is naturally isomorphic to
$\TP_{d-1}(H)\otimes\Z/2$, which is Lu-Shi-Lu's original conjecture~\cite[Conjecture 1]{LSL17}.
Xiong-Alexandradinata~\cite{XA18} also obtain this result using physics-based arguments.


\begin{proof}[Proof of \cref{glidethm}]
We calculate the $0^{\mathrm{th}}$ $\Z$-equivariant Borel-Moore $E$-homology of $\R^d$. As the $\Z$-action
is free, this is the $0^{\mathrm{th}}$ (nonequivariant) Borel-Moore $E$-homology of the fundamental domain
$X\coloneqq \R^d/\Z$. Since the one-point compactification $\overline X$ of $X$ is a finite CW complex, this
Borel-Moore homology is isomorphic to $\widetilde E_0(\overline X)$.

If $\sigma\to S^1$ denotes the Möbius bundle, then $X$ is diffeomorphic to the total space of
$\sigma\oplus\underline\R^{d-2}\to S^1$, so $\overline X$ is the Thom space
$(S^1)^{\sigma + d-2}$. The identification $(S^1)^\sigma\cong\RP^2$ induces $\overline X\cong\Sigma^{d-2}\RP^2$,
and therefore
\begin{equation}
	\Ki_*^\Z(\R^d; \underline E)\cong \tE_0(\Sigma^{d-2} \RP^2)\cong \tE_{2-d}(\RP^2).
\end{equation}
\begin{lem}
\label{isdeg2}
Let $p\colon S^2\to\RP^2$ be the double cover map and $s\colon \tE_k(S^1)\to\tE_{k+1}(S^2)$ be the suspension
isomorphism. The composition $p_*\circ\delta\circ s\colon \tE_{-1}(S^2)\to\tE_{-1}(S^2)$ is multiplication by $2$.
\end{lem}
\begin{proof}
This follows because the suspension is the cofiber of the cofiber; then one explicitly checks what happens on
mapping cylinders.
\end{proof}
\begin{lem}
\label{know_forget}
Under these isomorphisms, the forgetful map $\Ki_0^\Z(\R^d; \underline E)\to\Ki_0(\R^d; \underline E)$ is
identified with $\delta$.
\end{lem}
\begin{proof} 
Because $\Z$ acts freely on $\R^d$, $E_{0,\mathrm{BM}}^\Z(\R^d)$ is identified with $\tE_0$ of the one-point
compactification of $\R^d/\Z$, which we saw above is homeomorphic to $\Sigma^{d-2}\RP^2$. The codomain of the
forgetful map is $E_{0,\mathrm{BM}}(\R^d)\cong \tE_0(\Sigma^{d-2}S^2)$, so we have identified $\delta$ with a map
$\tE_0(\Sigma^{d-2}\RP^2)\to\tE_0(\Sigma^{d-2}S^2)$. But tracing through the construction in
\cref{change_symmetry}, this map comes from applying $\tE_0$ to an actual map
$\Sigma^{d-2}\RP^2\to\Sigma^{d-2}S^2$.

Next, precompose with $\Sigma^{d+2}p\colon \Sigma^{d-2}S^2\to\Sigma^{d-2}\RP^2$ and check that this map has degree
$2$, agreeing with \cref{isdeg2}. This suffices to identify the maps because $p^*\colon[\RP^2, S^2]\to [S^2, S^2]$
is injective.
\end{proof}
$\RP^2$ is homeomorphic to the cofiber of a degree-$2$ map $S^1\to S^1$. Hence there is a long exact
sequence in reduced $E$-homology
\begin{equation}
\label{glide_LES}
	\xymatrix{
		\dotsb\ar[r] & \tE_{2-d}(S^1)\ar[r]^-m & \tE_{2-d}(S^1)\ar[r]^r & \tE_{2-d}(\RP^2)\ar[r]^-\delta &
		\tE_{1-d}(S^1)\ar[r] & \dotsb
	}
\end{equation}
where $m$ is multiplication by $2$. Exactness implies $\ker(\delta) = \Im(r) = \coker(m)$. Using the suspension
isomorphism, $\tE_k(S^1)\cong\tE_{k-1}$, and therefore $\coker(m)\cong E_{-(d-1)}\otimes\Z/2$, and
\ref{know_forget} identifies $\delta$ with the forgetful map from equivariant to nonequivariant phase homology for
$\R^d$. In particular, $\widehat{\Ki}{}_0^\Z(\R^d; \underline E)\cong \ker(\delta)$, which we have naturally
identified with $E_{-(d-1)}\otimes\Z/2$.
\end{proof}
\begin{rem}
Using the long exact sequence~\eqref{glide_LES}, we observe that $\Ki_0^\Z(\R^d; \underline E)$ has exponent
$4$.
This is because for any long exact sequence of abelian groups
\begin{equation}
	\xymatrix{
		\dotsb\ar[r] & A\ar[r]^-{\cdot 2} & A\ar[r]^-f & B\ar[r]^-g & C\ar[r]^-{\cdot 2} & C\ar[r] & \dots
	}
\end{equation}
in which $A$ and $C$ are finitely generated, $\Im(f)\cong A/2$, hence has exponent $2$, and $\ker(g)$ is isomorphic
to the subgroup of order-$2$ elements of $C$, which also has exponent $2$. Since $B$ is an extension of $\ker(g)$
by $\Im(f)$, $B$ has exponent $4$.

Passing this observation through Freed-Hopkins' ansatz, this recovers an observation of
Xiong-Alexandradinata~\cite{XA18}: that \emph{any} invertible phase equivariant with respect to glide symmetry,
whether a glide SPT or not, has order dividing $4$.
\end{rem}
\begin{exm}
In Altland-Zirnbauer class AII, corresponding to the symmetry type pin\textsuperscript{$\tilde c+$}, the ansatz
predicts a unique nontrivial glide SPT in dimension $2+1$, coming from the classification
\begin{equation}
[\MTPin^{\tilde c+}, \Sigma^4 I_\Z] \otimes\Z/2 \cong\Z/2
\end{equation}
(the calculation of $[\MTPin^{\tilde c+}, \Sigma^4I_\Z]$ is due to Freed-Hopkins~\cite[\S 9.3]{FH16}). Physicists
are particularly interested in this nontrivial glide SPT phase, which is predicted to have unusual surface states
called ``hourglass fermions''~\cite{WACB16}, and which has been studied experimentally~\cite{MYLNS17}.
\end{exm}

\section{Conclusion and outlook}
	\label{conclusion}
We conclude by indicating a few directions of potential further research.
\subsection{From free fermions to interacting phases}
\label{free_ferm}
Free fermion phases are a rich source of examples of invertible phases in the physics literature, at least for
symmetry types spin, pin\textsuperscript{$\pm$}, \spinc, etc. The classification of free fermion systems uses
$K$-theory: see Kitaev~\cite{Kit09} for the original proposal and Freed-Moore~\cite{FM13} for a comprehensive
classification.  However, for a given dimension and symmetry type, the map from free fermion systems to invertible
phases of matter can in general have both kernel (as first observed by Fidkowski-Kitaev~\cite{FK10, FK11} and
Turner-Pollmann-Berg~\cite{TPB11}) and cokernel (as first observed by Wang-Potter-Senthil~\cite{WPS14} and
Wang-Senthil~\cite{WS14}). Researchers are also interested in the free-to-interacting map for phases with spatial
symmetries, and this map has been studied from a physics point of view for crystalline phases in several works,
including~\cite{Qi13, YR13, IF15, MFM15, LTH16, SS17, RL18, Zou18, LVK19, ACRFM21, RS20, ZYQG20, MCB23, HYZ24,
LSH24, MCB24, SAN24, BP25, SAC25}.

Freed-Hopkins~\cite[\FHrefllink{page.76}{\S 9.2}, \FHrefllink{page.88}{\S 9.3}]{FH16} mathematically model the map
from free to interacting systems using the Atiyah-Bott-Shapiro map $\MTSpin\to\KO$~\cite{ABS}, but they do not
consider spatial symmetries. In view of the large bodies of research on free fermions with spatial symmetries and
invertible phases with spatial symmetries, it would be nice to understand the map between them in the presence of
spatial symmetry from the low-energy field theory perspective, and to make contact with the work of Adem, Antolín
Camarena, Semenoff, and Sheinbaum~\cite{AACSS16}, Sheinbaum and Antolín Camarena~\cite{SC20, SAC24, SAC25},
Cornfeld-Carmeli~\cite{CC20}, and Sati-Schreiber~\cite{SS25} studying free fermion phases with spatial symmetries
using methods from homotopy theory. This is something we hope to tackle in future work.

Since the first version of this paper appeared online, the author, joint with Antolín Camarena, Krulewski,
Pacheco-Tallaj, Sheinbaum, and Stehouwer~\cite{ACDKPTSS24}, has proposed a generalization of Freed-Hopkins'
ansatz~\cite[\FHrefllink{page.76}{\S 9.2}, \FHrefllink{page.88}{\S 9.3}]{FH16} to weak SPT phases.
\subsection{Other symmetry types}
We investigated two of the ten Altland-Zirnbauer classes, and it would be interesting to know whether a
version of the FCEP holds for other classes. One starting point could be class C, corresponding to a
spin\textsuperscript{$h$} structure~\cite[\FHrefllink{page.77}{(9.25)}]{FH16};\footnote{Spin\textsuperscript{$h$}
is the symmetry type $\Spin\times_{\mu_2}\SU_2\to\O$. Freed-Hopkins~\cite[\FHrefllink{page.75}{Proposition
9.16}]{FH16} call this symmetry type $G^0$; it is sometimes also called spin-$\SU_2$, e.g.\ in~\cite{WWW19}.
Likewise, the symmetry types pin\textsuperscript{$h\pm$} we refer to later in this section are defined to be
$\Pin^\pm\times_{\mu_2}\SU_2$, and are called $G^\pm$ by Freed-Hopkins~\cite[\FHrefllink{page.75}{Proposition
9.16}]{FH16}.} the calculations in \S\ref{FCEP} could be applied to $\Spin_n^h$ to obtain a fermionic crystalline
equivalence principle for class C and hopefully phase homology calculations predicting the existence of additional
crystalline SPT phases.

Several teams of researchers have studied or classified interacting fermionic crystalline SPTs for other
Altland-Zirnbauer types, including~\cite{Qi13, YR13, YX14, CHMR15, LTH16, WF17, RL18, ZXXS20, CW18, ZYQG20, MCB23,
SXG18, ZMKB23, HABS24, HYZ24, MH19, RNQWG24, SAN24, BFKZ25, BP25, NRWQG25}. It would be good to compare their
computations with the predictions made by an FCEP in other symmetry types.

Since this paper first appeared online, Manjunath-Calvera-Barkeshli~\cite[\S III.A]{MCB23} proposed a general FCEP
that applies to all Altland-Zirnbauer classes, but does not use homotopy theory; this offers an answer to the
proposal in the previous paragraph. In example computations, our proposal and theirs appear to agree where they
overlap; it would be interesting to determine whether this is true in general, and to use this to extend our
homotopy-theoretic framework for the FCEP to the remaining eight Altland-Zirnbauer classes.

Another interesting potential connection with preexisting work is the case of class A phases with a spatial
reflection interacting with the internal $\T$ symmetry. Depending on how the symmetries mix,
Shiozaki-Shapourian-Gomi-Ryu~\cite[\S V.C, \S V.E]{SSGR18}, Thorngren-Else~\cite[\S VII.B]{TE18}, and
Ning-Mao-Li-Wang~\cite[\S III.B, \S V]{NMLW21}
obtain
classifications in terms of pin\textsuperscript{$\tilde c\pm$} bordism, and we would be interested in knowing
whether that can also be obtained from our ansatz. Relatedly, Kobayashi-Zhang-Wang-Barkeshli~\cite{KZWB25} consider
a symmetry group $Q\cong\Z/2$ acting by a combination of reflection and time-reversal, and classify 3d fermionic
SPTs with symmetry groups $Q$, $Q\times\T$, and $Q\ltimes\T$; in the latter two cases, fermion parity is $-1\in\T$.
Their classifications match the groups of deformation classes of 3d IFTs with $\xi$-structure for $\xi$ equal to
$\Spin\times\Z/2$, $\Spin^c\times\Z/2$, and $\Spin\times_{\set{\pm 1}}\Pin_2^+$ respectively.\footnote{References
for these groups of IFTs: $(I_\Z\MTSpin)^3(B\Z/2)\cong\Z\oplus\Z/8$, which follows from
\cref{2mod4_cyc_spin_bordism}, $(I_\Z\MTSpin^c)^3(B\Z/2)\cong\Z^2\oplus\Z/4$, which follows from
\cref{spinccyclicbordism}; and $I_\Z\mathit{MT}(\Spin\times_{\set{\pm 1}}\Pin_2^+)^3\cong\Z^2\oplus\Z/2$, which
follows from~\cite[\S 4.1]{Ste21}, \cite[\S IV.E.4, Appendix E.5]{MCB23}, or \cite[Proposition 3.47]{DYY25b}.} See
also Seiberg-Shao-Zhang~\cite{SSZ25}. It would be interesting to see whether it is possible to also obtain these
results from our ansatz.

Similarly, can one begin with class C
phases and a reflection acting on the internal $\SU_2$ symmetry and obtain a classification in terms of
pin\textsuperscript{$h\pm$} bordism? Since the first version of this paper appeared online,
Ning-Mao-Li-Wang~\cite[\S VII]{NMLW21} classified 4d invertible phases with these symmetries, starting with a
different ansatz, and obtained a classification mostly matching the corresponding pin\textsuperscript{$h\pm$}
bordism groups. It would be interesting to compare our approach with theirs.

\subsection{Crystallographic groups}
Though we discussed glide symmetries in \S\ref{glide_s}, we have barely touched upon the rich world of
crystallographic groups. Free-fermion phases equivariant for these groups have been studied, e.g.\
in~\cite{SMJZ13, KBWKS17, OSS19, SSG18}, but much less is known about the interacting case, even though the our
ansatz applies to it. There are some classifications by other methods for various classes of crystallographic
groups; for example,
\cite{OWGQ20, ZY21, MCB23, WQFCG23, ZMKB23, MCB24, RNQWG24, SAN24, BFKZ25}
study wallpaper group symmetries, and Sheinbaum-Antolín Camarena~\cite{SC20}
provide a general framework and a few examples. There is also work by
Wang-Alexandradinata-Cava-Bernevig~\cite{WACB16},
Guo-Ohmori-Putrov-Wan-Wang~\cite{GOPWW18},
Lee-Shiozaki-Hsieh~\cite[\S 6]{LSH24a},
Ning-Ren-Wang-Qi-Gu~\cite[\S V]{NRWQG25},
and Seiberg-Zhang~\cite{SZ26}
studying interacting
phases for various classes of crystallographic groups that are not point groups.

\subsection{Lattice realizations}
Modeling topological phases as lattice Hamiltonian systems can make any crystallographic symmetries acting on space
very explicit, using a lattice and Hamiltonian invariant under the symmetry of interest. Our predictions of
point group SPTs should correspond to actual lattice models of phases. We listed several specific predicted phases
of interest in \S\ref{interesting_to_study}, and these would make for good starting points for lattice
realizations.

Since this paper first appeared online, some of the phases predicted in this paper have been studied on the lattice
by Zhang-Ning-Qi-Gu~\cite{ZNQG25}, largely answering the question above.

%

\bibliographystyle{alpha}
\bibliography{references}{}

\end{document}